\documentclass[pdflatex,sn-mathphys-num]{sn-jnl}

\usepackage{graphicx}%
\usepackage{multirow}%
\usepackage{amsmath,amssymb,amsfonts}%
\usepackage{amsthm}%
\usepackage{mathrsfs}%
\usepackage{mathtools}%
\usepackage{bm}%
\usepackage[title]{appendix}%
\usepackage[table,dvipsnames]{xcolor}%
\usepackage{textcomp}%
\usepackage{manyfoot}%
\usepackage{booktabs}%
\usepackage{placeins}%
\usepackage{array}%
\usepackage{subcaption}%
\usepackage{algorithm}%
\usepackage{algorithmicx}%
\usepackage{algpseudocode}%
\usepackage[inline]{enumitem}%
\usepackage{cleveref}%

\makeatletter
\def\theHALG@line{\thealgorithm.\arabic{ALG@line}}
\makeatother

\algnewcommand{\LineComment}[1]{\Statex \(\triangleright\) \textcolor{gray}{#1}}

\theoremstyle{thmstyleone}%
\newtheorem{theorem}{Theorem}%
\newtheorem{lemma}{Lemma}%
\newtheorem{proposition}{Proposition}%
\newtheorem{corollary}{Corollary}%
\theoremstyle{thmstyletwo}%
\newtheorem{remark}{Remark}%
\theoremstyle{thmstylethree}%
\newtheorem{definition}{Definition}%

\crefname{item}{Item}{Items}
\Crefname{item}{Item}{Items}

\begin{document}

\title[SEC: One-Share Everywhere-Short Secret Sharing]{SEC: One-Share Everywhere-Short Secret Sharing from Lossless Expanders}

\author*[1]{\fnm{Muhammad} \sur{Arzaki}}\email{muhammad.arzaki@monash.edu}
\author[1]{\fnm{Muhammed F.} \sur{Esgin}}
\author[1]{\fnm{Amin} \sur{Sakzad}}
\author[1]{\fnm{Ron} \sur{Steinfeld}}

\affil*[1]{\orgdiv{Faculty of Information Technology}, \orgname{Monash University}, \orgaddress{\city{Melbourne}, \country{Australia}}}

\abstract{%
We introduce \(\mathsf{SEC}\), an expander-code-based ramp secret sharing scheme over \(\mathbb{Z}\)-modules, particularly suitable for lattice-based distributed cryptography.
It is everywhere-short: bounded secret coefficients yield bounded share coefficients, and the integer reconstruction vectors have bounded entries.
Moreover, each participant receives exactly one share, regardless of the number of parties and the privacy and reconstruction thresholds.
This improves on prior schemes with bounded share coefficients and bounded entries in the integer reconstruction vectors, in which the number of sub-shares or shares per participant grows either with the number of parties and the threshold or with the security parameter.
The construction rests on a coding-theoretic observation: in the bi-adjacency matrix of a lossless bipartite expander, unique-neighbor peeling completes every sufficiently small set of columns to a square unimodular submatrix.
We prove correctness, everywhere-shortness, and linearity, and give a publicly computable entrywise bound on the reconstruction vector our algorithm returns for each authorized set.
For privacy, we introduce hard-core local leakage resilience, which strengthens the leakage-resilience notion of Nielsen and Simkin (EUROCRYPT 2020) to protect a hard-core function of the secret, and we prove that \(\mathsf{SEC}\) satisfies it against correlated multi-query local leakage under the \(\mathsf{MatrixHint\text{-}MLWE}\) problem, which admits a reduction from the standard \(\mathsf{MLWE}\) problem.
For \(n=60\) participants and security parameter \(\kappa = 128\), \(\mathsf{SEC}\) achieves privacy threshold \(6\) against \(2^{64}\) leakage queries and empirical reconstruction threshold \(46\).
The ideal entropy-coded secret and share sizes are about \(3.5\)~KiB and \(6.4\)~KiB, respectively.
}



\maketitle

\section{Introduction}\label{Section:Introduction}

\subsection{Background and Related Work}\label{Section:Background-RelatedWork}

Secret sharing, introduced independently by Shamir and Blakley \cite{shamir1979share,blakley1979safeguarding}, distributes a secret among \(n\) parties so that authorized sets can reconstruct it while unauthorized sets learn nothing about it.
In a \(t\)-out-of-\(n\) threshold scheme, every set of at least \(t\) parties is authorized and every smaller set is unauthorized.
Ramp schemes permit a gap between a privacy threshold and a larger reconstruction threshold, enabling a better information rate than perfect threshold sharing \cite{blakley1985security,yamamoto1986secret}.

Secret sharing is closely connected to coding theory.
McEliece and Sarwate observed that Shamir's scheme is based on Reed-Solomon codes \cite{mceliece1981sharing}, and Massey characterized the access structure of a code-based scheme through the minimal codewords of the dual code \cite{massey1993minimal}.
Subsequent work constructs secret sharing schemes from random and algebraic codes \cite{chen2007secure,cramer2015linear} and from span programs over rings and modules \cite{fehr1998span}.

Black-box secret sharing, introduced by Desmedt and Frankel \cite{desmedt1989threshold}, works uniformly over finite abelian groups and reconstructs through group-independent integer linear combinations.
Threshold black-box schemes with asymptotically optimal \(O(\log n)\) expansion follow from number-theoretic and coding-theoretic constructions \cite{cramer2002optimal,cramer2005black,cramer2020blackbox}, while Cramer and Xing obtain near-threshold schemes with constant expansion \cite{cramer2020blackbox}.
Applebaum et al. construct near-threshold schemes with constant share size and \(O(n)\)-addition sharing and reconstruction by combining graph-based erasure codes with an extraction vector having small integer coefficients \cite{applebaum2023recover}.
The near-threshold schemes of \cite{cramer2020blackbox,applebaum2023recover} will be the closest relatives of \(\mathsf{SEC}\), and the construction of Applebaum et al.~\cite{applebaum2023recover} is the closest structurally because it also uses sparse bipartite-graph codes and reusable public parameters.
The requirement of Applebaum et al.~\cite{applebaum2023recover} that the extraction vector lie outside the row span of every unauthorized restriction modulo every prime is analogous to the integral column-module condition of \(\mathsf{SEC}\): both prevent an unauthorized coalition from isolating the secret through exact linear algebra over \(\mathbb{Z}\) or modulo an integer.
Black-box schemes use uniform masks and obtain perfect privacy, whereas \(\mathsf{SEC}\) uses the bounded-coefficient masks required in the lattice setting discussed next.

These coding-theoretic and black-box constructions provide the algebraic background for \(\mathsf{SEC}\), while lattice-based threshold cryptography supplies its principal application setting.
Large-scale quantum computers would break the factoring and discrete-logarithm assumptions underlying efficient classical threshold cryptography, and the NIST call for multi-party threshold schemes has made practical post-quantum threshold signatures a concrete target \cite{brandao2026nist}.
Classical schemes such as FROST provide partially non-interactive signing and native identifiable abort by verifying partial signatures directly against participant-specific public keys \cite{komlo2021frost}.
In a Fiat-Shamir lattice signature such as Raccoon, however, the response \(\bm{z}=c\bm{s}+\bm{r}\) must remain short \cite{del2024raccoon}.
Shamir sharing is unsuitable for directly generating such responses because its shares are uniform modulo \(q\) and its Lagrange reconstruction coefficients may be large.
Consequently, the resulting partial responses are neither short nor safe to reveal, as a bounded-coefficient mask cannot hide the large share-dependent term \cite{del2024threshold,borin2026hermine}. 
Threshold Raccoon therefore combines Shamir sharing with one-time additive masks \cite{del2024threshold}, an approach adopted by subsequent schemes \cite{katsumata2024adaptively,espitau2025two,zhu2025algebraic,boschini2025ringtail}.
Because masked partial responses cannot be verified directly, handling malicious signers requires robust, interactive, or proof-based mechanisms \cite{espitau2024flood,del2025unmasking,gur2025olingo}.

Small reconstruction coefficients have also been studied independently.
Ball et al. prove share-size lower bounds for linear threshold schemes with binary reconstruction, motivated by lattice-based cryptography \cite{ball2021linear}, and \(\{0,1\}\)-linear sharing has been used in threshold cryptosystems based on fully homomorphic encryption \cite{boneh2018threshold}.
Benaloh-Leichter sharing \cite{benaloh1990generalized}, instantiated with Boppana's monotone formulas \cite{boppana1985amplification}, gives reconstruction coefficients in \(\{0,1\}\) or \(\{0,\pm1\}\), depending on the presentation, with \(O(\min\{k_c,n-k_c\}^{4.3}\log n)\) shares per participant on average \cite{ball2021linear,chairattana2024partially}, where \(k_c\) is the reconstruction threshold.
Small reconstruction coefficients alone do not keep partial responses short, because the shares themselves must also have bounded coefficients.
This motivates \textit{everywhere-short} sharing: bounded secret coefficients yield bounded share coefficients, while the integer reconstruction vectors have bounded entries.
These properties keep partial responses short and permit direct verification against participant-specific public material \cite{del2024shortly,borin2026hermine}.

Existing everywhere-short constructions assign multiple atomic sub-shares with bounded coefficients to each participant (\Cref{Table:Intro-SS-Comparison}).
The \(\mathsf{sDKG}\) framework of \cite{del2024shortly} uses replicated sharing, with combinatorial share multiplicity, and random-assignment ramp sharing, which uses several parallel sharings.
Hermine uses Vandermonde sharing, which adapts the multiplicative construction of \cite{desmedt1994multiplicative} to the additive group \((\mathcal{R}^{\ell},+)\) and has a share count that grows with the threshold and \(\log n\).
For general thresholds, its best known per-participant upper bound is slightly superpolynomial, although it is claimed to remain concretely practical over Hermine's intended range \cite{battagliola2025enhancing,borin2026hermine}.
This multiplicity can substantially increase long-term secret-share storage and, in applications with participant-specific verification, the corresponding public material.
This issue is particularly relevant on constrained platforms, where post-quantum cryptography must operate under limited memory, computation, and bandwidth \cite{bormann2014rfc,kannwischer2019pqm4,bettale2022post,puch2026sok}.
We therefore study whether everywhere-short secret sharing over \(\mathbb{Z}\)-modules can be achieved with exactly one share per participant.

\subsection{Our Contributions}\label{Section:Contributions}

\begin{table}[!htp]
	\caption{Comparison of \(\mathsf{SEC}\) with secret sharing schemes with integer or short reconstruction used in lattice-based distributed cryptography}
	\label{Table:Intro-SS-Comparison}%
	\setlength{\tabcolsep}{2pt}\renewcommand{\arraystretch}{1.15}
	\begin{tabular}{@{}>{\raggedright\arraybackslash}p{2.05cm}>{\centering\arraybackslash}p{1.45cm}>{\centering\arraybackslash}p{1.9cm}*{2}{>{\centering\arraybackslash}p{1.5cm}}>{\centering\arraybackslash}p{2.3cm}>{\centering\arraybackslash}p{1.45cm}@{}}
		\toprule
		& Shamir \cite{shamir1979share} & Benaloh-Leichter \(+\) Boppana \cite{benaloh1990generalized,boppana1985amplification} & Replicated & Random-assignment ramp & Vandermonde \cite{desmedt1994multiplicative,battagliola2025enhancing} & \(\mathsf{SEC}\) (this work) \\
		\midrule
		Used in & \cite{del2024threshold,espitau2024flood,del2025unmasking,gur2025olingo} & \cite{chairattana2024partially,zhu2025algebraic} & \(\mathsf{sDKG}\) \cite{del2024shortly} & \(\mathsf{sDKG}\) \cite{del2024shortly} & Hermine \cite{borin2026hermine} & -- \\ \addlinespace[4pt]
		Access & threshold & threshold & threshold & ramp\({}^{c}\) & threshold & ramp\({}^{d}\) \\ \addlinespace[4pt]
		Shares per participant & \(1\) & \(O(\mu^{4.3}\log n)\)\({}^{a}\) & \(\tbinom{n-1}{k_c-1}\) & typically \(\kappa\) & \(O\big(\tbinom{k_c-1+\lceil\lg n\rceil}{k_c-1}\big)\) & \(1\) \\ \addlinespace[4pt]
		Bounded share coefficients & \(\times\) & \(\times\)\({}^{b}\) & \(\checkmark\) & \(\checkmark\) & \(\checkmark\) & \(\checkmark\) \\ \addlinespace[4pt]
		Reconstruction coefficients & large (Lagrange) & \(\{0,\pm1\}\) & binary & binary & binary & small \\ \addlinespace[4pt]
		Privacy & perfect & perfect\({}^{b}\) & comput. & comput. & comput. & comput. \\
		\midrule
		Secret size\({}^{e}\) & -- & -- & \(6.02\) KiB & \(6.28\) KiB & \(3.76\) KiB & \(3.54\) KiB \\ \addlinespace[4pt]
		Per-participant share storage\({}^{e}\) & -- & -- & \(259\) GiB & \(964\) KiB & \(699\) KiB & \(6.42\) KiB \\
		\botrule
	\end{tabular}
    \footnotetext{Note: \(k_p\) and \(k_c\) denote the privacy and reconstruction thresholds, and \(k_c=k_p+1\) for a threshold scheme.
    ``Binary'' means reconstruction coefficients in \(\{0,1\}\), while ``small'' permits bounded signed integers; ``comput.'' abbreviates computational.
    ``Shares per participant'' counts elements of \(\mathcal{R}^\ell\) for a module rank \(\ell\) determined by the security parameter and independent of \(n\); the size of each element, which may depend mildly on the parameters, is accounted for in the per-participant share storage.}
	\footnotetext[a]{Here \(\mu:=\min\{k_c,n-k_c\}\) \cite{ball2021linear,chairattana2024partially}.}
	\footnotetext[b]{As instantiated in \cite{chairattana2024partially,zhu2025algebraic}, the sharing randomness is uniform over \(\mathcal{R}_q\), so the share coefficients are not bounded and privacy is perfect; sampling bounded-coefficient pieces instead would bound the shares but make privacy computational.}
	\footnotetext[c]{The privacy and reconstruction thresholds hold with overwhelming probability over a public seed \cite{del2024shortly}.}
	\footnotetext[d]{The privacy threshold \(k_p\) holds for every coalition of at most \(k_p\) parties, given a lossless expander, while the reconstruction threshold \(k_c\) is selected empirically.}
    \footnotetext[e]{Ideal entropy-coded estimates of the secret and of all shares held by one participant at \(\kappa=128\) and \(n=60\) (Appendix~\ref{Appendix:Short-Share-Comparison}); \(\mathsf{SEC}\) uses \(Q=2^{64}\) and privacy threshold \(k_p=6\); the replicated and Vandermonde schemes use the privacy-matched threshold \(k_c=7\), so that they tolerate the same \(6\) corrupted parties as \(\mathsf{SEC}\), while the reported ramp-\(\mathsf{sDKG}\) storage is independent of the selected comparison threshold.
    The comparison schemes are extrapolated beyond their reported parameter ranges. 
    The \(\mathsf{sDKG}\) values include the error component \(\bm{e}\), as does the systematic \(\mathsf{SEC}\) secret. 
    The Vandermonde values replace \(\lg\sigma_\mathbf{s}\) by \(\lceil\lg\sigma_\mathbf{s}\rceil\), adding less than one bit per coefficient, and report the participant with the fewest sub-shares.}
\end{table}

We introduce \(\mathsf{SEC}\), an expander-code-based ramp secret sharing scheme over \(\mathbb{Z}\)-modules for lattice-based threshold cryptography.
Unlike prior everywhere-short constructions, \(\mathsf{SEC}\) assigns exactly one bounded-coefficient share to each participant, regardless of the number of parties and the privacy and reconstruction thresholds.
As summarized in \Cref{Table:Intro-SS-Comparison}, it is the only scheme in the comparison that combines one share per participant, bounded share coefficients, and bounded entries in the integer reconstruction vectors.
Privacy holds against every coalition of at most \(k_p\) parties.
The empirical reconstruction threshold \(k_c\) is the smallest tested coalition size such that, for every tested size \(t\geq k_c\), a uniformly random coalition of size \(t\) reconstructs with probability at least \(0.99\), where this probability is computed exactly or estimated by sampling.
A larger deterministic reconstruction threshold \(k_c^\mathsf{det}\) guarantees reconstruction for every coalition of at least that size.
All \(17\) evaluated instances, with \(10 \leq n \leq 60\), satisfy \(\tau_p:=k_p/n \geq 0.1\) and \(k/n < \tau_c:= k_c/n \leq (k+6)/n\), where \(k/n\) is the rate of the underlying code.

We prove correctness, everywhere-shortness, and \(\mathbb{Z}\)-linearity, and give a publicly computable entrywise bound on the reconstruction vector returned by our algorithm for each authorized set. 
Because bounded-coefficient shares can reveal partial information about a bounded secret, \(\mathsf{SEC}\) does not provide perfect information-theoretic privacy. 
Instead, we introduce a hard-core local leakage-resilience framework that strengthens \cite[Definition 5]{nielsen2020lower} for the lattice setting. 
Under suitable parameters, we prove that a linear hard-core function of the secret, corresponding to an \(\mathsf{MLWE}\) instance, remains computationally indistinguishable from uniform given the shares of a private coalition and prescribed noisy linear leakage from all shares, assuming the hardness of \(\mathsf{MatrixHint\text{-}MLWE}\), which in turn follows from the hardness of standard \(\mathsf{MLWE}\).
This leakage is correlated across shares and queries and includes the partial responses arising in threshold signing, making the resulting privacy guarantee suitable for lattice-based threshold cryptography. 
Informally, our main result is as follows.

\begin{theorem}[Main result, informal]\label{Theorem:Main-Informal} 
	Let \(\mathsf{SEC}\) be instantiated for \(n\) participants from a lossless bipartite expander, with privacy threshold \(k_p\). 
	Then the following hold.
    \begin{enumerate}
        \item Each participant holds exactly one share, whose coefficients are bounded whenever the coefficients of the secret and of the sharing randomness are bounded.\label{Item:Informal-Everywhere-Short}
		\item Every set for which the public reconstruction algorithm finds an integer reconstruction vector, called a certified set, recovers the secret exactly as an integer linear combination of its shares.
        The entries of the reconstruction vector satisfy a publicly computable bound. 
        In particular, every coalition of at least \(k_c^\mathsf{det}= n-b+1\) parties is certified, where \(b\geq\lfloor n/k\rfloor\) is the number of pairwise disjoint certified supports planted by the construction and \(k\) is the dimension of the underlying code. \label{Item:Informal-Reconstruction}
		\item For every coalition of at most \(k_p\) parties and every admissible choice of at most \(Q\) prescribed noisy linear leakage queries on all shares, a hard-core function of the secret remains pseudorandom under \(\mathsf{MLWE}\) with secret width \(\sigma\), provided that the widths of the sharing randomness and the leakage masks exceed \(\sigma\) by factors proportional to \(\sqrt{k_p}\) and \(\sqrt{Q}\), respectively, up to factors depending on \(n\) and the expander. \label{Item:Informal-SEC-HCLLR-multi}
    \end{enumerate}
    \Cref{Item:Informal-Everywhere-Short} is formalized in \Cref{Lemma:bound-for-si}, \Cref{Item:Informal-Reconstruction} in \Cref{Lemma:SEC-Exact-Reconstruction,Lemma:Deterministic-Correctness-Lower-Bound,Lemma:Certified-Coefficient-Bound}, and \Cref{Item:Informal-SEC-HCLLR-multi} in \Cref{Theorem:SEC-HCLLR-multi}. 
\end{theorem}

The main coding-theoretic ingredient may be of independent interest: in the bi-adjacency matrix of a lossless bipartite expander whose expansion exceeds half the left degree, unique-neighbor peeling completes every sufficiently small set of columns to a square unimodular submatrix (\Cref{Lemma:Expander-implies-UnimodularMinor}); we are not aware of a previous formulation of this unimodularity conclusion. 

\subsection{Technical Overview}\label{Section:Technical-Overview}

Following the coding-theoretic approach of McEliece and Sarwate \cite{mceliece1981sharing}, \(\mathsf{SEC}\) shares a secret through a public integer distribution matrix \(\mathbf{G}:=\begin{bmatrix}\mathbf{g} \ \mathbf{\hat{G}}\end{bmatrix}\in\mathbb{Z}^{n\times k}\), where \(\mathbf{\hat{G}}\) is a signed and permuted transpose of the bi-adjacency matrix of a lossless bipartite expander and \(\mathbf{g}\) is a secret-bearing column.
The dealer multiplies \(\mathbf{G}\) by a matrix whose first row is the secret and whose remaining rows are short Gaussian masks, giving one row of the product to each participant.
A set \(T\subseteq[n]\) is certified, and reconstructs exactly, when \(\mathbf{e}_1^\top\in\mathrm{Row}_{\mathbb{Z}}(\mathbf{G}_T)\), that is, when some \(\bm{\lambda}_T\in\mathbb{Z}^{|T|}\) satisfies \(\bm{\lambda}_T^\top\mathbf{G}_T=\mathbf{e}_1^\top\).
Applying the same integer combination to the shares recovers the secret; our reconstruction algorithm searches for such a vector with small entries, and these entries satisfy the publicly computable bound of \Cref{Lemma:Certified-Coefficient-Bound}.

Because the masks are short rather than uniform, privacy is not perfect.
Unlike sufficiently small sets of Shamir shares over a finite field, which are information-theoretically independent of the secret, shares generated from a sparse integer matrix offer no analogous guarantee: a coalition may attempt to cancel the masking contribution, possibly modulo an adversarially chosen integer \(q'\geq2\), while retaining a noiseless linear function of the secret.
We rule out such attacks through an integral column-module condition, which the lossless-expander construction satisfies deterministically for every coalition of at most \(k_p\) parties, owing to the unimodular submatrices obtained by unique-neighbor peeling.
This integral guarantee is stronger than sparsity or full rank over a fixed finite field and explains why an arbitrary sparse code does not suffice.

The structural condition excludes noiseless leakage but does not hide the noisy linear information carried by short shares.
We therefore formalize privacy through the hard-core leakage experiment \(\mathsf{HC\text{-}Leak}\), in which the adversary must distinguish a designated hard-core function of the secret from uniform, given the shares of a private coalition and prescribed local leakage from every share.
In our application, this function is the systematic \(\mathsf{MLWE}\) target \(f_{\bm{A}}(\bm{s}):=\bm{A}\bm{s}\bmod q\), where \(\bm{A}:=\begin{bmatrix}\bm{I} \ \bm{A}'\end{bmatrix}\), and the leakage consists of noisy linear functions of the shares, such as the partial responses revealed by a threshold signing protocol.

The main challenge is that every share is a linear function of the same masking matrix, so the corrupted shares and the leakage released across multiple queries are jointly correlated.
Treating them as independent hints would discard this correlation.
We instead express the complete adversarial view as one stacked linear hint with a jointly distributed perturbation and bound it uniformly over admissible coalitions and leakage patterns.
The integral image and kernel properties supplied by the unimodular submatrices allow us to apply the linear-transformation theorem of \cite{genise2020improved}, showing that the joint perturbation is statistically close to the elliptic discrete Gaussian required by \(\mathsf{MatrixHint\text{-}MLWE}\).
This enables a reduction of the multi-query \(\mathsf{HC\text{-}Leak}\) experiment to \(\mathsf{MatrixHint\text{-}MLWE}\), whose hardness in turn follows from that of standard \(\mathsf{MLWE}\).

\subsection{Organization}

\Cref{Section:Preliminaries} recalls the required background and introduces the unimodular-submatrix lemma for lossless expanders, the variant of \(\mathsf{MatrixHint\text{-}MLWE}\) used in our reduction, and the notions of everywhere-short secret sharing and hard-core local leakage resilience.
\Cref{Section:SEC} defines \(\mathsf{SEC}\) and constructs its distribution matrix.
\Cref{Section:Correctness-Shortness-Linearity} proves correctness, everywhere-shortness, and linearity, while \Cref{Section:Secret-Sharing-Privacy-Analysis} establishes multi-query privacy.
\Cref{Section:SEC-Empirical-Parameter} reports the concrete parameters and evaluation, and \Cref{Section:Discussion} discusses limitations and open problems.
The appendices provide supporting definitions, proofs, and analyses, as well as a size comparison with other short-share secret sharing schemes (Appendix~\ref{Appendix:Short-Share-Comparison}).

\section{Preliminaries}\label{Section:Preliminaries}

\subsection{General Notation}\label{Section:General-Notation}

For integers \(a\leq b\), let \([a,b]:=\{a,\dots,b\}\) and \([n]:=[1,n]\). 
We write \(x\gets\mathcal{D}\) for sampling from \(\mathcal{D}\) and \(x\stackrel{\$}{\gets}S\) for uniform sampling from a finite set \(S\). 
The notation \(T\stackrel{\$}{\gets}\binom{[n]}{t}\) means that \(T\) is sampled uniformly from the \(t\)-element subsets of \([n]\).
All algorithms are probabilistic polynomial time unless stated otherwise, and \(\mathsf{negl}(\kappa)\) denotes a negligible function of the security parameter \(\kappa\).

Let \(\mathcal{R}:=\mathbb{Z}[X]/(X^d+1)\) and \(\mathcal{R}_q:=\mathcal{R}/(q\mathcal{R})\), where \(d\) is a power of two.
Bold italic letters denote objects over \(\mathcal{R}\) or \(\mathcal{R}_q\), and bold upright letters denote coefficient embeddings or objects over \(\mathbb{Z}\), \(\mathbb{Q}\), \(\mathbb{R}\), or \(\mathbb{Z}_q\).
Thus, \(\bm{a}\in\mathcal{R}^n\) is identified with \(\mathbf{a}\in\mathbb{Z}^{dn}\), and \(\bm{A}\in\mathcal{R}^{m\times n}\) with its induced coefficient-embedded map \(\mathbf{A}\in\mathbb{Z}^{dm\times dn}\); analogously modulo \(q\).
Vectors are columns unless stated otherwise.

For a vector \(\mathbf{v}\), its \(i\)-th entry is \(\mathbf{v}[i]\), \((\mathbf{v})_T\) denotes its restriction to \(T\), and \(\mathrm{supp}(\mathbf{v}):=\{i:\mathbf{v}[i]\neq0\}\). 
For a matrix \(\mathbf{A}\), we use \(\mathbf{A}[i,j]\), \(\mathbf{A}[i,\cdot]\), and \(\mathbf{A}[\cdot,j]\) for its entries, rows, and columns, and \(\mathbf{A}[T,J]\) for a submatrix. 
We abbreviate \(\mathbf{A}_T:=\mathbf{A}[T,\cdot]\) and \(\mathbf{A}_{(J)}:=\mathbf{A}[\cdot,J]\). 
For a commutative ring or field \(\Sigma\), we write \(\mathrm{Col}_\Sigma(\mathbf{A})\), \(\mathrm{Row}_\Sigma(\mathbf{A})\), and \(\ker_\Sigma(\mathbf{A})\) for the column module, row module, and kernel of \(\mathbf{A}\). 
When \(\Sigma\) is a field, \(\mathrm{rank}_\Sigma(\mathbf{A})\) denotes its rank over \(\Sigma\). 
We retain the subscript when the scalar domain matters.

For a tuple of matrices \((\mathbf{A}_i)_{i\in I}\) indexed by a finite ordered set \(I\), \(\begin{bmatrix}(\mathbf{A}_i)_{i\in I}\end{bmatrix}\) denotes its ordered vertical stack; the same convention applies to vectors. 
Additional notation and linear-algebra conventions appear in Appendix~\ref{Appendix:Additional-Notation-Linear-Algebra}.

\subsection{Lattices and Discrete Gaussians}\label{Section:Lattices-Discrete-Gaussians}

A rank-\(r\) lattice generated by linearly independent vectors \(\mathbf{b}_1,\dots,\mathbf{b}_r\in\mathbb{R}^m\) is \(\Lambda := \left\{ \sum_{i=1}^{r}z_i\mathbf{b}_i : z_i\in\mathbb{Z} \right\} = \mathcal{L}(\mathbf{B}) = \mathbf{B}\mathbb{Z}^r\), where \(\mathbf{B} := \begin{bmatrix} \mathbf{b}_1 \ \cdots \ \mathbf{b}_r \end{bmatrix}\). 
Its dual within \(\mathrm{span}_\mathbb{R}(\Lambda)\) is \(\Lambda^* := \left\{ \mathbf{y}\in\mathrm{span}_\mathbb{R}(\Lambda) : \langle\mathbf{x},\mathbf{y}\rangle\in\mathbb{Z},\ \forall \mathbf{x}\in\Lambda \right\} = \mathbf{B}(\mathbf{B}^\top\mathbf{B})^{-1}\mathbb{Z}^r\). 
If \(\Lambda\) is full rank in \(\mathbb{R}^m\), then \(\Lambda^*=\mathbf{B}^{-\top}\mathbb{Z}^m\).

For \(s>0\), define \(\rho_s(\mathbf{x}) := \exp\left( -\pi\frac{\|\mathbf{x}\|_2^2}{s^2} \right)\) and \(\rho_s(S) := \sum_{\mathbf{x}\in S}\rho_s(\mathbf{x})\) for every countable set \(S\) for which the sum converges. 
For a countable set \(S\subseteq\mathbb{R}^m\) for which the normalizing sum below is finite, a center \(\mathbf{c}\in\mathbb{R}^m\), and a positive-definite parameter matrix \(\mathbf{\Sigma}\in\mathbb{R}^{m\times m}\), let \(\mathcal{D}_{S,\mathbf{c},\sqrt{\mathbf{\Sigma}}}\) denote the discrete Gaussian over \(S\) with probability mass function
\begin{equation}\label{Eq:DiscreteGaussianDist} 
	\mathcal{D}_{S,\mathbf{c},\sqrt{\mathbf{\Sigma}}}(\mathbf{x}) := \frac{ \exp\left( -\frac{1}{2} (\mathbf{x}-\mathbf{c})^\top \mathbf{\Sigma}^{-1} (\mathbf{x}-\mathbf{c}) \right) }{ \displaystyle \sum_{\mathbf{z}\in S} \exp\left( -\frac{1}{2} (\mathbf{z}-\mathbf{c})^\top \mathbf{\Sigma}^{-1} (\mathbf{z}-\mathbf{c}) \right) }, \qquad \mathbf{x}\in S. 
\end{equation} 
We omit the center when \(\mathbf{c}=\mathbf{0}\), write \(\mathcal{D}_{S,\mathbf{c},\sigma}\) when \(\mathbf{\Sigma}=\sigma^2\mathbf{I}_m\), and abbreviate this further to \(\mathcal{D}_\sigma\) when \(S=\mathbb{Z}^m\), \(\mathbf{c}=\mathbf{0}\), and the dimension is clear. 
The matrix \(\mathbf{\Sigma}\) is a Gaussian parameter matrix, not necessarily the exact covariance matrix of the resulting discrete distribution.

Using the standard-deviation convention of our Gaussians and the Matrix Hint-MLWE conditions in \cite{espitau2024flood}, we define the smoothing parameter of a nonzero lattice \(\Lambda\) for \(\epsilon>0\) by 
\begin{equation}\label{Eq:Smoothing-Parameter-SD} 
	\eta_\epsilon(\Lambda) := \inf\left\{ \sigma>0 : \sum_{\mathbf{y}\in\Lambda^*\setminus\{\mathbf{0}\}} \exp\left( -2\pi^2\sigma^2\|\mathbf{y}\|_2^2 \right) \leq\epsilon \right\}. 
\end{equation} 
For numerical estimates, we use \(\eta_\epsilon(\mathbb{Z}^N) \approx \sqrt{ \frac{\ln(2N/\epsilon)}{2\pi^2} }\), while the formal statements retain \(\eta_\epsilon(\mathbb{Z}^N)\) explicitly. 
The relation to the \(\rho_s\)-normalization of \cite{kim2023toward,genise2020improved} appears in Appendix~\ref{Appendix:MHMLWE-Proofs}.

\subsection{Linear Codes and Lossless Expanders}\label{Section:Linear-Codes-Lossless-Expanders}

\subsubsection{Linear Codes and Distance Transfer}\label{Section:Linear-Codes}

A linear code \(C\) of length \(n\) over a commutative ring or field \(\Sigma\) is a \(\Sigma\)-submodule of \(\Sigma^n\), whose codewords are written as columns. 
Its generator and parity-check matrices satisfy \(C = \mathrm{Col}_\Sigma(\mathbf{G}) = \ker_\Sigma(\mathbf{H})\), and its dual is \(C^\bot = \ker_\Sigma(\mathbf{G}^\top)\). 
For \(\mathbf{u}\in\Sigma^n\), let \(\mathrm{supp}(\mathbf{u})\) and \(wt_H(\mathbf{u}):=|\mathrm{supp}(\mathbf{u})|\) denote its support and Hamming weight, and let \(d(C)\) denote the minimum weight of a nonzero codeword. 
Over a field, \(d(C)\) equals the smallest number of linearly dependent columns of a parity-check matrix.

In our construction, a parity-check matrix over a finite field is also viewed over \(\mathbb{Z}\) and \(\mathbb{Q}\) through fixed integer representatives. 
The following lemma, whose short proof we include for completeness, shows that lifting the matrix to \(\mathbb{Q}\) does not decrease the associated code distance.

\begin{lemma}[Distance under lifting to \(\mathbb{Q}\)]\label{Lemma:Code-Distance-Preservation} 
	Let \(p\) be prime and let \(\mathbf{H}\in\mathbb{F}_p^{m\times n}\). 
	Fix an integer lift of \(\mathbf{H}\), still denoted by \(\mathbf{H}\), whose reduction modulo \(p\) is the original matrix. 
	Define \(C_p := \ker_{\mathbb{F}_p}(\mathbf{H}) \subseteq \mathbb{F}_p^n\) and \(C_\mathbb{Q} := \ker_\mathbb{Q}(\mathbf{H}) \subseteq \mathbb{Q}^n\). 
	Then \(d(C_\mathbb{Q}) \geq d(C_p)\). 
\end{lemma}

The proof clears denominators and removes the largest common power of \(p\); it appears in Appendix~\ref{Appendix:Deferred-Proofs}. 
Moreover, \(\ker_\mathbb{Q}(\mathbf{U}\mathbf{H}) = \ker_\mathbb{Q}(\mathbf{H})\) for every \(\mathbf{U}\in GL_m(\mathbb{Z})\). 
Hence, unimodular row transformations, including signed permutations, preserve the rational code and its distance.

\subsubsection{Lossless Bipartite Expanders and Expander Codes}\label{Section:Lossless-Expanders}

Let \(G=(L,R,E)\) be a simple bipartite graph with \(|L|=n\) and \(|R|=m\), whose vertices we identify with \([n]\) and \([m]\), respectively. 
The graph is \(c_\mathrm{exp}\)-left-regular if every left vertex has degree \(c_\mathrm{exp}\). 
For \(S\subseteq L\), let \(N(S)\subseteq R\) be its neighbor set; a vertex of \(R\) adjacent to exactly one vertex in \(S\) is called a \textit{unique neighbor} of \(S\).

\begin{definition}[Lossless bipartite expander and expander code, cf.~{\cite{sipser1996expander}, \cite[Chapter 11]{guruswami2023essential}}]\label{Definition:Lossless-Expander}
    Let \(0<\epsilon_\mathrm{exp}<1/2\) and \(\gamma\in(0,1]\). 
	An \(\left( n,m,c_\mathrm{exp},\gamma, (1-\epsilon_\mathrm{exp})c_\mathrm{exp} \right)\) \emph{lossless bipartite expander} is a \(c_\mathrm{exp}\)-left-regular bipartite graph \(G=(L,R,E)\) with \(|L|=n\) and \(|R|=m\) such that \(|N(S)| \geq (1-\epsilon_\mathrm{exp})c_\mathrm{exp}|S|\) for every \(S\subseteq L\) with \(|S|\leq\gamma n\). 
    Its bi-adjacency matrix \(\mathbf{H}\in\{0,1\}^{m\times n}\) has rows indexed by \(R\), columns indexed by \(L\), and \(\mathbf{H}[i,j]=1\) exactly when \(i\) and \(j\) are adjacent. 
    The associated \emph{binary expander code} is \(C(G) := \ker_{\mathbb{F}_2}(\mathbf{H}) \subseteq \mathbb{F}_2^n\).
\end{definition}

Since \(\mathbf{H}\) has entries in \(\{0,1\}\), it can be read over any ring.
The binary code \(C(G)\) is the standard expander code of \cite{sipser1996expander} and is used only for the distance bound below, whereas our construction uses \(\mathbf{H}\) over \(\mathbb{Z}\), and \Cref{Lemma:Code-Distance-Preservation} transfers distance bounds from \(\mathbb{F}_p\) to the rational kernel.

The following theorem restates known existence results for convenience: its first part is \cite[Lemma 1.9]{hoory2006expander} and \cite[Theorem 11.2.3]{guruswami2023essential}, and its second part, which bounds the failure probability of a random construction, is \cite[Lemma 2]{xie2022orion}.
We use it only to show that constant-rate families of lossless bipartite expanders with the required asymptotic parameters exist.
Our concrete instances do not rely on it and are certified directly in \Cref{Section:SEC-Search-Certification}.

\begin{theorem}[Existence of lossless bipartite expanders {\cite[Lemma 1.9]{hoory2006expander}, \cite[Theorem 11.2.3]{guruswami2023essential}}]\label{Theorem:ExpanderExistence}
	Fix \(0<\epsilon_\mathrm{exp}<1/2\).
    For sufficiently large integers \(n\geq m\), there exists an \(\left(n,m,c_\mathrm{exp},\gamma,(1-\epsilon_\mathrm{exp})c_\mathrm{exp}\right)\) lossless bipartite expander with \(c_\mathrm{exp}=\Theta\left(\frac{\log(n/m)+\log(1/\epsilon_\mathrm{exp})}{\epsilon_\mathrm{exp}}\right)\) and \(\gamma=\Theta\left(\frac{\epsilon_\mathrm{exp}m}{c_\mathrm{exp}n}\right)\).
	Moreover, if \(m=\Theta(n)\), then by \cite[Lemma 2]{xie2022orion}, for every fixed integer \(a\geq1\), one can choose \(c_\mathrm{exp} = \Theta\left( \frac{ \log(n/m)+\log(1/\epsilon_\mathrm{exp})+a }{ \epsilon_\mathrm{exp} } \right) \) such that a random \(c_\mathrm{exp}\)-left-regular simple bipartite graph, obtained by letting each left vertex independently choose a uniform \(c_\mathrm{exp}\)-subset of \(R\), has the required expansion with probability at least \(1-n^{-a}\). 
    The hidden constants may depend on the fixed rate, expansion gap, and failure exponent.
\end{theorem}

\paragraph{Unique-Neighbor Properties}

For \(S\subseteq L\), let \(U(S) := \left\{ v\in R: v \text{ is adjacent to exactly one vertex in }S \right\}\) denote its unique-neighbor set. 
Lossless expansion implies that every sufficiently small set of left vertices has many unique neighbors.

\begin{lemma}[Unique-neighbor bound {\cite[Lemma 11.3.4]{guruswami2023essential}}]\label{Lemma:Expander-Graph-Unique-Neighbor}
	Let \(G=(L,R,E)\) be an \(\left( n,m,c_\mathrm{exp},\gamma, (1-\epsilon_\mathrm{exp})c_\mathrm{exp} \right)\) lossless bipartite expander with \(0<\epsilon_\mathrm{exp}<1/2\).
	Then, for every \(S\subseteq L\) satisfying \(|S|\leq\gamma n\), \(|U(S)| \geq c_\mathrm{exp} (1-2\epsilon_\mathrm{exp}) |S|\).
\end{lemma}

We extend \Cref{Lemma:Expander-Graph-Unique-Neighbor} to a larger range, with the weaker conclusion that every such set has a unique neighbor.

\begin{lemma}[Extended unique-neighbor property]\label{Lemma:Expander-Graph-Unique-Neighbor-extended}
	Let \(G=(L,R,E)\) be an \(\big( n,m,c_\mathrm{exp},\gamma, (1-\epsilon_\mathrm{exp})c_\mathrm{exp} \big)\) lossless bipartite expander with \(0<\epsilon_\mathrm{exp}<1/2\) and \(\lfloor\gamma n\rfloor\geq1\).
	Then every nonempty set \(S\subseteq L\) satisfying \(|S| < 2(1-\epsilon_\mathrm{exp}) \lfloor\gamma n\rfloor\) has a unique neighbor.
\end{lemma}

The proof applies \Cref{Lemma:Expander-Graph-Unique-Neighbor} to a subset of size \(\lfloor\gamma n\rfloor\) and counts the neighbors of the remaining vertices; see Appendix~\ref{Appendix:Deferred-Proofs}. 
The same unique-neighbor argument gives \(d(C(G)) \geq 2(1-\epsilon_\mathrm{exp})\gamma n\) for every \(\left( n,m,c_\mathrm{exp},\gamma, (1-\epsilon_\mathrm{exp})c_\mathrm{exp} \right)\) lossless bipartite expander \(G\) \cite[Theorem 11.3.8]{guruswami2023essential}.

\paragraph{Unimodular Submatrices}

An integer matrix contains a \(t\times t\) \textit{unimodular submatrix} if some choice of \(t\) rows and \(t\) columns yields a submatrix with determinant \(\pm1\).
Unique-neighbor peeling repeatedly removes a column that is the only remaining neighbor of some row.
As in iterative erasure decoding, it stops before removing all columns exactly when the remaining columns form a stopping set, a nonempty set without unique neighbors \cite{di2002finite}, which \Cref{Lemma:Expander-Graph-Unique-Neighbor-extended} rules out for sufficiently small sets.
Applying the domain-independent peeling method of \cite{applebaum2023recover} to a column restriction \(\mathbf{H}_{(T)}\) and reversing the peeling order produces a triangular submatrix with unit diagonal. 
It is therefore invertible over \(\mathbb{Z}\) and modulo every prime, properties used in both the structural and computational privacy analyses. 
Although unique-neighbor peeling and stopping-set arguments are standard, we are not aware of a previous formulation that extracts this unimodular-submatrix conclusion over \(\mathbb{Z}\).
We later use the following lemma to show that the necessary condition for privacy holds with probability \(1\) for every coalition \(T\) of size at most \(k_p\).

\begin{lemma}[Unimodular submatrices from lossless expansion]\label{Lemma:Expander-implies-UnimodularMinor}
	Let \(\mathbf{H}\in\{0,1\}^{m\times n}\) be the bi-adjacency matrix of an \(\left( n,m,c_\mathrm{exp},\gamma, (1-\epsilon_\mathrm{exp})c_\mathrm{exp} \right)\) lossless bipartite expander, where \(0<\epsilon_\mathrm{exp}<1/2\) and \(\lfloor\gamma n\rfloor\geq1\).
	Then, for every nonempty set \(T\subseteq[n]\) satisfying \(|T| \leq \left\lceil 2(1-\epsilon_\mathrm{exp}) \lfloor\gamma n\rfloor \right\rceil-1\), the column restriction \(\mathbf{H}_{(T)}\) contains a \(|T|\times|T|\) unimodular submatrix.
\end{lemma}

\begin{proof} 
	Fix a nonempty \(T\subseteq[n]\), and let \(\tau := |T| \leq \left\lceil 2(1-\epsilon_\mathrm{exp}) \lfloor\gamma n\rfloor \right\rceil-1\). 
	Since \(\tau \in \mathbb{Z}\), this implies \(\tau < 2(1-\epsilon_\mathrm{exp}) \lfloor\gamma n\rfloor\). 
	Set \(T_0:=T\). 
	For each \(i\in[\tau]\), \Cref{Lemma:Expander-Graph-Unique-Neighbor-extended} gives a unique neighbor \(r_i\) of the nonempty set \(T_{i-1}\). 
    Let \(t_i\) be the unique vertex of \(T_{i-1}\) adjacent to \(r_i\), and set \(T_i := T_{i-1}\setminus\{t_i\}\).
	The indices \(t_1,\dots,t_\tau\) are distinct by construction. 
	The indices \(r_1,\dots,r_\tau\) are also distinct because, after \(t_i\) is removed, \(r_i\) has no neighbors in \(T_i\). 
	Let \(\mathbf{M}_0\) be the submatrix whose rows and columns are ordered as \(r_\tau,\dots,r_1\) and \(t_\tau,\dots,t_1\), respectively. 
	For each \(i\), the row \(r_i\) is adjacent to \(t_i\) and to no other vertex in \(T_{i-1} = \{t_i,t_{i+1},\dots,t_\tau\}\). 
	Hence \(\mathbf{M}_0\) is triangular with unit diagonal and \(\det(\mathbf{M}_0)=1\). 
	Reordering the rows and columns changes the determinant only by a sign, so \(\mathbf{H}_{(T)}\) contains a \(|T|\times|T|\) unimodular submatrix. 
\end{proof}

\paragraph{Deterministic Peeling}

The proof of \Cref{Lemma:Expander-implies-UnimodularMinor} is constructive.
The scheme \(\mathsf{SEC}\) defined in \Cref{Section:SEC} uses \(\mathbf{\hat{G}} := (\mathbf{P}\mathbf{H})^\top\), where \(\mathbf{P}\) is a signed permutation matrix, so every nonzero entry of \(\mathbf{\hat{G}}\) lies in \(\{-1,1\}\).
For \(U\subseteq[n]\), we peel the row restriction \(\mathbf{\hat{G}}_U\) deterministically as follows.
Initially, all rows of \(\mathbf{\hat{G}}_U\) are active. 
Each step selects the smallest-index column, among those not yet selected, that has exactly one nonzero entry in the active rows, records that column together with the row containing this entry, and then deactivates the row. 
When \(|U| \leq \left\lceil 2(1-\epsilon_\mathrm{exp}) \lfloor\gamma n\rfloor \right\rceil-1\), such a column exists at every step by \Cref{Lemma:Expander-Graph-Unique-Neighbor-extended}, exactly as in the proof of \Cref{Lemma:Expander-implies-UnimodularMinor}. 
The selected columns then form a matrix \(\mathbf{M}_U\in GL_{|U|}(\mathbb{Z})\); after ordering its rows according to the peeling order, it is triangular with diagonal entries in \(\{-1,1\}\).
The smallest-index rule makes \(\mathbf{M}_U\) a deterministic function of \(\mathbf{\hat{G}}\) and \(U\); \Cref{Algorithm:Construct-MU} in Appendix~\ref{Appendix:Deferred-Proofs} states the procedure as pseudocode. 
We call \(\mathbf{M}_U\) the \textit{peeled unimodular submatrix}; its conditioning enters the kernel-smoothing bound of \Cref{Section:Uniform-Kernel-Smoothing}.

\subsection{Hardness Assumptions}\label{Section:Hardness-Assumption}

Our privacy analysis relies on decisional Module-LWE in systematic form and Matrix Hint-MLWE, which augments an MLWE secret with noisy linear leakage.

\subsubsection{Module Learning with Errors}\label{Section:MLWE}

\begin{definition}[Module-LWE in systematic form, cf.~{\cite{langlois2015worst}}]\label{Definition:MLWE}
	Let \(n,m,q,d\in\mathbb{Z}^+\), and let \(\psi\) be a distribution over \(\mathbb{Z}\), also denoting its coefficient-wise extension to \(\mathcal{R}_q\). 
	Sample \(\bm{A}\stackrel{\$}{\gets}\mathcal{R}_q^{n\times m}\) and define \(\overline{\bm{A}} := \begin{bmatrix} \bm{I}_n \ \bm{A} \end{bmatrix}\). 
	The decisional \(\mathsf{MLWE}_{n,m,q,d,\psi}\) problem is to distinguish \((\bm{A},\overline{\bm{A}}\bm{s})\) from \((\bm{A},\bm{u})\), where \(\bm{s}\gets\psi^{n+m}\) and \(\bm{u}\stackrel{\$}{\gets}\mathcal{R}_q^n\). 
	The advantage of an adversary \(\mathcal{A}\) is \(\mathsf{Adv}_\mathcal{A}^\mathsf{MLWE}(\kappa) := \left| \Pr\!\left[ 1\gets \mathcal{A} (\bm{A},\overline{\bm{A}}\bm{s}) \right] - \Pr\!\left[ 1\gets \mathcal{A}(\bm{A},\bm{u}) \right] \right|\). 
\end{definition}

Writing \(\bm{s}=\begin{bmatrix}\bm{e}\\\bm{s}'\end{bmatrix}\) with \(\bm{e}\gets\psi^n\) and \(\bm{s}'\gets\psi^m\) gives \(\overline{\bm{A}}\bm{s}=\bm{A}\bm{s}'+\bm{e}\); thus, the systematic secret combines the conventional MLWE error and secret. 
When \(\psi=\mathcal{D}_{\mathbb{Z},\sigma}\), we abbreviate \(\mathsf{MLWE}_{n,m,q,d,\psi}\) as \(\mathsf{MLWE}_{n,m,q,d,\sigma}\), where \(\sigma\) is the source width.

\subsubsection{Matrix Hint-MLWE}\label{Section:Matrix-Hint-MLWE}

\paragraph{Problem Definition}

Hint-MLWE augments an MLWE sample with noisy scalar multiples of its secret \cite{kim2023toward}, while Matrix Hint-MLWE permits noisy linear images \cite{espitau2024flood}. 
Following the rectangular single-hint formulation of \cite{wang2025post}, we allow multiple rectangular hints with possibly different output dimensions. 
As in \Cref{Definition:MLWE}, let \(\overline{\bm{A}} := \begin{bmatrix} \bm{I}_n \ \bm{A} \end{bmatrix}\).

\begin{definition}[Matrix Hint-MLWE with varying hint dimensions]\label{Definition:MHMLWE}
	Let \(n,m,q,d,h\in\mathbb{Z}^{+}\) and \((n_i)_{i\in[h]}\in(\mathbb{Z}^{+})^h\).
	Let \(\chi_{\bm{s}}\) be a distribution over \(\mathcal{R}^{n+m}\).
	For every \(i\in[h]\), let \(\chi_{\bm{y}_i}\) be a distribution over \(\mathcal{R}^{n_i}\), and let \(\mathcal{M}_i\) be a distribution over \(\mathcal{R}^{n_i\times(n+m)}\).
	The problem \(\mathsf{MatrixHint\text{-}MLWE}_{ (n_i)_{i\in[h]},n,m,q,d,\chi_{\bm{s}} }^{ h, (\chi_{\bm{y}_i})_{i\in[h]}, (\mathcal{M}_i)_{i\in[h]} }\) is to distinguish:
	\begin{enumerate*}[label=(\arabic*)]
		\item (\(\mathsf{Real}\)) \(\left( \bm{A}, \overline{\bm{A}}\bm{s}, (\bm{M}_i)_{i\in[h]}, (\bm{z}_i)_{i\in[h]} \right)\), and
		\item (\(\mathsf{Random}\)) \(\left( \bm{A}, \bm{u}, (\bm{M}_i)_{i\in[h]}, (\bm{z}_i)_{i\in[h]} \right)\),
	\end{enumerate*}
	where \(\bm{A} \stackrel{\$}{\gets} \mathcal{R}_q^{n\times m}\), \(\bm{u} \stackrel{\$}{\gets} \mathcal{R}_q^n\), \(\bm{s}\gets\chi_{\bm{s}}\), and, independently for every \(i\in[h]\), \(\bm{M}_i\gets\mathcal{M}_i\), \(\bm{y}_i\gets\chi_{\bm{y}_i}\), and \(\bm{z}_i := \bm{M}_i\bm{s}+\bm{y}_i\).
	The secret is reduced coefficient-wise modulo \(q\) when forming \(\overline{\bm{A}}\bm{s}\in\mathcal{R}_q^n\), while the hints are computed over \(\mathcal{R}\), or equivalently over their integer coefficient embeddings.
	The distinguishing advantage of an adversary \(\mathcal{A}\) is \(\mathsf{Adv}_\mathcal{A}^\mathsf{MHMLWE}(\kappa) := \left| \Pr[\mathcal{A}(\mathsf{Real})=1] - \Pr[\mathcal{A}(\mathsf{Random})=1] \right|\), where the probabilities are taken over the stated sampling and the internal randomness of \(\mathcal{A}\).
\end{definition}

\paragraph{Joint-Precision Hardness Bound}

We use the elliptic posterior-resampling argument of \cite{kim2023toward,espitau2024flood,wang2025post} and prove the following variant for multiple hints.
Rather than separately bounding \(\|\mathbf{M}_i^\top\mathbf{M}_i\|_2 \|\mathbf{\Sigma}_{\mathbf{y}_i}^{-1}\|_2\), it retains the tighter joint term \(\|\mathbf{M}_i^\top \mathbf{\Sigma}_{\mathbf{y}_i}^{-1}\mathbf{M}_i\|_2\).
Under coefficient embedding, let \(N:=d(n+m)\) and \(N_i:=dn_i\) for \(i \in [h]\).

\begin{theorem}[Joint-precision hardness of Matrix Hint-MLWE]\label{Theorem:Hardness-MHMLWE-Tight}
	Consider the setting of \Cref{Definition:MHMLWE}.
	Suppose that, under coefficient embedding, \(\mathbf{s}\gets \mathcal{D}_{\mathbb{Z}^{N},\mathbf{c}_\mathbf{s}, \sqrt{\mathbf{\Sigma}_\mathbf{s}}}\), and, for every \(i\in[h]\), the \(i\)-th perturbation is distributed as \(\mathbf{y}_i\gets \mathcal{D}_{\mathbb{Z}^{N_i},\mathbf{c}_{\mathbf{y}_i}, \sqrt{\mathbf{\Sigma}_{\mathbf{y}_i}}}\), where \(\mathbf{c}_\mathbf{s}\in\mathbb{R}^{N}\), \(\mathbf{\Sigma}_\mathbf{s}\succ0\), and, for every \(i\in[h]\), \(\mathbf{c}_{\mathbf{y}_i}\in\mathbb{R}^{N_i}\) and \(\mathbf{\Sigma}_{\mathbf{y}_i}\succ0\).
	Fix a tuple of hint matrices in the support of \((\mathcal{M}_i)_{i\in[h]}\), and let \(\mathbf{M}_i\in\mathbb{Z}^{N_i\times N}\) be the coefficient embedding of its \(i\)-th component.
	Define the corresponding effective source width \(\sigma>0\) by \(1/\sigma^2 := 2\big(\|\mathbf{\Sigma}_\mathbf{s}^{-1}\|_2 + \sum_{i=1}^{h} \|\mathbf{M}_i^\top \mathbf{\Sigma}_{\mathbf{y}_i}^{-1} \mathbf{M}_i\|_2 \big)\).
	If \(\sigma \geq \sqrt{2}\eta_\epsilon(\mathbb{Z}^{N})\) for some \(0<\epsilon\leq1/2\), then there exists an efficient reduction from \(\mathsf{MLWE}_{n,m,q,d,\sigma}\) to the corresponding fixed-matrix Matrix Hint-MLWE problem, with additive distinguishing loss at most \(2\epsilon\).
\end{theorem}

The posterior-resampling proof uses \(\mathbf{\Sigma}_\mathbf{s}^{-1}+\sum_{i}\mathbf{M}_i^\top\mathbf{\Sigma}_{\mathbf{y}_i}^{-1}\mathbf{M}_i\) as the precision of the secret conditioned on the hints and requires only an upper bound on its norm; see Appendix~\ref{Appendix:MHMLWE-Proofs}.

\begin{remark}[Interpreting the effective source width]\label{Remark:MHMLWE-Width-Interpretation}
	For an isotropic secret, \(\mathbf{\Sigma}_\mathbf{s}=\sigma_\mathbf{s}^2\mathbf{I}_N\), the effective source width of \Cref{Theorem:Hardness-MHMLWE-Tight} satisfies \(1/\sigma^2 = 2\left(1/\sigma_\mathbf{s}^2+\Xi\right)\), where \(\Xi := \sum_{i=1}^{h}\|\mathbf{M}_i^\top\mathbf{\Sigma}_{\mathbf{y}_i}^{-1}\mathbf{M}_i\|_2\) is the leakage term. 
	Hence \(\sigma\leq\sigma_\mathbf{s}/\sqrt{2}\), with equality in the absence of hints, and the hints reduce \(\sigma\) further. 
	If the hint perturbations are wide enough that \(\Xi\leq c/\sigma_\mathbf{s}^2\) for a constant \(c\geq0\), then \(\sigma\geq\sigma_\mathbf{s}/\sqrt{2(1+c)}\), so the reduction loses only a constant factor in the \(\mathsf{MLWE}\) source width; for example, \(\Xi\leq1/\sigma_\mathbf{s}^2\) gives \(\sigma\geq\sigma_\mathbf{s}/2\). 
	In \(\mathsf{SEC}\), the hint perturbations are the sharing and leakage masks, so \(\Xi\) decreases as \(\sigma_\mathbf{R}\) and \(\sigma_{\bm{r}}\) grow; the concrete parameters of \Cref{Section:SEC-Concrete-Privacy} directly require the resulting width \(\sigma_\mathrm{MQ}\) to meet the \(\mathsf{MLWE}\) security target.
\end{remark}

The reduction remains valid from a smaller source width that still satisfies the smoothing condition.

\begin{corollary}[Reduction from a smaller source width]\label{Corollary:Hardness-MHMLWE-Tight-Smaller-Sigma}
	Fix a tuple of hint matrices as in \Cref{Theorem:Hardness-MHMLWE-Tight}, and let \(\sigma\) be its effective source width.
	If, for some \(0<\epsilon\leq1/2\), \(\sqrt{2}\, \eta_\epsilon(\mathbb{Z}^{d(n+m)}) \leq \sigma' \leq \sigma\), then there is an efficient reduction from \(\mathsf{MLWE}_{n,m,q,d,\sigma'}\) to the corresponding Matrix Hint-MLWE problem, with additive distinguishing loss at most \(2\epsilon\).
\end{corollary}

The proof verifies positive definiteness of the correction parameter matrix and smoothing of the convolution step; see Appendix~\ref{Appendix:MHMLWE-Proofs}.

\subsection{Secret Sharing}\label{Section:General-Secret-Sharing}

We recall the access structures, syntax, and correctness of secret sharing, and then formalize the two additional properties used in this work: everywhere-shortness and hard-core local leakage resilience.

\subsubsection{Basic Definitions}\label{Section:Secret-Sharing-Syntax}

We recall the following definition of partial and ramp access structures.

\begin{definition}[Partial and ramp access structures, cf.~{\cite[Definition 1]{applebaum2023recover}}]\label{Definition:Partial-Ramp-Access}
	A partial access structure over \(n\) parties is a pair \(\Gamma=(\Gamma_0,\Gamma_1)\) of nonempty families \(\Gamma_0,\Gamma_1\subseteq 2^{[n]}\) such that \(B\not\subseteq A\) for every \(A\in\Gamma_0\) and \(B\in\Gamma_1\).
	Sets in \(\Gamma_0\) are called \emph{private}, whereas sets in \(\Gamma_1\) are called \emph{authorized}.
	For \(0<\tau_p<\tau_c\leq1\), define \(k_p:=\lceil\tau_p n\rceil\) and \(k_c:=\lfloor\tau_c n\rfloor\), and assume \(k_p<k_c\).
	The \((\tau_p,\tau_c)\)-ramp access structure over \(n\) parties is the partial access structure \(\Gamma=(\Gamma_0,\Gamma_1)\) given by \(\Gamma_0:=\{U\subseteq [n]: |U|\leq k_p\}\) and \(\Gamma_1:=\{T\subseteq [n]: |T|\geq k_c\}\).
\end{definition}

\begin{remark}[Privacy vs. reconstruction]\label{Remark:Privacy-vs-Reconstruction} 
	In a ramp scheme, the privacy and reconstruction thresholds need not coincide, and no guarantee is prescribed for sets in neither \(\Gamma_0\) nor \(\Gamma_1\): for a \((k_p,k_c)\)-ramp access structure, sets with \(k_p<|U|<k_c\) may learn partial information, and some may even reconstruct the secret. 
	Accordingly, our privacy statements are always quantified over \(U\in\Gamma_0\), whereas our correctness statements are quantified over \(T\in\Gamma_1\). 
\end{remark}

We use the following secret-sharing syntax, which allows the access structure to depend on the public parameters.

\begin{definition}[Secret sharing syntax, cf.~{\cite[Section 2.1]{applebaum2023recover}}]\label{Definition:SecretSharing-Formal}
	Let \(n\in\mathbb{Z}^+\), let \(S\) be a secret space, and let \((S_i)_{i\in[n]}\) be the share spaces.
	A secret sharing scheme \(\mathsf{Sharing}=(\mathsf{Setup},\mathsf{Deal},\mathsf{Rec})\) consists of the following efficient algorithms: \begin{enumerate}
		\item \(\mathsf{Setup}(1^\kappa)\) is a randomized algorithm that outputs public parameters \(\mathsf{pp}\) and determines a partial access structure \(\Gamma(\mathsf{pp})=(\Gamma_0(\mathsf{pp}),\Gamma_1(\mathsf{pp}))\) over \(n\) parties.
		\item \(\mathsf{Deal}_\mathsf{pp}(s)\), on input \(s\in S\), is a randomized algorithm that outputs shares \((s_i)_{i\in[n]}\in\prod_{i=1}^n S_i\).
		\item \(\mathsf{Rec}_\mathsf{pp}(T,(s_i)_{i\in T})\), on input \(T\subseteq[n]\) and the corresponding shares, is a deterministic algorithm that outputs an element of \(S\cup\{\bot\}\).
	\end{enumerate}
\end{definition}

\begin{definition}[Correctness, cf.~{\cite[Definition 2]{applebaum2023recover}}]\label{Definition:SecretSharing-Correctness}
	Let \(\mathsf{Sharing}=(\mathsf{Setup},\mathsf{Deal},\mathsf{Rec})\) be a secret sharing scheme.
	The scheme is \(\epsilon_c\)-correct if, for every \(\mathsf{pp}\) in the support of \(\mathsf{Setup}(1^\kappa)\), every authorized set \(T\in\Gamma_1(\mathsf{pp})\), and every secret \(s\in S\), \( \Pr\!\left[ \mathsf{Rec}_\mathsf{pp}\!\left(T,(s_i)_{i\in T}\right)\neq s \;\middle|\; (s_i)_{i\in[n]}\gets\mathsf{Deal}_\mathsf{pp}(s) \right] \leq\epsilon_c\), where the probability is over the randomness of \(\mathsf{Deal}\).
	The scheme is \emph{perfectly correct} if \(\epsilon_c=0\).
\end{definition}

The conventional pairwise-secret privacy notion is recalled in Appendix~\ref{Appendix:Conventional-Privacy}.
The privacy property used in this work is instead the distributional leakage-resilience notion defined in \Cref{Section:HC-LLR}.

\subsubsection{Everywhere-Short Secret Sharing}\label{Section:Everywhere-Short-Secret-Sharing}

Everywhere-short secret sharing strengthens the syntax and correctness of \Cref{Definition:SecretSharing-Formal,Definition:SecretSharing-Correctness} by requiring that, for every secret of bounded infinity norm, the shares have bounded infinity norm, up to a permitted shortness error, and the reconstruction coefficients have bounded infinity norm. 
Following prior work, we call such objects \textit{short}. 
The term was introduced informally in \cite{borin2026hermine} for schemes whose secrets, shares, and reconstruction coefficients are simultaneously short. 
Related coefficient bounds appear in \cite{chairattana2024partially}, without short shares, and in \cite{del2024shortly}, as properties of distributed key generation. 
We formalize the notion below, building on the integer-linear reconstruction of black-box secret sharing \cite{cramer2002optimal,cramer2020blackbox}.

\begin{definition}[Everywhere-short secret sharing]\label{Definition:Everywhere-Short-Secret-Sharing}
	Let \(\mathsf{Sharing}=(\mathsf{Setup},\mathsf{Deal},\mathsf{Rec})\) be a secret sharing scheme over \(n\) parties, and for \(\mathsf{pp}\) in the support of \(\mathsf{Setup}(1^\kappa)\) let \(\Gamma(\mathsf{pp})=(\Gamma_0(\mathsf{pp}),\Gamma_1(\mathsf{pp}))\) be the associated partial access structure.
    Let \(L\) denote the total number of atomic sub-shares, indexed by \([L]\).
    Suppose that the secret space \(S\) and the atomic share spaces \((\widehat{S}_j)_{j\in[L]}\) are contained in a common \(\mathbb{Z}\)-module with a fixed integer coefficient embedding, and let \(\|\cdot\|_\infty\) denote the infinity norm, namely the largest absolute value of a coefficient.
    Let \((I_i)_{i\in[n]}\) be a partition of \([L]\), where party \(i\) holds the aggregate share \((\bm{s}_j)_{j\in I_i}\in S_i:=\prod_{j\in I_i}\widehat{S}_j\).
	We say that \(\mathsf{Sharing}\) is \emph{everywhere-short} with secret bound \(\beta_\mathsf{sec}\), share bound \(\beta_\mathsf{sh}\), reconstruction bound \(\beta_{\bm{\lambda}}\), and shortness error \(\epsilon_\mathsf{sh}\) if the following hold.
	\begin{enumerate}
		\item \textbf{Short shares:} for every \(\mathsf{pp}\) in the support of \(\mathsf{Setup}(1^\kappa)\) and every \(\bm{s}\in S\) with \(\|\bm{s}\|_\infty\leq\beta_\mathsf{sec}\), if \(\big((\bm{s}_j)_{j\in I_i}\big)_{i\in[n]}\gets\mathsf{Deal}_\mathsf{pp}(\bm{s})\), then \(\Pr\!\big[\max_{j\in[L]}\|\bm{s}_j\|_\infty>\beta_\mathsf{sh}\big]\leq\epsilon_\mathsf{sh}\), where the probability is over the randomness of \(\mathsf{Deal}\).
        \item \textbf{Short linear reconstruction:} for every \(\mathsf{pp}\) in the support of \(\mathsf{Setup}(1^\kappa)\) and every \(T\in\Gamma_1(\mathsf{pp})\), \(\mathsf{Rec}_\mathsf{pp}(T,\cdot)\) is defined by \(\big((\bm{s}_j)_{j\in I_i}\big)_{i\in T}\mapsto\sum_{j\in\bigcup_{i\in T}I_i}\lambda_{T,j}\bm{s}_j\) for a public reconstruction vector \(\bm{\lambda}_T\in\mathbb{Z}^{L}\) satisfying \(\mathrm{supp}(\bm{\lambda}_T)\subseteq\bigcup_{i\in T}I_i\), depending only on \(\mathsf{pp}\) and \(T\), hence independent of the shares, with \(\|\bm{\lambda}_T\|_\infty\leq\beta_{\bm{\lambda}}\).
        Moreover, this map takes the value \(\bm{s}\) on every sharing in the support of \(\mathsf{Deal}_\mathsf{pp}(\bm{s})\), for every \(\bm{s}\in S\) with \(\|\bm{s}\|_\infty\leq\beta_\mathsf{sec}\). \label{Item:Short-Linear-Recon}
	\end{enumerate}
	The \emph{expansion factor} of the scheme is \(L/n\).
\end{definition}

All coefficient-magnitude bounds in \Cref{Definition:Everywhere-Short-Secret-Sharing} use the infinity norm, as in \cite{chairattana2024partially,del2024shortly,borin2026hermine}.
Euclidean bounds follow from \(\|\mathbf{x}\|_2\leq\sqrt{N}\|\mathbf{x}\|_\infty\) for \(\mathbf{x}\in\mathbb{Z}^N\).
Moreover, whenever \(\max_{j\in[L]}\|\bm{s}_j\|_\infty\leq\beta_\mathsf{sh}\), every reconstructed combination satisfies \(\left\|\sum_{j\in\bigcup_{i\in T}I_i}\lambda_{T,j}\bm{s}_j\right\|_\infty\leq wt_H(\bm{\lambda}_T)\,\beta_{\bm{\lambda}}\,\beta_\mathsf{sh}\) for every \(T\in\Gamma_1(\mathsf{pp})\).

\begin{remark}\label{Remark:Lambda-Public}
    Since \(\bm{\lambda}_T\) may depend only on \(\mathsf{pp}\) and \(T\), it may be published without affecting privacy. 
	The reconstruction vector need not be unique: if \(\bm{\lambda}_T\) and \(\bm{\lambda}_T'\) both satisfy \Cref{Item:Short-Linear-Recon} in \Cref{Definition:Everywhere-Short-Secret-Sharing} for the same \((\mathsf{pp},T)\), then their induced linear combinations agree on every sharing in the support of \(\mathsf{Deal}_\mathsf{pp}(\bm{s})\), for every \(\bm{s}\in S\) with \(\|\bm{s}\|_\infty\leq\beta_\mathsf{sec}\), since both recover the shared secret. 
\end{remark}

Binary reconstruction \cite{boneh2018threshold,ball2021linear} is recovered by restricting the reconstruction vector in \Cref{Item:Short-Linear-Recon} of \Cref{Definition:Everywhere-Short-Secret-Sharing} to \(\bm{\lambda}_T\in\{0,1\}^L\), in which case \(\beta_{\bm{\lambda}}=1\).
The preceding definition more generally permits short signed integer reconstruction coefficients.

\subsubsection{Hard-Core Local Leakage Resilience}\label{Section:HC-LLR}

Everywhere-short shares may reveal partial information about a short secret, so conventional privacy is generally too strong for our setting. 
We instead protect a prescribed hard-core function of a secret sampled from a designated distribution, even given the shares of a private set and prescribed local leakage from all shares. 
Our model follows the local-leakage setting, in which each share is processed by its own leakage function, studied for non-malleable secret sharing in \cite{goyal2018non} and for linear secret sharing in \cite{benhamouda2021local}.

\begin{definition}[Local leakage function]\label{Definition:Leakage-Function} 
	Let \(\mathsf{Sharing}\) have share spaces \((S_i)_{i\in[n]}\). 
	A \emph{local leakage function} is a tuple \(L=(L_i)_{i\in[n]}\), where \(L_i:S_i\to D_i\), acting component-wise as \(L((s_i)_{i\in[n]}) := (L_i(s_i))_{i\in[n]}\). 
	We write \(\mathcal{L}\) for a family of such functions and \(\mathcal{D}_\mathcal{L}\) for a distribution over \(\mathcal{L}\). 
\end{definition}

The public description of \(L\) may include linear maps and fresh randomness. 
In our analysis, \(L_i(\mathbf{s}_i) := \mathbf{L}_i\mathbf{s}_i+\mathbf{r}_i\), where \(\mathbf{L}_i\) is public and \(\mathbf{r}_i\) is fresh noise, possibly across several queries, as when a threshold signing protocol releases noisy linear functions of the shares. 
We omit the publicly known \(L\) from the adversary's displayed input below.
We now define the experiment underlying our privacy notion.

\begin{definition}[HC-Leak experiment]\label{Definition:HC-Leak} 
	Let \(\mathsf{Sharing}\) be a secret sharing scheme over \(n\) parties with secret space \(S\).
	Let \(\mathcal{D}_S\) be a distribution over \(S\), \(\mathcal{D}_{\mathcal{L}}\) a distribution over a family \(\mathcal{L}\) of local leakage functions, and \(\mathcal{D}_{\mathcal{F}}\) a distribution over a family \(\mathcal{F}\) of functions \(f:S\to Z\), where \(Z\) is finite. 
	Fix \(\mathsf{pp}\) in the support of \(\mathsf{Setup}(1^\kappa)\) and a private set \(U\subseteq[n]\) with \(|U|\leq k_p\). 
	The experiment \(\mathsf{HC\text{-}Leak}_{\mathcal{A},\mathsf{Sharing}}^{\mathsf{pp},U}(\kappa)\) proceeds as follows: 
	\begin{enumerate} 
		\item Sample \(f\gets\mathcal{D}_{\mathcal{F}}\), \(L\gets\mathcal{D}_{\mathcal{L}}\), and \(s\gets\mathcal{D}_S\), and generate \( (s_i)_{i\in[n]} \gets \mathsf{Deal}_{\mathsf{pp}}(s)\).  
		\item Sample \(z_0\stackrel{\$}{\gets}Z\) and \(b\stackrel{\$}{\gets}\{0,1\}\), and set \(z_1:=f(s)\). 
		\item Give \(\big( \mathsf{pp}, f, (s_i)_{i\in U}, L((s_i)_{i\in[n]}), z_b \big)\) to \(\mathcal{A}\), which outputs \(b'\in\{0,1\}\). 
	\end{enumerate} 
	The advantage of \(\mathcal{A}\) is \(\mathsf{Adv}_{ \mathcal{A},\mathsf{HC\text{-}Leak} }^{\mathsf{pp},U}(\kappa) := 2\left| \Pr[b'=b]-1/2 \right|\). 
\end{definition}

In our application, the hard-core function is the systematic \(\mathsf{MLWE}\) target \(f_{\bm{A}}(\bm{s}) := \bm{A}\bm{s}\bmod q\), where \(\bm{A} := \begin{bmatrix} \bm{I}_p \ \bm{A}' \end{bmatrix} \in \mathcal{R}_q^{p\times\ell}\) and \(\bm{A}'\) is sampled as part of the function description. 
Thus, \(Z=\mathcal{R}_q^p\), and the experiment asks whether \(f_{\bm{A}}(\bm{s})\) is indistinguishable from an independent uniform element of \(\mathcal{R}_q^p\). 
The following definition strengthens the leakage-resilience notion of \cite[Definition 5]{nielsen2020lower}.

\begin{definition}[Hard-core function for local leakage resilience]\label{Definition:HC-LLR}
    Let \(\mathsf{Sharing}=(\mathsf{Setup},\mathsf{Deal},\mathsf{Rec})\) be an \(\epsilon_c\)-correct secret sharing scheme with secret space \(S\) and privacy threshold \(k_p\). 
	Let \(\mathcal{D}_S\), \(\mathcal{D}_\mathcal{L}\), and \(\mathcal{D}_\mathcal{F}\) be as in \Cref{Definition:HC-Leak}, let \(\delta_p\in[0,1]\), and let \(\epsilon_p\) assign a bound \(\epsilon_p(\kappa,\mathcal{A},U)\in[0,1]\) to every security parameter \(\kappa\), PPT adversary \(\mathcal{A}\), and coalition \(U\).
    We say that \(\mathcal{F}\) is a \((\delta_p,\epsilon_p)\)-private hard-core function family for local leakage resilience with respect to \(\mathsf{Sharing}\), \(\mathcal{D}_S\), \(\mathcal{D}_\mathcal{L}\), and \(\mathcal{D}_\mathcal{F}\) if, for every \(\kappa\), 
    \begin{align*} 
		\Pr_{\mathsf{pp}\gets\mathsf{Setup}(1^\kappa)} \Big[ \forall\,\textnormal{PPT }\mathcal{A},\ \forall\,U\subseteq[n],|U|\leq k_p: \mathsf{Adv}_{ \mathcal{A},\mathsf{HC\text{-}Leak} }^{\mathsf{pp},U}(\kappa) \leq \epsilon_p(\kappa,\mathcal{A},U) \Big] \geq 1-\delta_p. 
	\end{align*} 
    In words, except with probability at most \(\delta_p\) over \(\mathsf{pp}\gets\mathsf{Setup}(1^\kappa)\), every PPT adversary has advantage at most \(\epsilon_p(\kappa,\mathcal{A},U)\) in the \(\mathsf{HC\text{-}Leak}\) experiment for every coalition \(U\subseteq[n]\) of size at most \(k_p\). 
	If, for every PPT adversary \(\mathcal{A}\) and every such coalition \(U\), the function \(\epsilon_p(\kappa,\mathcal{A},U)\) is negligible in \(\kappa\), we say that \(\mathcal{F}\) is \(\delta_p\)-private. 
\end{definition}

Thus, \(\delta_p\) bounds the probability of exceptional setup outputs, which are determined before the adversary and the coalition are chosen; outside them, every PPT adversary has advantage at most \(\epsilon_p\) against every coalition of at most \(k_p\) parties. 
Unlike conventional privacy, this notion protects a prescribed hard-core function of a secret sampled from \(\mathcal{D}_S\), rather than requiring indistinguishability of the complete views induced by arbitrary fixed secrets.

\section{The \(\mathsf{SEC}\) Construction}\label{Section:SEC}

This section explains our ramp secret sharing scheme \(\mathsf{SEC}\) along with the construction of its distribution matrix. 
Correctness, everywhere-shortness, and linearity are established in \Cref{Section:Correctness-Shortness-Linearity}, and privacy in \Cref{Section:Secret-Sharing-Privacy-Analysis}.

\subsection{Overview}\label{Section:SEC-Overview}

Our construction follows the coding-theoretic approach to secret sharing initiated by McEliece and Sarwate \cite{mceliece1981sharing} and developed through linear secret sharing, span programs, and black-box secret sharing \cite{karchmer1993span,massey1993minimal,massey1994some,cramer2002optimal,cramer2005black,cramer2015linear,cramer2020blackbox,applebaum2023recover}.
Unlike conventional finite-field schemes, \(\mathsf{SEC}\) operates over a \(\mathbb{Z}\)-module and is targeted primarily for lattice-based distributed cryptography, where a common requirement is for both share coefficients and reconstruction coefficients to be bounded.

A designated dealer samples an \((n,m,c_\mathrm{exp},\gamma,(1-\epsilon_\mathrm{exp})c_\mathrm{exp})\) lossless bipartite expander (\Cref{Definition:Lossless-Expander}) with bi-adjacency matrix \(\mathbf{H}\in\{0,1\}^{m\times n}\), viewed over \(\mathbb{Z}\).
It samples a signed permutation matrix \(\mathbf{P}\in GL_m(\mathbb{Z})\), sets \(\mathbf{\hat{G}}:=(\mathbf{P}\mathbf{H})^\top\), and constructs a secret-bearing column \(\mathbf{g}\in\mathbb{Z}^n\) that supports reconstruction by sufficiently large sets.
The distribution matrix is \(\mathbf{G} := \begin{bmatrix} \mathbf{g} \ \mathbf{\hat{G}} \end{bmatrix} \in\mathbb{Z}^{n\times k}\), where \(k:=m+1\).
The column \(\mathbf{g}\) determines how the secret enters the shares, while the sparse columns of \(\mathbf{\hat{G}}\) distribute short masking randomness and keep the shares short.

Let the secret be \(\bm{s}\in\mathcal{R}^\ell\), where \(\ell\) is the module rank and \(\mathcal{R}=\mathbb{Z}[X]/(X^d+1)\) has degree \(d\), and let \(\mathbf{s}\in\mathbb{Z}^D\), with \(D:=d\ell\), denote its coefficient embedding. 
To share \(\mathbf{s}\), the dealer samples a short matrix \(\mathbf{R}\in\mathbb{Z}^{(k-1)\times D}\) and computes \(\mathbf{S} := \mathbf{G} \begin{bmatrix} \mathbf{s}^{\top}\\ \mathbf{R} \end{bmatrix} \in\mathbb{Z}^{n\times D}\).
Party \(i\) receives the row \(\mathbf{s}_i^\top:=\mathbf{S}[i,\cdot]\).
For a participant index list \(T \subseteq [n]\), any \(\bm{\lambda}_T\in\mathbb{Z}^{|T|}\) satisfying \(\bm{\lambda}_T^\top\mathbf{G}_T = \mathbf{e}_1^\top\) yields \(\bm{\lambda}_T^\top\mathbf{S}_T=\mathbf{s}^\top\), so reconstruction by \(T\subseteq[n]\) reduces to finding such a vector with small \(\ell_\infty\)-norm.
For every private set \(U\), equivalently every \(U\subseteq[n]\) with \(|U|\leq k_p\), we additionally require \((\mathbf{g})_U \in \mathrm{Col}_\mathbb{Z}(\mathbf{\hat{G}}_U)\), which rules out exact cancellation of the masking term over the integers and modulo any integer. 
This structural condition does not by itself give privacy, which we prove computationally in \Cref{Section:Secret-Sharing-Privacy-Analysis}.

\subsection{Setup, Sharing, and Reconstruction}\label{Section:SEC-formal}

We first characterize reconstruction and give a deterministic procedure for finding a short reconstruction vector.
We then formally define the setup, sharing, and reconstruction algorithms.

The following standard linear-reconstruction criterion underlies span-program, linear, and black-box secret sharing \cite{beimel1996secure,karchmer1993span,cramer2002optimal,applebaum2023recover}.

\begin{lemma}[Authorized reconstruction]\label{Lemma:Authorized-Reconstruction}
	Let \(\mathbf{G}\in\mathbb{Z}^{n\times k}\), \(\mathbf{s}\in\mathbb{Z}^{D}\), \(\mathbf{R}\in\mathbb{Z}^{(k-1)\times D}\), and define \(\mathbf{S}:=\mathbf{G}\begin{bmatrix}\mathbf{s}^\top \\ \mathbf{R}\end{bmatrix}\).
	For any \(T\subseteq[n]\), if \(\mathbf{e}_1^\top\in\mathrm{Row}_\mathbb{Z}(\mathbf{G}_T)\), then there exists \(\bm{\lambda}_T\in\mathbb{Z}^{|T|}\) such that \(\bm{\lambda}_T^\top\mathbf{G}_T=\mathbf{e}_1^\top\), and every such vector satisfies \((\bm{\lambda}_T^\top\mathbf{S}_T)^\top=\mathbf{s}\).
\end{lemma}

\begin{proof}
	Suppose that \(\mathbf{e}_1^\top\in\mathrm{Row}_\mathbb{Z}(\mathbf{G}_T)\).
	By the definition of the integer row module, there exists \(\bm{\lambda}_T\in\mathbb{Z}^{|T|}\) such that \(\bm{\lambda}_T^\top\mathbf{G}_T=\mathbf{e}_1^\top\).
	For every such vector, \(\bm{\lambda}_T^\top\mathbf{S}_T = \bm{\lambda}_T^\top\mathbf{G}_T \begin{bmatrix} \mathbf{s}^\top\\ \mathbf{R} \end{bmatrix} = \mathbf{e}_1^\top \begin{bmatrix} \mathbf{s}^\top\\ \mathbf{R} \end{bmatrix} = \mathbf{s}^\top\).
	Transposing both sides gives \((\bm{\lambda}_T^\top\mathbf{S}_T)^\top=\mathbf{s}\).
\end{proof}

Conversely, if a share-independent \(\bm{\lambda}_T\in\mathbb{Z}^{|T|}\) satisfies \((\bm{\lambda}_T^\top\mathbf{S}_T)^\top=\mathbf{s}\) for all \(\mathbf{s}\) and \(\mathbf{R}\), then \(\bm{\lambda}_T^\top\mathbf{G}_T=\mathbf{e}_1^\top\), since \(\begin{bmatrix}\mathbf{s}^\top\\ \mathbf{R}\end{bmatrix}\) ranges over all of \(\mathbb{Z}^{k\times D}\).
Hence \(\mathbf{e}_1^\top\in\mathrm{Row}_\mathbb{Z}(\mathbf{G}_T)\) characterizes the sets that admit integer-linear reconstruction.
The row Hermite normal form both tests this condition and extracts a reconstruction vector.

\begin{lemma}[Row-HNF characterization of authorization]\label{Lemma:e1-row-span-HNF-e1}
	Let \(\mathbf{A}\in\mathbb{Z}^{a\times b}\) where \(a,b \in \mathbb{Z}^+\), and let \(\mathbf{U}_\mathbf{A}\in GL_a(\mathbb{Z})\) be such that \(\mathbf{R}_\mathbf{A}:=\mathbf{U}_\mathbf{A}\mathbf{A}\) is the row-HNF of \(\mathbf{A}\).
	That is, for some \(r\geq0\), the nonzero rows of \(\mathbf{R}_\mathbf{A}\) are its first \(r\) rows, the first nonzero entry \(p_i>0\) of row \(i\in[r]\) lies in column \(j_i\) with \(j_1<\dots<j_r\), and \(0\leq\mathbf{R}_\mathbf{A}[\iota,j_i]<p_i\) for all \(\iota<i\).
	Then \(\mathbf{e}_1^\top\in\mathrm{Row}_\mathbb{Z}(\mathbf{A})\) if and only if \(\mathbf{R}_\mathbf{A}[1,\cdot]=\mathbf{e}_1^\top\).
	In this case, \(\bm{\lambda}_0:=(\mathbf{U}_\mathbf{A}[1,\cdot])^\top\) satisfies \(\bm{\lambda}_0^\top\mathbf{A}=\mathbf{e}_1^\top\).
\end{lemma}

We prove \Cref{Lemma:e1-row-span-HNF-e1} in Appendix~\ref{Appendix:Deferred-Proofs}; the proof uses only that unimodular row operations preserve \(\mathrm{Row}_\mathbb{Z}\) and that the row-HNF pivots reduce the entries above them.

Although \(\mathsf{Rec}_\mathsf{pp}\) ultimately computes only the vector-matrix product \((\bm{\lambda}_T^\top\mathbf{S}_T)^\top\), one must first select an integer vector \(\bm{\lambda}_T\) satisfying \(\bm{\lambda}_T^\top\mathbf{G}_T=\mathbf{e}_1^\top\).
By \Cref{Lemma:e1-row-span-HNF-e1}, the row-HNF determines whether such a vector exists and, when it does, yields the particular solution \(\bm{\lambda}_{T,0}:=(\mathbf{U}_{\mathbf{G}_T}[1,\cdot])^\top\).
The transformation matrix \(\mathbf{U}_{\mathbf{G}_T}\) need not be unique, and the resulting vector \(\bm{\lambda}_{T,0}\) need not be short.
Since adding any \(\bm{\nu}\in K_T:=\ker_\mathbb{Z}(\mathbf{G}_T^\top)\) preserves the reconstruction equation, all integer solutions form the affine coset \(\bm{\lambda}_{T,0}+K_T\), and \Cref{Algorithm:Construct-Small-Lambda} fixes a deterministic procedure for selecting a short representative of it.

\begin{algorithm}[!htp]\caption{Constructing a small reconstruction vector \(\bm{\lambda}_T\) for \(\mathbf{G}_T\).}
	\label{Algorithm:Construct-Small-Lambda}
	\begin{algorithmic}[1]
        \small
		\Require A submatrix \(\mathbf{G}_T\in\mathbb{Z}^{|T|\times k}\); a fixed Lov\'asz parameter \(\delta_\mathsf{LLL}\in(1/4,1)\); and fixed small greedy parameters \(\mathit{passes},\mathit{window}\in\mathbb{Z}^+\).
		\Ensure A vector \(\bm{\lambda}_T\in\mathbb{Z}^{|T|}\) satisfying \(\bm{\lambda}_T^\top\mathbf{G}_T=\mathbf{e}_1^\top\), or \(\bot\) if \(\mathbf{e}_1^\top\notin\mathrm{Row}_\mathbb{Z}(\mathbf{G}_T)\).
		\State Compute the row-HNF \(\mathbf{R}_{\mathbf{G}_T}\) of \(\mathbf{G}_T\) together with a \(\mathbf{U}_{\mathbf{G}_T}\in GL_{|T|}(\mathbb{Z})\) such that \(\mathbf{U}_{\mathbf{G}_T}\mathbf{G}_T=\mathbf{R}_{\mathbf{G}_T}\).
		\If{\(\mathbf{R}_{\mathbf{G}_T}[1,\cdot]\neq\mathbf{e}_1^\top\)}
			\State \Return \(\bot\).
		\EndIf
		\State Set \(\bm{\lambda}_{T,0}:=(\mathbf{U}_{\mathbf{G}_T}[1,\cdot])^\top\).
		\State Let \(\mathbf{B}_K\in\mathbb{Z}^{r\times|T|}\) be the row-HNF basis of \(K_T:=\ker_\mathbb{Z}(\mathbf{G}_T^\top)\), where \(r:=\mathrm{rank}(K_T)\).
		\If{\(r=0\)}
			\State \Return \(\bm{\lambda}_{T,0}\).
		\EndIf
		\State Compute \(\mathbf{B}_K':=\mathsf{LLL}_{\delta_\mathsf{LLL}}(\mathbf{B}_K)\) in exact rational arithmetic using the fixed reduction and tie-breaking conventions.
		\State Compute \(\bm{\lambda}_T\gets\mathsf{BabaiGreedy}(\bm{\lambda}_{T,0},\mathbf{B}_K',\mathit{passes},\mathit{window})\) according to \Cref{Algorithm:Construct-small-lambda-Babai-greedy}.
		\If{\(\bm{\lambda}_T^\top\mathbf{G}_T\neq\mathbf{e}_1^\top\)}
			\State \Return \(\bot\). \Comment{exact certification}
		\EndIf
		\State \Return \(\bm{\lambda}_T\).
	\end{algorithmic}
\end{algorithm}

Finding a representative of minimum \(\ell_\infty\)-norm is a closest-vector problem and is NP-hard in general \cite{van1981another,micciancio2002complexity}; we therefore use exact-arithmetic LLL reduction and Babai decoding \cite{babai1986lovasz,hhan2026hybrid}, followed by a deterministic greedy search that directly targets the \(\ell_\infty\)-norm. 
Since \(\mathsf{BabaiGreedy}\) (\Cref{Algorithm:Construct-small-lambda-Babai-greedy}) modifies \(\bm{\lambda}_{T,0}\) only by vectors in \(K_T\), its output remains in the affine coset \(\bm{\lambda}_{T,0}+K_T\) and therefore preserves the reconstruction equation. 
The final equality test certifies this equation exactly. 
The resulting coefficient size is analyzed in \Cref{Section:Bounding-Reconstruction-Coefficients} and evaluated empirically in \Cref{Section:SEC-Reconstruction-Coefficient-Evaluation}. 
The detailed Babai and greedy procedures are given in \Cref{Algorithm:Construct-small-lambda-Babai-greedy} in Appendix~\ref{Appendix:Babai-Greedy}.

We formally define \(\mathsf{SEC}\), an expander-code-based secret sharing scheme, as follows.

\begin{definition}[Expander-code-based secret sharing]\label{Definition:SEC}
    Fix \(d,\ell,n,m,c_\mathrm{exp}\in\mathbb{Z}^+\) with \(n > m\), let \(D:=d\ell\) and \(k:=m+1\), and fix \(\gamma\in(0,1]\), \(\epsilon_\mathrm{exp}\in(0,1/2)\), \(\sigma_\mathbf{s},\sigma_\mathbf{R}\in\mathbb{R}^+\), a privacy threshold \(k_p\in[n-1]\) satisfying \(k_p\leq\left\lceil 2(1-\epsilon_\mathrm{exp})\lfloor\gamma n\rfloor \right\rceil-1\), a target reconstruction-coverage error \(\epsilon_c\in[0,0.01]\), and the fixed public parameters used by \Cref{Algorithm:Construct-Small-Lambda}.
	Let the secret space be \(S:=\mathbb{Z}^D\), let each party \(i\in[n]\) have share space \(S_i:=\mathbb{Z}^D\), and let \(\mathcal{D}_S:=\mathcal{D}_{\sigma_\mathbf{s}}^D\) be the designated secret distribution for the everywhere-shortness and privacy analyses.
	The expander-code-based secret sharing scheme \(\mathsf{SEC}=(\mathsf{Setup},\mathsf{Deal},\mathsf{Rec})\) consists of the following algorithms.
	\begin{enumerate}
		\item \(\mathsf{Setup}(1^\kappa)\).
		Sample a lossless bipartite expander graph \((n,m,c_\mathrm{exp},\gamma,(1-\epsilon_\mathrm{exp})c_\mathrm{exp})\), and let \(\mathbf{H}\in\{0,1\}^{m\times n}\) be its bi-adjacency matrix, viewed over \(\mathbb{Z}\).
        Sample a signed permutation matrix \(\mathbf{P}\in GL_m(\mathbb{Z})\), set \(\mathbf{\hat{G}}:=(\mathbf{P}\mathbf{H})^\top\in\mathbb{Z}^{n\times m}\), construct \(\mathbf{g}\in\mathbb{Z}^n\) according to \Cref{Section:Constructing-First-Column-G}, and define \(\mathbf{G}:=\begin{bmatrix}\mathbf{g} \ \mathbf{\hat{G}}\end{bmatrix}\in\mathbb{Z}^{n\times k}\).
        By the restrictions on \(\epsilon_\mathrm{exp}\) and \(k_p\) above, \(\mathbf{\hat{G}}\) satisfies \(\mathrm{Col}_\mathbb{Z}(\mathbf{\hat{G}}_U)=\mathbb{Z}^{|U|}\), and hence \((\mathbf{g})_U\in \mathrm{Col}_\mathbb{Z}(\mathbf{\hat{G}}_U)\), for every \(U\subseteq[n]\) with \(|U|\leq k_p\).
        Let \(k_c\) be the smallest integer in \([k_p+1,n]\) such that, for every \(t\in[k_c,n]\), \(\Pr_{T\stackrel{\$}{\gets}\binom{[n]}{t}} \left[ \mathbf{e}_1^\top \notin \mathrm{Row}_\mathbb{Z}(\mathbf{G}_T) \right] \leq\epsilon_c\). 
		Define \(\Gamma_0(\mathsf{pp}):=\{U\subseteq[n]:|U|\leq k_p\}\) and \(\Gamma_1(\mathsf{pp}) := \{T\subseteq[n]:|T|\geq k_c \, \land \, \mathbf{e}_1^\top\in\mathrm{Row}_\mathbb{Z}(\mathbf{G}_T)\}\), and output the public parameters \(\mathsf{pp}\), including \(\mathbf{G}\), \(k_c\), and the public sampling and reconstruction parameters.
		\item \(\mathsf{Deal}_\mathsf{pp}(\mathbf{s})\).
		On input \(\mathbf{s}\in\mathbb{Z}^{D}\), sample a masking matrix \(\mathbf{R}\in\mathbb{Z}^{(k-1)\times D}\) with independent entries from \(\mathcal{D}_{\sigma_\mathbf{R}}\), compute
        \begin{align}\label{Eq:SEC-Deal}
            \mathbf{S}:= \mathbf{G}  \begin{bmatrix}\mathbf{s}^\top \\ \mathbf{R}\end{bmatrix},
        \end{align}
        and output the shares \((\mathbf{s}_i)_{i\in[n]}\), where \(\mathbf{s}_i:=(\mathbf{S}[i,\cdot])^\top\in\mathbb{Z}^{D}\).
		\item \(\mathsf{Rec}_\mathsf{pp}(T,(\mathbf{s}_i)_{i\in T})\).
		If \(|T|<k_c\), output \(\bot\).
		Otherwise, let \(\mathbf{S}_T\in\mathbb{Z}^{|T|\times D}\) be the matrix whose rows are the transposes of the shares indexed by \(T\), ordered according to the indices in \(T\), and compute \(\bm{\lambda}_T\) by running \Cref{Algorithm:Construct-Small-Lambda} on input \(\mathbf{G}_T\).
		If the algorithm returns \(\bot\), output \(\bot\); otherwise, output \((\bm{\lambda}_T^\top\mathbf{S}_T)^\top\).
	\end{enumerate}
\end{definition}

By \Cref{Lemma:e1-row-span-HNF-e1,Lemma:Authorized-Reconstruction}, \(\mathsf{Rec}_\mathsf{pp}\) exactly recovers the secret for every \(T\in\Gamma_1(\mathsf{pp})\) (\Cref{Lemma:SEC-Exact-Reconstruction}).
The private-set condition asserted in \Cref{Definition:SEC} follows from the expander structure under the restrictions on \(\epsilon_\mathrm{exp}\) and \(k_p\) (\Cref{Lemma:Expander-implies-UnimodularMinor,Lemma:Column-Space-Ghat-U-Expander}).

\begin{remark}[Deterministic selection of reconstruction vectors]\label{Remark:Canonicity-of-Construct-Small-Lambda} 
	Although the reconstruction equation may admit multiple integer solutions, \Cref{Algorithm:Construct-Small-Lambda} fixes a deterministic output for each \(\mathbf{G}_T\) and its stated parameters. 
	This follows from the unique row-HNF basis of \(K_T\), the fixed exact-arithmetic LLL and Babai conventions, and the deterministic greedy refinement; in particular, the final output is independent of the transformation matrix used to obtain \(\bm{\lambda}_{T,0}\). 
	The decoding and tie-breaking rules, together with the determinism justification, are given in \Cref{Algorithm:Construct-small-lambda-Babai-greedy} in Appendix~\ref{Appendix:Babai-Greedy}. 
	Once the resulting public vector \(\bm{\lambda}_T\) is fixed, reconstruction is the vector-matrix multiplication \(\mathbf{S}_T\mapsto (\bm{\lambda}_T^\top\mathbf{S}_T)^\top\), and is therefore efficient and deterministic. 
\end{remark}

\subsection{Constructing the Distribution Matrix}\label{Section:Constructing-First-Column-G}

We describe how \(\mathsf{Setup}\) constructs the distribution matrix \(\mathbf{G}=\begin{bmatrix}\mathbf{g}\ \mathbf{\hat{G}}\end{bmatrix}\).
The expander-derived masking block \(\mathbf{\hat{G}}\) alone guarantees the private-set condition, allowing the secret-bearing column \(\mathbf{g}\) to be chosen to promote reconstruction.
We construct \(\mathbf{g}\) by planting pairwise disjoint certified supports, thereby obtaining a deterministic reconstruction threshold while keeping its entries small.

The expander structure ensures that every restriction of \(\mathbf{\hat{G}}\) to at most \(k_p\) rows is integrally surjective.

\begin{lemma}[Integral surjectivity of row restrictions to private sets]\label{Lemma:Column-Space-Ghat-U-Expander}
	Let \(\mathbf{\hat{G}}=(\mathbf{P}\mathbf{H})^\top\in\{-1,0,1\}^{n\times m}\), where \(\mathbf{P}\in GL_m(\mathbb{Z})\) is a signed permutation matrix and \(\mathbf{H}\in\{0,1\}^{m\times n}\) is the bi-adjacency matrix of an \((n,m,c_\mathrm{exp},\gamma,(1-\epsilon_\mathrm{exp})c_\mathrm{exp})\) lossless bipartite expander with \(0<\epsilon_\mathrm{exp}<1/2\).
	If \(k_p\leq\left\lceil2(1-\epsilon_\mathrm{exp})\lfloor\gamma n\rfloor\right\rceil-1\), then \(\mathrm{Col}_\mathbb{Z}(\mathbf{\hat{G}}_U)=\mathbb{Z}^{|U|}\) for every \(U\subseteq[n]\) with \(|U|\leq k_p\).
\end{lemma}

\begin{proof}
	The claim is immediate if \(U=\emptyset\).
	Now let \(\emptyset\neq U\subseteq[n]\) with \(|U|\leq k_p\).
	By \Cref{Lemma:Expander-implies-UnimodularMinor}, the restricted matrix \(\mathbf{H}_{(U)}\in\{0,1\}^{m\times|U|}\) contains a \(|U|\times|U|\) unimodular submatrix, and hence \(\mathrm{Row}_\mathbb{Z}(\mathbf{H}_{(U)})=\mathbb{Z}^{|U|}\).
	Since \(\mathbf{P}\in GL_m(\mathbb{Z})\), left multiplication by \(\mathbf{P}\) preserves the integer row module, so \(\mathrm{Row}_\mathbb{Z}((\mathbf{P}\mathbf{H})_{(U)})=\mathbb{Z}^{|U|}\).
	Finally, \(\mathbf{\hat{G}}_U=((\mathbf{P}\mathbf{H})_{(U)})^\top\), and therefore \(\mathrm{Col}_\mathbb{Z}(\mathbf{\hat{G}}_U)=\mathrm{Row}_\mathbb{Z}((\mathbf{P}\mathbf{H})_{(U)})=\mathbb{Z}^{|U|}\).
\end{proof}

Consequently, \((\mathbf{g})_U\in\mathrm{Col}_\mathbb{Z}(\mathbf{\hat{G}}_U)\) for every \(\mathbf{g}\in\mathbb{Z}^n\) and every \(U\subseteq[n]\) with \(|U|\leq k_p\).
Thus, after fixing \(\mathbf{\hat{G}}\), the construction of \(\mathbf{g}\) can focus on promoting reconstruction without affecting the private-set condition required in \Cref{Definition:SEC}.

To motivate the construction of \(\mathbf{g}\), write \(\mathbf{H}':=\mathbf{P}\mathbf{H}\), so that \(\mathbf{\hat{G}}_T=(\mathbf{H}'_{(T)})^\top\). 
By \Cref{Lemma:Authorized-Reconstruction}, a set \(T\subseteq[n]\) reconstructs whenever \(\mathbf{e}_1^\top\in\mathrm{Row}_\mathbb{Z}(\mathbf{G}_T)\). 
Since \(\mathbf{G}_T=\begin{bmatrix}(\mathbf{g})_T\ \mathbf{\hat{G}}_T\end{bmatrix}\), this condition can be rewritten as
\begin{align}\label{Eq:Correctness-criterion-T} 
	\mathbf{e}_1^\top\in\mathrm{Row}_\mathbb{Z}(\mathbf{G}_T) \quad\Longleftrightarrow\quad \exists\bm{\lambda}_T \in\ker_\mathbb{Z}(\mathbf{H}'_{(T)}) : \bm{\lambda}_T^\top(\mathbf{g})_T=1. 
\end{align} 
If \(\ker_\mathbb{Z}(\mathbf{H}'_{(T)})\neq\{\mathbf{0}\}\), then dividing any nonzero vector in this kernel by the greatest common divisor of its entries gives a primitive vector \(\bm{\lambda}_T\).  
The generalized B\'ezout identity then gives \((\mathbf{g})_T\in\mathbb{Z}^{|T|}\) satisfying \(\bm{\lambda}_T^\top(\mathbf{g})_T=1\).
The remaining challenge is to construct a single global vector \(\mathbf{g}\in\mathbb{Z}^n\) that satisfies this condition for many sets \(T\), while keeping its entries small. 

We address this challenge by planting pairwise disjoint supports that satisfy \eqref{Eq:Correctness-criterion-T}.
This guarantees that every set of size at least a deterministic threshold \(k_c^\mathsf{det}\) reconstructs.
Smaller sets are not guaranteed to reconstruct, but \Cref{Section:SEC-Reconstruction-Thresholds} shows empirically that most sets of size at least the smaller threshold \(k_c\) of \Cref{Definition:SEC} do.

We call \(T\subseteq[n]\) a \textit{certified support} if \(\mathbf{e}_1^\top\in\mathrm{Row}_\mathbb{Z}(\mathbf{G}_T)\).
Certification is monotone: if \(T\) is certified, then every \(S\supseteq T\) is also certified.
Our construction plants pairwise disjoint certified supports so that every sufficiently large set contains at least one of them.

\begin{lemma}[Deterministic reconstruction threshold]\label{Lemma:Deterministic-Correctness-Lower-Bound}
	If \(T_1,\dots,T_b\subseteq[n]\) are pairwise disjoint certified supports, then every \(S\subseteq[n]\) with \(|S|\geq n-b+1\) is certified.
\end{lemma}

\begin{proof}
	Let \(R:=[n]\setminus\bigcup_{i=1}^bT_i\).
	If \(S\subseteq[n]\) contains none of \(T_1,\dots,T_b\), then it omits at least one element from each \(T_i\), so \(|S|\leq|R|+\sum_{i=1}^b(|T_i|-1)=n-b\).
	Therefore, every \(S\subseteq[n]\) with \(|S|\geq n-b+1\) contains some \(T_i\).
	Since certification is monotone, such an \(S\) is certified.
\end{proof}

As a result, planting \(b\) pairwise disjoint certified supports gives the deterministic reconstruction threshold \(k_c^\mathsf{det}:=n-b+1\).
The construction therefore seeks many small pairwise disjoint sets \(T_i\) for which \(\ker_\mathbb{Z}(\mathbf{H}'_{(T_i)})\neq\{\mathbf{0}\}\), chooses a short primitive vector in each kernel, and selects \((\mathbf{g})_{T_i}\) using the generalized B\'ezout identity so that \eqref{Eq:Correctness-criterion-T} holds. 
The following rank argument gives a deterministic baseline for the number of such supports.

\begin{lemma}[Rank-based disjoint supports]\label{Lemma:DisjointBlocksRankLeftover}
	Let \(\mathbf{H}'\in\mathbb{Z}^{m\times n}\), and let \(r:=\mathrm{rank}_\mathbb{Q}(\mathbf{H}')\).
	Then there exist pairwise disjoint sets \(T_1,\dots,T_b\subseteq[n]\) with \(|T_i|=r+1\) and \(\ker_\mathbb{Z}(\mathbf{H}'_{(T_i)})\neq\{\mathbf{0}\}\) for every \(i\in[b]\), such that the leftover set \(R:=[n]\setminus\bigcup_{i=1}^bT_i\) satisfies \(|R|\leq r\).
\end{lemma}

Indeed, any \(r+1\) columns of \(\mathbf{H}'\) are linearly dependent over \(\mathbb{Q}\), so partitioning \([n]\) into \(\lfloor n/(r+1)\rfloor\) blocks of size \(r+1\), together with a leftover set of size at most \(r\), proves the claim (Appendix~\ref{Appendix:Deferred-Proofs}). 
Making each block certified as above gives \(b=\lfloor n/(r+1)\rfloor\), and hence \(k_c^\mathsf{det} \leq n-\lfloor n/(r+1)\rfloor+1\).
In practice, the construction searches for smaller kernel supports, thereby increasing \(b\) and lowering the deterministic reconstruction threshold. 
The complete construction is given in \Cref{Algorithm:Construct-g-disjoint-approach}; it prefers B\'ezout solutions in \(\{-1,0,1\}^{|T_i|}\) when possible and assigns small values to the remaining coordinates of \(\mathbf{g}\).

\begin{algorithm}[!htp]
	\caption{Constructing \(\mathbf{g}\) from disjoint certified supports.}
	\label{Algorithm:Construct-g-disjoint-approach}
    \small
	\begin{algorithmic}[1]
        \Require A matrix \(\mathbf{H}'=\mathbf{P}\mathbf{H}\in\mathbb{Z}^{m\times n}\) with \(n>m\); a target starting size \(t_0\in[t_\mathrm{max}]\), where \(t_\mathrm{max}:=\mathrm{rank}_\mathbb{Q}(\mathbf{H}')+1\); and a fixed bound \(\mathit{maxTrials}\in\mathbb{Z}^+\) on the number of sampled subsets per size.
		\Ensure A vector \(\mathbf{g}\in\mathbb{Z}^n\) and a family \(\mathcal{T}\) of pairwise disjoint certified supports.
		\State Initialize \(\Omega:=[n]\), \(\mathbf{g}:=\mathbf{0}\), and \(\mathcal{T}:=\emptyset\).
		\While{\(|\Omega|\geq t_0\)}
			\State Set \(\mathit{found}:=\mathbf{False}\).
			\For{\(t=t_0\) \textbf{ to } \(\min\{t_\mathrm{max},|\Omega|\}\)}
				\If{\(t<t_\mathrm{max}\)}
					\State Sample and examine up to \(\mathit{maxTrials}\) subsets \(T\subseteq\Omega\) with \(|T|=t\), using the fixed sampling and tie-breaking conventions.
					\If{some \(T\) satisfies \(\ker_\mathbb{Z}(\mathbf{H}'_{(T)})\neq\{\mathbf{0}\}\)}\label{Line:Kernel-Test}
						\State Record \(T\), set \(\mathit{found}:=\mathbf{True}\), and \textbf{break}.
					\EndIf
				\Else
					\State Choose the lexicographically first \(T\subseteq\Omega\) with \(|T|=t\).
					\State Set \(\mathit{found}:=\mathbf{True}\) and \textbf{break}.
				\EndIf
			\EndFor
			\If{\textbf{not} \(\mathit{found}\)}
				\State \textbf{break}. \Comment{no further certified support is found}
			\EndIf
			\State Compute a short primitive vector \(\bm{\lambda}_T\in\ker_\mathbb{Z}(\mathbf{H}'_{(T)})\). \label{Line:Heuristic-Small-Primitive-Lambda}
			\State Compute a small solution \((\mathbf{g})_T\in\mathbb{Z}^{|T|}\) to \(\bm{\lambda}_T^\top(\mathbf{g})_T=1\) using the generalized B\'ezout identity.
			\State Write \(T=\{t_1,\dots,t_{|T|}\}\) with \(t_1<\cdots<t_{|T|}\), and set \(\mathbf{g}[t_a]=(\mathbf{g})_T[a]\) for every \(a\in[|T|]\).
			\State Set \(\mathcal{T}:=\mathcal{T}\cup\{T\}\) and \(\Omega:=\Omega\setminus T\).
		\EndWhile
		\State Assign each remaining coordinate \(\mathbf{g}[j]\), for \(j\in\Omega\), a fixed value in \(\{-1,0,1\}\), taking \(\mathbf{g}[j]=0\) by default.
		\State \Return \((\mathbf{g},\mathcal{T})\).
	\end{algorithmic}
\end{algorithm}

For every support \(T\in\mathcal{T}\), the algorithm produces \(\bm{\lambda}_T\in\ker_\mathbb{Z}(\mathbf{H}'_{(T)})\) and \((\mathbf{g})_T\) satisfying \(\bm{\lambda}_T^\top(\mathbf{g})_T=1\), so \(T\) is certified by \eqref{Eq:Correctness-criterion-T}.
Every selected support is removed from \(\Omega\) before the next iteration, so the supports are pairwise disjoint, and \Cref{Lemma:Deterministic-Correctness-Lower-Bound} guarantees reconstruction for every set of size at least \(k_c^\mathsf{det}=n-b+1\), where \(b:=|\mathcal{T}|\).

The branch \(t=t_\mathrm{max}\) always admits a nonzero integer kernel vector whenever \(|\Omega|\geq t_\mathrm{max}\).
Indeed, \(t_\mathrm{max}=\mathrm{rank}_\mathbb{Q}(\mathbf{H}')+1\), so any \(t_\mathrm{max}\) columns of \(\mathbf{H}'\) are linearly dependent over \(\mathbb{Q}\), and clearing denominators yields a nonzero integer kernel vector.
Thus, the deterministic baseline of \Cref{Lemma:DisjointBlocksRankLeftover} is recovered even when the searches for smaller supports are unsuccessful.

The choices that keep \(\mathbf{g}\) short do not affect certification, which depends only on the exact relations \(\bm{\lambda}_T\in\ker_\mathbb{Z}(\mathbf{H}'_{(T)})\) and \(\bm{\lambda}_T^\top(\mathbf{g})_T=1\); the resulting value of \(\|\mathbf{g}\|_\infty\) enters the individual-share bound of \Cref{Section:SEC-Share-Shortness}.

For fixed search parameters, the construction has polynomial bit complexity; the operation count is given in Appendix~\ref{Appendix:Deferred-Proofs}.
The existence of a certified support also ensures that the secret-bearing column is not contained in the integer column module generated by the masking columns.

\begin{lemma}\label{Lemma:g-not-in-ColZ-hatG}
	If \(\mathbf{e}_1^\top\in\mathrm{Row}_\mathbb{Z}(\mathbf{G}_T)\) for some \(T\subseteq[n]\), then \(\mathbf{g}\notin\mathrm{Col}_\mathbb{Z}(\mathbf{\hat{G}})\).
\end{lemma}

\begin{proof}
	Suppose, toward a contradiction, that \(\mathbf{g}\in\mathrm{Col}_\mathbb{Z}(\mathbf{\hat{G}})\).
	Then there exists \(\mathbf{a}\in\mathbb{Z}^{k-1}\) such that \(\mathbf{g}=\mathbf{\hat{G}}\mathbf{a}=(\mathbf{H}')^\top\mathbf{a}\).
	Let \(T\subseteq[n]\) satisfy \(\mathbf{e}_1^\top\in\mathrm{Row}_\mathbb{Z}(\mathbf{G}_T)\).
	By \eqref{Eq:Correctness-criterion-T}, there exists \(\bm{\lambda}_T\in\ker_\mathbb{Z}(\mathbf{H}'_{(T)})\) satisfying \(\bm{\lambda}_T^\top(\mathbf{g})_T=1\).
	Restricting \(\mathbf{g}=(\mathbf{H}')^\top\mathbf{a}\) to \(T\) gives \((\mathbf{g})_T=(\mathbf{H}'_{(T)})^\top\mathbf{a}\).
	Therefore, \(\bm{\lambda}_T^\top(\mathbf{g})_T=(\mathbf{H}'_{(T)}\bm{\lambda}_T)^\top\mathbf{a}=0\), contradicting \(\bm{\lambda}_T^\top(\mathbf{g})_T=1\).
	Hence \(\mathbf{g}\notin\mathrm{Col}_\mathbb{Z}(\mathbf{\hat{G}})\).
\end{proof}

Accordingly, whenever the construction produces at least one certified support, the resulting distribution matrix admits nontrivial secret reconstruction while retaining the private-set condition established in \Cref{Lemma:Column-Space-Ghat-U-Expander}.

\section{Correctness, Everywhere-Shortness, and Linearity}\label{Section:Correctness-Shortness-Linearity}

This section establishes the functional properties of \(\mathsf{SEC}\).
\Cref{Section:SEC-Correctness} shows that \(\mathsf{SEC}\) is perfectly correct in the sense of \Cref{Definition:SecretSharing-Correctness} and analyzes the reconstruction coverage measured by \(\epsilon_c\).
\Cref{Section:SEC-Everywhere-Shortness} shows that \(\mathsf{SEC}\) is everywhere-short in the sense of \Cref{Definition:Everywhere-Short-Secret-Sharing} by bounding the shares and the reconstruction coefficients separately; since each party holds a single share, the expansion factor is \(1\).
Finally, \Cref{Section:SEC-Linearity} proves the linearity of reconstruction and shows that public linear operations may be applied locally to the shares before reconstruction.

\subsection{Correctness}\label{Section:SEC-Correctness}

\subsubsection{Exact Reconstruction}\label{Section:SEC-Exact-Reconstruction}

The correctness analysis of \(\mathsf{SEC}\) separates exact reconstruction from reconstruction coverage: every authorized set reconstructs exactly, while the reconstruction error \(\epsilon_c\) measures only how often a sampled set fails the integer row-span criterion.

\begin{lemma}[Exact reconstruction for authorized sets]\label{Lemma:SEC-Exact-Reconstruction}
	Fix public parameters \(\mathsf{pp}\) generated by \(\mathsf{Setup}(1^\kappa)\).
	For every \(T\subseteq[n]\) satisfying \(|T|\geq k_c\) and \(\mathbf{e}_1^\top\in\mathrm{Row}_\mathbb{Z}(\mathbf{G}_T)\), every secret \(\mathbf{s}\in\mathbb{Z}^D\), and every \((\mathbf{s}_i)_{i\in[n]}\) in the support of \(\mathsf{Deal}_\mathsf{pp}(\mathbf{s})\), we have \(\mathsf{Rec}_\mathsf{pp}(T,(\mathbf{s}_i)_{i\in T})=\mathbf{s}\).
\end{lemma}

\begin{proof}
	Since \(|T|\geq k_c\), \(\mathsf{Rec}_\mathsf{pp}\) does not reject \(T\) based on its cardinality, and \Cref{Algorithm:Construct-Small-Lambda} returns a vector \(\bm{\lambda}_T\in\mathbb{Z}^{|T|}\) satisfying \(\bm{\lambda}_T^\top\mathbf{G}_T=\mathbf{e}_1^\top\).
	By \Cref{Lemma:Authorized-Reconstruction}, we have \((\bm{\lambda}_T^\top\mathbf{S}_T)^\top=\mathbf{s}\), and hence \(\mathsf{Rec}_\mathsf{pp}(T,(\mathbf{s}_i)_{i\in T})=\mathbf{s}\).
\end{proof}

For a fixed instantiation of \(\mathbf{G}\), the reconstruction threshold \(k_c\) is selected so that, for every \(t\geq k_c\), a uniformly sampled set \(T\subseteq[n]\) of size \(t\) satisfies \(\mathbf{e}_1^\top\in\mathrm{Row}_\mathbb{Z}(\mathbf{G}_T)\) except with probability at most \(\epsilon_c\), and we require \(\epsilon_c\leq 0.01\); for every certified set of size at least \(k_c\), reconstruction is exact by \Cref{Lemma:SEC-Exact-Reconstruction}.
The planted certified supports of \Cref{Algorithm:Construct-g-disjoint-approach} additionally guarantee reconstruction, with zero failure probability, for every set of size at least \(k_c^\mathsf{det}=n-b+1\) (\Cref{Lemma:Deterministic-Correctness-Lower-Bound}).
Accordingly, the correctness error of \Cref{Definition:SecretSharing-Correctness} is zero for authorized sets, while throughout the remainder of the paper \(\epsilon_c\) denotes the reconstruction-coverage error defined above.

Both \(k_c\) and \(k_c^\mathsf{det}\) depend on the instantiated generator matrix \(\mathbf{G}\).
We derive analytical lower bounds on the certification probability induced by the planted certified supports in \Cref{Section:SEC-Coverage}, and report the instantiated reconstruction thresholds in \Cref{Section:SEC-Reconstruction-Thresholds}.

\subsubsection{Reconstruction-Coverage Analysis}\label{Section:SEC-Coverage}

Reconstruction coverage measures the probability that a set sampled uniformly from all subsets of \([n]\) of a fixed size satisfies the integer row-span criterion.

Fix \(\mathbf{G}\) and pairwise disjoint certified supports \(T_1,\dots,T_b\subseteq[n]\), and let \(S\stackrel{\$}{\gets}\binom{[n]}{s}\).
Since every superset of a certified support is certified, \(\Pr[\mathbf{e}_1^\top\in\mathrm{Row}_\mathbb{Z}(\mathbf{G}_S)]\geq\Pr[\bigcup_{i=1}^b(T_i\subseteq S)]\).
The bound improves if each \(T_i\) is replaced by the support \(W_i:=\mathrm{supp}(\bm{\lambda}_{T_i})\subseteq T_i\) of its planted reconstruction vector \(\bm{\lambda}_{T_i}\in \ker_\mathbb{Z}(\mathbf{H}'_{(T_i)})\) satisfying \(\bm{\lambda}_{T_i}^\top(\mathbf{g})_{T_i}=1\). 
Indeed, restricting \(\bm{\lambda}_{T_i}\) to its support shows that \(W_i\) is itself certified, so \(W_i\subseteq S\) implies certification.
Here and below, \(\binom{a}{c}:=0\) whenever \(c<0\) or \(c>a\).

\begin{corollary}[Coverage from planted supports]\label{Corollary:Correctness-LB-T}
	Let \(t_i:=|T_i|\). For \(S\stackrel{\$}{\gets}\binom{[n]}{s}\),
	\begin{align}
		\Pr\left[\mathbf{e}_1^\top\in\mathrm{Row}_\mathbb{Z}(\mathbf{G}_S)\right] &\geq \frac{1}{\binom{n}{s}} \sum_{\emptyset\neq J\subseteq[b]} (-1)^{|J|+1} \binom{n-\sum_{i\in J}t_i}{s-\sum_{i\in J}t_i} \label{Eq:IE-Correctness-Hypergeometric}\\
		&\geq \sum_{i=1}^b \frac{\binom{n-t_i}{s-t_i}}{\binom{n}{s}} - \sum_{1\leq i<j\leq b} \frac{\binom{n-t_i-t_j}{s-t_i-t_j}}{\binom{n}{s}}. \label{Eq:IE-Correctness-Bonferroni}
	\end{align}
\end{corollary}

\begin{corollary}[Coverage from witness supports]\label{Corollary:Correctness-LB-Witness}
	The bounds \eqref{Eq:IE-Correctness-Hypergeometric} and \eqref{Eq:IE-Correctness-Bonferroni} remain valid with \(t_i\) replaced by \(w_i:=|W_i|\leq t_i\).
\end{corollary}

The right-hand side of \eqref{Eq:IE-Correctness-Hypergeometric} is the exact probability of containing at least one planted support, by inclusion-exclusion over disjoint sets, and \eqref{Eq:IE-Correctness-Bonferroni} is the second-order Bonferroni bound; the witness version and the heuristic design rules \(k_c\approx(\ln(1/\epsilon_c)/b)^{1/t}n\) and \(k_c \approx (\ln(1/\epsilon_c)/b)^{1/w}n\) derived from them are given in Appendix~\ref{Appendix:Deferred-Proofs}.
Containing a planted or witness support is sufficient but not necessary for certification, so the certification probability of an instantiated matrix may be substantially larger than these bounds.
The empirical evaluation in \Cref{Section:SEC-Reconstruction-Thresholds} therefore computes \(k_c\) exactly when feasible and otherwise estimates it from the observed certification rate.

\subsection{Everywhere-Shortness}\label{Section:SEC-Everywhere-Shortness}

\subsubsection{Share Shortness}\label{Section:SEC-Share-Shortness}

We first bound the individual shares produced by \(\mathsf{Deal}_\mathsf{pp}\).
By \eqref{Eq:SEC-Deal}, each share \(\mathbf{s}_i\) is a linear function of the secret \(\mathbf{s}\) and the masking matrix \(\mathbf{R}\):
\begin{align}
	\mathbf{s}_i^\top &= \mathbf{G}[i,\cdot] \begin{bmatrix} \mathbf{s}^\top\\ \mathbf{R} \end{bmatrix} = \begin{bmatrix} \mathbf{g}[i] & \mathbf{\hat{G}}[i,\cdot] \end{bmatrix} \begin{bmatrix} \mathbf{s}^\top\\ \mathbf{R} \end{bmatrix} \nonumber\\
	&= \mathbf{g}[i]\mathbf{s}^\top + \mathbf{\hat{G}}[i,\cdot]\mathbf{R}. \label{Eq:s_i-linear-function-s+R}
\end{align}
\begin{lemma}[Boundedness of the shares]\label{Lemma:bound-for-si}
	Let \(\beta_\mathbf{s},\beta_\mathbf{R}\in\mathbb{R}^+\), and suppose that \(\|\mathbf{s}\|_\infty\leq\beta_\mathbf{s}\) and \(\|\mathbf{R}\|_{\max}\leq\beta_\mathbf{R}\).
	Then, for every \(i\in[n]\), \(\|\mathbf{s}_i\|_\infty \leq |\mathbf{g}[i]|\beta_\mathbf{s} + \|\mathbf{\hat{G}}[i,\cdot]\|_1\beta_\mathbf{R}\).
	In particular, if \(\mathbf{\hat{G}}=(\mathbf{P}\mathbf{H})^\top\), where \(\mathbf{P}\) is a signed permutation matrix and \(\mathbf{H}\) is the bi-adjacency matrix of a \(c_\mathrm{exp}\)-left-regular bipartite lossless expander, then \(\|\mathbf{s}_i\|_\infty \leq |\mathbf{g}[i]|\beta_\mathbf{s} + c_\mathrm{exp}\beta_\mathbf{R}\).
	If, moreover, \(\|\mathbf{g}\|_\infty\leq\beta_\mathbf{g}\) for some \(\beta_\mathbf{g}\in\mathbb{R}^+\), then \(\|\mathbf{s}_i\|_\infty \leq \beta_\mathbf{g}\beta_\mathbf{s} + c_\mathrm{exp}\beta_\mathbf{R}\).
\end{lemma}

\begin{proof}
	Coordinate-wise, \eqref{Eq:s_i-linear-function-s+R} gives \(|\mathbf{s}_i[j]| \leq |\mathbf{g}[i]||\mathbf{s}[j]| + \sum_{t}|\mathbf{\hat{G}}[i,t]||\mathbf{R}[t,j]| \leq |\mathbf{g}[i]|\beta_\mathbf{s} + \|\mathbf{\hat{G}}[i,\cdot]\|_1 \beta_\mathbf{R}\).
	For \(\mathbf{\hat{G}}=(\mathbf{P}\mathbf{H})^\top\), \(\|\mathbf{\hat{G}}[i,\cdot]\|_1 = \|\mathbf{H}[\cdot,i]\|_1 = c_\mathrm{exp}\) by left-regularity, and \(|\mathbf{g}[i]|\leq\beta_\mathbf{g}\) gives the last claim.
\end{proof}

Under the designated secret distribution \(\mathcal{D}_S=\mathcal{D}_{\sigma_\mathbf{s}}^D\) and the entrywise discrete Gaussian sampling of \(\mathbf{R}\), we choose \(\beta_\mathbf{s}\) and \(\beta_\mathbf{R}\) so that the event \(\|\mathbf{s}\|_\infty\leq\beta_\mathbf{s}\) and \(\|\mathbf{R}\|_{\max}\leq\beta_\mathbf{R}\) holds except with probability at most \(\epsilon_\mathsf{sh}\).
Consequently, if \(\|\mathbf{g}\|_\infty\leq\beta_\mathbf{g}\), then \(\max_{i\in[n]}\|\mathbf{s}_i\|_\infty\leq\beta_\mathbf{g}\beta_\mathbf{s}+c_\mathrm{exp}\beta_\mathbf{R}\) except with probability at most \(\epsilon_\mathsf{sh}\).
Thus, \(\mathsf{SEC}\) satisfies the individual-share shortness requirement of \Cref{Definition:Everywhere-Short-Secret-Sharing} with share bound \(\beta_\mathsf{sh}:=\beta_\mathbf{g}\beta_\mathbf{s}+c_\mathrm{exp}\beta_\mathbf{R}\) and failure probability \(\epsilon_\mathsf{sh}\).
The tail parameters in \Cref{Section:SEC-Gaussian-Tail} choose \(\beta_\mathbf{s}\) and \(\beta_\mathbf{R}\) so that \(\epsilon_\mathsf{sh}\leq2\cdot2^{-\kappa}\).

\subsubsection{Reconstruction-Coefficient Bounds}\label{Section:Bounding-Reconstruction-Coefficients}

We next bound the reconstruction coefficients produced by \Cref{Algorithm:Construct-Small-Lambda}.
Fix an authorized set \(T\in\Gamma_1(\mathsf{pp})\), and write \(\mathbf{A}_T:=\mathbf{G}_T^\top\), a local abbreviation rather than a row restriction.
The reconstruction equation \(\bm{\lambda}^\top\mathbf{G}_T=\mathbf{e}_1^\top\) is equivalent to \(\mathbf{A}_T\bm{\lambda}=\mathbf{e}_1\).
We first decompose every real solution into the minimum-\(\ell_2\)-norm solution and a real-kernel component, and then bound the kernel component of the integer vector returned by \Cref{Algorithm:Construct-Small-Lambda}.

\paragraph{Affine Reconstruction Cosets}

\begin{lemma}[Affine coset of reconstruction vectors]\label{Lemma:Affine-Coset-Reconstruction-Vectors}
	Let \(T\in\Gamma_1(\mathsf{pp})\), let \(\mathbf{A}_T:=\mathbf{G}_T^\top\), and let \(\bm{\lambda}_{T,0}\in\mathbb{Z}^{|T|}\) satisfy \(\mathbf{A}_T\bm{\lambda}_{T,0}=\mathbf{e}_1\).
	Then the set of integer reconstruction vectors for \(T\) is the affine lattice coset \(\mathcal{L}_T:=\bm{\lambda}_{T,0}+K_T\), where \(K_T:=\ker_\mathbb{Z}(\mathbf{A}_T)\).
\end{lemma}

\begin{proof}
	For every \(\bm{\nu}\in K_T\), \(\mathbf{A}_T(\bm{\lambda}_{T,0}+\bm{\nu}) = \mathbf{e}_1\).
    Conversely, every \(\bm{\lambda}\in\mathbb{Z}^{|T|}\) with \(\mathbf{A}_T\bm{\lambda}=\mathbf{e}_1\) satisfies \(\mathbf{A}_T(\bm{\lambda}-\bm{\lambda}_{T,0}) = \mathbf{0}\), so \(\bm{\lambda}-\bm{\lambda}_{T,0}\in K_T\).
\end{proof}

Although the representative \(\bm{\lambda}_{T,0}\) obtained from a transformation matrix in the row-HNF computation need not be unique, the coset \(\mathcal{L}_T\) depends only on \(\mathbf{G}_T\).

\paragraph{Orthogonal Decomposition of Reconstruction Vectors}

We next identify the unavoidable real least-norm component of every reconstruction vector.
Let \(r_T:=\mathrm{rank}_\mathbb{R}(\mathbf{A}_T)=\mathrm{rank}_\mathbb{R}(\mathbf{G}_T)\), and let \(d_T:=\dim(\ker_\mathbb{R}(\mathbf{A}_T))=|T|-r_T\).
We write \(\sigma_{\min}^+(\mathbf{G}_T)\) for the smallest positive singular value of \(\mathbf{G}_T\).

\begin{lemma}[Minimum-norm real reconstruction vector]\label{Lemma:Real-Least-Norm-Reconstruction}
	Let \(T\in\Gamma_1(\mathsf{pp})\), let \(\mathbf{A}_T:=\mathbf{G}_T^\top\), and define \(\bm{\lambda}_{T,\mathbb{R}} := \mathbf{A}_T^+\mathbf{e}_1\), where \((\cdot)^+\) denotes the Moore-Penrose pseudoinverse.
	Then \(\bm{\lambda}_{T,\mathbb{R}}\) is the unique solution of \(\mathbf{A}_T\bm{\lambda}=\mathbf{e}_1\) having minimum Euclidean norm.
	Moreover, \(\|\bm{\lambda}_{T,\mathbb{R}}\|_2^2 = \mathbf{e}_1^\top (\mathbf{A}_T\mathbf{A}_T^\top)^+ \mathbf{e}_1 = \mathbf{e}_1^\top (\mathbf{G}_T^\top\mathbf{G}_T)^+ \mathbf{e}_1\), and \(\|\bm{\lambda}_{T,\mathbb{R}}\|_2 \leq \frac{1}{\sigma_{\min}^+(\mathbf{G}_T)}\).
\end{lemma}

\begin{proof}
	Since \(T\in\Gamma_1(\mathsf{pp})\), we have \(\mathbf{e}_1\in\mathrm{Col}_\mathbb{R}(\mathbf{A}_T)\), so the system \(\mathbf{A}_T\bm{\lambda}=\mathbf{e}_1\) is consistent.
	The Moore-Penrose pseudoinverse gives its unique minimum-Euclidean-norm solution, and hence \(\bm{\lambda}_{T,\mathbb{R}}=\mathbf{A}_T^+\mathbf{e}_1\).
	Using the pseudoinverse identity \((\mathbf{A}_T^+)^\top\mathbf{A}_T^+=(\mathbf{A}_T\mathbf{A}_T^\top)^+\), we obtain
	\begin{align*}
		\|\bm{\lambda}_{T,\mathbb{R}}\|_2^2
		= (\mathbf{A}_T^+\mathbf{e}_1)^\top (\mathbf{A}_T^+\mathbf{e}_1)
		= \mathbf{e}_1^\top (\mathbf{A}_T^+)^\top \mathbf{A}_T^+ \mathbf{e}_1
		= \mathbf{e}_1^\top (\mathbf{A}_T\mathbf{A}_T^\top)^+ \mathbf{e}_1
		= \mathbf{e}_1^\top (\mathbf{G}_T^\top\mathbf{G}_T)^+ \mathbf{e}_1.
	\end{align*}
	Furthermore, \( \|\bm{\lambda}_{T,\mathbb{R}}\|_2 \leq \|\mathbf{A}_T^+\|_2\|\mathbf{e}_1\|_2 = \frac{1}{\sigma_{\min}^+(\mathbf{A}_T)} = \frac{1}{\sigma_{\min}^+(\mathbf{G}_T)}\), where we used \(\|\mathbf{e}_1\|_2=1\) and the fact that a matrix and its transpose have the same positive singular values.
\end{proof}

The following decomposition separates the kernel component of a reconstruction vector from the unavoidable minimum-\(\ell_2\)-norm real component.

\begin{lemma}[Orthogonal decomposition of reconstruction vectors]\label{Lemma:Orthogonal-Decomposition-RecCoeff}
    Let \(T\in\Gamma_1(\mathsf{pp})\).
	For every \(\bm{\lambda}\in\mathcal{L}_T\), \(\|\bm{\lambda}\|_2^2 = \|\bm{\lambda}_{T,\mathbb{R}}\|_2^2 + \|\bm{\lambda}-\bm{\lambda}_{T,\mathbb{R}}\|_2^2\).
	In particular, fix any \(\bm{\lambda}_{T,0}\in\mathcal{L}_T\), define \(\mathbf{t}_T := \bm{\lambda}_{T,\mathbb{R}} - \bm{\lambda}_{T,0}\), and write \(\bm{\lambda}=\bm{\lambda}_{T,0}+\bm{\nu}\) for some \(\bm{\nu}\in K_T\).
	Then \(\|\bm{\lambda}_{T,0}+\bm{\nu}\|_2^2 = \|\bm{\lambda}_{T,\mathbb{R}}\|_2^2 + \|\bm{\nu}-\mathbf{t}_T\|_2^2\).
\end{lemma}

\begin{proof}
	The minimum-Euclidean-norm solution satisfies \(\bm{\lambda}_{T,\mathbb{R}} \in \mathrm{Row}_\mathbb{R}(\mathbf{A}_T) = \mathrm{Col}_\mathbb{R}(\mathbf{A}_T^\top) = (\ker_\mathbb{R}(\mathbf{A}_T))^\bot\).
	Since \(\bm{\lambda}\) and \(\bm{\lambda}_{T,\mathbb{R}}\) satisfy \(\mathbf{A}_T\bm{\lambda}=\mathbf{e}_1\) and \(\mathbf{A}_T\bm{\lambda}_{T,\mathbb{R}}=\mathbf{e}_1\), respectively, their difference lies in \(\ker_\mathbb{R}(\mathbf{A}_T)\).
	Hence \(\bm{\lambda}_{T,\mathbb{R}}\) and \(\bm{\lambda}-\bm{\lambda}_{T,\mathbb{R}}\) are orthogonal, and the first claim follows from the Pythagorean identity.
	For the second claim, observe that \(\bm{\lambda}-\bm{\lambda}_{T,\mathbb{R}} = \bm{\lambda}_{T,0}+\bm{\nu} - \bm{\lambda}_{T,\mathbb{R}} = \bm{\nu}-\mathbf{t}_T\), and substitute this equality into the first identity.
\end{proof}

Replacing \(\bm{\lambda}_{T,0}\) by another element of \(\mathcal{L}_T\) translates \(\mathbf{t}_T\) by an element of \(K_T\), so the lattice-decoding problem is independent of the HNF transformation matrix used to obtain the initial representative. 
By \Cref{Lemma:Orthogonal-Decomposition-RecCoeff}, minimizing \(\|\bm{\lambda}_{T,0}+\bm{\nu}\|_2^2\) over \(\bm{\nu}\in K_T\) is equivalent to the closest-vector problem \(\min_{\bm{\nu}\in K_T}\|\bm{\nu}-\mathbf{t}_T\|_2^2\), since the remaining term \(\|\bm{\lambda}_{T,\mathbb{R}}\|_2^2\) is fixed.

\paragraph{Basis-Dependent Babai Certificate}

We write \(B_{\bm{\lambda}}^*(T) := \min_{\bm{\lambda}\in\mathcal{L}_T}\|\bm{\lambda}\|_\infty\) for the optimal reconstruction-coefficient size and \(B_{\bm{\lambda}}(T):=\|\bm{\lambda}_T\|_\infty\) for the value achieved by \Cref{Algorithm:Construct-Small-Lambda}. 
Since \(\bm{\lambda}_T\in\mathcal{L}_T\), we have \(B_{\bm{\lambda}}^*(T)\leq B_{\bm{\lambda}}(T)\). 
Computing \(B_{\bm{\lambda}}^*(T)\) is an \(\ell_\infty\)-closest-vector problem in \(K_T\), which is hard in general. 
The algorithm instead applies Babai's nearest-plane procedure to an LLL-reduced basis of \(K_T\), followed by a greedy refinement that accepts only updates strictly decreasing the \(\ell_\infty\)-norm. 
The following lemma gives a Euclidean certificate for the resulting vector.

\begin{lemma}[Certified coefficient bound]\label{Lemma:Certified-Coefficient-Bound} 
	Let \(T\in\Gamma_1(\mathsf{pp})\), let \(d_T:=\dim(\ker_\mathbb{R}(\mathbf{G}_T^\top))\), and let \(\sigma_{\min}^+(\mathbf{G}_T)\) denote the smallest positive singular value of \(\mathbf{G}_T\). 
	Let \(\mathbf{B}_K'\in\mathbb{Z}^{d_T\times|T|}\), with rows \(\mathbf{b}_1,\dots,\mathbf{b}_{d_T}\), be the LLL-reduced basis of \(K_T:=\ker_\mathbb{Z}(\mathbf{G}_T^\top)\) used by \Cref{Algorithm:Construct-Small-Lambda}, and let \(\mathbf{b}_1^*,\dots,\mathbf{b}_{d_T}^*\) be its Gram-Schmidt vectors. 
	If \(\bm{\lambda}_T\) is the output of \Cref{Algorithm:Construct-Small-Lambda}, then 
	\begin{align}\label{Eq:Certified-Coefficient-Bound} 
		B_{\bm{\lambda}}(T) := \|\bm{\lambda}_T\|_\infty \leq \left( \frac{1}{ \left(\sigma_{\min}^+(\mathbf{G}_T)\right)^2 } + \frac{1}{4} \sum_{i=1}^{d_T}\|\mathbf{b}_i^*\|_2^2 \right)^{1/2}. 
	\end{align} 
	When \(d_T=0\), the sum is interpreted as zero. 
\end{lemma}

\begin{proof} 
	If \(d_T=0\), then \(K_T=\{\mathbf{0}\}\), so \(\mathcal{L}_T\) contains a unique vector.
    This vector equals \(\bm{\lambda}_{T,\mathbb{R}}\), since \(\ker_\mathbb{R}(\mathbf{A}_T)=\{\mathbf{0}\}\) makes the real solution unique, and hence \(\|\bm{\lambda}_T\|_\infty \leq \|\bm{\lambda}_{T,\mathbb{R}}\|_2 \leq \frac{1}{\sigma_{\min}^+(\mathbf{G}_T)}\) by \Cref{Lemma:Real-Least-Norm-Reconstruction}. 
    This is \eqref{Eq:Certified-Coefficient-Bound} with an empty Gram-Schmidt sum. 
	
	Suppose now that \(d_T>0\). 
    Let \(\bm{\lambda}_\mathrm{dec}\) be the output of Stage~1 of \Cref{Algorithm:Construct-small-lambda-Babai-greedy}, and let \(\bm{\nu}_\mathsf{Babai}:=\bm{\lambda}_\mathrm{dec}-\bm{\lambda}_{T,0}\in K_T\). 
	By \Cref{Lemma:Orthogonal-Decomposition-RecCoeff} and \(\|\cdot\|_\infty\leq\|\cdot\|_2\), 
	\begin{align}\label{Eq:decoding-certificate} 
		\|\bm{\lambda}_\mathrm{dec}\|_\infty \leq \|\bm{\lambda}_\mathrm{dec}\|_2 \leq \left( \|\bm{\lambda}_{T,\mathbb{R}}\|_2^2 + \|\bm{\nu}_\mathsf{Babai}-\mathbf{t}_T\|_2^2 \right)^{1/2}. 
	\end{align} 
    Since \(\mathbf{A}_T\mathbf{t}_T=\mathbf{A}_T\bm{\lambda}_{T,\mathbb{R}}-\mathbf{A}_T\bm{\lambda}_{T,0}=\mathbf{0}\), the target \(\mathbf{t}_T\) lies in \(\ker_\mathbb{R}(\mathbf{A}_T)\), which equals \(\mathrm{span}_\mathbb{R}(K_T)=\mathrm{span}_\mathbb{R}(\mathbf{b}_1^*,\dots,\mathbf{b}_{d_T}^*)\) because \(\mathbf{A}_T\) has integer entries.
    Stage~1 of \Cref{Algorithm:Construct-small-lambda-Babai-greedy} starts from \(\bm{\lambda}=\bm{\lambda}_{T,0}\) and processes \(i=d_T,\dots,1\) in turn, following Babai's nearest-plane procedure \cite{babai1986lovasz}.
	At step \(i\), it subtracts the integer multiple of \(\mathbf{b}_i\) that places the coefficient of the current vector \(\bm{\lambda}\) along \(\mathbf{b}_i^*\) in \([-1/2,1/2)\).
	Later steps subtract only multiples of \(\mathbf{b}_j\) with \(j<i\), which are orthogonal to \(\mathbf{b}_i^*\), so this coefficient is not changed afterward.
	Since \(\bm{\lambda}_{T,\mathbb{R}}\in(\ker_\mathbb{R}(\mathbf{A}_T))^\perp\) is orthogonal to every \(\mathbf{b}_i^*\), the coefficients of \(\bm{\lambda}_\mathrm{dec}-\bm{\lambda}_{T,\mathbb{R}}=\bm{\nu}_\mathsf{Babai}-\mathbf{t}_T\) along the vectors \(\mathbf{b}_i^*\) coincide with those of \(\bm{\lambda}_\mathrm{dec}\).
	This difference lies in \(\mathrm{span}_\mathbb{R}(\mathbf{b}_1^*,\dots,\mathbf{b}_{d_T}^*)\), so \(\bm{\nu}_\mathsf{Babai}-\mathbf{t}_T=\sum_{i=1}^{d_T}c_i'\mathbf{b}_i^*\) with \(c_i'\in[-1/2,1/2)\), and since the Gram-Schmidt vectors are mutually orthogonal,
    \begin{align}\label{Eq:Babai-radius} 
		\|\bm{\nu}_\mathsf{Babai}-\mathbf{t}_T\|_2 \leq R_\mathsf{Babai}(\mathbf{B}_K') := \frac{1}{2} \left( \sum_{i=1}^{d_T}\|\mathbf{b}_i^*\|_2^2 \right)^{1/2}. 
	\end{align}
	Combining \eqref{Eq:decoding-certificate} and \eqref{Eq:Babai-radius} gives \(\|\bm{\lambda}_\mathrm{dec}\|_\infty \leq \left( \|\bm{\lambda}_{T,\mathbb{R}}\|_2^2 + \frac{1}{4} \sum_{i=1}^{d_T}\|\mathbf{b}_i^*\|_2^2 \right)^{1/2}\). 
	Stage~2 of \Cref{Algorithm:Construct-small-lambda-Babai-greedy} accepts only updates that strictly decrease the \(\ell_\infty\)-norm, so \(\|\bm{\lambda}_T\|_\infty \leq\|\bm{\lambda}_\mathrm{dec}\|_\infty\). 
	Finally, \Cref{Lemma:Real-Least-Norm-Reconstruction} gives \(\|\bm{\lambda}_{T,\mathbb{R}}\|_2 \leq \frac{1}{\sigma_{\min}^+(\mathbf{G}_T)}\). 
	Substituting this bound proves \eqref{Eq:Certified-Coefficient-Bound}. 
\end{proof}

The certificate in \eqref{Eq:Certified-Coefficient-Bound} is computable from \(\mathbf{G}_T\) and the reduced kernel basis used by the reconstruction algorithm, and applies to the vector actually returned. 
Unlike the initial HNF representative \(\bm{\lambda}_{T,0}\), it is independent of the transformation matrix used in the row-HNF computation. 
The certificate may nevertheless be conservative, since it does not capture the full improvement from the greedy refinement.
The proof in fact bounds the Euclidean norm of the Stage-1 Babai vector \(\bm{\lambda}_\mathrm{dec}\); it is stated for the \(\ell_\infty\)-norm because the greedy refinement guarantees only \(\|\bm{\lambda}_T\|_\infty\leq\|\bm{\lambda}_\mathrm{dec}\|_\infty\) and may increase the Euclidean norm.
Heuristically, if the coordinates of \(\bm{\lambda}_\mathrm{dec}\) have comparable magnitudes, its \(\ell_\infty\)-norm is about \(\sqrt{|T|}\) times smaller than its Euclidean norm, so \eqref{Eq:Certified-Coefficient-Bound} may overestimate \(B_{\bm{\lambda}}(T)\) by roughly this factor.
Appendix~\ref{Appendix:Heuristic-Coefficients} relates the terms of the certificate to the sparsity of \(\mathbf{G}\) through the kernel-lattice determinant and gives heuristic estimates under the geometric-series assumption.

\begin{corollary}[Coefficient bound from the LLL root-Hermite factor]\label{Corollary:Certified-Coefficient-RHF}
	Let \(T\in\Gamma_1(\mathsf{pp})\) with \(d_T>0\), and let \(D_T:=\det(K_T)^{1/d_T}\).
	Let \(\delta_\mathsf{LLL}\in(1/4,1)\) be the Lov\'asz parameter used by \Cref{Algorithm:Construct-Small-Lambda}, and let \(\delta_\mathrm{RHF}^\mathrm{wc}:=(\delta_\mathsf{LLL}-1/4)^{-1/4}\) be the corresponding worst-case root-Hermite factor of LLL.
	If the Gram-Schmidt vectors in \Cref{Lemma:Certified-Coefficient-Bound} satisfy \(\|\mathbf{b}_i^*\|_2\leq\|\mathbf{b}_1^*\|_2\) for every \(i\in[d_T]\), then
	\begin{align}\label{Eq:Certified-Coefficient-RHF}
		B_{\bm{\lambda}}(T) \leq \left( \frac{1}{\left(\sigma_{\min}^+(\mathbf{G}_T)\right)^2} + \frac{d_T}{4}\,\big(\delta_\mathrm{RHF}^\mathrm{wc}\big)^{2(d_T-1)}D_T^2 \right)^{1/2}.
	\end{align}
\end{corollary}

\begin{proof}
    The Lov\'asz condition and size reduction give \(\|\mathbf{b}_{i+1}^*\|_2^2\geq(\delta_\mathsf{LLL}-1/4)\|\mathbf{b}_i^*\|_2^2\) for every \(i\in[d_T-1]\), and hence \(\|\mathbf{b}_1^*\|_2^2\leq(\delta_\mathsf{LLL}-1/4)^{-(i-1)}\|\mathbf{b}_i^*\|_2^2\) for every \(i\in[d_T]\).
	Multiplying these \(d_T\) inequalities and using \(\prod_{i=1}^{d_T}\|\mathbf{b}_i^*\|_2=\det(K_T)\) gives \(\|\mathbf{b}_1\|_2=\|\mathbf{b}_1^*\|_2\leq(\delta_\mathsf{LLL}-1/4)^{-(d_T-1)/4}\det(K_T)^{1/d_T}=\big(\delta_\mathrm{RHF}^\mathrm{wc}\big)^{d_T-1}D_T\) \cite{lenstra1982factoring}.
	By the hypothesis, \(\sum_{i=1}^{d_T}\|\mathbf{b}_i^*\|_2^2\leq d_T\|\mathbf{b}_1^*\|_2^2\leq d_T\big(\delta_\mathrm{RHF}^\mathrm{wc}\big)^{2(d_T-1)}D_T^2\).
	Substituting this bound into \eqref{Eq:Certified-Coefficient-Bound} proves the claim.
\end{proof}

Here the root-Hermite factor is normalized by \(\|\mathbf{b}_1\|_2=\delta^{d_T-1}\det(K_T)^{1/d_T}\), consistently with \eqref{Eq:GS-norm-profile}.
The hypothesis holds whenever the Gram-Schmidt norms are non-increasing, as in the geometric-series model of Appendix~\ref{Appendix:Heuristic-Coefficients}; LLL does not guarantee it, but it can be checked directly from \(\mathbf{B}_K'\).
For \(\delta_\mathsf{LLL}=0.99\), as used in \Cref{Section:SEC-Reconstruction-Coefficient-Evaluation}, \(\delta_\mathrm{RHF}^\mathrm{wc}\approx1.078\), whereas LLL typically achieves a root-Hermite factor of about \(1.02\) in practice \cite{nguyen2006lll,gama2008predicting}; substituting the latter for \(\delta_\mathrm{RHF}^\mathrm{wc}\) gives a heuristic estimate.
Bounds on \(D_T\) in terms of the sparsity of \(\mathbf{G}\) follow from \eqref{Eq:det-KT_bound} and \eqref{Eq:det-KT_bound-full-rank}.

Since \(\Gamma_1(\mathsf{pp})\) is finite, the maximum of the right-hand side of \eqref{Eq:Certified-Coefficient-Bound} over all \(T\in\Gamma_1(\mathsf{pp})\) is a valid choice of the reconstruction bound in \Cref{Item:Short-Linear-Recon} of \Cref{Definition:Everywhere-Short-Secret-Sharing}, so \(\mathsf{SEC}\) is everywhere-short with this bound.
Evaluating this maximum exactly requires the certificate for every authorized set, which is infeasible for the instances considered in \Cref{Section:SEC-Empirical-Parameter}.

For the concrete instantiations, we evaluate \(B_{\bm{\lambda}}(T)=\|\bm{\lambda}_T\|_\infty\) over the authorized sets examined during the parameter search. 
We choose \(\beta_{\bm{\lambda}}\) slightly above the largest observed value \(\hat{\beta}_{\bm{\lambda}}\), and similarly choose \(\omega_{\bm{\lambda}}\) above the largest observed weight \(\hat{\omega}_{\bm{\lambda}}\). 
Thus, \((\beta_{\bm{\lambda}},\omega_{\bm{\lambda}})\) are empirically selected parameters rather than worst-case consequences of \eqref{Eq:Certified-Coefficient-Bound} over all \(T\in\Gamma_1(\mathsf{pp})\). 
An application may impose both as additional admissibility checks; the selected margins then affect how many candidate sets are rejected, rather than the guarantees conditioned on satisfying the bounds. 
The sampling methodology, observed values, and selected margins are reported in \Cref{Section:SEC-Reconstruction-Coefficient-Evaluation}.

If \(\|\bm{\lambda}_T\|_\infty\leq\beta_{\bm{\lambda}}\) and \(wt_H(\bm{\lambda}_T)\leq\omega_{\bm{\lambda}}\), then every family \((\mathbf{x}_i)_{i\in T}\) with \(\|\mathbf{x}_i\|_\infty\leq\beta_\mathsf{sh}\), such as shares bounded by \Cref{Lemma:bound-for-si}, satisfies \(\|\sum_{i\in T}\lambda_{T,i}\mathbf{x}_i\|_\infty \leq \omega_{\bm{\lambda}}\beta_{\bm{\lambda}}\beta_\mathsf{sh}\leq|T|\beta_{\bm{\lambda}}\beta_\mathsf{sh}\).
This bounds every linear combination computed with a reconstruction vector satisfying the selected coefficient and weight bounds, as required by applications such as threshold signing.

\subsection{Linearity and Public-Map Compatibility}\label{Section:SEC-Linearity}

For fixed public parameters \(\mathsf{pp}\) and an authorized set \(T\in\Gamma_1(\mathsf{pp})\), let \(\bm{\lambda}_T\in\mathbb{Z}^{|T|}\) be the deterministic reconstruction vector produced by \Cref{Algorithm:Construct-Small-Lambda}. 
For vectors \((\mathbf{x}_i)_{i\in T}\) of any common dimension, let \(\mathbf{X}_T\) denote the matrix whose rows are their transposes, ordered according to the indices in \(T\). 
We use the same notation \(\mathsf{Rec}_\mathsf{pp}\) for the dimension-independent reconstruction rule \(\mathsf{Rec}_\mathsf{pp} \left(T,(\mathbf{x}_i)_{i\in T}\right) := (\bm{\lambda}_T^\top\mathbf{X}_T)^\top\). 
We record two consequences that make \(\mathsf{SEC}\) compatible with public linear computations on shares.

\begin{lemma}[Linearity of reconstruction]\label{Lemma:Linear-Combination-Rec}
	Fix public parameters \(\mathsf{pp}\) and an authorized set \(T\in\Gamma_1(\mathsf{pp})\).
	For every \(\alpha\in\mathbb{Z}\) and every \(\mathbf{x}_i,\mathbf{y}_i\in\mathbb{Z}^D\), \(i\in T\), we have \(\mathsf{Rec}_\mathsf{pp} \left( T, (\alpha\mathbf{x}_i+\mathbf{y}_i)_{i\in T} \right) = \alpha\, \mathsf{Rec}_\mathsf{pp} \left( T, (\mathbf{x}_i)_{i\in T} \right) + \mathsf{Rec}_\mathsf{pp} \left( T, (\mathbf{y}_i)_{i\in T} \right)\).
\end{lemma}

\begin{lemma}[Compatibility with public matrix multiplication]\label{Lemma:Matrix-Multiplication-Rec}
	Fix public parameters \(\mathsf{pp}\) and an authorized set \(T\in\Gamma_1(\mathsf{pp})\).
	Let \(\mathbf{M}\in\mathbb{Z}^{p\times D}\) and let \(\mathbf{x}_i\in\mathbb{Z}^D\) for every \(i\in T\).
	Then \(\mathbf{M}\, \mathsf{Rec}_\mathsf{pp} \left( T, (\mathbf{x}_i)_{i\in T} \right) = \mathsf{Rec}_\mathsf{pp} \left( T, (\mathbf{M}\mathbf{x}_i)_{i\in T} \right)\).
\end{lemma}

Both lemmas follow directly from \(\mathsf{Rec}_\mathsf{pp}(T,(\mathbf{x}_i)_{i\in T})=(\bm{\lambda}_T^\top\mathbf{X}_T)^\top\); the short computations are given in Appendix~\ref{Appendix:Deferred-Proofs}.
The first property permits local linear combinations of shares to be reconstructed as the corresponding linear combinations of their reconstructed values, and the second permits a public matrix to be applied locally to each share before reconstruction.

\section{Privacy of \(\mathsf{SEC}\)}\label{Section:Secret-Sharing-Privacy-Analysis}

The privacy analysis follows the five-step roadmap in \Cref{Table:Privacy-Roadmap}: we exclude exact cancellation attacks; express the correlated multi-query view as a stacked linear hint; bound its joint precision and perturbation parameters; derive uniform bounds and the required Gaussian approximation; and reduce the resulting HC-Leak experiment to Matrix Hint-MLWE and standard MLWE.
The notation is summarized in \Cref{Table:Privacy-Notation}, and lengthy proofs are deferred to Appendix~\ref{Appendix:Privacy-Proofs}.

\begin{table}[!htp]
	\centering
	\small
    \caption{Roadmap of the privacy analysis.}
	\label{Table:Privacy-Roadmap}
	\begin{tabular}{@{}l l l l@{}}
		\toprule
		Step & Section & Main statement & Output \\
		\midrule
		Structural condition & \ref{Section:Structural-Obstructions} & \Cref{Proposition:No-Noise-Cancellation-Summary} & no exact cancellation for \(|U|\leq k_p\) \\
		Stacked representation & \ref{Section:Multi-Query-HC-Leak-Representation} & \eqref{Eq:Multi-Query-Stacked-Matrix-Hint}, \eqref{Eq:Joint-Perturbation-Formula} & \(\mathbf{z}=\mathbf{M}\mathbf{s}+\mathbf{\Phi}\mathbf{r}\) \\
		Joint precision & \ref{Section:Bounding-Joint-Precision-Term} & \Cref{Lemma:Final-Joint-Bound} & pointwise precision bound \\
		Uniform parameters & \ref{Section:Uniform-Parameters-Multi-Query-Privacy} & \Cref{Lemma:Uniform-MQ-Precision,Lemma:Uniform-Gaussian-Joint-Perturbation} & \(\Xi_\mathrm{MQ}\), \(\sigma_\mathrm{MQ}\), loss \(2\epsilon'\) \\
		Reduction & \ref{Section:Multi-Query-HCLLR-Privacy} & \Cref{Theorem:SEC-HCLLR-multi} & HC-LLR from \(\mathsf{MLWE}\) \\
		\botrule
	\end{tabular}
\end{table}

\begin{table}[!htp]
	\centering
	\small
	\caption{Notation for the privacy analysis.}
	\label{Table:Privacy-Notation}
	\begin{tabular}{@{}l >{\raggedright\arraybackslash}p{0.31\linewidth} @{\hspace{1.5em}}l >{\raggedright\arraybackslash}p{0.31\linewidth}@{}}
		\toprule
		\text{Symbol} & Meaning & \text{Symbol} & Meaning \\
		\midrule
		\(D=d\ell\) & coefficient dimension of a secret or share & \(Q\) & number of leakage queries \\
		\(U_\mathrm{raw}\) & parties whose raw shares are exposed & \(V_s\) & parties whose responses leak in query \(s\) \\
		\(c_s,\ \mathbf{L}_s\) & leakage description and its local map & \(\omega_c\) & uniform bound on \(\|\mathbf{L}_s\|_2\) \\
		\(\mathcal{S}\) & leakage-block indices \((s,i)\), \(i\in V_s\) & \(\mathcal{S}_\mathrm{in},\mathcal{S}_\mathrm{out}\) & blocks with \(i\in U_\mathrm{raw}\), blocks with \(i\notin U_\mathrm{raw}\) \\
		\(\mathbf{r}_\mathbf{R}\) & \(\mathrm{vec}(\mathbf{R}^\top)\), sharing mask, width \(\sigma_\mathbf{R}\) & \(\mathbf{r}_{i,s}\) & fresh response mask, width \(\sigma_\mathbf{r}\) \\
		\(\mathbf{\hat{G}}_\mathrm{raw}\) & \(\mathbf{\hat{G}}_{U_\mathrm{raw}}\otimes\mathbf{I}_D\) & \(\mathbf{T}_\mathcal{S}\) & \(\left(\mathbf{\hat{G}}[i,\cdot]\otimes\mathbf{L}_s\right)_{(s,i) \in \mathcal{S}}\) (vertical stack of \(\mathbf{\hat{G}}[i,\cdot]\otimes \mathbf{L}_s\) for \((s,i) \in \mathcal{S}\)) \\
		\(\mathbf{M}_{\mathcal{U},\mathbf{c}},\ \mathbf{y}_{\mathcal{U},\mathbf{c}}\) & stacked hint matrix and perturbation & \(\mathbf{\Sigma}_{\mathbf{y}_{\mathcal{U},\mathbf{c}}}\) & target parameter matrix of \(\mathbf{y}_{\mathcal{U},\mathbf{c}}\) \\
		\(\mathbf{W}\) & public elimination transformation & \(\mathbf{\Sigma}_\mathrm{cor},\mathbf{B},\mathbf{C}\) & correlated block, transfer matrix, Schur complement \\
		\(\alpha_{k_p}\) & \(\min_{|V|=k_p}\sigma_{\min}(\mathbf{\hat{G}}_V)\) & \(\beta_{k_p}\) & \(\max_{|U|\leq k_p}\|\mathbf{\hat{G}}_U\|_2\) \\
		\(\gamma_{k_p}\) & \(\min_{|U|\leq k_p}\sigma_{\min}\) of peeled submatrix & \(\gamma_{k_p}^{\ker}\) & uniform kernel Gram-Schmidt bound \\
		\(\sigma_\mathrm{eff}\) & effective perturbation width & \(\Xi_\mathrm{MQ},\ \sigma_\mathrm{MQ}\) & uniform precision and MLWE source width \\
		\botrule
	\end{tabular}
\end{table}

\subsection{Structural Setup Condition}\label{Section:Structural-Obstructions}

Fix a private coalition \(U\in\Gamma_0\), whose share matrix is \(\mathbf{S}_U=(\mathbf{g})_U\mathbf{s}^\top+ \mathbf{\hat{G}}_U\mathbf{R}\). 
Before reducing privacy to \(\mathsf{MLWE}\), we must exclude exact linear combinations that annihilate the masking term while retaining a nonzero function of the secret. 
Because the shares are integral, an adversary may also reduce them modulo any \(q'\geq2\) and attempt cancellation over \(\mathbb{Z}_{q'}\). 
By \Cref{Lemma:ComputationalPrivacy-SEC-Column-Z} in Appendix~\ref{Appendix:SEC-Structural-Obstructions}, if \((\mathbf{g})_U\notin \mathrm{Col}_\mathbb{Z}(\mathbf{\hat{G}}_U)\), then some modular linear combination cancels the masking term and reveals a noiseless nonzero linear function of \(\mathbf{s}\bmod q'\). 
We therefore impose the structural setup condition \((\mathbf{g})_U\in \mathrm{Col}_\mathbb{Z}(\mathbf{\hat{G}}_U)\) for every \(U\in\Gamma_0\).

\begin{proposition}[Exclusion of structural noise cancellation]\label{Proposition:No-Noise-Cancellation-Summary}
	Fix public parameters \(\mathsf{pp}\) and a private coalition \(U\in\Gamma_0\).
	If \((\mathbf{g})_U\in\mathrm{Col}_\mathbb{Z}(\mathbf{\hat{G}}_U)\), then:
	\begin{enumerate*}[label=(\arabic*)]
		\item \((\mathbf{g})_U\in\mathrm{Col}_\mathbb{Q}(\mathbf{\hat{G}}_U)\), and hence rational noise cancellation is impossible;
		\item \(\mathrm{Col}_\mathbb{Z}(\mathbf{G}_U)=\mathrm{Col}_\mathbb{Z}(\mathbf{\hat{G}}_U)\) and \(\ker_\mathbb{Z}(\mathbf{G}_U^\top)=\ker_\mathbb{Z}(\mathbf{\hat{G}}_U^\top)\), and hence integer noise cancellation is impossible;
		\item for every \(q'\geq2\), \(\overline{(\mathbf{g})_U}\in\mathrm{Col}_{\mathbb{Z}_{q'}}(\overline{\mathbf{\hat{G}}_U})\), and hence modular noise cancellation is impossible.
	\end{enumerate*}
\end{proposition}

Writing \((\mathbf{g})_U=\mathbf{\hat{G}}_U\mathbf{a}\) for some integral \(\mathbf{a}\) proves all three claims; the details are given in Appendix~\ref{Appendix:SEC-Structural-Obstructions}. 
The rational condition also has a coding-theoretic interpretation through the dual distance of the masking code.

\begin{lemma}\label{Lemma:DualDistance-NoiseCancellation-Q}
	Let \(\hat{C}_\mathbb{Q} := \mathrm{Col}_\mathbb{Q}(\mathbf{\hat{G}}) \subseteq \mathbb{Q}^n\), so that \(\hat{C}_\mathbb{Q}^\bot = \ker_\mathbb{Q}(\mathbf{\hat{G}}^\top)\), and let \(\hat{d}_\mathbb{Q}^\bot\) be the minimum Hamming distance of \(\hat{C}_\mathbb{Q}^\bot\).
	Then, for every \(U\subseteq[n]\) satisfying \(|U| \leq \hat{d}_\mathbb{Q}^\bot-1\), we have \(\ker_\mathbb{Q} (\mathbf{\hat{G}}_U^\top) = \{\mathbf{0}\}\).
	Consequently, rational noise cancellation is impossible for every such \(U\).
\end{lemma}

\begin{proof}
	A nonzero \(\bm{\alpha}_U \in \ker_\mathbb{Q}(\mathbf{\hat{G}}_U^\top)\), extended by zero outside \(U\), would be a nonzero codeword of \(\hat{C}_\mathbb{Q}^\bot\) of weight at most \(|U| \leq \hat{d}_\mathbb{Q}^\bot-1\), a contradiction.
	The final claim follows from \Cref{Lemma:Noise-Cancellation-SEC}.
\end{proof}

When \(\mathbf{H}\) is a lifted finite-field parity-check matrix, \Cref{Lemma:Code-Distance-Preservation} transfers the finite-field distance bound to the rational kernel code, which is unchanged by multiplication with the signed permutation \(\mathbf{P}\). 
Dual distance, however, neither implies the integral condition nor excludes cancellation modulo an adversarially chosen modulus. 
The lossless-expander construction gives the stronger guarantee \(\mathrm{Col}_\mathbb{Z}(\mathbf{\hat{G}}_U) = \mathbb{Z}^{|U|}\) for every \(|U|\leq k_p\) by \Cref{Lemma:Column-Space-Ghat-U-Expander}, so the structural condition holds for every private coalition independently of \(\mathbf{g}\).

\subsection{Multi-Query HC-Leak Representation}\label{Section:Multi-Query-HC-Leak-Representation}

Fix \(d,\ell\in\mathbb{Z}^+\) and write \(D:=d\ell\).
Let \(U_\mathrm{raw}\subseteq[n]\) with \(|U_\mathrm{raw}|\leq k_p\) be the set of parties whose raw shares are exposed.
The raw-share view is
\begin{align}\label{Eq:S_U-full-SEC}
	\mathbf{S}_{U_\mathrm{raw}} = (\mathbf{g})_{U_\mathrm{raw}}\mathbf{s}^\top + \mathbf{\hat{G}}_{U_\mathrm{raw}}\mathbf{R},
\end{align}
where \(\mathbf{s}\in\mathbb{Z}^D\) is the coefficient embedding of the shared secret and \(\mathbf{R}\in\mathbb{Z}^{(k-1)\times D}\) is the sharing mask.
By \Cref{Section:Structural-Obstructions}, \(\mathbf{S}_{U_\mathrm{raw}} = \mathbf{\hat{G}}_{U_\mathrm{raw}} ( \mathbf{R} + \mathbf{a}\mathbf{s}^\top )\) for some \(\mathbf{a}\in\mathbb{Z}^{k-1}\), so the secret enters the raw-share view only through a shift of the common mask.

The adversary also observes noisy linear local leakage across at most \(Q\) queries. 
This model is motivated by lattice-based threshold signatures, in which party \(i\) answers query \(s\) with a partial response \(\bm{z}_{i,s}:=c_s\bm{s}_i+\bm{r}_{i,s}\in\mathcal{R}^\ell\), where \(c_s\) is the challenge and \(\bm{r}_{i,s}\) is a fresh mask. %(see \Cref{Section:Application}). 
Each query \(s\in[Q]\) is therefore specified by a public \(c_s\in\mathcal{C}\), where \(\mathcal{C}\subseteq\mathcal{R}\) is a fixed set of bounded-norm ring elements. 
Under the coefficient embedding, multiplication by \(c_s\) on \(\mathcal{R}^\ell\) is represented by \(\mathbf{L}_s := \mathbf{I}_\ell\otimes\mathrm{Rot}(c_s) \in\mathbb{Z}^{D\times D}\). 
We assume \(\|\mathbf{L}_s\|_2\leq\omega_c\) for every \(c_s\in\mathcal{C}\). 
For example, if every element of \(\mathcal{C}\) has exactly \(\omega_c\) nonzero coefficients in \(\{-1,1\}\), then \(\|\mathbf{L}_s\|_2\leq\|c_s\|_1=\omega_c\).

Let \(V_s\subseteq[n]\) index the response blocks exposed in query \(s\). 
For every \(i\in V_s\), the exposed coefficient-embedded block is 
\begin{equation}\label{Eq:Multi-Query-Noisy-Linear-Leakage} 
	\mathbf{z}_{i,s} := \mathbf{L}_s\mathbf{s}_i+\mathbf{r}_{i,s}, 
\end{equation} where \(\mathbf{r}_{i,s}\gets \mathcal{D}_{\mathbb{Z}^D,\sigma_\mathbf{r}}\) are mutually independent and independent of \(\mathbf{R}\) and \(\mathbf{s}\). 
We write \(\mathcal{U} := \bigl(U_\mathrm{raw},(V_s)_{s\in[Q]}\bigr)\) for the multi-query leakage pattern and \(\mathcal{S} := \{(s,i):s\in[Q],\,i\in V_s\}\) for its leakage-block index set.

\subsubsection{Fixed-Pattern HC-Leak View}\label{Section:Fixed-Pattern-View}

Fix an admissible pattern \(\mathcal{U}\), a public leakage-description sequence \(\mathbf{c}:=(c_s)_{s\in[Q]}\), and a public ordering of \(\mathcal{S}\).
Let \(\bm{A}'\stackrel{\$}{\gets}\mathcal{R}_q^{p\times(\ell-p)}\), set \(\bm{A} := \begin{bmatrix} \bm{I}_p \ \bm{A}' \end{bmatrix}\), and define the hard-core map \(f_{\bm{A}}(\bm{s}) := \bm{A}\bm{s} \bmod q \in \mathcal{R}_q^p\).
We encode the auxiliary view \(\mathsf{Leak}_{\mathcal{U},\mathbf{c}} := ( \mathbf{S}_{U_\mathrm{raw}}, (\mathbf{z}_{i,s})_{(s,i)\in\mathcal{S}} )\) by
\begin{equation}\label{Eq:z-Uc-Structured-Encoding}
	\mathbf{z}_{\mathcal{U},\mathbf{c}} := \begin{bmatrix} \mathrm{vec}\left(\mathbf{S}_{U_\mathrm{raw}}^\top\right)\\ (\mathbf{z}_{i,s})_{(s,i)\in\mathcal{S}} \end{bmatrix} \in \mathbb{Z}^{M'},
\end{equation}
where \(M':=(|U_\mathrm{raw}|+|\mathcal{S}|)D\).
In the fixed-pattern \(\mathsf{HC\text{-}Leak}\) experiment, the real branch supplies \(f_{\bm{A}}(\bm{s})\) together with \(\mathsf{Leak}_{\mathcal{U},\mathbf{c}}\), whereas the random branch replaces \(f_{\bm{A}}(\bm{s})\) by an independently uniform element of \(\mathcal{R}_q^p\); the auxiliary leakage has the same marginal distribution in both branches.

\subsubsection{Stacked Matrix-Hint Representation}\label{Section:Stacked-Matrix-Hint}

Write \(\mathbf{r}_\mathbf{R} := \mathrm{vec}(\mathbf{R}^\top) \in \mathbb{Z}^{(k-1)D}\) and \(\mathbf{\hat{g}}_i^\top:=\mathbf{\hat{G}}[i,\cdot]\) for \(i\in[n]\), and define
\begin{align}
	\mathbf{\hat{G}}_\mathrm{raw} &:= \mathbf{\hat{G}}_{U_\mathrm{raw}} \otimes \mathbf{I}_D \in \mathbb{Z}^{|U_\mathrm{raw}|D\times(k-1)D}, \label{Eq:Ghat-raw-definition}\\
	\mathbf{T}_\mathcal{S} &:= \left( \mathbf{\hat{g}}_i^\top \otimes \mathbf{L}_s \right)_{(s,i)\in\mathcal{S}} \in \mathbb{Z}^{|\mathcal{S}|D\times(k-1)D}, \label{Eq:T-S-definition}
\end{align}
where the block rows of \(\mathbf{T}_\mathcal{S}\) are stacked in the fixed ordering of \(\mathcal{S}\).
Applying \(\mathrm{vec}(\mathbf{X}\mathbf{Y}\mathbf{Z})=(\mathbf{Z}^\top\otimes\mathbf{X})\mathrm{vec}(\mathbf{Y})\) to \eqref{Eq:S_U-full-SEC} and to \(\mathbf{s}_i=\mathbf{g}[i]\mathbf{s}+\mathbf{R}^\top\mathbf{\hat{g}}_i\) gives
\begin{align}
	\mathrm{vec} \left( \mathbf{S}_{U_\mathrm{raw}}^\top \right) &= \mathbf{M}_\mathrm{raw}\mathbf{s} + \mathbf{y}_\mathrm{raw}, \label{Eq:Multi-Query-Raw-Hint}\\
	\mathbf{z}_{i,s} &= \mathbf{M}_{i,s}\mathbf{s} + \mathbf{y}_{i,s}, \qquad (s,i)\in\mathcal{S}, \label{Eq:Multi-Query-Individual-Hint}
\end{align}
where \(\mathbf{M}_\mathrm{raw} := (\mathbf{g})_{U_\mathrm{raw}} \otimes \mathbf{I}_D\), \(\mathbf{y}_\mathrm{raw} := \mathbf{\hat{G}}_\mathrm{raw}\mathbf{r}_\mathbf{R}\), \(\mathbf{M}_{i,s} := \mathbf{g}[i]\mathbf{L}_s\), and \(\mathbf{y}_{i,s} := ( \mathbf{\hat{g}}_i^\top \otimes \mathbf{L}_s ) \mathbf{r}_\mathbf{R} + \mathbf{r}_{i,s}\).
Stacking these blocks in the fixed ordering, define
\begin{equation}\label{Eq:z_Uc-y_Uc-M_Uc}
	\mathbf{M}_{\mathcal{U},\mathbf{c}} := \begin{bmatrix} \mathbf{M}_\mathrm{raw}\\ (\mathbf{M}_{i,s})_{(s,i)\in\mathcal{S}} \end{bmatrix} \in \mathbb{Z}^{M'\times D},
	\qquad
	\mathbf{y}_{\mathcal{U},\mathbf{c}} := \begin{bmatrix} \mathbf{y}_\mathrm{raw}\\ (\mathbf{y}_{i,s})_{(s,i)\in\mathcal{S}} \end{bmatrix} \in \mathbb{Z}^{M'}.
\end{equation}
The complete auxiliary view is then the single stacked linear hint
\begin{equation}\label{Eq:Multi-Query-Stacked-Matrix-Hint}
	\mathbf{z}_{\mathcal{U},\mathbf{c}} = \mathbf{M}_{\mathcal{U},\mathbf{c}}\mathbf{s} + \mathbf{y}_{\mathcal{U},\mathbf{c}}.
\end{equation}
Its perturbation is one integer linear image of the independent masks: with \(\mathbf{r}_\mathcal{S}:=(\mathbf{r}_{i,s})_{(s,i)\in\mathcal{S}}\),
\begin{equation}\label{Eq:Joint-Perturbation-Formula}
	\mathbf{y}_{\mathcal{U},\mathbf{c}} = \mathbf{\Phi}_{\mathcal{U},\mathbf{c}} \begin{bmatrix} \mathbf{r}_\mathbf{R}\\ \mathbf{r}_\mathcal{S} \end{bmatrix},
	\qquad
	\mathbf{\Phi}_{\mathcal{U},\mathbf{c}} := \begin{bmatrix} \mathbf{\hat{G}}_\mathrm{raw} & \mathbf{O}\\ \mathbf{T}_\mathcal{S} & \mathbf{I}_{|\mathcal{S}|D} \end{bmatrix}.
\end{equation}
Accordingly, we define the target parameter matrix of the joint perturbation by
\begin{align}
	\mathbf{\Sigma}_{\mathbf{y}_{\mathcal{U},\mathbf{c}}} &:= \mathbf{\Phi}_{\mathcal{U},\mathbf{c}} \mathbf{\Sigma}_0 \mathbf{\Phi}_{\mathcal{U},\mathbf{c}}^\top, \label{Eq:Sigma-y-definition}\\
	\mathbf{\Sigma}_0 &:= \mathrm{diag} \left( \sigma_\mathbf{R}^2\mathbf{I}_{(k-1)D}, \sigma_\mathbf{r}^2\mathbf{I}_{|\mathcal{S}|D} \right), \label{Eq:Sigma-0-Pushforward}
\end{align}
where \(\mathbf{\Sigma}_0\) is the parameter matrix of the independent masks \((\mathbf{r}_\mathbf{R},\mathbf{r}_\mathcal{S})\).
Because the blocks of \(\mathbf{y}_{\mathcal{U},\mathbf{c}}\) share the common mask \(\mathbf{r}_\mathbf{R}\), the matrix \(\mathbf{\Sigma}_{\mathbf{y}_{\mathcal{U},\mathbf{c}}}\) need not be block diagonal, and the view must be treated as one Matrix-Hint instance rather than as independent per-block hints.
\Cref{Section:Bounding-Joint-Precision-Term} bounds the resulting precision term, and \Cref{Section:Uniform-Parameters-Multi-Query-Privacy} shows that \(\mathbf{y}_{\mathcal{U},\mathbf{c}}\) is statistically close to \(\mathcal{D}_{\mathbb{Z}^{M'},\sqrt{\mathbf{\Sigma}_{\mathbf{y}_{\mathcal{U},\mathbf{c}}}}}\).

\subsection{Bounding the Joint Precision Term}\label{Section:Bounding-Joint-Precision-Term}

To apply \Cref{Theorem:Hardness-MHMLWE-Tight}, we bound \( \| \mathbf{M}_{\mathcal{U},\mathbf{c}}^\top \mathbf{\Sigma}_{\mathbf{y}_{\mathcal{U},\mathbf{c}}}^{-1} \mathbf{M}_{\mathcal{U},\mathbf{c}} \|_2\).
The response blocks are divided according to whether the same party's raw share is exposed:
\begin{align}\label{Eq:S_in-S_out}
	\mathcal{S}_\mathrm{in} := \{(s,i)\in\mathcal{S}:i\in U_\mathrm{raw}\}, \qquad \mathcal{S}_\mathrm{out} := \{(s,i)\in\mathcal{S}:i\notin U_\mathrm{raw}\}.
\end{align}
For \((s,i)\in\mathcal{S}_\mathrm{in}\), the adversary can compute \(\mathbf{z}_{i,s}-\mathbf{L}_s\mathbf{s}_i = \mathbf{r}_{i,s}\), so a public transformation isolates these blocks as secret-independent fresh masks.
After this elimination, a Schur-complement decomposition separates the raw-share precision from the conditional precision of the unpaired blocks, and bounding both components gives the pointwise estimate of \Cref{Lemma:Final-Joint-Bound}.

\subsubsection{Transformation and Block Structure}\label{Section:Joint-Transformation}

We reorder the block rows of \(\mathbf{M}_{\mathcal{U},\mathbf{c}}\) and \(\mathbf{y}_{\mathcal{U},\mathbf{c}}\) as raw-share blocks, then \(\mathcal{S}_\mathrm{in}\), then \(\mathcal{S}_\mathrm{out}\), and write the corresponding blocks as \(\mathbf{M}_\mathrm{raw},\mathbf{M}_\mathrm{in},\mathbf{M}_\mathrm{out}\) and \(\mathbf{y}_\mathrm{raw},\mathbf{y}_\mathrm{in},\mathbf{y}_\mathrm{out}\).
A common permutation of the hint and perturbation preserves the precision matrix, so we keep the notation; empty blocks are omitted throughout.
Let \(\mathbf{L}_{\mathcal{S}_\mathrm{in}}\) be the block-sparse matrix whose block in row \((s,i)\in\mathcal{S}_\mathrm{in}\) and column \(i\in U_\mathrm{raw}\) equals \(\mathbf{L}_s\).
Then
\begin{align}
	\mathbf{L}_{\mathcal{S}_\mathrm{in}} \mathbf{M}_\mathrm{raw} &= \mathbf{M}_\mathrm{in},\label{Eq:L-Sin-Mraw-Min}\\
	\mathbf{y}_\mathrm{in} - \mathbf{L}_{\mathcal{S}_\mathrm{in}} \mathbf{y}_\mathrm{raw} &= (\mathbf{r}_{i,s})_{(s,i)\in\mathcal{S}_\mathrm{in}} =: \mathbf{r}_\mathrm{in}.\label{Eq:L-Sin-yraw-yin}
\end{align}
Define the public transformation
\begin{equation}\label{Eq:W-public-transformation}
	\mathbf{W} := \begin{bmatrix} \mathbf{I} & \mathbf{O} & \mathbf{O} \\ -\mathbf{L}_{\mathcal{S}_\mathrm{in}} & \mathbf{I} & \mathbf{O} \\ \mathbf{O} & \mathbf{O} & \mathbf{I} \end{bmatrix} \in GL_{M'}(\mathbb{Z}),
\end{equation}
which is unimodular because it is block lower triangular with identity diagonal blocks.
Writing \(\mathbf{\widetilde{M}}_{\mathcal{U},\mathbf{c}} := \mathbf{W}\mathbf{M}_{\mathcal{U},\mathbf{c}}\) and \(\mathbf{\widetilde{y}}_{\mathcal{U},\mathbf{c}} := \mathbf{W}\mathbf{y}_{\mathcal{U},\mathbf{c}}\), \eqref{Eq:L-Sin-Mraw-Min} and \eqref{Eq:L-Sin-yraw-yin} yield the zero block:
\begin{align}\label{Eq:Transformed-M-and-y}
	\mathbf{\widetilde{M}}_{\mathcal{U},\mathbf{c}} &= \begin{bmatrix} \mathbf{M}_\mathrm{raw}\\ \mathbf{O}\\ \mathbf{M}_\mathrm{out} \end{bmatrix}, & \mathbf{\widetilde{y}}_{\mathcal{U},\mathbf{c}} &= \begin{bmatrix} \mathbf{y}_\mathrm{raw}\\ \mathbf{r}_\mathrm{in}\\ \mathbf{y}_\mathrm{out} \end{bmatrix}.
\end{align}
Since \(\mathbf{\Sigma}_{\mathbf{\widetilde{y}}_{\mathcal{U},\mathbf{c}}} = \mathbf{W} \mathbf{\Sigma}_{\mathbf{y}_{\mathcal{U},\mathbf{c}}} \mathbf{W}^\top\) and \(\mathbf{W}\) is invertible,
\begin{align}\label{Eq:Precision-Invariance-under-W}
	\mathbf{M}_{\mathcal{U},\mathbf{c}}^\top \mathbf{\Sigma}_{\mathbf{y}_{\mathcal{U},\mathbf{c}}}^{-1} \mathbf{M}_{\mathcal{U},\mathbf{c}} = \mathbf{\widetilde{M}}_{\mathcal{U},\mathbf{c}}^\top \mathbf{\Sigma}_{\mathbf{\widetilde{y}}_{\mathcal{U},\mathbf{c}}}^{-1} \mathbf{\widetilde{M}}_{\mathcal{U},\mathbf{c}}.
\end{align}
Let \(\mathbf{\Sigma}_\mathrm{raw}\), \(\mathbf{\Sigma}_\mathrm{out}\), and \(\mathbf{\Sigma}_{\mathrm{raw},\mathrm{out}}\) be the parameter blocks of \(\mathbf{y}_\mathrm{raw}\), of \(\mathbf{y}_\mathrm{out}\), and between them, and define
\begin{equation}\label{Eq:Sigma-cor-definition}
	\mathbf{\Sigma}_\mathrm{cor} := \begin{bmatrix} \mathbf{\Sigma}_\mathrm{raw} & \mathbf{\Sigma}_{\mathrm{raw},\mathrm{out}} \\ \mathbf{\Sigma}_{\mathrm{raw},\mathrm{out}}^\top & \mathbf{\Sigma}_\mathrm{out} \end{bmatrix}.
\end{equation}

\begin{lemma}[Block structure of the transformed perturbation]\label{Lemma:Block-Diagonal-Reduction}
	Assume that \(\sigma_\mathbf{r}>0\) and \(\mathbf{\Sigma}_\mathrm{cor}\succ0\).
	Write \(\mathbf{\Sigma}_\mathrm{cor}^{-1}[a,b]\), \(a,b\in\{1,2\}\), for the blocks of \(\mathbf{\Sigma}_\mathrm{cor}^{-1}\) in the raw-share and \(\mathcal{S}_\mathrm{out}\) coordinates.
	Then, in the \((U_\mathrm{raw},\mathcal{S}_\mathrm{in},\mathcal{S}_\mathrm{out})\) ordering,
	\begin{equation}\label{Eq:Sigma-ytilde-inverse-unpermuted}
		\mathbf{\Sigma}_{\mathbf{\widetilde{y}}_{\mathcal{U},\mathbf{c}}}^{-1} = \begin{bmatrix} \mathbf{\Sigma}_\mathrm{cor}^{-1}[1,1] & \mathbf{O} & \mathbf{\Sigma}_\mathrm{cor}^{-1}[1,2] \\ \mathbf{O} & \sigma_\mathbf{r}^{-2} \mathbf{I} & \mathbf{O} \\ \mathbf{\Sigma}_\mathrm{cor}^{-1}[2,1] & \mathbf{O} & \mathbf{\Sigma}_\mathrm{cor}^{-1}[2,2] \end{bmatrix},
	\end{equation}
	and \(\| \mathbf{\Sigma}_{\mathbf{\widetilde{y}}_{\mathcal{U},\mathbf{c}}}^{-1} \|_2 = \max \{ \|\mathbf{\Sigma}_\mathrm{cor}^{-1}\|_2, \sigma_\mathbf{r}^{-2} \}\).
\end{lemma}

The lemma holds because \(\mathbf{r}_\mathrm{in}\) is independent of \(\mathbf{y}_\mathrm{raw}\) and \(\mathbf{y}_\mathrm{out}\); the proof is given in Appendix~\ref{Appendix:Privacy-Proofs-Joint}.
Thus, all remaining dependence induced by the common sharing mask is confined to the raw-share and \(\mathcal{S}_\mathrm{out}\) coordinates.

\subsubsection{Elimination and Schur-Complement Decomposition}\label{Section:Joint-Schur}

The isolated fresh-mask block is multiplied on both sides by the zero block of \(\mathbf{\widetilde{M}}_{\mathcal{U},\mathbf{c}}\) and therefore contributes nothing to the precision.
Define \(\mathbf{M}_\mathrm{cor} := \begin{bmatrix} \mathbf{M}_\mathrm{raw}^\top & \mathbf{M}_\mathrm{out}^\top \end{bmatrix}^\top\).

\begin{lemma}[Elimination of paired response blocks]\label{Lemma:Exact-Reduction-Mcor}
	Under the conditions of \Cref{Lemma:Block-Diagonal-Reduction},
	\begin{equation}\label{Eq:Exact-Reduction-Mcor}
		\mathbf{M}_{\mathcal{U},\mathbf{c}}^\top \mathbf{\Sigma}_{\mathbf{y}_{\mathcal{U},\mathbf{c}}}^{-1} \mathbf{M}_{\mathcal{U},\mathbf{c}} = \mathbf{M}_\mathrm{cor}^\top \mathbf{\Sigma}_\mathrm{cor}^{-1} \mathbf{M}_\mathrm{cor}.
	\end{equation}
\end{lemma}

Assume that \(\mathbf{\Sigma}_\mathrm{raw}\succ0\), and define the parameter-transfer matrix and the Schur complement of \(\mathbf{\Sigma}_\mathrm{raw}\) in \(\mathbf{\Sigma}_\mathrm{cor}\) by
\begin{align}
	\mathbf{B} &:= \mathbf{\Sigma}_\mathrm{raw}^{-1} \mathbf{\Sigma}_{\mathrm{raw},\mathrm{out}}, \label{Eq:B-cross-definition} \\
	\mathbf{C} &:= \mathbf{\Sigma}_\mathrm{out} - \mathbf{\Sigma}_{\mathrm{raw},\mathrm{out}}^\top \mathbf{\Sigma}_\mathrm{raw}^{-1} \mathbf{\Sigma}_{\mathrm{raw},\mathrm{out}}. \label{Eq:C-Schur-complement-definition}
\end{align}

\begin{lemma}[Schur-complement precision decomposition]\label{Lemma:Splitting-M-cor}
	Assume that \(\mathbf{\Sigma}_\mathrm{raw}\succ0\) and \(\mathbf{C}\succ0\). Then
	\begin{align}\label{Eq:Splitting-M-cor}
		\mathbf{M}_\mathrm{cor}^\top \mathbf{\Sigma}_\mathrm{cor}^{-1} \mathbf{M}_\mathrm{cor} &= \mathbf{M}_\mathrm{raw}^\top \mathbf{\Sigma}_\mathrm{raw}^{-1} \mathbf{M}_\mathrm{raw} + \left( \mathbf{M}_\mathrm{out} - \mathbf{B}^\top\mathbf{M}_\mathrm{raw} \right)^\top \mathbf{C}^{-1} \left( \mathbf{M}_\mathrm{out} - \mathbf{B}^\top\mathbf{M}_\mathrm{raw} \right),
	\end{align}
	where the second term is omitted if \(\mathcal{S}_\mathrm{out}=\emptyset\).
\end{lemma}

Both lemmas follow from \eqref{Eq:Precision-Invariance-under-W}, \eqref{Eq:Sigma-ytilde-inverse-unpermuted}, and the block \(\mathbf{L}\mathbf{D}\mathbf{L}^\top\) factorization of \(\mathbf{\Sigma}_\mathrm{cor}\); see Appendix~\ref{Appendix:Privacy-Proofs-Joint}.

\subsubsection{Bounds on the Parameter and Hint Blocks}\label{Section:Joint-Bounds}

For \(1\leq|U_\mathrm{raw}|\leq k_p\), \Cref{Lemma:Column-Space-Ghat-U-Expander} implies that \(\mathbf{\hat{G}}_{U_\mathrm{raw}}\) has full row rank, so \(\sigma_{\min}(\mathbf{\hat{G}}_{U_\mathrm{raw}})>0\).
Recall \(\mathbf{\hat{G}}_\mathrm{raw}\) and \(\mathbf{T}_\mathcal{S}\) from \eqref{Eq:Ghat-raw-definition}, and let
\begin{equation}\label{Eq:T-out-definition}
	\mathbf{T}_\mathrm{out} := \left( \mathbf{\hat{g}}_i^\top \otimes \mathbf{L}_s \right)_{(s,i)\in\mathcal{S}_\mathrm{out}}
\end{equation}
be the rows of \(\mathbf{T}_\mathcal{S}\) indexed by \(\mathcal{S}_\mathrm{out}\), so that \(\mathbf{y}_\mathrm{out} = \mathbf{T}_\mathrm{out}\mathbf{r}_\mathbf{R} + \mathbf{r}_\mathrm{out}\) with \(\mathbf{r}_\mathrm{out}:=(\mathbf{r}_{i,s})_{(s,i)\in\mathcal{S}_\mathrm{out}}\).
By \eqref{Eq:Sigma-y-definition},
\begin{align}
	\mathbf{\Sigma}_\mathrm{raw} &= \sigma_\mathbf{R}^2 \mathbf{\hat{G}}_\mathrm{raw} \mathbf{\hat{G}}_\mathrm{raw}^\top, \label{Eq:Covariance-Block-Identities-raw} \\
	\mathbf{\Sigma}_{\mathrm{raw},\mathrm{out}} &= \sigma_\mathbf{R}^2 \mathbf{\hat{G}}_\mathrm{raw} \mathbf{T}_\mathrm{out}^\top, \label{Eq:Covariance-Block-Identities-rawout}\\
	\mathbf{\Sigma}_\mathrm{out} &= \sigma_\mathbf{R}^2 \mathbf{T}_\mathrm{out} \mathbf{T}_\mathrm{out}^\top + \sigma_\mathbf{r}^2 \mathbf{I}. \label{Eq:Covariance-Block-Identities-out}
\end{align}

\begin{lemma}[Raw parameter and parameter-transfer bounds]\label{Lemma:Sigma-raw-Tout-B-bound}
	Suppose that \(1\leq|U_\mathrm{raw}|\leq k_p\) and \(\sigma_\mathbf{R}>0\).
	Then \(\mathbf{\Sigma}_\mathrm{raw}\succ0\) and \(\| \mathbf{\Sigma}_\mathrm{raw}^{-1} \|_2 = 1/( \sigma_\mathbf{R}^2 \sigma_{\min}^2 (\mathbf{\hat{G}}_{U_\mathrm{raw}}) )\).
	If \(\mathcal{S}_\mathrm{out}\neq\emptyset\), then \(\|\mathbf{B}\|_2 \leq \omega_c\sqrt{Q} \|\mathbf{\hat{G}}\|_2 / \sigma_{\min} (\mathbf{\hat{G}}_{U_\mathrm{raw}})\).
	Moreover, for the expander-based construction, \(\|\mathbf{\hat{G}}\|_2 \leq \sqrt{c_\mathrm{exp}n}\).
\end{lemma}

\begin{lemma}[Conditional response-parameter bound]\label{Lemma:C-inverse-bound}
	Suppose that \(1\leq|U_\mathrm{raw}|\leq k_p\), \(\mathcal{S}_\mathrm{out}\neq\emptyset\), and \(\sigma_\mathbf{R},\sigma_\mathbf{r}>0\).
	Then \(\mathbf{C} \succeq \sigma_\mathbf{r}^2 \mathbf{I} \succ0\), and hence \(\|\mathbf{C}^{-1}\|_2 \leq \sigma_\mathbf{r}^{-2}\).
\end{lemma}

\begin{lemma}[Raw, outside, and residual hint bounds]\label{Lemma:M-raw-out-norms}
	Suppose that \(1\leq|U_\mathrm{raw}|\leq k_p\) and \(\|\mathbf{g}\|_\infty\leq\beta_\mathbf{g}\).
	Then \(\|\mathbf{M}_\mathrm{raw}\|_2 \leq \beta_\mathbf{g}\sqrt{k_p}\).
	If \(\mathcal{S}_\mathrm{out}\neq\emptyset\), then \(\|\mathbf{M}_\mathrm{out}\|_2 \leq \beta_\mathbf{g} \omega_c \sqrt{nQ}\) and
	\begin{align}\label{Eq:Residual-M-bound}
		\left\| \mathbf{M}_\mathrm{out} - \mathbf{B}^\top\mathbf{M}_\mathrm{raw} \right\|_2 &\leq \beta_\mathbf{g} \omega_c \sqrt{Q} \left( \sqrt{n} + \sqrt{k_p}\, \frac{ \|\mathbf{\hat{G}}\|_2 }{ \sigma_{\min} (\mathbf{\hat{G}}_{U_\mathrm{raw}}) } \right).
	\end{align}
\end{lemma}

The first lemma uses the Kronecker structure of \(\mathbf{\Sigma}_\mathrm{raw}\) and the factorization \(\mathbf{T}_\mathrm{out} =\mathbf{L}_{\mathcal{S}_\mathrm{out}} (\mathbf{\hat{G}}_{\mathcal{S}_\mathrm{out}}\otimes\mathbf{I}_D)\), in which each party appears in at most \(Q\) block rows. 
The second follows because \(\mathbf{C}-\sigma_\mathbf{r}^2\mathbf{I}\) is a congruence of an orthogonal projector. 
The third follows from the block structure of the hint matrices, the operator-norm bounds on the local maps, and the triangle inequality. 
The proofs are given in Appendix~\ref{Appendix:Privacy-Proofs-Joint}.

\subsubsection{Final Joint Precision Bound}\label{Section:Joint-Final}

\begin{lemma}[Joint precision bound]\label{Lemma:Final-Joint-Bound}
	Suppose that \(1\leq|U_\mathrm{raw}|\leq k_p\), \(\sigma_\mathbf{R},\sigma_\mathbf{r}>0\), and \(\|\mathbf{g}\|_\infty\leq\beta_\mathbf{g}\).
	Then
	\begin{align}\label{Eq:Final-Joint-Precision-Bound}
		\left\| \mathbf{M}_{\mathcal{U},\mathbf{c}}^\top \mathbf{\Sigma}_{\mathbf{y}_{\mathcal{U},\mathbf{c}}}^{-1} \mathbf{M}_{\mathcal{U},\mathbf{c}} \right\|_2 &\leq \frac{ \beta_\mathbf{g}^2 k_p }{ \sigma_\mathbf{R}^2 \sigma_{\min}^2 ( \mathbf{\hat{G}}_{U_\mathrm{raw}} ) } + \frac{ \beta_\mathbf{g}^2 \omega_c^2 Q }{ \sigma_\mathbf{r}^2 } \left( \sqrt{n} + \sqrt{k_p}\, \frac{ \|\mathbf{\hat{G}}\|_2 }{ \sigma_{\min} ( \mathbf{\hat{G}}_{U_\mathrm{raw}} ) } \right)^2.
	\end{align}
\end{lemma}

\begin{proof}[Proof sketch]
	By \Cref{Lemma:Exact-Reduction-Mcor,Lemma:Splitting-M-cor}, the precision matrix is the sum of \(\mathbf{M}_\mathrm{raw}^\top \mathbf{\Sigma}_\mathrm{raw}^{-1} \mathbf{M}_\mathrm{raw}\) and the term involving \(\mathbf{C}^{-1}\).
	Bounding the first by \Cref{Lemma:Sigma-raw-Tout-B-bound,Lemma:M-raw-out-norms} and the second by \Cref{Lemma:C-inverse-bound} and \eqref{Eq:Residual-M-bound} gives the two terms.
	The full argument, including the case \(\mathcal{S}_\mathrm{out}=\emptyset\), is in Appendix~\ref{Appendix:Privacy-Proofs-Joint}.
\end{proof}

The first term bounds the precision contributed by the exposed raw shares, and the second the conditional precision contributed by the response blocks whose raw shares are not exposed; paired response blocks contribute no separate term.
By \Cref{Remark:MHMLWE-Width-Interpretation}, the reduction loses only a constant factor in the \(\mathsf{MLWE}\) source width when the right-hand side of \eqref{Eq:Final-Joint-Precision-Bound} is at most \(c/\sigma_\mathbf{s}^2\) for a constant \(c>0\).
It suffices that each of its two terms is at most \(c/(2\sigma_\mathbf{s}^2)\), that is, \(\sigma_\mathbf{R}/\sigma_\mathbf{s}\geq\sqrt{2/c}\,\beta_\mathbf{g}\sqrt{k_p}/\sigma_{\min}(\mathbf{\hat{G}}_{U_\mathrm{raw}})\) and \(\sigma_\mathbf{r}/\sigma_\mathbf{s}\geq\sqrt{2/c}\,\beta_\mathbf{g}\omega_c\sqrt{Q}\,\big(\sqrt{n}+\sqrt{k_p}\,\|\mathbf{\hat{G}}\|_2/\sigma_{\min}(\mathbf{\hat{G}}_{U_\mathrm{raw}})\big)\).
Under this sufficient allocation, the sharing-mask width exceeds the secret width by a factor proportional to \(\sqrt{k_p}\), and the leakage-mask width exceeds it by a factor proportional to \(\sqrt{Q}\).
The concrete parameters of \Cref{Section:SEC-Concrete-Privacy} do not fix \(c=\sigma_\mathbf{s}^2\,\Xi_\mathrm{MQ}\) in advance, where \(\Xi_\mathrm{MQ}\) is the uniform precision parameter in \eqref{Eq:Xi-MQ}.
Instead, they impose the uniform version of this bound through \(\sigma_\mathrm{MQ}\) in \eqref{Eq:Sigma-MQ-Uniform-Parameter}, and the selected widths give \(c\) between about \(0.5\) and \(8.5\), with \(\sigma_\mathbf{r}/\sigma_\mathbf{s}\approx2^{40}\) to \(2^{42.5}\) for \(Q=2^{64}\).

\begin{remark}[Boundary leakage patterns]\label{Remark:Boundary-Leakage-Pattern}
	If \(U_\mathrm{raw}=\emptyset\) and \(\mathcal{S}\neq\emptyset\), then \(\| \mathbf{M}_{\mathcal{U},\mathbf{c}}^\top \mathbf{\Sigma}_{\mathbf{y}_{\mathcal{U},\mathbf{c}}}^{-1} \mathbf{M}_{\mathcal{U},\mathbf{c}} \|_2 \leq \beta_\mathbf{g}^2 \omega_c^2 nQ / \sigma_\mathbf{r}^2\).
	If \(U_\mathrm{raw}\neq\emptyset\) and \(\mathcal{S}_\mathrm{out}=\emptyset\), only the first term of \eqref{Eq:Final-Joint-Precision-Bound} remains.
	If \(U_\mathrm{raw}=\emptyset\) and \(\mathcal{S}=\emptyset\), no hint is present and the precision term is defined to be \(0\).
	These cases are verified in Appendix~\ref{Appendix:Privacy-Proofs-Joint}.
\end{remark}

The joint bound is tighter than the estimate obtained by bounding the two factors separately, namely \(\| \mathbf{M}_{\mathcal{U},\mathbf{c}}^\top \mathbf{M}_{\mathcal{U},\mathbf{c}} \|_2 \| \mathbf{\Sigma}_{ \mathbf{y}_{\mathcal{U},\mathbf{c}} }^{-1} \|_2\), which still pays for the \(\mathcal{S}_\mathrm{in}\) blocks whose hint-matrix block vanishes after the transformation \(\mathbf{W}\).
With the remaining parameters fixed, the separate estimate grows cubically in \(Q\), whereas the leakage-dependent term in \eqref{Eq:Final-Joint-Precision-Bound} grows linearly.
We therefore apply \Cref{Theorem:Hardness-MHMLWE-Tight} in the reduction.

\subsection{Uniform Parameters for Multi-Query Privacy}\label{Section:Uniform-Parameters-Multi-Query-Privacy}

We now replace the pattern-dependent quantities in \Cref{Lemma:Final-Joint-Bound} by parameters valid uniformly over every admissible leakage pattern and leakage-description sequence, that is, in the worst case over every set \(U_\mathrm{raw}\subseteq[n]\) of exposed raw shares with \(|U_\mathrm{raw}|\leq k_p\), all response sets \(V_s\subseteq[n]\), and all \(\mathbf{c}\in\mathcal{C}^Q\).
We then establish the distributional condition required by Matrix Hint-MLWE.

\subsubsection{Uniform Conditioning and Precision}\label{Section:Uniform-Conditioning}

Define
\begin{equation}\label{Eq:alpha-kp-definition}
	\alpha_{k_p} := \min\left\{ \sigma_{\min}(\mathbf{\hat{G}}_V): V\subseteq[n],\, |V|=k_p \right\}.
\end{equation}

\begin{lemma}[Uniform smallest-singular-value bound]\label{Lemma:alpha-kp-uniform}
	Suppose that \(\mathbf{\hat{G}}_U\) has full row rank for every \(U\subseteq[n]\) with \(1\leq|U|\leq k_p\).
	Then \(\alpha_{k_p}>0\), and \(\sigma_{\min}(\mathbf{\hat{G}}_U)\geq\alpha_{k_p}\) for every nonempty \(U\subseteq[n]\) with \(|U|\leq k_p\).
\end{lemma}

The second claim follows from eigenvalue interlacing, since \(\mathbf{\hat{G}}_U\mathbf{\hat{G}}_U^\top\) is a principal submatrix of \(\mathbf{\hat{G}}_V\mathbf{\hat{G}}_V^\top\) for every \(V\supseteq U\) with \(|V|=k_p\).
For the expander-based instantiation, the full-row-rank hypothesis follows from \Cref{Lemma:Column-Space-Ghat-U-Expander}.
Assume the hypothesis of \Cref{Lemma:alpha-kp-uniform}, so that \(\alpha_{k_p}>0\), and that \(\sigma_\mathbf{R},\sigma_\mathbf{r}>0\) and \(\|\mathbf{g}\|_\infty\leq\beta_\mathbf{g}\). 
Define the uniform multi-query precision parameter
\begin{align}\label{Eq:Xi-MQ}
	\Xi_\mathrm{MQ} := \frac{ \beta_\mathbf{g}^2 k_p }{ \sigma_\mathbf{R}^2\alpha_{k_p}^2 } + \frac{ \beta_\mathbf{g}^2\omega_c^2Q }{ \sigma_\mathbf{r}^2 } \left( \sqrt{n} + \sqrt{k_p}\, \frac{ \|\mathbf{\hat{G}}\|_2 }{ \alpha_{k_p} } \right)^2.
\end{align}

\begin{lemma}[Uniform multi-query precision bound]\label{Lemma:Uniform-MQ-Precision}
	For every admissible pattern \(\mathcal{U}\) with \(|U_\mathrm{raw}|\leq k_p\) and every \(\mathbf{c}\in\mathcal{C}^{Q}\),
	\begin{align}\label{Eq:Uniform-Joint-Precision-Bound}
		\left\| \mathbf{M}_{\mathcal{U},\mathbf{c}}^\top \mathbf{\Sigma}_{\mathbf{y}_{\mathcal{U},\mathbf{c}}}^{-1} \mathbf{M}_{\mathcal{U},\mathbf{c}} \right\|_2 \leq \Xi_\mathrm{MQ},
	\end{align}
	where the left-hand side is interpreted as \(0\) when the auxiliary view is empty.
\end{lemma}

The bound follows from \Cref{Lemma:Final-Joint-Bound,Lemma:alpha-kp-uniform} and \Cref{Remark:Boundary-Leakage-Pattern}; it is independent of \(\mathbf{c}\) because each \(c_s\) enters only through \(\|\mathbf{L}_s\|_2\leq\omega_c\).
The proofs of both lemmas are given in Appendix~\ref{Appendix:Privacy-Proofs-Uniform}.

\subsubsection{Linear-Image and Kernel Structure}\label{Section:Linear-Image-Kernel}

The precision bound alone does not establish the distributional condition of Matrix Hint-MLWE: by \eqref{Eq:Joint-Perturbation-Formula}, \(\mathbf{y}_{\mathcal{U},\mathbf{c}}\) is an integer linear image of independent discrete Gaussians and need not itself be an elliptic discrete Gaussian.
This section, together with the subsequent \Cref{Section:Uniform-Kernel-Smoothing,Section:Uniform-Gaussian-Approximation}, derives kernel-smoothing conditions under which \(\mathbf{y}_{\mathcal{U},\mathbf{c}}\) is, uniformly over admissible realizations, within statistical distance \(\epsilon'\) of the elliptic discrete Gaussian assumed in \Cref{Theorem:Hardness-MHMLWE-Tight}.
Splitting \(\mathbf{r}_\mathcal{S}\) into \(\mathbf{r}_\mathrm{in}\) and \(\mathbf{r}_\mathrm{out}\) and applying \(\mathbf{W}\) to \eqref{Eq:Joint-Perturbation-Formula} gives
\begin{equation}\label{Eq:Phi-tilde-definition}
	\mathbf{\widetilde{y}}_{\mathcal{U},\mathbf{c}} = \mathbf{\widetilde{\Phi}}_{\mathcal{U},\mathbf{c}}\, \mathbf{r}_0,
	\qquad \mathbf{\widetilde{\Phi}}_{\mathcal{U},\mathbf{c}} := \mathbf{W}\mathbf{\Phi}_{\mathcal{U},\mathbf{c}} = \begin{bmatrix} \mathbf{\hat{G}}_\mathrm{raw} & \mathbf{O} & \mathbf{O}\\ \mathbf{O} & \mathbf{I} & \mathbf{O}\\ \mathbf{T}_\mathrm{out} & \mathbf{O} & \mathbf{I} \end{bmatrix},
\end{equation}
where \(\mathbf{r}_0 := \begin{bmatrix} \mathbf{r}_\mathbf{R}^\top \ \mathbf{r}_\mathrm{in}^\top \ \mathbf{r}_\mathrm{out}^\top \end{bmatrix}^\top\) in agreement with \eqref{Eq:Transformed-M-and-y}.
Let \(M:=(k-1+|\mathcal{S}|)D\), so that \(\mathbf{\widetilde{\Phi}}_{\mathcal{U},\mathbf{c}} \in\mathbb{Z}^{M'\times M}\) and \(\mathbf{r}_0\gets \mathcal{D}_{\mathbb{Z}^M,\sqrt{\mathbf{\Sigma}_0}}\), where \(\mathbf{\Sigma}_0\) is as in \eqref{Eq:Sigma-0-Pushforward}, with the fresh-mask coordinates ordered as \((\mathcal{S}_\mathrm{in},\mathcal{S}_\mathrm{out})\).

\begin{lemma}[Integral surjectivity of the transformed leakage map]\label{Lemma:Phi-tilde-Surjective}
	For every admissible fixed public realization \((\mathcal{U},\mathbf{c})\), \(\mathbf{\widetilde{\Phi}}_{\mathcal{U},\mathbf{c}} \mathbb{Z}^{M} = \mathbb{Z}^{M'}\).
\end{lemma}

Surjectivity follows from \(\mathbf{\hat{G}}_{U_\mathrm{raw}}\mathbb{Z}^{k-1} = \mathbb{Z}^{|U_\mathrm{raw}|}\) (\Cref{Lemma:Column-Space-Ghat-U-Expander}) and the identity blocks in \eqref{Eq:Phi-tilde-definition}; see Appendix~\ref{Appendix:Privacy-Proofs-Uniform}.
The block equations in \eqref{Eq:Phi-tilde-definition} also give the integer kernel
\begin{equation}\label{Eq:Phi-tilde-kernel}
	\ker_\mathbb{Z} \left( \mathbf{\widetilde{\Phi}}_{\mathcal{U},\mathbf{c}} \right) = \left\{ \begin{bmatrix} \mathbf{v}\\ \mathbf{0}\\ -\mathbf{T}_\mathrm{out}\mathbf{v} \end{bmatrix} : \mathbf{v} \in \ker_\mathbb{Z} \left( \mathbf{\hat{G}}_\mathrm{raw} \right) \right\},
\end{equation}
so the \(\mathcal{S}_\mathrm{in}\) blocks do not contribute to the kernel geometry.
Finally, the target parameter matrix of the transformed perturbation is
\begin{equation}\label{Eq:Sigma-ytilde-Phi-tilde}
	\mathbf{\Sigma}_{\mathbf{\widetilde{y}}_{\mathcal{U},\mathbf{c}}} = \mathbf{\widetilde{\Phi}}_{\mathcal{U},\mathbf{c}} \mathbf{\Sigma}_0 \mathbf{\widetilde{\Phi}}_{\mathcal{U},\mathbf{c}}^\top.
\end{equation}
For every nonempty auxiliary view, \Cref{Lemma:Phi-tilde-Surjective} implies that \(\mathbf{\widetilde{\Phi}}_{\mathcal{U},\mathbf{c}}\) has full row rank; since \(\mathbf{\Sigma}_0\succ0\) and \(\mathbf{W}\) is invertible, both \(\mathbf{\Sigma}_{\mathbf{\widetilde{y}}_{\mathcal{U},\mathbf{c}}}\) and \(\mathbf{\Sigma}_{\mathbf{y}_{\mathcal{U},\mathbf{c}}}\) are positive definite.

\subsubsection{Uniform Kernel-Smoothing Condition}\label{Section:Uniform-Kernel-Smoothing}

For an integer lattice \(\Lambda\), let \(\widetilde{bl}(\Lambda)\) denote the minimum, over all ordered bases of \(\Lambda\), of the largest Euclidean norm of a Gram-Schmidt vector, with \(\widetilde{bl}(\{\mathbf{0}\}):=0\).
Define the uniform kernel-geometry parameter and the effective perturbation width
\begin{align}
	\gamma_{k_p}^{\ker} &:= \max \left\{ 1,\, \max_{\substack{ U\subseteq[n]\\ 1\leq|U|\leq k_p }} \widetilde{bl} \left( \ker_\mathbb{Z}(\mathbf{\hat{G}}_U) \right) \right\},\label{Eq:gamma-kernel-definition}\\
	\sigma_\mathrm{eff} &:= \left( \frac{1}{\sigma_\mathbf{R}^2} + \frac{ \omega_c^2Q \|\mathbf{\hat{G}}\|_2^2 }{ \sigma_\mathbf{r}^2 } \right)^{-1/2}. \label{Eq:sigma-eff-definition}
\end{align}
For \(0<\epsilon'\leq1/2\), we impose the uniform kernel-smoothing condition
\begin{equation}\label{Eq:sigma-eff-condition}
	\sigma_\mathrm{eff} \geq \gamma_{k_p}^{\ker} \sqrt{ \frac{ \ln \left( 2(k-1)D \left( 2+1/\epsilon' \right) \right) }{ 2\pi^2 } }.
\end{equation}
All quantities in \eqref{Eq:sigma-eff-condition} depend only on the public sharing matrix, the distribution parameters, and the global bounds \(k_p\), \(\omega_c\), and \(Q\).

\begin{lemma}[Uniform whitened-kernel smoothing]\label{Lemma:Whitened-Kernel-Smoothing}
	Assume \eqref{Eq:sigma-eff-condition}, and let \(\epsilon_0:=\epsilon'/(1+\epsilon')\).
	Then, for every admissible fixed public realization \((\mathcal{U},\mathbf{c})\) with a nonempty auxiliary view and \(\ker_\mathbb{Z}(\mathbf{\widetilde{\Phi}}_{\mathcal{U},\mathbf{c}})\neq\{\mathbf{0}\}\),
	\begin{equation}\label{Eq:Whitened-Kernel-Smoothing-Condition}
		\eta_{\epsilon_0} \left( \mathbf{\Sigma}_0^{-1/2} \ker_\mathbb{Z} \left( \mathbf{\widetilde{\Phi}}_{\mathcal{U},\mathbf{c}} \right) \right) \leq 1.
	\end{equation}
\end{lemma}

The proof, in Appendix~\ref{Appendix:Privacy-Proofs-Uniform}, represents the whitened kernel as the image of \(\ker_\mathbb{Z}(\mathbf{\hat{G}}_\mathrm{raw})\) under a linear map of norm at most \(1/\sigma_\mathrm{eff}\), bounds its Gram-Schmidt profile by \(\gamma_{k_p}^{\ker}/\sigma_\mathrm{eff}\), and applies the basis-dependent smoothing estimate of \cite[Lemma 2.8]{genise2020improved}.
To make \(\gamma_{k_p}^{\ker}\) explicit, define
\begin{align}
	\beta_{k_p} &:= \max_{\substack{ U\subseteq[n]\\ 1\leq|U|\leq k_p }} \|\mathbf{\hat{G}}_U\|_2, \label{Eq:beta-kp-definition}\\
	\gamma_{k_p} &:= \min_{\substack{ U\subseteq[n]\\ 1\leq|U|\leq k_p }} \sigma_{\min}(\mathbf{M}_U), \label{Eq:gamma-kp-definition}
\end{align}
where \(\mathbf{M}_U\) is the peeled unimodular submatrix of \(\mathbf{\hat{G}}_U\) from \Cref{Section:Lossless-Expanders}.
Since every \(\mathbf{M}_U\) is invertible, \(\gamma_{k_p}>0\).

\begin{lemma}[Explicit kernel basis and bound on \(\gamma_{k_p}^{\ker}\)]\label{Lemma:Kernel-GS-Bound}
	Let \(U\subseteq[n]\) satisfy \(1\leq|U|\leq k_p\), write \(r:=k-1\), and let \(\mathbf{M}_U=\mathbf{\hat{G}}_U[\cdot,J_U]\) be the peeled unimodular submatrix, where \(|J_U|=|U|\).
	Set \(t:=r-|U|\), \(\overline{J_U}:=[r]\setminus J_U\), \(\mathbf{N}_U := \mathbf{\hat{G}}_U[\cdot,\overline{J_U}]\), and let \(\mathbf{P}_U\) be the permutation matrix with \(\mathbf{\hat{G}}_U\mathbf{P}_U = \begin{bmatrix} \mathbf{M}_U \ \mathbf{N}_U \end{bmatrix}\).
	Then the vectors
	\begin{align}\label{Eq:Explicit-Kernel-Basis-Vector}
		\mathbf{b}_a := \mathbf{P}_U\begin{bmatrix} -\mathbf{M}_U^{-1} \mathbf{N}_U\mathbf{e}_a\\ \mathbf{e}_a \end{bmatrix} \in \mathbb{Z}^{r}, \qquad a\in[t],
	\end{align}
	form a basis of \(\ker_\mathbb{Z}(\mathbf{\hat{G}}_U)\).
	Consequently,
	\begin{equation}\label{Eq:Kernel-GS-Bound-Explicit}
		\gamma_{k_p}^{\ker} \leq \sqrt{ 1+ \min\left\{ k_p,\, \beta_{k_p}^2 \right\} / \gamma_{k_p}^2 }.
	\end{equation}
\end{lemma}

The basis is integral because \(\mathbf{M}_U^{-1}\) is integral, and each \(\|\mathbf{b}_a\|_2^2\) is at most \(1+\|\mathbf{N}_U\mathbf{e}_a\|_2^2/\sigma_{\min}^2(\mathbf{M}_U)\); the proof is given in Appendix~\ref{Appendix:Privacy-Proofs-Uniform}.
Combining \Cref{Lemma:Kernel-GS-Bound} with \eqref{Eq:sigma-eff-condition} gives the computable sufficient condition
\begin{equation}\label{Eq:sigma-eff-condition-explicit-practical}
	\sigma_\mathrm{eff} \geq \sqrt{ 1+ \frac{ \min\left\{ k_p,\, \beta_{k_p}^2 \right\} }{ \gamma_{k_p}^2 } } \sqrt{ \frac{ \ln\left( 2(k-1)D \left( 2+1/\epsilon' \right) \right) }{ 2\pi^2 } }.
\end{equation}
This is the only point of the privacy analysis at which the peeling procedure enters quantitatively; the values of \(\beta_{k_p}\) and \(\gamma_{k_p}\) for our instances are reported in \Cref{Table:SEC-Selected-Privacy-Statistics}.

\subsubsection{Uniform Gaussian Approximation}\label{Section:Uniform-Gaussian-Approximation}

\begin{lemma}[Uniform Gaussian approximation of the joint perturbation]\label{Lemma:Uniform-Gaussian-Joint-Perturbation}
	Suppose that \(\sigma_\mathbf{R},\sigma_\mathbf{r}>0\) and that \eqref{Eq:sigma-eff-condition} holds.
	Then, for every admissible fixed public realization \((\mathcal{U},\mathbf{c})\) inducing a nonempty auxiliary view, \(\Delta ( \mathbf{y}_{\mathcal{U},\mathbf{c}}, \mathcal{D}_{ \mathbb{Z}^{M'}, \sqrt{ \mathbf{\Sigma}_{ \mathbf{y}_{\mathcal{U},\mathbf{c}} } } } ) \leq \epsilon'\).
\end{lemma}

\begin{proof}[Proof sketch] 
	If \(\ker_\mathbb{Z} (\mathbf{\widetilde{\Phi}}_{\mathcal{U},\mathbf{c}}) =\{\mathbf{0}\}\), then integral surjectivity makes \(\mathbf{\widetilde{\Phi}}_{\mathcal{U},\mathbf{c}}\) unimodular, so the pushforward is exact. 
	Otherwise, \Cref{Lemma:Phi-tilde-Surjective,Lemma:Whitened-Kernel-Smoothing} and the linear-transformation theorem of \cite[Theorem 3.1]{genise2020improved} give statistical distance at most \(\epsilon_0/(1-\epsilon_0)=\epsilon'\). 
	Applying the bijection \(\mathbf{W}^{-1}\) preserves this distance. 
	The details are given in Appendix~\ref{Appendix:Privacy-Proofs-Uniform}. 
\end{proof}

In the fixed-pattern \(\mathsf{HC\text{-}Leak}\) experiment, replacing the actual perturbation by its target elliptic discrete Gaussian therefore costs at most \(\epsilon'\) in each branch and at most \(2\epsilon'\) in total; the empty view incurs no loss.

\subsubsection{Uniform Matrix Hint-MLWE Source Parameter}\label{Section:Uniform-Source-Parameter}

Suppose that the coefficient-embedded secret has parameter matrix \(\sigma_\mathbf{s}^2\mathbf{I}_D\).
For a fixed nonempty realization \((\mathcal{U},\mathbf{c})\), the effective source width of \Cref{Theorem:Hardness-MHMLWE-Tight} and its uniform replacement are
\begin{align}
	\sigma_{\mathcal{U},\mathbf{c}} &:= \left( 2 \left( \frac{1}{\sigma_\mathbf{s}^2} + \left\| \mathbf{M}_{\mathcal{U},\mathbf{c}}^\top \mathbf{\Sigma}_{\mathbf{y}_{\mathcal{U},\mathbf{c}}}^{-1} \mathbf{M}_{\mathcal{U},\mathbf{c}} \right\|_2 \right) \right)^{-1/2}, \label{Eq:Sigma-Pattern-Dependent-MQ}\\
    \sigma_\mathrm{MQ} &:= \left( 2 \left( \frac{1}{\sigma_\mathbf{s}^2} + \Xi_\mathrm{MQ} \right) \right)^{-1/2}. \label{Eq:Sigma-MQ-Uniform-Parameter}
\end{align}
By \Cref{Lemma:Uniform-MQ-Precision}, \(\sigma_{\mathcal{U},\mathbf{c}} \geq \sigma_\mathrm{MQ}\) for every admissible nonempty realization.
Hence, if \(\sigma_\mathrm{MQ} \geq \sqrt{2}\eta_\epsilon(\mathbb{Z}^{D})\) for some \(0<\epsilon\leq1/2\), \Cref{Corollary:Hardness-MHMLWE-Tight-Smaller-Sigma} lets every such instance be reduced from \(\mathsf{MLWE}_{p,\ell-p,q,d,\sigma_\mathrm{MQ}}\) with loss \(2\epsilon\).
If \(U_\mathrm{raw}=\emptyset\) and \(\mathcal{S}=\emptyset\), then the auxiliary view is empty and contains no hint; this case is handled directly under \(\mathsf{MLWE}_{p,\ell-p,q,d,\sigma_\mathbf{s}}\).

\subsection{Multi-Query HC-LLR Privacy}\label{Section:Multi-Query-HCLLR-Privacy}

\subsubsection{Fixed-Pattern Reduction to MLWE}\label{Section:Fixed-Pattern-Reduction}

Fix an admissible public realization \((\mathcal{U},\mathbf{c})\) inducing a nonempty auxiliary view.
Define the Gaussianized hint and perturbation
\begin{align}\label{Eq:z-breve-Fixed-Pattern}
	\mathbf{\breve{z}}_{\mathcal{U},\mathbf{c}} := \mathbf{M}_{\mathcal{U},\mathbf{c}}\mathbf{s} + \mathbf{\breve{y}}_{\mathcal{U},\mathbf{c}},
    \qquad
    \mathbf{\breve{y}}_{\mathcal{U},\mathbf{c}} \gets \mathcal{D}_{ \mathbb{Z}^{M'}, \sqrt{ \mathbf{\Sigma}_{ \mathbf{y}_{\mathcal{U},\mathbf{c}} } } }
\end{align}
where \(\mathbf{\breve{y}}_{\mathcal{U},\mathbf{c}}\) is sampled independently of the secret.

\begin{lemma}[Fixed-pattern reduction to Matrix Hint-MLWE]\label{Lemma:HCLeak-to-MHMLWE-multi}
	Suppose that the hypotheses of \Cref{Lemma:Uniform-Gaussian-Joint-Perturbation} hold for some \(0<\epsilon'\leq1/2\).
	Then, for every PPT adversary \(\mathcal{A}\) against the fixed-pattern \(\mathsf{HC\text{-}Leak}\) experiment for \((\mathcal{U},\mathbf{c})\), there exists a PPT adversary \(\mathcal{B}\) against the one-hint Matrix Hint-MLWE instance with hint matrix \(\mathbf{M}_{\mathcal{U},\mathbf{c}}\) and hint \(\mathbf{\breve{z}}_{\mathcal{U},\mathbf{c}}\) in \eqref{Eq:z-breve-Fixed-Pattern} such that \(\mathsf{Adv}_{\mathcal{A},\mathsf{HC\text{-}Leak}}^{(\mathsf{pp},\mathcal{U},\mathbf{c})}(\kappa) \leq \mathsf{Adv}_\mathcal{B}^\mathsf{MHMLWE}(\kappa) + 2\epsilon'\).
\end{lemma}

\begin{lemma}[Fixed-pattern reduction to MLWE]\label{Lemma:Multi-HCLeak-to-MLWE}
	Suppose that the hypotheses of \Cref{Lemma:Uniform-MQ-Precision,Lemma:Uniform-Gaussian-Joint-Perturbation} hold and that \(\sigma_\mathrm{MQ} \geq \sqrt{2}\eta_\epsilon(\mathbb{Z}^D)\) for some \(0<\epsilon\leq1/2\).
	Then, for every PPT adversary \(\mathcal{A}\) against the fixed-pattern \(\mathsf{HC\text{-}Leak}\) experiment for \((\mathcal{U},\mathbf{c})\), there exists a PPT adversary \(\mathcal{B}\) against \(\mathsf{MLWE}_{p,\ell-p,q,d,\sigma_\mathrm{MQ}}\) such that \(\mathsf{Adv}_{\mathcal{A},\mathsf{HC\text{-}Leak}}^{(\mathsf{pp},\mathcal{U},\mathbf{c})}(\kappa) \leq \mathsf{Adv}_{\mathcal{B},\mathsf{MLWE}_{p,\ell-p,q,d,\sigma_\mathrm{MQ}}}(\kappa) + 2\epsilon' + 2\epsilon\).
\end{lemma}

The first lemma replaces the actual joint perturbation by its target elliptic discrete Gaussian, parses \(\mathbf{\breve{z}}_{\mathcal{U},\mathbf{c}}\) according to \eqref{Eq:z-Uc-Structured-Encoding}, and loses at most \(\epsilon'\) in each branch by \Cref{Lemma:Uniform-Gaussian-Joint-Perturbation}. 
The second composes this reduction with \Cref{Corollary:Hardness-MHMLWE-Tight-Smaller-Sigma}, using \(\sigma_\mathrm{MQ}\leq\sigma_{\mathcal{U},\mathbf{c}}\), and incurs the additional loss \(2\epsilon\). 
The proofs are given in Appendix~\ref{Appendix:Privacy-Proofs-Reduction}.

\subsubsection{Admissible Multi-Query Leakage Patterns}\label{Section:Admissible-Patterns}

Define \(\mathcal{U}_{k_p,Q} := \left\{ \left( U_\mathrm{raw}, (V_s)_{s\in[Q]} \right): U_\mathrm{raw}\in\Gamma_0,\ |U_\mathrm{raw}|\leq k_p,\ V_s\subseteq[n]\ \forall s\in[Q] \right\}\).
No disjointness is required between \(U_\mathrm{raw}\) and the sets \(V_s\), the same party may occur in several queries, and empty sets are permitted, including the completely empty auxiliary view.
For every \(\mathbf{c}\in\mathcal{C}^{Q}\), the induced local maps satisfy \(\|\mathbf{L}_s\|_2\leq\omega_c\), so \Cref{Lemma:Multi-HCLeak-to-MLWE} applies pointwise to every \((\mathcal{U},\mathbf{c})\) with a nonempty view, with the same source width \(\sigma_\mathrm{MQ}\).

\subsubsection{Expander Setup Event}\label{Section:Expander-Setup-Event}

Let \(\mathsf{Good}_\mathrm{MQ}(\mathsf{pp})\) be the event that, for every nonempty \(U\subseteq[n]\) with \(|U|\leq k_p\), the matrix \(\mathbf{\hat{G}}_U\) contains a unimodular \(|U|\times|U|\) column submatrix, and let
\begin{align}\label{Eq:delta-MQ}
	\delta_\mathrm{MQ} := \Pr_{\mathsf{pp}\gets\mathsf{Setup}(1^\kappa)} \left[ \neg\mathsf{Good}_\mathrm{MQ}(\mathsf{pp}) \right].
\end{align} 
On \(\mathsf{Good}_\mathrm{MQ}(\mathsf{pp})\), every such \(\mathbf{\hat{G}}_U\) is integrally surjective and has full row rank, so the uniform conditioning, integral-surjectivity, and kernel-geometry conclusions of \Cref{Section:Uniform-Parameters-Multi-Query-Privacy} hold simultaneously over \(\mathcal{U}_{k_p,Q}\). 
The probability \(\delta_\mathrm{MQ}\) is taken only over the setup output; the conditions \eqref{Eq:sigma-eff-condition} and \(\sigma_\mathrm{MQ}\geq \sqrt{2}\eta_\epsilon(\mathbb{Z}^D)\) are deterministic, and their statistical errors \(\epsilon'\) and \(\epsilon\) are accounted for separately. 
For the certified expander-based construction, \Cref{Lemma:Expander-implies-UnimodularMinor} shows that \(\mathsf{Good}_\mathrm{MQ}(\mathsf{pp})\) holds for every setup output, so \(\delta_\mathrm{MQ}=0\).

\subsubsection{Multi-Query Privacy Theorem}\label{Section:MQ-Privacy-Theorem}

For fixed \(\mathsf{pp}\) and \(\mathcal{U}\in\mathcal{U}_{k_p,Q}\), write \(\mathsf{Adv}_{\mathcal{A},\mathsf{HC\text{-}Leak}}^{(\mathsf{pp},\mathcal{U})}(\kappa)\) for the advantage in the experiment in which the leakage-description sequence is sampled internally as \(\mathbf{c}\gets\mathcal{C}^{Q}\).
The theorem instantiates \Cref{Definition:HC-LLR} with coalitions replaced by patterns \(\mathcal{U}\in\mathcal{U}_{k_p,Q}\); its exceptional setup outputs are those violating \(\mathsf{Good}_\mathrm{MQ}\), which depends neither on the adversary nor on the pattern.

Let \(\mathcal{F}_\mathrm{MQ} := \{f_{\bm{A}}: \bm{s}\mapsto\bm{A}\bm{s}\bmod q \mid \bm{A} = \begin{bmatrix} \bm{I}_p \ \bm{A}' \end{bmatrix},\ \bm{A}' \in \mathcal{R}_q^{p\times(\ell-p)} \}\), with \(\bm{A}'\) uniform. 
For each \(\kappa\), let \(\mathsf{Good}_\kappa\) be the set of setup outputs \(\mathsf{pp}\) in the support of \(\mathsf{Setup}(1^\kappa)\) satisfying \(\mathsf{Good}_\mathrm{MQ}(\mathsf{pp})\). 
For a PPT adversary \(\mathcal{A}\), a pattern \(\mathcal{U}\in\mathcal{U}_{k_p,Q}\), and \(\mathsf{pp}\in\mathsf{Good}_\kappa\), let \(\mathcal{B}_\mathsf{pp}\) be obtained from \(\mathcal{A}\), \(\mathcal{U}\), and \(\mathsf{pp}\) as follows: if the auxiliary view of \(\mathcal{U}\) is nonempty, \(\mathcal{B}_\mathsf{pp}\) samples \(\mathbf{c}\gets\mathcal{C}^Q\) and runs the reduction of \Cref{Lemma:Multi-HCLeak-to-MLWE} against \(\mathsf{MLWE}_{p,\ell-p,q,d,\sigma_\mathrm{MQ}}\); otherwise, \(\mathcal{B}_\mathsf{pp}\) is the direct reduction to \(\mathsf{MLWE}_{p,\ell-p,q,d,\sigma_\mathbf{s}}\) given in the proof in Appendix~\ref{Appendix:Privacy-Proofs-Reduction}.
Define
\begin{align}\label{Eq:epsilon-MQ}
	\epsilon_\mathrm{MQ}(\kappa,\mathcal{A},\mathcal{U}) := \max_{\mathsf{pp}\in\mathsf{Good}_\kappa}\mathsf{Adv}_{\mathcal{B}_\mathsf{pp},\mathsf{MLWE}}(\kappa)+2\epsilon'+2\epsilon,
\end{align}
where the maximum is over a finite set and is taken to be \(0\) if \(\mathsf{Good}_\kappa=\emptyset\).
Since \(\mathcal{B}_\mathsf{pp}\) uses \(\mathsf{pp}\) only as polynomial-size advice, \(\epsilon_\mathrm{MQ}\) is negligible whenever the relevant \(\mathsf{MLWE}\) problems are hard against non-uniform PPT adversaries and \(2\epsilon'+2\epsilon\) is negligible.

\begin{theorem}[Multi-query HC-LLR privacy from \(\mathsf{MLWE}\)]\label{Theorem:SEC-HCLLR-multi}
    Suppose that the following hold:
	\begin{enumerate}
		\item the secret, sharing masks, and response masks satisfy the distributional and independence conditions of \Cref{Section:Multi-Query-HC-Leak-Representation};
		\item \eqref{Eq:sigma-eff-condition} holds for some \(0<\epsilon'\leq1/2\), and \(\sigma_\mathrm{MQ}\geq\sqrt{2}\eta_\epsilon(\mathbb{Z}^D)\) for some \(0<\epsilon\leq1/2\), where \(\sigma_\mathrm{MQ}\) is defined in \eqref{Eq:Sigma-MQ-Uniform-Parameter};
        \item \(\mathsf{MLWE}_{p,\ell-p,q,d,\sigma_\mathrm{MQ}}\) and \(\mathsf{MLWE}_{p,\ell-p,q,d,\sigma_\mathbf{s}}\) are hard against non-uniform PPT adversaries, and \(2\epsilon'+2\epsilon\) is negligible in \(\kappa\).
	\end{enumerate}
	Then, for every \(\kappa\), with \(\delta_\mathrm{MQ}\) and \(\epsilon_\mathrm{MQ}\) as in \eqref{Eq:delta-MQ} and \eqref{Eq:epsilon-MQ}, respectively, 
	\begin{align*}
		\Pr_{\mathsf{pp}\gets\mathsf{Setup}(1^\kappa)} \Big[ \forall\,\textnormal{PPT }\mathcal{A},\ \forall\,\mathcal{U}\in\mathcal{U}_{k_p,Q}: \mathsf{Adv}_{\mathcal{A},\mathsf{HC\text{-}Leak}}^{(\mathsf{pp},\mathcal{U})}(\kappa) \leq \epsilon_\mathrm{MQ}(\kappa,\mathcal{A},\mathcal{U}) \Big] \geq 1-\delta_\mathrm{MQ}.
	\end{align*}
    That is, except with probability at most \(\delta_\mathrm{MQ}\) over \(\mathsf{pp}\gets\mathsf{Setup}(1^\kappa)\), every PPT adversary \(\mathcal{A}\) has advantage at most \(\epsilon_\mathrm{MQ}(\kappa,\mathcal{A},\mathcal{U})\) for every admissible pattern \(\mathcal{U}\). 
	Consequently, \(\mathcal{F}_\mathrm{MQ}\) is \((\delta_\mathrm{MQ},\epsilon_\mathrm{MQ})\)-private for \(\mathsf{SEC}\) in the multi-query \(\mathsf{HC\text{-}LLR}\) sense of \Cref{Definition:HC-LLR}, and hence \(\delta_\mathrm{MQ}\)-private because \(\epsilon_\mathrm{MQ}(\kappa,\mathcal{A},\mathcal{U})\) is negligible in \(\kappa\) for every PPT adversary \(\mathcal{A}\) and every admissible pattern \(\mathcal{U}\).
    For the certified expander-based construction, \(\delta_\mathrm{MQ}=0\).
\end{theorem}

\begin{proof}[Proof sketch] 
	Fix \(\mathsf{pp}\in\mathsf{Good}_\kappa\), a PPT adversary \(\mathcal{A}\), and a pattern \(\mathcal{U}\in\mathcal{U}_{k_p,Q}\). 
	If the auxiliary view of \(\mathcal{U}\) is empty, \(\mathcal{B}_\mathsf{pp}\) directly distinguishes \(\mathsf{MLWE}_{p,\ell-p,q,d,\sigma_\mathbf{s}}\), so \(\mathsf{Adv}_{\mathcal{A},\mathsf{HC\text{-}Leak}}^{(\mathsf{pp},\mathcal{U})}(\kappa)\leq\mathsf{Adv}_{\mathcal{B}_\mathsf{pp},\mathsf{MLWE}}(\kappa)\). 
	Otherwise, \(\mathsf{Good}_\mathrm{MQ}(\mathsf{pp})\) ensures that \Cref{Lemma:Multi-HCLeak-to-MLWE} applies uniformly over \(\mathbf{c}\in\mathcal{C}^Q\), with source width \(\sigma_\mathrm{MQ}\) and loss \(2\epsilon'+2\epsilon\), so \(\mathsf{Adv}_{\mathcal{A},\mathsf{HC\text{-}Leak}}^{(\mathsf{pp},\mathcal{U})}(\kappa)\leq\mathsf{Adv}_{\mathcal{B}_\mathsf{pp},\mathsf{MLWE}}(\kappa)+2\epsilon'+2\epsilon\). 
	In both cases, the advantage is at most \(\epsilon_\mathrm{MQ}(\kappa,\mathcal{A},\mathcal{U})\) by its definition as a maximum over \(\mathsf{Good}_\kappa\). 
	The result follows because \(\mathsf{Good}_\kappa\) has probability \(1-\delta_\mathrm{MQ}\) and depends neither on \(\mathcal{A}\) nor on \(\mathcal{U}\). 
	The complete proof is given in Appendix~\ref{Appendix:Privacy-Proofs-Reduction}. 
\end{proof}

The two MLWE assumptions correspond to different cases: \(\sigma_\mathrm{MQ}\) applies uniformly to every nonempty auxiliary view, whereas the empty view uses the original secret width \(\sigma_\mathbf{s}\). 
By \eqref{Eq:Sigma-MQ-Uniform-Parameter}, \(\sigma_\mathrm{MQ}\leq\sigma_\mathbf{s}/\sqrt{2}\). 
The theorem states both assumptions separately, but for our concrete instantiations it suffices in practice to evaluate hardness at \(\sigma_\mathrm{MQ}\); see \Cref{Section:SEC-Concrete-Requirements}.

\section{Parameters and Evaluation}\label{Section:SEC-Empirical-Parameter}

We empirically evaluate concrete \(\mathsf{SEC}\) distribution matrices for party counts \(n\in\{10,15,\dots,60\}\). 
Our purpose is not to characterize random expander-code instances, but to identify concrete matrices satisfying the structural, privacy, reconstruction, and shortness properties established in \Cref{Section:SEC,Section:Correctness-Shortness-Linearity,Section:Secret-Sharing-Privacy-Analysis}. 
We report all \(17\) selected matrices, including multiple selected matrices for some party counts. 
Among the selected matrices with \(n\geq25\), the \(n=40\) instance achieves the largest observed ratio \(\tau_p/\tau_c\approx0.192\), while the \(n=60\) instance is the largest configuration considered. 
We use these two instances for the concrete privacy instantiations.
One should be able to obtain results for larger \(n\) values by expending more computation.

\subsection{Search and Certification Methodology}\label{Section:SEC-Search-Certification}

\subsubsection{Search Space}\label{Section:SEC-Search-Space}

For each party count, we consider masking dimensions in the regimes \(m\approx n/2\) and \(m\approx2n/3\), whenever applicable.
For each pair \((n,m)\), we consider candidate parameter tuples \((n,m,c_\mathrm{exp},\gamma,\epsilon_\mathrm{exp})\), where \(\epsilon_\mathrm{exp}\in(0,1/2)\) and \(\gamma \geq 0.1\).
We search increasing values of \(c_\mathrm{exp}\) and retain the smallest left degree for which the search finds a suitable certified lossless-expander instance.
The selected instances use \(c_\mathrm{exp}\in\{2,3\}\).

The principal constraint is obtaining a sufficiently large certified value of \(\gamma\), since the certified expansion range determines the private-coalition threshold \(k_p:=\left\lceil2(1-\epsilon_\mathrm{exp})\lfloor\gamma n\rfloor\right\rceil-1\).
For every reported instance, \(\gamma\) is chosen so that \(\gamma n\in\mathbb{Z}\), and the underlying full-precision parameters satisfy \(\left\lceil2(1-\epsilon_\mathrm{exp})\lfloor\gamma n\rfloor\right\rceil-1=\lfloor\gamma n\rfloor\).
Hence \(k_p=\gamma n\) and \(\tau_p:=k_p/n=\gamma\).

The values of \(\gamma\), \(\epsilon_\mathrm{exp}\), \(\alpha_{k_p}\), \(\|\mathbf{\hat{G}}\|_2\), and \(\|\mathbf{G}\|_2\) are reported to three decimal places.
Expansion verification and the computation of \(k_p\) use the underlying full-precision parameters rather than the rounded table entries.

\subsubsection{Expander Sampling and Certification}\label{Section:SEC-Expander-Sampling}

For each candidate tuple, we sample a \(c_\mathrm{exp}\)-left-regular bipartite graph with left vertices \(L=[n]\) and right vertices \(R=[m]\) by letting each left vertex independently choose a uniformly random \(c_\mathrm{exp}\)-subset of \(R\), so that \(\mathbf{H} \in \{0,1\}^{m\times n}\) has exactly \(c_\mathrm{exp}\) nonzero entries in each column.
Every sampled candidate is then certified by exhaustive verification, which the regime \(n\leq60\) permits: for every \(S\subseteq L\) with \(|S| \leq \lfloor\gamma n\rfloor\), we verify \(|N(S)| \geq (1-\epsilon_\mathrm{exp}) c_\mathrm{exp}|S|\), and only candidates passing every check are selected.
Thus, the concrete matrices rely on exhaustive certification rather than \Cref{Theorem:ExpanderExistence}, and \Cref{Lemma:Expander-implies-UnimodularMinor} converts the certified expansion property into the unimodular submatrices required by the structural privacy condition.

\subsubsection{Construction of the Distribution Matrix}\label{Section:SEC-Distribution-Matrix-Construction}

For each selected \(\mathbf{H}\), we construct \(\mathbf{G} = \begin{bmatrix} \mathbf{g} \ \mathbf{\hat{G}} \end{bmatrix} \) as in \Cref{Definition:SEC}, with \(\mathbf{g}\) from \Cref{Algorithm:Construct-g-disjoint-approach}; for every selected instance, \( \|\mathbf{g}\|_\infty = 1\), and hence \(\beta_\mathbf{g}=1\).
Among the candidates satisfying the certified expansion and generator-matrix construction requirements, we select representative instances based primarily on their empirical reconstruction thresholds and observed reconstruction-vector magnitudes, while also accounting for their privacy-side matrix statistics.
For some party counts, more than one instance is selected to compare different choices of \(m\) or \(c_\mathrm{exp}\).

\subsection{Structural and Reconstruction Parameters}\label{Section:SEC-Structural-Reconstruction-Parameters}

\subsubsection{Selected Instances and Privacy-Side Parameters}\label{Section:SEC-Privacy-Side-Parameters}

The privacy-side statistics of all selected instances are reported in \Cref{Table:SEC-Selected-Privacy-Statistics}.
The threshold \(\hat{d}^{\bot}-1\) is the certified rational no-cancellation threshold associated with the code having parity-check matrix \(\mathbf{P}\mathbf{H}\) (\Cref{Lemma:DualDistance-NoiseCancellation-Q}).
The quantities \(\alpha_{k_p}\) and \(\|\mathbf{\hat{G}}\|_2\) enter the uniform precision parameter \(\Xi_\mathrm{MQ}\) in \eqref{Eq:Xi-MQ}, while \(\beta_{k_p}\) and \(\gamma_{k_p}\) enter the explicit kernel-smoothing condition \eqref{Eq:sigma-eff-condition-explicit-practical}.
The quantities \(\alpha_{k_p}\), \(\beta_{k_p}\), and \(\gamma_{k_p}\) are evaluated exhaustively over their subset domains.

To compare the integral and rational no-cancellation thresholds independently of \(\mathbf{g}\), let \(\hat{d}_\mathbb{Z}^\bot:=\min_p\hat{d}_p^\bot\), where \(\hat{d}_p^\bot\) is the minimum distance of \(\hat{C}_p^\bot:=\ker_{\mathbb{F}_p}(\overline{\mathbf{H}})\), and the minimum ranges over all primes \(p\).
By the Smith normal form, \(\mathrm{Col}_\mathbb{Z}(\mathbf{\hat{G}}_U) =\mathbb{Z}^{|U|}\) for every \(|U|\leq\hat{d}_\mathbb{Z}^\bot-1\). Moreover, \(k_p \leq \hat{d}_\mathbb{Z}^\bot-1 \leq \hat{d}_\mathbb{Q}^\bot-1\), where the first inequality follows from \Cref{Lemma:Column-Space-Ghat-U-Expander} and the second from \Cref{Lemma:Code-Distance-Preservation}. 
Thus, a strict gap between \(k_p\) and \(\hat{d}^\bot-1\) may reflect both slack in the expander-based certificate and a genuine gap between the integral and rational no-cancellation thresholds. 
For coalitions satisfying \(k_p<|U|\leq\hat{d}^\bot-1\), rational cancellation remains impossible, but the expander certificate no longer guarantees resistance to modular cancellation. 
If \((\mathbf{g})_U\notin \mathrm{Col}_\mathbb{Z}(\mathbf{\hat{G}}_U)\), such a coalition admits a modular noise-cancellation attack by \Cref{Lemma:ComputationalPrivacy-SEC-Column-Z}.
Coalitions beyond \(k_p\) lie outside the stated privacy guarantee.

\begin{table}[!htp]
	\caption{Selected \(\mathsf{SEC}\) instances and their privacy-side matrix statistics}
	\label{Table:SEC-Selected-Privacy-Statistics}%
	\setlength{\tabcolsep}{3.5pt}
    \renewcommand{\arraystretch}{1.15}
	\begin{tabular}{@{}cccccccccccc@{}}
		\toprule
		No. & \(n\) & \(k=m+1\) & \(c_\mathrm{exp}\) & \(\gamma\) (\(=\tau_p\)) & \(\epsilon_\mathrm{exp}\) & \(k_p\) & \(\hat{d}^\bot-1\) & \(\alpha_{k_p}\) & \(\|\mathbf{\hat{G}}\|_2\) & \(\beta_{k_p}\) & \(\gamma_{k_p}\) \\
		\midrule
		1 & \(10\) & \(6\) & \(2\) & \(0.200\) & \(0.333\) & \(2\) & \(3\) & \(1.000\) & \(2.828\) & \(1.732\) & \(0.618\) \\
		2 & \(10\) & \(6\) & \(3\) & \(0.200\) & \(0.333\) & \(2\) & \(3\) & \(1.000\) & \(4.243\) & \(2.236\) & \(0.618\) \\
		3 & \(15\) & \(8\) & \(3\) & \(0.200\) & \(0.450\) & \(3\) & \(3\) & \(0.792\) & \(4.440\) & \(2.646\) & \(0.445\) \\
		4 & \(20\) & \(10\) & \(2\) & \(0.150\) & \(0.333\) & \(3\) & \(3\) & \(0.765\) & \(3.000\) & \(2.000\) & \(0.445\) \\
		5 & \(20\) & \(10\) & \(3\) & \(0.150\) & \(0.490\) & \(3\) & \(3\) & \(0.792\) & \(4.634\) & \(2.646\) & \(0.445\) \\
		6 & \(25\) & \(13\) & \(2\) & \(0.120\) & \(0.333\) & \(3\) & \(3\) & \(0.765\) & \(3.143\) & \(2.000\) & \(0.445\) \\
		7 & \(25\) & \(13\) & \(3\) & \(0.120\) & \(0.450\) & \(3\) & \(3\) & \(0.792\) & \(4.433\) & \(2.646\) & \(0.445\) \\
		8 & \(30\) & \(15\) & \(2\) & \(0.100\) & \(0.333\) & \(3\) & \(3\) & \(0.765\) & \(3.298\) & \(2.000\) & \(0.445\) \\
		9 & \(30\) & \(15\) & \(3\) & \(0.100\) & \(0.450\) & \(3\) & \(3\) & \(0.792\) & \(4.628\) & \(2.646\) & \(0.445\) \\
		10 & \(30\) & \(20\) & \(2\) & \(0.133\) & \(0.450\) & \(4\) & \(5\) & \(0.618\) & \(2.769\) & \(2.236\) & \(0.347\) \\
		11 & \(30\) & \(20\) & \(3\) & \(0.133\) & \(0.450\) & \(4\) & \(5\) & \(0.618\) & \(4.053\) & \(2.665\) & \(0.295\) \\
		12 & \(35\) & \(18\) & \(3\) & \(0.114\) & \(0.450\) & \(4\) & \(5\) & \(0.618\) & \(4.459\) & \(2.665\) & \(0.295\) \\
		13 & \(40\) & \(20\) & \(3\) & \(0.125\) & \(0.480\) & \(5\) & \(5\) & \(0.464\) & \(4.432\) & \(2.940\) & \(0.193\) \\
		14 & \(45\) & \(23\) & \(3\) & \(0.111\) & \(0.470\) & \(5\) & \(5\) & \(0.464\) & \(4.427\) & \(2.868\) & \(0.193\) \\
		15 & \(50\) & \(25\) & \(3\) & \(0.100\) & \(0.470\) & \(5\) & \(5\) & \(0.464\) & \(4.529\) & \(2.940\) & \(0.193\) \\
		16 & \(55\) & \(37\) & \(3\) & \(0.109\) & \(0.450\) & \(6\) & \(\geq 8\) & \(0.445\) & \(3.863\) & \(2.893\) & \(0.152\) \\
		17 & \(60\) & \(40\) & \(3\) & \(0.100\) & \(0.450\) & \(6\) & \(\geq 8\) & \(0.518\) & \(3.937\) & \(2.893\) & \(0.152\) \\
		\botrule
	\end{tabular}
	\footnotetext{Note: the entries \(\hat{d}^\bot-1 \geq 8\) are certified lower bounds.
	Here \(\beta_{k_p}:=\max_{U}\|\mathbf{\hat{G}}_U\|_2\) and \(\gamma_{k_p}:=\min_U\sigma_{\min}(\mathbf{M}_U)\) over \(1\leq|U|\leq k_p\), where \(\mathbf{M}_U\) is the peeled unimodular submatrix.}
\end{table}

\subsubsection{Reconstruction Thresholds and Certification Probability}\label{Section:SEC-Reconstruction-Thresholds}

\paragraph{Deterministic Reconstruction Threshold}

A deterministic correctness threshold \(k_c^\mathsf{det}\) satisfies \(\mathbf{e}_1^\top \in \mathrm{Row}_\mathbb{Z}(\mathbf{G}_T)\) for every \(T\subseteq[n]\) with \(|T|\geq k_c^\mathsf{det}\).
Since certification is preserved under enlarging a coalition, it is sufficient to verify this condition for subsets of size exactly \(k_c^\mathsf{det}\).
In the empirical search, the deterministic threshold is obtained either from the planted certified supports through \Cref{Lemma:Deterministic-Correctness-Lower-Bound} or by exhaustive verification whenever the number of relevant subsets is sufficiently small.
In the exhaustive case, we report the smallest verified threshold.
When the planted-support argument is used, the reported value is a certified deterministic threshold but need not be the smallest possible one.

\paragraph{Certification Probability and Empirical Correctness}

\begin{definition}[Certification probability and average-case correctness]\label{Definition:Certification-Probability}
	Fix public parameters \(\mathsf{pp}\) containing \(\mathbf{G}\in\mathbb{Z}^{n\times k}\).
	For \(t\in[n]\), define
	\begin{align}\label{Eq:Certification-Probability}
		\Pr_\mathrm{cert}(t,\mathbf{G}) := \Pr_{T\stackrel{\$}{\gets}\binom{[n]}{t}} \left[ \mathbf{e}_1^\top \in \mathrm{Row}_\mathbb{Z}(\mathbf{G}_T) \right].
	\end{align}
	For \(k_c\in[n]\) and \(\epsilon_c\in[0,1]\), the fixed instance \((\mathsf{SEC},\mathsf{pp})\) is \(\epsilon_c\)-correct for random coalitions above threshold \(k_c\) if \(\Pr_\mathrm{cert}(t,\mathbf{G})\geq1-\epsilon_c\) for every \(t\geq k_c\).
\end{definition}

Thus, \(\Pr_\mathrm{cert}(t,\mathbf{G})\) is the fraction of size-\(t\) coalitions certified by the integer row-span condition, and it is nondecreasing in \(t\) because certification is monotone.
We compute it exactly when \(\binom{n}{t}\leq10^6\), and otherwise estimate it from \(N_t\) distinct uniformly sampled subsets, with \(10^5\leq N_t\leq10^6\) and \(N_t=10^6\) near \(t=n/2\).
The empirical threshold \(k_c\) is the smallest tested value for which the exact or estimated probability is at least \(0.99\) for every tested \(t\geq k_c\), that is, \(\epsilon_c\leq0.01\); when sampling is used, \(k_c\) is an empirical estimate rather than an exhaustive certification.

\begin{figure}[!htp]
    \centering
    \begin{subfigure}[b]{0.70\linewidth}
        \centering
        \includegraphics[width=\linewidth]{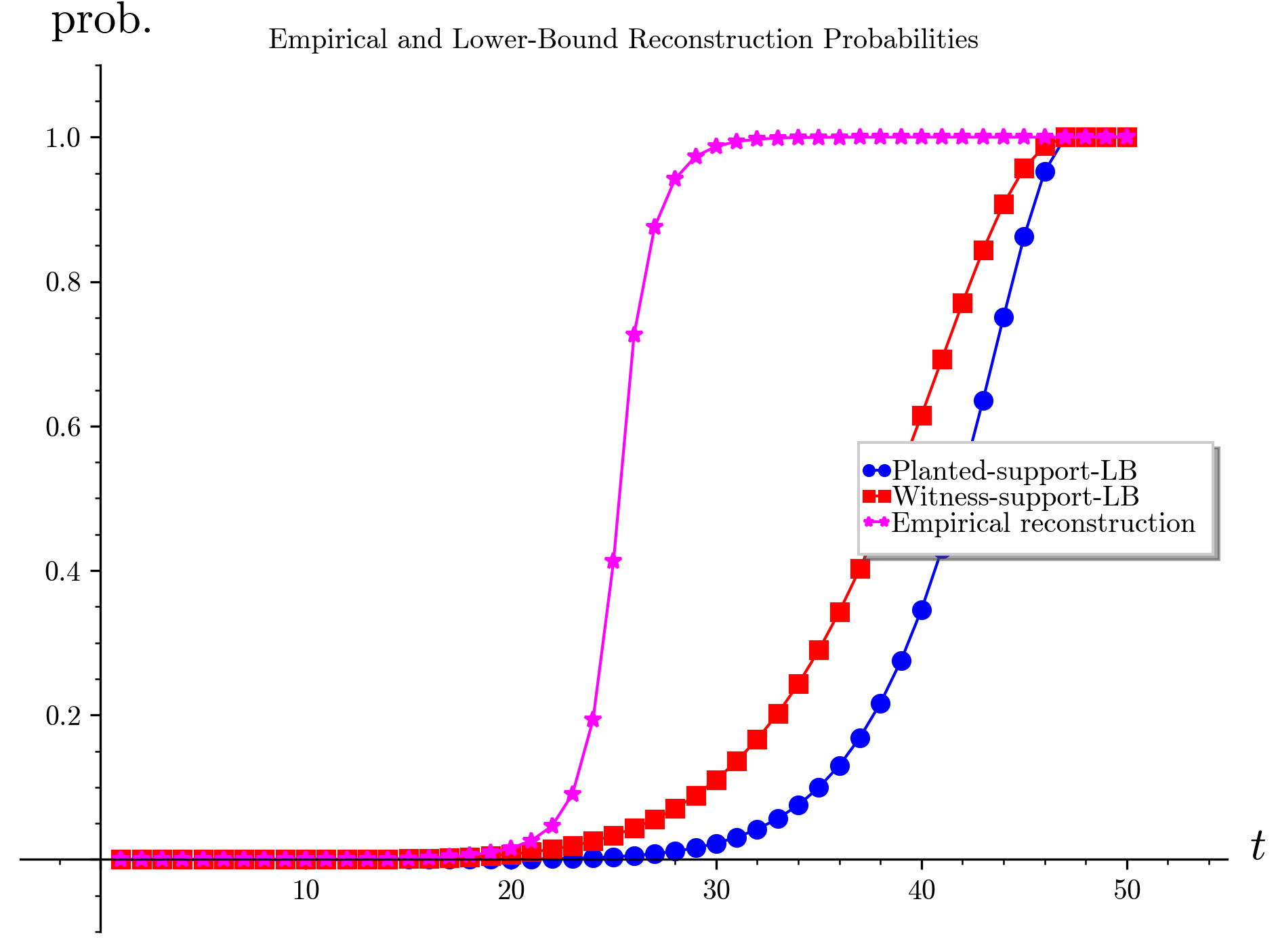}
        \caption{\(n=50\)}
        \label{Figure:Correctness-50}
    \end{subfigure}
    \hfill
    \begin{subfigure}[b]{0.70\linewidth}
        \centering
        \includegraphics[width=\linewidth]{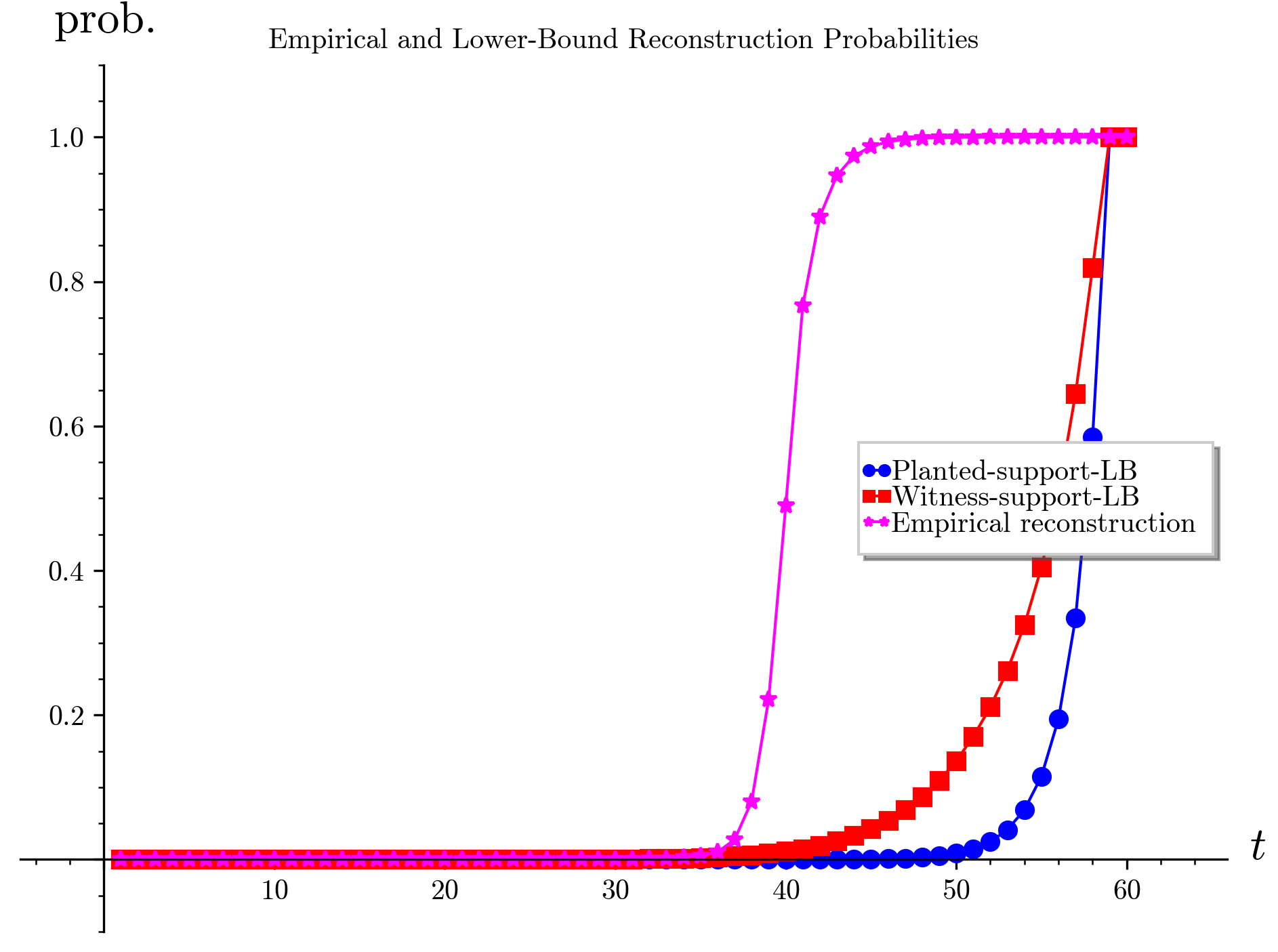}
        \caption{\(n=60\)}
        \label{Figure:Correctness-60}
    \end{subfigure}
    \caption{Empirical certification probability and the planted- and witness-support lower bounds for the selected instances with \(n=50\) and \(n=60\).}
    \label{Figure:Correctness-50-60}
\end{figure}

\Cref{Figure:Correctness-50-60} compares the exact or estimated certification probability of the selected instances with \(n=50\) and \(n=60\) with the planted-support and witness-support lower bounds of \Cref{Corollary:Correctness-LB-T,Corollary:Correctness-LB-Witness}, which describe only sufficient conditions for certification.
For \(n=50\), the empirical threshold is \(k_c=31\), whereas the deterministic threshold reported in \Cref{Table:SEC-Reconstruction-Statistics} is \(k_c^\mathsf{det}=45\).
For \(n=60\), the corresponding thresholds are \(k_c=46\) and \(k_c^\mathsf{det}=56\).
These results illustrate that random coalitions typically reconstruct substantially earlier than the conservative worst-case deterministic threshold.

\begin{figure}[!htp]
    \centering
    \includegraphics[width=0.85\linewidth]{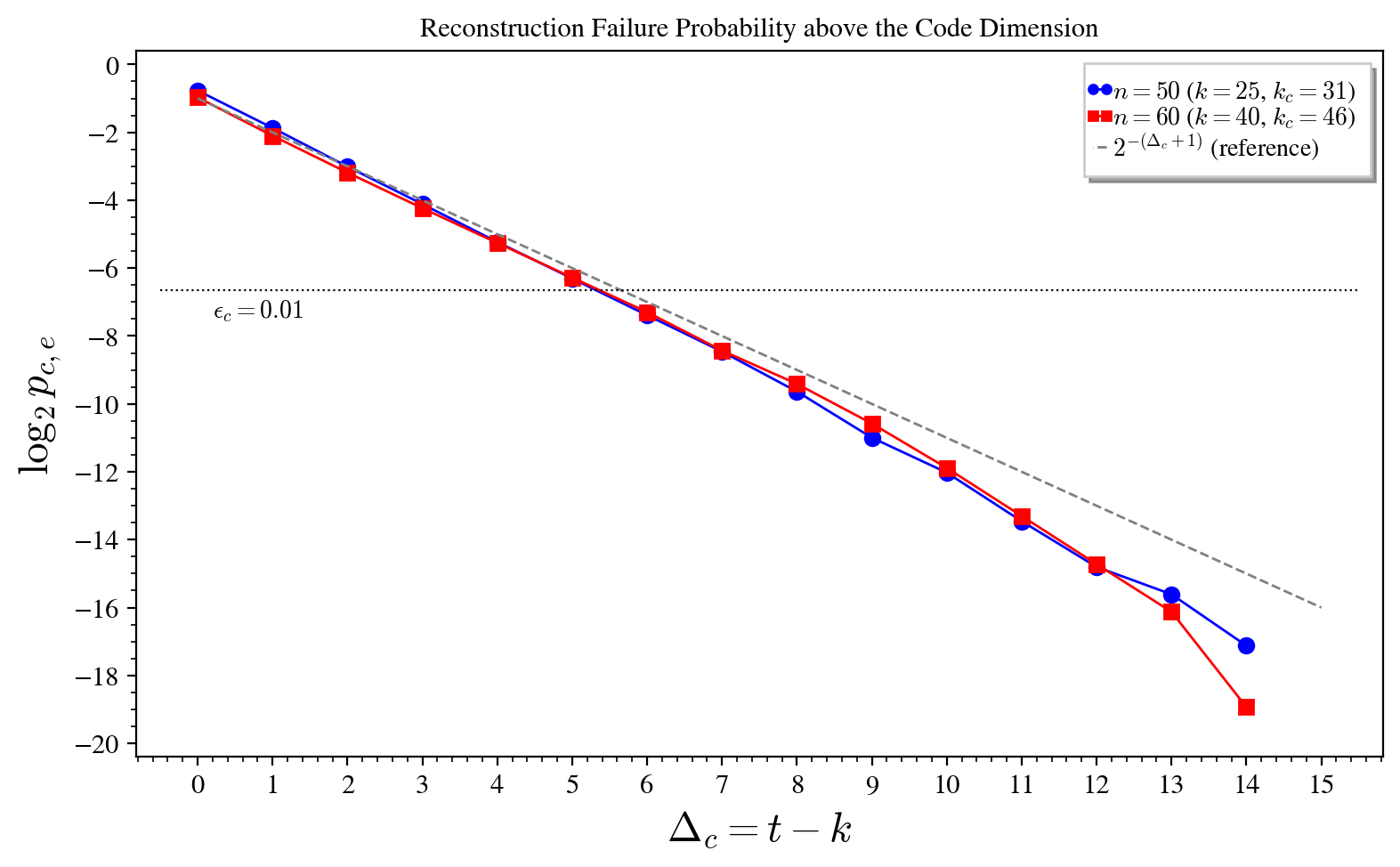}
    \caption{Reconstruction failure probability \(p_{c,e}(t)=1-\Pr_\mathrm{cert}(t,\mathbf{G})\) against the excess coalition size \(\Delta_c=t-k\) for the selected instances with \(n=50\) and \(n=60\). 
    The dashed line is \(2^{-(\Delta_c+1)}\), and the dotted line marks \(\epsilon_c=0.01\).}
    \label{Figure:Failure-Tail-50-60}
\end{figure}

To examine how the reconstruction failure probability decreases as the coalition size exceeds the code dimension, let \(p_{c,e}(t):=1-\Pr_\mathrm{cert}(t,\mathbf{G})\), and let \(\Delta_c:=t-k\) denote the excess coalition size over \(k=m+1\). 
\Cref{Figure:Failure-Tail-50-60} plots \(\lg p_{c,e}\) against \(\Delta_c\in[0,15]\) for the same two instances. 
For both instances, \(\lg p_{c,e}\) is approximately linear in \(\Delta_c\): a least-squares fit over \(0\leq\Delta_c\leq12\) gives slopes of \(-1.15\) for \(n=50\) and \(-1.11\) for \(n=60\), so each additional party reduces the failure probability by a factor of approximately \(2^{1.1}\). 
The two curves nearly coincide and remain close to \(2^{-(\Delta_c+1)}\), suggesting that, for these instances, the failure probability depends primarily on \(\Delta_c\), rather than separately on \(n\) or \(k\). 
This empirical tail is consistent with the observed thresholds: the failure probability first drops below \(\epsilon_c=0.01\) at \(\Delta_c=6\), giving \(k_c=k+6\) in both cases. 
It also suggests that each halving of \(\epsilon_c\) requires roughly one additional party, until \(k_c^\mathsf{det}\) guarantees reconstruction for every coalition. 
The points with \(\Delta_c\geq13\) are based on only a few observed failures among at most \(10^6\) samples and are therefore excluded from the fit.

\subsection{Reconstruction-Coefficient Evaluation}\label{Section:SEC-Reconstruction-Coefficient-Evaluation}

\subsubsection{Reconstruction-Vector Statistics}\label{Section:SEC-Reconstruction-Vector-Statistics}

The reconstruction-vector experiment considers coalition sizes \(t\geq k_c\), corresponding to the empirical high-certification regime of \Cref{Section:SEC-Reconstruction-Thresholds}.
Computing a reconstruction vector requires the row-HNF of \(\mathbf{G}_T\), LLL reduction of a basis of \(\ker_\mathbb{Z}(\mathbf{G}_T^\top)\), and the decoding and refinement steps of \Cref{Algorithm:Construct-Small-Lambda}, which is considerably more expensive than the row-span test used for the certification probability in \Cref{Section:SEC-Reconstruction-Thresholds}.
We therefore use a smaller sample per coalition size: for each tested size \(t\), let \(\mathcal{T}_\mathrm{test}^{t}(\mathbf{G})\) be the set of tested size-\(t\) subsets, obtained by enumerating all size-\(t\) subsets when \(\binom{n}{t}\leq30000\), and otherwise by sampling at most \(30000\) distinct size-\(t\) subsets uniformly without replacement.
Every tested subset \(T\) is first checked for certification, and only certified subsets contribute reconstruction vectors; when sampling is used, the number contributing at a given size may therefore be smaller than \(30000\).
For each certified subset, we apply \Cref{Algorithm:Construct-Small-Lambda} with \(\delta_\mathsf{LLL}=0.99\), obtaining a vector \(\bm{\lambda}_T\) that satisfies \(\bm{\lambda}_T^\top\mathbf{G}_T=\mathbf{e}_1^\top\) by the exact certification step.
We report \(\hat{\beta}_{\bm{\lambda}} := \max_T\|\bm{\lambda}_T\|_\infty\) and \(\hat{\omega}_{\bm{\lambda}} := \max_T wt_H(\bm{\lambda}_T)\), where both maxima range over the tested certified subsets with \(|T|\geq k_c\), not over every certified coalition of size at least \(k_c\).
The reconstruction statistics of all selected instances are reported in \Cref{Table:SEC-Reconstruction-Statistics}.

\begin{table}[!htp]
	\caption{Deterministic and empirical reconstruction statistics for the selected \(\mathsf{SEC}\) instances}
	\label{Table:SEC-Reconstruction-Statistics}%
    \renewcommand{\arraystretch}{1.15}
	\begin{tabular}{@{}cccccccc@{}}
		\toprule
		No. & \(n\) & \(k=m+1\) & \(k_c^\mathsf{det}\) & \(k_c\) & \(\tau_c=k_c/n\) & \(\hat{\beta}_{\bm{\lambda}}\) & \(\hat{\omega}_{\bm{\lambda}}\) \\
		\midrule
		1 & \(10\) & \(6\) & \(9\) & \(8\) & \(0.800\) & \(1\) & \(4\) \\
		2 & \(10\) & \(6\) & \(9\) & \(8\) & \(0.800\) & \(1\) & \(4\) \\
		3 & \(15\) & \(8\) & \(13\) & \(12\) & \(0.800\) & \(1\) & \(6\) \\
		4 & \(20\) & \(10\) & \(17\) & \(14\) & \(0.700\) & \(1\) & \(6\) \\
		5 & \(20\) & \(10\) & \(16\) & \(14\) & \(0.700\) & \(2\) & \(12\) \\
		6 & \(25\) & \(13\) & \(21\) & \(16\) & \(0.640\) & \(1\) & \(10\) \\
		7 & \(25\) & \(13\) & \(22\) & \(18\) & \(0.720\) & \(4\) & \(17\) \\
		8 & \(30\) & \(15\) & \(25\) & \(19\) & \(0.633\) & \(1\) & \(8\) \\
		9 & \(30\) & \(15\) & \(26\) & \(20\) & \(0.667\) & \(3\) & \(17\) \\
		10 & \(30\) & \(20\) & \(28\) & \(25\) & \(0.833\) & \(1\) & \(12\) \\
		11 & \(30\) & \(20\) & \(28\) & \(25\) & \(0.833\) & \(7\) & \(25\) \\
		12 & \(35\) & \(18\) & \(31\) & \(24\) & \(0.686\) & \(4\) & \(22\) \\
		13 & \(40\) & \(20\) & \(35\) & \(26\) & \(0.650\) & \(4\) & \(24\) \\
		14 & \(45\) & \(23\) & \(41\) & \(29\) & \(0.644\) & \(7\) & \(27\) \\
		15 & \(50\) & \(25\) & \(45\) & \(31\) & \(0.620\) & \(4\) & \(28\) \\
		16 & \(55\) & \(37\) & \(51\) & \(43\) & \(0.782\) & \(6\) & \(39\) \\
		17 & \(60\) & \(40\) & \(56\) & \(46\) & \(0.767\) & \(8\) & \(42\) \\
		\botrule
	\end{tabular}
	\footnotetext{Note: \(\hat{\beta}_{\bm{\lambda}}\) and \(\hat{\omega}_{\bm{\lambda}}\) are observed maxima over the tested certified subsets of sizes at least \(k_c\).}
\end{table}

\subsubsection{Observed Reconstruction-Coefficient Bounds}\label{Section:SEC-Observed-Reconstruction-Coefficient-Bounds}

To examine how representative the largest observed coefficient is, we also consider the empirical distribution of \(\|\bm{\lambda}_T\|_\infty\).
For each coalition size \(t\), let \(\mathcal{T}_\mathrm{cert}^{t}(\mathbf{G}) := \big\{ T\in \mathcal{T}_\mathrm{test}^{t}(\mathbf{G}): \mathbf{e}_1^\top \in \mathrm{Row}_\mathbb{Z}(\mathbf{G}_T) \big\}\) denote the tested certified subsets of size \(t\).
For a range \(I\subseteq[k_c,n]\), define \(F_\infty^I(b) := \sum_{t\in I} \left| \left\{ T\in \mathcal{T}_\mathrm{cert}^{t}(\mathbf{G}): \|\bm{\lambda}_T\|_\infty = b \right\} \right|\) and \(\widehat{F}_\infty^I(b) := \frac{ F_\infty^I(b) }{ \sum_{b'\geq0} F_\infty^I(b') }\).
Thus, \(\widehat{F}_\infty^I\) is the pooled empirical distribution of \(\|\bm{\lambda}_T\|_\infty\) over the tested certified subsets whose sizes lie in \(I\).

\begin{figure}[!htp]
    \centering
    \includegraphics[width=1\linewidth]{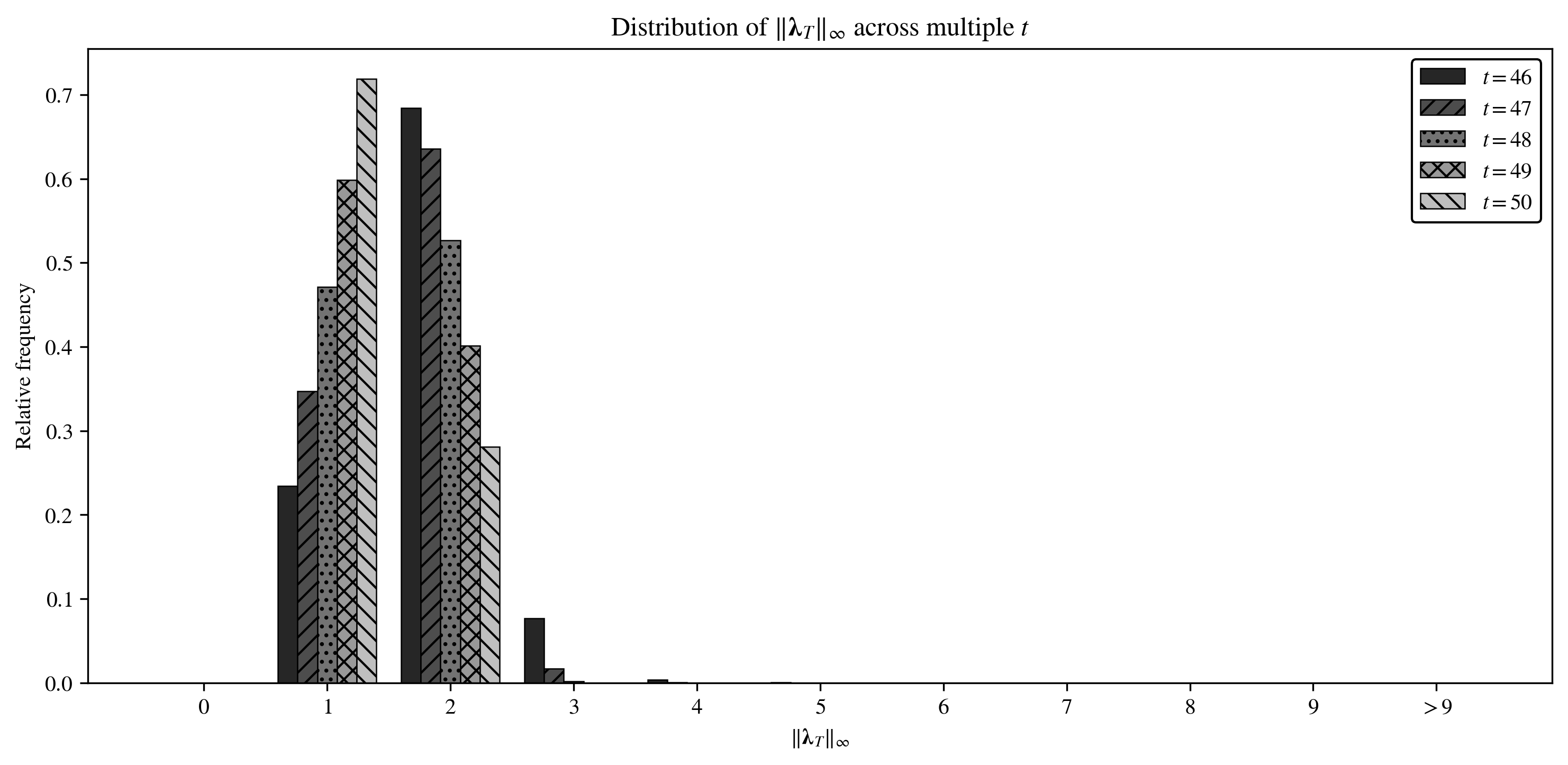}
	\caption{Pooled relative-frequency distribution of the observed values of \(\|\bm{\lambda}_T\|_\infty\) over the tested certified subsets with \(|T|\in[46,50]\) for the selected \(n=60\) instance.}
    \label{Figure:Relative-Frequency-46-50}
\end{figure}

For the selected \(n=60\) instance, we report the range \(I=[46,50]\), which begins at the empirical threshold \(k_c=46\), in \Cref{Figure:Relative-Frequency-46-50}.
For this instance, \(\|\bm{\lambda}_T\|_\infty\leq2\) for \(99.18\%\) of the tested certified subsets, and the distribution shifts toward smaller values as the coalition size increases, consistent with the additional freedom provided by larger reconstructing coalitions.
In particular, \(\|\bm{\lambda}_T\|_\infty \leq 2\) for all observed \(T\) such that \(|T| \in [51,60]\).
The observed maximum \(\hat{\beta}_{\bm{\lambda}}=8\) is attained by a single subset, of the smallest size \(|T|=46\).

The \(n=40\) instance is even more concentrated: \(\|\bm{\lambda}_T\|_\infty\leq2\) for \(99.99\%\) of the tested certified subsets, and the observed maximum \(\hat{\beta}_{\bm{\lambda}}=4\) is attained by only three subsets, all of the smallest size \(|T|=26\).

We therefore set \(\beta_{\bm{\lambda}}:=\hat{\beta}_{\bm{\lambda}}+1\), that is, \(\beta_{\bm{\lambda}}=5\) for \(n=40\) and \(\beta_{\bm{\lambda}}=9\) for \(n=60\).
For the Hamming weight, which is trivially bounded by \(|T|\leq n\), we use \(\omega_{\bm{\lambda}}=25\) and \(\omega_{\bm{\lambda}}=45\), slightly above the observed maxima \(\hat{\omega}_{\bm{\lambda}}=24\) and \(42\), respectively.
If an application enforces both bounds, as discussed in \Cref{Section:Bounding-Reconstruction-Coefficients}, they become additional admissibility conditions for authorized sets; none of the tested subsets is rejected under the selected values.

\subsection{Concrete Privacy Instantiations}\label{Section:SEC-Concrete-Privacy}

\subsubsection{Gaussian Tail Parameters}\label{Section:SEC-Gaussian-Tail}

The share bound of \Cref{Section:SEC-Share-Shortness} and the requirements below use explicit tail bounds for the three families of Gaussian samples in \(\mathsf{SEC}\): the secret \(\bm{s}\), the masking matrix \(\mathbf{R}\), and the leakage masks \(\bm{r}_{i,s}\).
We use the following discrete Gaussian tail bound \cite[Lemma 4.4]{lyubashevsky2012lattice}, stated for the standard-deviation convention of \eqref{Eq:DiscreteGaussianDist}, together with an explicit sufficient condition on the tail factor.
We use \(\sigma_{\bm{s}}=\sigma_\mathbf{s}\) and \(\beta_{\bm{s}}=\beta_\mathbf{s}\), since \(\bm{s}\in\mathcal{R}^\ell\) and its coefficient embedding \(\mathbf{s}\in\mathbb{Z}^D\) represent the same secret.
The choice of notation indicates only the representation.

\begin{lemma}[Discrete Gaussian tail bound]\label{Lemma:Discrete-Gaussian-Tail-IA}
	Let \(\sigma>0\), let \(N\geq1\), and let \(\mathbf{x}\gets\mathcal{D}_{\mathbb{Z}^N,\sigma}\).
	For \(\theta>1\), define \(C_\mathrm{tail}(\theta) := \theta \exp \left( \frac{1-\theta^2}{2} \right)\).
	Then \(C_\mathrm{tail}(\theta) < 1\) and
	\begin{equation}\label{Eq:Discrete-Gaussian-Tail}
		\Pr \left[ \|\mathbf{x}\|_2 > \theta\sigma\sqrt{N} \right] < C_\mathrm{tail}(\theta)^N.
	\end{equation}
	Moreover, for any \(\kappa_\mathrm{tail}>0\), if \(\theta \geq 1+\sqrt{2\kappa_\mathrm{tail}\ln 2/N}\), then \(C_\mathrm{tail}(\theta)^N \leq 2^{-\kappa_\mathrm{tail}}\).
\end{lemma}

The second claim follows from \(-\ln C_\mathrm{tail}(1+x)\geq x^2/2\) for \(x>0\); the proof is given in Appendix~\ref{Appendix:Gaussian-Tail-Size}.
Fix the security parameter \(\kappa\).
For the secret, we choose
\begin{equation}\label{Eq:theta_s}
	\theta_{\bm{s}} \geq 1+\sqrt{\frac{2\kappa\ln 2}{D}}.
\end{equation}
For the masking matrix, let \(\kappa_\mathbf{R} := \kappa + \left\lceil \log_2((k-1)D) \right\rceil\) and choose
\begin{equation}\label{Eq:theta_R-single}
	\theta_\mathbf{R} \geq 1+\sqrt{2\kappa_\mathbf{R}\ln 2}.
\end{equation}
For the leakage masks, when \(Q\geq1\), let \(\kappa_{\bm{r}} := \kappa + \left\lceil \log_2(nQ) \right\rceil \) and choose
\begin{equation}\label{Eq:theta_r}
	\theta_{\bm{r}} \geq 1+\sqrt{\frac{2\kappa_{\bm{r}}\ln 2}{D}}.
\end{equation}
We define \(\beta_{\bm{s}}:=\theta_{\bm{s}}\sigma_{\bm{s}}\sqrt{D}\) and \(\beta_\mathbf{R}:=\theta_\mathbf{R}\sigma_\mathbf{R}\).
If \(Q\geq1\), we additionally define \(\beta_{\bm{r}}:=\theta_{\bm{r}}\sigma_{\bm{r}}\sqrt{D}\).
The secret parameter \(\theta_{\bm{s}}\) controls a single sample in \(\mathbb{Z}^D\), whereas \(\theta_\mathbf{R}\) absorbs a union bound over the \((k-1)D\) entries of \(\mathbf{R}\).
If \(Q\geq1\), \(\theta_{\bm{r}}\) absorbs a union bound over at most \(nQ\) leakage masks.
\Cref{Lemma:Discrete-Gaussian-Tail-IA} then gives
\begin{align}
	\Pr \left[ \|\bm{s}\|_\infty > \beta_{\bm{s}} \right] &\leq 2^{-\kappa}, \label{Eq:Good-tail-bound-kappa-s}\\
	\Pr \left[ \exists (i',j')\in[k-1]\times[D]: \lvert\mathbf{R}[i',j']\rvert > \beta_\mathbf{R} \right] &\leq 2^{-\kappa}, \label{Eq:Good-tail-bound-kappa_bigR} \\
	\Pr \left[ \exists s\in[Q],\ \exists i\in[n]: \|\bm{r}_{i,s}\|_2 > \beta_{\bm{r}} \right] &\leq 2^{-\kappa} \text{ (when \(Q\geq 1\))}, \label{Eq:Good-tail-bound-kappa_smallr}
\end{align}
where \eqref{Eq:Good-tail-bound-kappa-s} uses \(\|\bm{s}\|_\infty\leq\|\bm{s}\|_2\), and the derivations are given in Appendix~\ref{Appendix:Gaussian-Tail-Size}.
We collect these bounds in the following event.

\begin{definition}[Good-tail event]\label{Definition:Good-tail}
	Let \(\mathsf{Good}_\mathsf{tail}\) denote the event that
	\begin{enumerate*}[label=(\arabic*)]
		\item the secret satisfies \(\|\bm{s}\|_\infty\leq\beta_{\bm{s}}\);
		\item every entry of the masking matrix satisfies \(\lvert\mathbf{R}[i',j']\rvert\leq\beta_\mathbf{R}\); and
		\item when \(Q\geq1\), for every query \(s\in[Q]\) and every \(i\in[n]\), the leakage mask satisfies \(\|\bm{r}_{i,s}\|_2\leq\beta_{\bm{r}}\).
	\end{enumerate*}
\end{definition}

\begin{lemma}[Failure probability of the good-tail event]\label{Lemma:Good-tail-Failure} 
	For the choices of \(\theta_{\bm{s}}\) and \(\theta_\mathbf{R}\) in \eqref{Eq:theta_s} and \eqref{Eq:theta_R-single}, and, when \(Q\geq1\), of \(\theta_{\bm{r}}\) in \eqref{Eq:theta_r}, we have \( \Pr[\neg\mathsf{Good}_\mathsf{tail}] \leq 3\cdot2^{-\kappa}\). 
    When \(Q=0\), the leakage-mask condition in \Cref{Definition:Good-tail} is vacuous, and \(\Pr[\neg\mathsf{Good}_\mathsf{tail}] \leq2\cdot2^{-\kappa}\).
\end{lemma}

The lemma follows from \eqref{Eq:Good-tail-bound-kappa-s}--\eqref{Eq:Good-tail-bound-kappa_smallr} by a union bound; the leakage-mask condition is vacuous when \(Q=0\). 
The secret and masking-matrix conditions, together with \Cref{Lemma:bound-for-si}, give \(\max_{i\in[n]}\|\bm{s}_i\|_\infty\leq\beta_\mathsf{sh}=\beta_\mathbf{g}\beta_{\bm{s}}+c_\mathrm{exp}\beta_\mathbf{R}\), so the shortness error of \Cref{Definition:Everywhere-Short-Secret-Sharing} satisfies \(\epsilon_\mathsf{sh}\leq2\cdot2^{-\kappa}\).
The leakage-mask condition is not needed for the privacy or shortness of \(\mathsf{SEC}\); when \(Q\geq1\), it bounds the released leakage blocks by \(\|\bm{z}_{i,s}\|_2\leq\omega_c\sqrt{D}\beta_\mathsf{sh}+\beta_{\bm{r}}\), which an application that verifies partial responses must accept.

\subsubsection{Instantiation Methodology}\label{Section:SEC-Instantiation-Methodology}

We instantiate the selected instances with \(n\in\{40,60\}\) from \Cref{Table:SEC-Selected-Privacy-Statistics}, for \(\kappa\in\{128,192,256\}\).
We fix the ring degree \(d=512\) and the systematic hard-core matrix \(\bm{A}=\begin{bmatrix}\bm{I}_p \ \bm{A}'\end{bmatrix}\in\mathcal{R}_q^{p\times\ell}\), and write \(\ell=p+\ell'\), so that the hardness assumptions are \(\mathsf{MLWE}_{p,\ell',q,d,\sigma}\) for the relevant widths \(\sigma\).
We consider two query budgets, \(Q\in\{0,2^{64}\}\).
The budget \(Q=0\) corresponds to \(\mathsf{SEC}\) as a standalone sharing scheme, in which the adversary observes only the shares of at most \(k_p\) parties; the budget \(Q=2^{64}\) matches the signing-query budgets targeted by lattice-based threshold signatures.
The leakage-description set is \(\mathcal{C} := \{ c\in\mathcal{R}: \|c\|_\infty=1,\ \|c\|_1=\omega_c \}\), so that \(|\mathcal{C}| = 2^{\omega_c}\binom{d}{\omega_c}\) and \(\|\mathbf{L}_s\|_2\leq\omega_c\).
Following the convention of \cite[Section 7]{borin2026hermine}, we set \(\omega_c := \min\{\omega\in[d]: 2^{\omega}\binom{d}{\omega}\geq2^\kappa\}\), which gives \(\omega_c=19\), \(31\), and \(44\) for \(\kappa=128\), \(192\), and \(256\), respectively.
When \(Q=0\), the second term of \(\Xi_\mathrm{MQ}\) in \eqref{Eq:Xi-MQ} vanishes, \(\sigma_\mathrm{eff}=\sigma_\mathbf{R}\) in \eqref{Eq:sigma-eff-definition}, and the width \(\sigma_{\bm{r}}\) plays no role. 
For each of these instances, each pair \((\kappa,Q)\), and each candidate \(\ell\), we choose widths \(\sigma_{\bm{s}}\), \(\sigma_\mathbf{R}\), and \(\sigma_{\bm{r}}\) satisfying the requirements of \Cref{Section:SEC-Concrete-Requirements} so as to minimize the effective share width \(\sigma_\mathsf{sh}\), defined in Appendix~\ref{Appendix:Gaussian-Tail-Size}, which satisfies \(\sigma_\mathsf{sh}\geq \beta_\mathbf{g}\sigma_{\bm{s}}\) and determines the size of one share. 
When \(Q\geq1\), minimizing \(\sigma_\mathsf{sh}\) alone yields no finite optimizer, because increasing \(\sigma_{\bm{r}}\) continually weakens the query-dependent restriction and drives the objective toward its \(Q=0\) limit. 
We therefore minimize \(\sigma_\mathsf{sh}\sigma_{\bm{r}}\), which balances share size against the size of the released leakage masks.
The resulting parameter sets and their sizes are reported in \Cref{Table:SEC-Selected-Parameters,Table:SEC-Selected-Sizes} and discussed in \Cref{Section:SEC-Security-Estimates}.

\begin{table}[!htp]
	\caption{Selected \(\mathsf{SEC}\) parameter sets for the application-driven and fixed \(32\)-bit modulus regimes}
	\label{Table:SEC-Selected-Parameters}%
	\setlength{\tabcolsep}{4pt}
    \renewcommand{\arraystretch}{1.15}
    \begin{tabular}{@{}ccccccccccc@{}} 
		\toprule 
		\((n,k)\) & \(\kappa\) & \((k_p,k_c,k_c^\mathsf{det})\) & \(Q\) & \(\omega_c\) & \(\lceil\lg q\rceil\) & \(\ell\) & \(p=\ell'\) & \(\lg\sigma_{\bm{s}}\) & \(\lg\sigma_\mathbf{R}\) & \(\lg\sigma_{\bm{r}}\) \\ 
		\midrule \multicolumn{11}{c}{application-driven threshold-signature moduli} \\ 
		\midrule 
		\((40,20)\) & \(128\) & \((5,26,35)\) & \(0\) & \(19\) & \(61\) & \(12\) & \(6\) & \(2.957\) & \(4.736\) & -- \\ 
		\((40,20)\) & \(128\) & \((5,26,35)\) & \(2^{64}\) & \(19\) & \(61\) & \(12\) & \(6\) & \(3.730\) & \(5.222\) & \(43.646\) \\ 
		\((40,20)\) & \(192\) & \((5,26,35)\) & \(0\) & \(31\) & \(63\) & \(16\) & \(8\) & \(3.236\) & \(4.995\) & -- \\ 
		\((40,20)\) & \(192\) & \((5,26,35)\) & \(2^{64}\) & \(31\) & \(63\) & \(16\) & \(8\) & \(4.039\) & \(5.478\) & \(44.615\) \\ 
		\((40,20)\) & \(256\) & \((5,26,35)\) & \(0\) & \(44\) & \(64\) & \(20\) & \(10\) & \(3.436\) & \(5.186\) & -- \\ 
		\((40,20)\) & \(256\) & \((5,26,35)\) & \(2^{64}\) & \(44\) & \(64\) & \(20\) & \(10\) & \(4.257\) & \(5.666\) & \(45.313\) \\ 
        \addlinespace 
		\((60,40)\) & \(128\) & \((6,46,56)\) & \(0\) & \(19\) & \(64\) & \(12\) & \(6\) & \(2.463\) & \(5.210\) & -- \\ 
		\((60,40)\) & \(128\) & \((6,46,56)\) & \(2^{64}\) & \(19\) & \(64\) & \(12\) & \(6\) & \(2.670\) & \(5.711\) & \(43.934\) \\ 
		\((60,40)\) & \(192\) & \((6,46,56)\) & \(0\) & \(31\) & \(66\) & \(16\) & \(8\) & \(2.733\) & \(5.468\) & -- \\ 
		\((60,40)\) & \(192\) & \((6,46,56)\) & \(2^{64}\) & \(31\) & \(66\) & \(16\) & \(8\) & \(2.944\) & \(5.969\) & \(44.898\) \\ 
		\((60,40)\) & \(256\) & \((6,46,56)\) & \(0\) & \(44\) & \(66\) & \(20\) & \(10\) & \(2.928\) & \(5.657\) & -- \\ 
		\((60,40)\) & \(256\) & \((6,46,56)\) & \(2^{64}\) & \(44\) & \(66\) & \(20\) & \(10\) & \(3.142\) & \(6.158\) & \(45.593\) \\ 
		\midrule 
		\multicolumn{11}{c}{fixed modulus \(q=2^{32}-10239\)} \\ 
		\midrule 
		\((40,20)\) & \(128\) & \((5,26,35)\) & \(0\) & \(19\) & \(32\) & \(6\) & \(3\) & \(2.952\) & \(4.731\) & -- \\ 
		\((40,20)\) & \(128\) & \((5,26,35)\) & \(2^{64}\) & \(19\) & \(32\) & \(6\) & \(3\) & \(3.724\) & \(5.217\) & \(43.641\) \\ 
		\((40,20)\) & \(192\) & \((5,26,35)\) & \(0\) & \(31\) & \(32\) & \(8\) & \(4\) & \(3.232\) & \(4.992\) & -- \\ 
		\((40,20)\) & \(192\) & \((5,26,35)\) & \(2^{64}\) & \(31\) & \(32\) & \(8\) & \(4\) & \(4.035\) & \(5.475\) & \(44.612\) \\ 
		\((40,20)\) & \(256\) & \((5,26,35)\) & \(0\) & \(44\) & \(32\) & \(10\) & \(5\) & \(3.434\) & \(5.183\) & -- \\ 
		\((40,20)\) & \(256\) & \((5,26,35)\) & \(2^{64}\) & \(44\) & \(32\) & \(10\) & \(5\) & \(4.254\) & \(5.664\) & \(45.310\) \\ 
        \addlinespace 
		\((60,40)\) & \(128\) & \((6,46,56)\) & \(0\) & \(19\) & \(32\) & \(6\) & \(3\) & \(2.458\) & \(5.205\) & -- \\ 
		\((60,40)\) & \(128\) & \((6,46,56)\) & \(2^{64}\) & \(19\) & \(32\) & \(6\) & \(3\) & \(2.665\) & \(5.706\) & \(43.929\) \\ 
		\((60,40)\) & \(192\) & \((6,46,56)\) & \(0\) & \(31\) & \(32\) & \(8\) & \(4\) & \(2.729\) & \(5.465\) & -- \\ 
		\((60,40)\) & \(192\) & \((6,46,56)\) & \(2^{64}\) & \(31\) & \(32\) & \(8\) & \(4\) & \(2.941\) & \(5.966\) & \(44.895\) \\ 
		\((60,40)\) & \(256\) & \((6,46,56)\) & \(0\) & \(44\) & \(32\) & \(10\) & \(5\) & \(2.926\) & \(5.655\) & -- \\ 
		\((60,40)\) & \(256\) & \((6,46,56)\) & \(2^{64}\) & \(44\) & \(32\) & \(10\) & \(5\) & \(3.139\) & \(6.156\) & \(45.590\) \\ 
		\botrule 
	\end{tabular}
    \footnotetext{Note: \(\ell=p+\ell'\) and \(p=\ell'=\ell/2\). 
	All logarithms are base two and are computed from the full-precision parameter values. 
	The application-driven rows use the exact threshold-signature prime moduli corresponding to the displayed bit lengths, while the fixed \(32\)-bit rows use \(q=2^{32}-10239\). 
	For \(Q=0\), no leakage mask is sampled. 
	For \(Q=2^{64}\), the leakage blocks are released over \(\mathcal{R}^\ell\), not modulo \(q\), so \(\sigma_{\bm{r}}\) may exceed \(q\).}
\end{table}

\subsubsection{Concrete Parameter Requirements}\label{Section:SEC-Concrete-Requirements}

For every admissible pattern, \Cref{Theorem:SEC-HCLLR-multi} bounds the distinguishing advantage by \(\epsilon_\mathrm{MQ}=\mathsf{Adv}_\mathsf{MLWE}+2\epsilon'+2\epsilon\) (see \eqref{Eq:epsilon-MQ}), where \(\mathsf{Adv}_\mathsf{MLWE}\) denotes the maximum advantage, over the good setup outputs, of the corresponding non-uniform reduction against the relevant \(\mathsf{MLWE}\) problem, and \(2\epsilon'+2\epsilon\) is the statistical loss of the Gaussian approximation and posterior resampling (\Cref{Lemma:Multi-HCLeak-to-MLWE}).
For a concrete instantiation, we target a total advantage of at most \(2^{-\kappa}\).
For the privacy reduction, we set \(\epsilon'=\epsilon:=2^{-(\kappa+3)}\) and \(\kappa_\mathsf{MLWE}:=\kappa+1\), so that its statistical and computational contributions are each bounded by \(2^{-(\kappa+1)}\).
Here, following the usual convention, \(\kappa_\mathsf{MLWE}\) bits of \(\mathsf{MLWE}\) security are read as \(\mathsf{Adv}_\mathsf{MLWE}\leq2^{-\kappa_\mathsf{MLWE}}\).
A parameter set \((q,p,\ell',\sigma_{\bm{s}},\sigma_\mathbf{R},\sigma_{\bm{r}})\), with \(\sigma_{\bm{r}}\) omitted when \(Q = 0\), is admissible for an instance and a pair \((\kappa,Q)\) if it satisfies the following requirements, which are the hypotheses of \Cref{Theorem:SEC-HCLLR-multi} together with the tail requirements of \Cref{Section:SEC-Gaussian-Tail}.
\begin{enumerate}
	\item \textbf{Kernel smoothing.}
    The computable condition \eqref{Eq:sigma-eff-condition-explicit-practical} holds with target \(\epsilon'\), using the underlying full-precision values of \(\beta_{k_p}\) and \(\gamma_{k_p}\) reported in rounded form in \Cref{Table:SEC-Selected-Privacy-Statistics}.
	\item \textbf{Posterior resampling.}
    The width \(\sigma_\mathrm{MQ}\) in \eqref{Eq:Sigma-MQ-Uniform-Parameter} satisfies \(\sigma_\mathrm{MQ}\geq \sqrt{2}\eta_\epsilon(\mathbb{Z}^{D})\), evaluated using the estimate of \(\eta_\epsilon(\mathbb{Z}^N)\) from \Cref{Section:Lattices-Discrete-Gaussians} and the underlying full-precision values of \(\alpha_{k_p}\) and \(\|\mathbf{\hat{G}}\|_2\) reported in rounded form in \Cref{Table:SEC-Selected-Privacy-Statistics}.
    \item \textbf{Computational hardness.} 
	Both \(\mathsf{MLWE}_{p,\ell',q,d,\sigma_\mathrm{MQ}}\), for nonempty auxiliary views, and \(\mathsf{MLWE}_{p,\ell',q,d,\sigma_{\bm{s}}}\), for the empty view, provide at least \(\kappa_\mathsf{MLWE}\) bits of security against the known attacks evaluated by the lattice estimator \cite{albrecht2015concrete} under the cost model of \cite{alkim2016post}. 
	For this concrete evaluation, we assume that the polynomial-size advice \(\mathsf{pp}\), which is independent of the random \(\mathsf{MLWE}\) matrix \(\bm{A}'\), does not improve these attacks; the formal non-uniform hardness assumption remains that of \Cref{Theorem:SEC-HCLLR-multi}. 
	Since \(\sigma_\mathrm{MQ}\leq\sigma_{\bm{s}}/\sqrt{2}\), the requirement for \(\mathsf{MLWE}_{p,\ell',q,d,\sigma_{\bm{s}}}\) is implied in practice by the one for \(\mathsf{MLWE}_{p,\ell',q,d,\sigma_\mathrm{MQ}}\), because the estimated security does not decrease with the width.
	\item \textbf{Tail bounds and shortness.}
    The tail parameters satisfy \eqref{Eq:theta_s}, \eqref{Eq:theta_R-single}, and, when \(Q\geq1\), \eqref{Eq:theta_r}; the resulting share bound is \(\beta_\mathsf{sh}=\beta_\mathbf{g}\beta_{\bm{s}}+c_\mathrm{exp}\beta_\mathbf{R}\) from \Cref{Lemma:bound-for-si}.
\end{enumerate}

\begin{table}[!htp]
    \caption{Ideal entropy-coded sizes, in KiB, of the secret and of one share for the selected parameter sets}
	\label{Table:SEC-Selected-Sizes}%
    \renewcommand{\arraystretch}{1.15}
	\begin{tabular}{@{}cccccccc@{}}
		\toprule
		& & & & \multicolumn{2}{c}{threshold-signature moduli} & \multicolumn{2}{c}{\(\lceil\lg q\rceil=32\)} \\
		\cmidrule(lr){5-6}\cmidrule(l){7-8}
		\((n,k)\) & \((k_p,k_c,k_c^\mathsf{det})\) & \(\kappa\) & \(Q\) & Secret \(\bm{s}\) & Share \(\bm{s}_i\) & Secret \(\bm{s}\) & Share \(\bm{s}_i\) \\
		\midrule
        \((40,20)\) & \((5,26,35)\) & \(128\) & \(0\) & \(3.753\) & \(5.697\) & \(1.875\) & \(2.847\) \\
        \((40,20)\) & \((5,26,35)\) & \(128\) & \(2^{64}\) & \(4.333\) & \(6.068\) & \(2.164\) & \(3.032\) \\
        \((40,20)\) & \((5,26,35)\) & \(192\) & \(0\) & \(5.283\) & \(7.856\) & \(2.640\) & \(3.926\) \\
        \((40,20)\) & \((5,26,35)\) & \(192\) & \(2^{64}\) & \(6.087\) & \(8.350\) & \(3.041\) & \(4.173\) \\
        \((40,20)\) & \((5,26,35)\) & \(256\) & \(0\) & \(6.854\) & \(10.058\) & \(3.425\) & \(5.027\) \\
        \((40,20)\) & \((5,26,35)\) & \(256\) & \(2^{64}\) & \(7.880\) & \(10.674\) & \(3.938\) & \(5.335\) \\
        \midrule
        \((60,40)\) & \((6,46,56)\) & \(128\) & \(0\) & \(3.383\) & \(6.041\) & \(1.689\) & \(3.019\) \\
        \((60,40)\) & \((6,46,56)\) &\(128\) & \(2^{64}\) & \(3.538\) & \(6.416\) & \(1.767\) & \(3.206\) \\
        \((60,40)\) & \((6,46,56)\) &\(192\) & \(0\) & \(4.780\) & \(8.313\) & \(2.388\) & \(4.155\) \\
        \((60,40)\) & \((6,46,56)\) &\(192\) & \(2^{64}\) & \(4.991\) & \(8.812\) & \(2.494\) & \(4.404\) \\
        \((60,40)\) & \((6,46,56)\) &\(256\) & \(0\) & \(6.219\) & \(10.628\) & \(3.108\) & \(5.312\) \\
        \((60,40)\) & \((6,46,56)\) &\(256\) & \(2^{64}\) & \(6.486\) & \(11.252\) & \(3.242\) & \(5.624\) \\
		\botrule
	\end{tabular}
	\footnotetext{Note: sizes exclude implementation-specific coding, framing, and byte-alignment overheads; each participant stores exactly one share.}
\end{table}

\subsubsection{Security Estimates and Selected Parameter Sets}\label{Section:SEC-Security-Estimates}

Privacy imposes no lower bound on the hard-core modulus \(q\), and the requirements of \Cref{Section:SEC-Concrete-Requirements} are independent of it.
We therefore evaluate two modulus regimes: the application-driven moduli obtained from the \(\mathsf{MSIS}\) estimates for the corresponding threshold-signature instantiations, and the fixed smaller prime \(q=2^{32}-10239\), for which \(\lceil\lg q\rceil=32\) and \(q\equiv1\pmod{2d}\).
For each pair \((n,\kappa)\) and each modulus regime, we select the smallest even \(\ell\), with \(p=\ell'=\ell/2\), for which \(\mathsf{MLWE}_{p,\ell',q,d,\sigma_\mathrm{MQ}}\) provides at least \(\kappa_\mathsf{MLWE}\) bits of security. 
This choice also minimizes the sizes, which increase with \(\ell\). 
At the selected widths, the posterior-resampling requirement holds with equality, so \(\sigma_\mathrm{MQ}=\sqrt{2}\eta_\epsilon(\mathbb{Z}^{D})\) depends only on \(\kappa\) and \(\ell\), and the same security estimate applies to both query budgets. 
Every retained tuple is evaluated at its exact prime modulus using the full \(\mathsf{LWE}\) attack set of the \textit{lattice estimator} introduced in \cite{albrecht2015concrete}, including primal, BDD, dual, and hybrid attacks under the reduction-cost model of \cite{alkim2016post}.

For the threshold-signature moduli, the selected parameters remain private for every hard-core modulus satisfying \(\lceil\lg q\rceil \leq 68\) when \(\kappa\in\{128,192\}\) and \(\lceil\lg q\rceil \leq 69\) when \(\kappa=256\). 
For the fixed \(32\)-bit modulus, the selected module dimensions are \(\ell=6\), \(8\), and \(10\) for \(\kappa=128\), \(192\), and \(256\), respectively, for \(n \in \{40,60\}\). 
These parameter sets remain private for every hard-core modulus satisfying \(\lceil\lg q\rceil \leq 35\), and their secrets and shares are approximately half the size of those obtained with the threshold-signature moduli.
\Cref{Table:SEC-Selected-Parameters} reports the parameter sets for both modulus regimes. 
The corresponding ideal entropy-coded sizes are reported in \Cref{Table:SEC-Selected-Sizes} and computed as described in Appendix~\ref{Appendix:Gaussian-Tail-Size}.

\FloatBarrier
\section{Discussion and Conclusion}\label{Section:Discussion}

\(\mathsf{SEC}\) shows that lossless bipartite expanders yield everywhere-short secret sharing over \(\mathbb{Z}\)-modules in which each participant receives exactly one share, with bounded share coefficients and bounded entries in the integer reconstruction vectors.
Its correctness and shortness follow from integer linear algebra over the expander-derived distribution matrix, while its hard-core privacy against correlated multi-query local leakage reduces to standard \(\mathsf{MLWE}\).
We conclude by discussing the ramp gap, limitations, and open problems.

\subsection{The Ramp Gap}\label{Section:Ramp-Gap}

The principal trade-off in assigning exactly one bounded-coefficient share to each party is the gap between the privacy threshold \(k_p\) and the reconstruction threshold \(k_c\), which remains the main limitation of the construction. 
The privacy threshold is governed by the expansion of the underlying bipartite graph, whereas the reconstruction threshold depends on the resulting distribution matrix and the construction of its secret-bearing column. 
For the selected instances in \Cref{Table:SEC-Selected-Privacy-Statistics,Table:SEC-Reconstruction-Statistics}, \(\tau_p\in[0.100,0.200]\) and \(\tau_c\in[0.620,0.833]\). 
The near-threshold black-box schemes of \cite{cramer2020blackbox,applebaum2023recover} and the random-assignment ramp sharing of \cite{del2024shortly} likewise trade a gap between the thresholds for compact shares, although their reconstruction coefficients are either integers without a stated bound \cite{cramer2020blackbox} or binary \cite{applebaum2023recover,del2024shortly}, rather than bounded signed integers.

\subsection{Open Problems and Future Directions}\label{Section:Open-Problems}

The main open problem is to construct an everywhere-short secret sharing scheme over \(\mathbb{Z}\)-modules with one share per participant, or more generally \(O(1)\) shares per participant, that realizes a perfect (non-ramp) threshold access structure, or at least a substantially narrower ramp gap than \(\mathsf{SEC}\).
A further question is how to instantiate \(\mathsf{SEC}\) for larger committees, beyond the exhaustively certified regime considered here.
Finally, applications of \(\mathsf{SEC}\) to lattice-based distributed cryptography, such as threshold-signing protocols, are left for future work.

\backmatter

\section*{Statements and Declarations}

\paragraph{Funding}

This work was partially supported by Australian Research Council Discovery Grants DP250100229 and DP260100245.
M. A. is supported by a scholarship from the Indonesia Endowment Fund for Education (LPDP), grant no.~SKPB9126/LPDP/LPDP.3/2023.

\paragraph{Competing interests}

The authors have no competing interests to declare that are relevant to the content of this article.

\paragraph{Data availability}

The SageMath scripts used for the expander search, certification, and evaluation in \Cref{Section:SEC-Empirical-Parameter} are publicly available at \url{https://github.com/arzaki-cstheory/SEC-LaTSEC} under the GPL-3.0 license.

\paragraph{Author contributions}

M. A. contributed to the construction, developed the proofs, implemented the experiments and parameter search, and wrote the initial manuscript.
A. S. and R. S. contributed to the conception of the work and the interpretation of the results, supervised the project, and revised the manuscript.
M. F. E. contributed to the interpretation of the results, supervised the project, and revised the manuscript.
All authors read and approved the final manuscript.

\paragraph{Use of AI tools}

During the preparation of this manuscript, the authors used Claude (Anthropic) and Microsoft Copilot to assist with editing the text, improving the technical writing, checking the consistency of proofs and notation, drafting revisions of some passages, and preparing and debugging the SageMath scripts used for the experiments and parameter search.
All such content and code were reviewed, verified, and revised by the authors, who take full responsibility for the content of the publication.

\begin{appendices}

\counterwithin*{equation}{section}% equation numbers restart in each appendix: (A1), (B1), ...
\renewcommand{\theHequation}{app.\Alph{section}.\arabic{equation}}% unique hyperlink anchors
\renewcommand{\theHtable}{app.\Alph{section}.\arabic{table}}%
\renewcommand{\theHfigure}{app.\Alph{section}.\arabic{figure}}%

\section{Additional Notation and Linear-Algebra Conventions}\label{Appendix:Additional-Notation-Linear-Algebra}

For distributions \(X\) and \(Y\), their statistical distance is denoted by \(\Delta(X,Y)\).
For real vectors \(\mathbf{u}\) and \(\mathbf{v}\) of the same dimension, let \(\langle\mathbf{u},\mathbf{v}\rangle:=\mathbf{u}^\top\mathbf{v}\).
We write \(\mathbf{A}^\top\) for transpose, and \(\mathbf{I}\) and \(\mathbf{O}\) for identity and zero matrices of the required dimensions.
The symbols \(\otimes\) and \(\oplus\) denote the Kronecker product and block-diagonal direct sum, respectively, and \(\mathrm{vec}(\mathbf{A})\) denotes column-wise vectorization.
A matrix of the form \(\begin{bmatrix}\mathbf{I}\ \mathbf{A}\end{bmatrix}\) is in systematic form.

For vectors, \(\|\cdot\|_1\), \(\|\cdot\|_2\), and \(\|\cdot\|_\infty\) denote the \(1\)-norm, Euclidean norm, and infinity norm, respectively, applied after coefficient embedding when necessary. 
For real matrices, \(\|\cdot\|_2\), \(\|\cdot\|_F\), \(\sigma_{\min}(\cdot)\), and \(\sigma_{\min}^{+}(\cdot)\) denote the spectral norm, Frobenius norm, smallest singular value, and smallest positive singular value, respectively, and \(\|\mathbf{A}\|_{\max}:=\max_{i,j}|\mathbf{A}[i,j]|\).
We write \(\mathbf{A}\succeq\mathbf{B}\) (resp. \(\mathbf{A}\succ\mathbf{B}\)) when \(\mathbf{A}-\mathbf{B}\) is positive semidefinite (resp. positive definite), and \(\mathbf{A}^+\) for the Moore-Penrose pseudoinverse; if \(\mathbf{A}\) has full row rank, then \(\mathbf{A}^+ = \mathbf{A}^\top (\mathbf{A}\mathbf{A}^\top)^{-1}\) and \(\|\mathbf{A}^+\|_2 = 1/\sigma_{\min}(\mathbf{A})\).
Recall that \(\mathbf{A}\succeq\mathbf{B}\succ0\) implies \(\mathbf{A}^{-1}\preceq\mathbf{B}^{-1}\).

\section{Gaussian Normalizations and Matrix Hint-MLWE Proofs}\label{Appendix:MHMLWE-Proofs}

\subsection{Normalizations}

Since \(\rho_{1/s}(\mathbf{y}) =\exp(-\pi s^2\|\mathbf{y}\|_2^2)\), substituting \(s=\sqrt{2\pi}\sigma\) in \eqref{Eq:Smoothing-Parameter-SD} gives \(\eta_\epsilon(\Lambda) = \frac{ \eta^\pi_\epsilon(\Lambda) }{ \sqrt{2\pi} }\), where \(\eta^\pi_\epsilon(\Lambda) := \inf\left\{ s>0: \rho_{1/s} (\Lambda^*\setminus\{\mathbf{0}\}) \leq\epsilon \right\} \) is the normalization of \cite[Definition 5]{kim2023toward}. 
Similarly, for \(\mathbf{\Sigma} \succ 0\) and \(\mathbf{S}:=\sqrt{2\pi}\mathbf{\Sigma}^{1/2}\), \(\exp\left( -\pi\|\mathbf{S}^{-1}\mathbf{x}\|_2^2 \right) = \exp\left( -\frac{1}{2} \mathbf{x}^{\top} \mathbf{\Sigma}^{-1} \mathbf{x} \right)\). 
Thus, the parameter matrix \(\mathbf{\Sigma}\) in \eqref{Eq:DiscreteGaussianDist} corresponds to the matrix parameter \(\sqrt{2\pi}\mathbf{\Sigma}^{1/2}\) under the convention of \cite{genise2020improved}.

\subsection{Deferred Proofs}

\begin{proof}[Proof of \Cref{Theorem:Hardness-MHMLWE-Tight}]
	Fix the tuple of hint matrices from the theorem statement and define \(\mathbf{\Xi}:=\mathbf{\Sigma}_\mathbf{s}^{-1}+\sum_{i=1}^{h}\mathbf{M}_i^\top\mathbf{\Sigma}_{\mathbf{y}_i}^{-1}\mathbf{M}_i\).
	The posterior-resampling proofs underlying the Matrix Hint-MLWE hardness results of \cite{espitau2024flood,wang2025post} use \(\mathbf{\Xi}\) as the precision matrix of the secret conditioned on the hints.
	The argument requires an upper bound on \(\|\mathbf{\Xi}\|_2\).
	By the triangle inequality and the definition of the effective source width, \(\|\mathbf{\Xi}\|_2 \leq \left\|\mathbf{\Sigma}_\mathbf{s}^{-1}\right\|_2 + \sum_{i=1}^{h} \left\| \mathbf{M}_i^\top \mathbf{\Sigma}_{\mathbf{y}_i}^{-1} \mathbf{M}_i \right\|_2 = \frac{1}{2\sigma^2}\).
	Thus, the precision bound required by the correction and smoothing steps holds directly.
	In contrast, the standard separate-product formulation further applies \(\|\mathbf{M}_i^\top\mathbf{\Sigma}_{\mathbf{y}_i}^{-1}\mathbf{M}_i\|_2 \leq \|\mathbf{M}_i^\top\mathbf{M}_i\|_2 \|\mathbf{\Sigma}_{\mathbf{y}_i}^{-1}\|_2\).
	Retaining the joint term yields the condition stated in \Cref{Theorem:Hardness-MHMLWE-Tight}.
	The remaining posterior-resampling, convolution, uniform-target, and advantage-accounting arguments are unchanged and incur additive distinguishing loss at most \(2\epsilon\).
\end{proof}

\begin{proof}[Proof of \Cref{Corollary:Hardness-MHMLWE-Tight-Smaller-Sigma}]
	Fix the tuple of hint matrices and write \(\mathbf{\Xi}:=\mathbf{\Sigma}_\mathbf{s}^{-1}+\sum_{i=1}^{h}\mathbf{M}_i^\top\mathbf{\Sigma}_{\mathbf{y}_i}^{-1}\mathbf{M}_i\) and \(\mathbf{\Sigma}_0:=\mathbf{\Xi}^{-1}\).
	By the definition of \(\sigma\), we have \(\|\mathbf{\Xi}\|_2\leq1/(2\sigma^2)\), and hence \(\lambda_{\min}(\mathbf{\Sigma}_0)=1/\|\mathbf{\Xi}\|_2\geq2\sigma^2\).
	Run the posterior-resampling reduction underlying \Cref{Theorem:Hardness-MHMLWE-Tight} with source width \(\sigma'\) in place of \(\sigma\).
	The corresponding correction parameter matrix is \(\mathbf{\Sigma}_0-\sigma'^2\mathbf{I}_{N}\), which is positive definite because \(\lambda_{\min}\!\left( \mathbf{\Sigma}_0-\sigma'^2\mathbf{I}_{N} \right) \geq 2\sigma^2-\sigma'^2 \geq \sigma'^2 > 0\).
	It remains to verify the smoothing condition in the convolution step.
	Set \(\mathbf{\Sigma}_1:=\sigma'^2\mathbf{I}_{N}\), \(\mathbf{\Sigma}_2:=\mathbf{\Sigma}_0-\sigma'^2\mathbf{I}_{N}\), and define \(\mathbf{\Sigma}_3^{-1}:=\mathbf{\Sigma}_1^{-1}+\mathbf{\Sigma}_2^{-1}\).
	The preceding bound gives \(\lambda_{\min}(\mathbf{\Sigma}_2)\geq\sigma'^2\), and therefore \(\left\|\mathbf{\Sigma}_3^{-1}\right\|_2 \leq \frac{1}{\sigma'^2} + \frac{1}{\lambda_{\min}(\mathbf{\Sigma}_2)} \leq \frac{2}{\sigma'^2} \leq \frac{1}{\eta_\epsilon(\mathbb{Z}^{N})^2}\), 	where the last inequality follows from \(\sigma'\geq\sqrt{2}\eta_\epsilon(\mathbb{Z}^{N})\).
    Since \(\exp(-\frac{1}{2}\mathbf{x}^\top\mathbf{\Sigma}^{-1}\mathbf{x}) = \exp(-\pi\,\mathbf{x}^\top(2\pi\mathbf{\Sigma})^{-1}\mathbf{x})\) and \(\eta^\pi_\epsilon = \sqrt{2\pi}\,\eta_\epsilon\), this is exactly the hypothesis of the convolution lemma \cite[Lemmas 3 and 5]{kim2023toward} in the normalization \eqref{Eq:Smoothing-Parameter-SD}.
    Thus, the same convolution argument applies with statistical loss at most \(2\epsilon\).
 	The posterior-resampling identity, the uniform-target branch, and the advantage accounting are independent of the choice between \(\sigma\) and \(\sigma'\), so the remainder of the reduction is unchanged.
\end{proof}

\section{Conventional Privacy}\label{Appendix:Conventional-Privacy}

Conventional secret-sharing privacy compares the share views induced by two arbitrarily fixed secrets.

\begin{definition}[Conventional privacy]\label{Definition:SecretSharing-Privacy}
	Let \(\mathsf{Sharing}=(\mathsf{Setup},\mathsf{Deal},\mathsf{Rec})\) be a secret sharing scheme with secret space \(S\).
	Fix \(\mathsf{pp}\) in the support of \(\mathsf{Setup}(1^\kappa)\).
	For \(U\in\Gamma_0(\mathsf{pp})\), secrets \(s,s'\in S\), and an adversary \(\mathcal{A}\), define
	\begin{align*}
		\mathsf{Adv}_{\mathcal{A},\mathsf{Sharing}}^\mathsf{priv}(\mathsf{pp},U,s,s') := \Big| &\Pr\!\left[ \mathcal{A}\!\left(\mathsf{pp},(s_i)_{i\in U}\right)=1 \,\middle|\, (s_i)_{i\in[n]}\gets\mathsf{Deal}_\mathsf{pp}(s) \right] \\ -&\Pr\!\left[ \mathcal{A}\!\left(\mathsf{pp},(s_i')_{i\in U}\right)=1 \,\middle|\, (s_i')_{i\in[n]}\gets\mathsf{Deal}_\mathsf{pp}(s') \right] \Big|,
	\end{align*}
	where the probabilities are over the randomness of \(\mathsf{Deal}\) and \(\mathcal{A}\).
	The scheme is \emph{computationally \(\delta\)-private} for \(\mathsf{pp}\) if \(\mathsf{Adv}_{\mathcal{A},\mathsf{Sharing}}^\mathsf{priv}(\mathsf{pp},U,s,s')\leq\delta\) for every PPT adversary \(\mathcal{A}\), every \(U\in\Gamma_0(\mathsf{pp})\), and every \(s,s'\in S\).
	It is \emph{statistically \(\delta\)-private} if the same condition holds for every possibly unbounded adversary, and it is \emph{perfectly private} if it is statistically \(0\)-private.
\end{definition}

\section{Deferred Proofs for Codes, Construction, and Correctness}\label{Appendix:Deferred-Proofs}

\subsection{Codes and Expanders}

\begin{proof}[Proof of \Cref{Lemma:Code-Distance-Preservation}]
	Let \(\mathbf{c}\in C_\mathbb{Q}\setminus\{\mathbf{0}\}\), and let \(\mathbf{a} := D\mathbf{c} \in \mathbb{Z}^n\) for a positive common denominator \(D\) of its entries, so that \(\mathbf{H}\mathbf{a} = \mathbf{0}\) over \(\mathbb{Z}\) and \(\mathrm{supp}(\mathbf{a}) = \mathrm{supp}(\mathbf{c})\).
	With \(t := \min_{j\in\mathrm{supp}(\mathbf{a})}v_p(\mathbf{a}[j])\), where \(v_p\) denotes the \(p\)-adic valuation, the vector \(\mathbf{b} := p^{-t}\mathbf{a} \in \mathbb{Z}^n\) satisfies \(\mathbf{H}\mathbf{b}=\mathbf{0}\) and has an entry that is nonzero modulo \(p\).
	Hence \(\overline{\mathbf{b}} := \mathbf{b}\bmod p\) is a nonzero codeword of \(C_p\) with \(\mathrm{supp}(\overline{\mathbf{b}}) \subseteq \mathrm{supp}(\mathbf{c})\), so \(d(C_p) \leq wt_H(\overline{\mathbf{b}}) \leq wt_H(\mathbf{c})\).
	Taking the minimum over all nonzero \(\mathbf{c}\in C_\mathbb{Q}\) gives \(d(C_\mathbb{Q}) \geq d(C_p)\).
\end{proof}

\begin{proof}[Proof of \Cref{Lemma:Expander-Graph-Unique-Neighbor-extended}]
	Let \(S\subseteq L\) be nonempty and satisfy \(|S| < 2(1-\epsilon_\mathrm{exp}) \lfloor\gamma n\rfloor\).
	If \(|S|\leq\gamma n\), then \Cref{Lemma:Expander-Graph-Unique-Neighbor} gives \(|U(S)| \geq c_\mathrm{exp} (1-2\epsilon_\mathrm{exp}) |S| > 0\), where the strict inequality uses \(|S|\geq1\), that is, the nonemptiness of \(S\), together with \(c_\mathrm{exp}\geq1\) and \(\epsilon_\mathrm{exp}<1/2\).
	Otherwise, \(|S|>\gamma n\geq\lfloor\gamma n\rfloor\geq1\), so we may choose \(T\subsetneq S\) with \(|T| = \lfloor\gamma n\rfloor\).
    Then \(|U(T)| \geq c_\mathrm{exp} (1-2\epsilon_\mathrm{exp}) \lfloor\gamma n\rfloor\) by \Cref{Lemma:Expander-Graph-Unique-Neighbor}, while \(|S\setminus T| < 2(1-\epsilon_\mathrm{exp})\lfloor\gamma n\rfloor-\lfloor\gamma n\rfloor = (1-2\epsilon_\mathrm{exp}) \lfloor\gamma n\rfloor\).
	By left-regularity, \(|N(S\setminus T)| \leq c_\mathrm{exp}|S\setminus T| < |U(T)|\), so some vertex of \(U(T)\setminus N(S\setminus T)\) has exactly one neighbor in \(T\) and none in \(S\setminus T\), and hence exactly one neighbor in \(S\).
\end{proof}

\paragraph{Deterministic Peeling Pseudocode}

\Cref{Algorithm:Construct-MU} states the deterministic peeling procedure of \Cref{Section:Lossless-Expanders}. 
The rows of \(\mathbf{\hat{G}}_U\) are indexed by \([r]\), where \(r:=|U|\), in increasing order of the elements of \(U\). 

\begin{algorithm}[!htp]
	\caption{Peeling-based construction of a unimodular submatrix.}
	\label{Algorithm:Construct-MU}
	\begin{algorithmic}[1]
	\small
	\Require A matrix \(\mathbf{\hat{G}}_U\in\mathbb{Z}^{r\times(k-1)}\), where \(r:=|U|\), satisfying the unit-valued unique-neighbor property: for every nonempty \(S\subseteq[r]\), there exists \(j\in[k-1]\) such that \(\left| \left\{ i\in S: \mathbf{\hat{G}}_U[i,j]\neq0 \right\} \right| = 1\), and the unique nonzero entry belongs to \(\{-1,1\}\).
	\Ensure An ordered column tuple \(J_U=(j_1,\dots,j_r)\) such that \(\mathbf{M}_U:=\mathbf{\hat{G}}_U[\cdot,J_U]\) is unimodular.
	\State \(S:=[r]\), \(J_U:=()\), and \(I_U:=()\).
	\For{\(t=1\) \textbf{to} \(r\)}
		\For{\(j\in[k-1]\setminus\{j_1,\ldots,j_{t-1}\}\)}
			\State \(\mathrm{supp}_S(j) := \{i\in S:\mathbf{\hat{G}}_U[i,j]\neq0\}\).
		\EndFor
		\State \(C := \left\{ j\in[k-1]\setminus\{j_1,\dots,j_{t-1}\}: \begin{array}{l} |\mathrm{supp}_S(j)|=1,\\ \mathbf{\hat{G}}_U[i(j),j]\in\{-1,1\} \end{array} \right\}\), where \(i(j)\) is the unique row in \(\mathrm{supp}_S(j)\).
		\If{\(C=\emptyset\)}
			\State \Return \(\bot\). \Comment{This cannot occur under the assumed property.}
		\EndIf
		\State \(j_t:=\min C\).
		\State Let \(i_t\) be the unique row in \(\mathrm{supp}_S(j_t)\).
		\State \(J_U:=(j_1,\dots,j_t)\) and \(I_U:=(i_1,\dots,i_t)\).
		\State \(S:=S\setminus\{i_t\}\).
	\EndFor
	\State \(\mathbf{M}_U:=\mathbf{\hat{G}}_U[\cdot,J_U]\).
	\State \Return \((\mathbf{M}_U,J_U,I_U)\).
	\end{algorithmic}
\end{algorithm}

For \(1\leq|U|\leq \left\lceil 2(1-\epsilon_\mathrm{exp})\lfloor\gamma n\rfloor \right\rceil-1\), the row restriction \(\mathbf{\hat{G}}_U\) of \(\mathbf{\hat{G}}=(\mathbf{P}\mathbf{H})^\top\) satisfies the unit-valued unique-neighbor property: a column with exactly one nonzero entry in \(S\) corresponds to a unique neighbor of the corresponding nonempty set of left vertices, which exists by \Cref{Lemma:Expander-Graph-Unique-Neighbor-extended}, and every nonzero entry of \(\mathbf{\hat{G}}\) lies in \(\{-1,1\}\) because \(\mathbf{P}\) is a signed permutation matrix. 
Moreover, once a column \(j_t\) and its unique active row \(i_t\) are selected, the column \(j_t\) has no nonzero entry among the remaining active rows. 
Thus, no previously selected column can have singleton support in a later iteration, so excluding the previously selected columns does not affect the nonemptiness of \(C\). 
Hence, the candidate set \(C\) in \Cref{Algorithm:Construct-MU} is nonempty at every iteration.

Under the peeling row order \(I_U=(i_1,\dots,i_r)\), the matrix \(\mathbf{M}_U^\triangle :=\mathbf{\hat{G}}_U[I_U,J_U]\) is triangular with diagonal entries in \(\{-1,1\}\), so \(\left|\det\left(\mathbf{M}_U^\triangle\right)\right|=1\). 
Since \(\mathbf{M}_U\) differs from \(\mathbf{M}_U^\triangle\) only by a row permutation, \(|\det(\mathbf{M}_U)|=1\), and thus \(\mathbf{M}_U\in GL_r(\mathbb{Z})\). 
The deterministic choice \(j_t:=\min C\) and the fixed ordering conventions make \((\mathbf{M}_U,J_U,I_U)\) uniquely determined by \(\mathbf{\hat{G}}\) and \(U\).

\subsection{Construction}

\begin{proof}[Proof of \Cref{Lemma:e1-row-span-HNF-e1}]
	Since \(\mathbf{U}_\mathbf{A}\) is unimodular, \(\mathrm{Row}_\mathbb{Z}(\mathbf{R}_\mathbf{A})=\mathrm{Row}_\mathbb{Z}(\mathbf{A})\).
	If \(\mathbf{R}_\mathbf{A}[1,\cdot]=\mathbf{e}_1^\top\), then \(\mathbf{e}_1^\top\in\mathrm{Row}_\mathbb{Z}(\mathbf{R}_\mathbf{A})=\mathrm{Row}_\mathbb{Z}(\mathbf{A})\).
	Conversely, suppose that \(\mathbf{e}_1^\top=\sum_{i=1}^r\alpha_i\mathbf{R}_\mathbf{A}[i,\cdot]\) with \(\alpha_1,\dots,\alpha_r\in\mathbb{Z}\).
	Since \(\mathbf{e}_1^\top\) has a nonzero first coordinate, some row of \(\mathbf{R}_\mathbf{A}\) is nonzero in column \(1\), so \(j_1=1\), and since \(j_1<\dots<j_r\), row \(1\) is the only such row.
	Comparing first coordinates gives \(1=\alpha_1p_1\), so \(\alpha_1=p_1=1\).
	Suppose, for contradiction, that \(\alpha_i\neq0\) for some \(i\geq2\), and let \(i\) be the smallest such index.
	Rows \(\iota>i\) vanish in column \(j_i\), and \(\mathbf{e}_1^\top\) vanishes in column \(j_i>1\), so comparing coordinates \(j_i\) gives \(0=\mathbf{R}_\mathbf{A}[1,j_i]+\alpha_ip_i\).
	Since \(0\leq\mathbf{R}_\mathbf{A}[1,j_i]<p_i\), this forces \(\alpha_i=0\), a contradiction.
	Hence \(\alpha_2=\dots=\alpha_r=0\), and therefore \(\mathbf{R}_\mathbf{A}[1,\cdot]=\mathbf{e}_1^\top\).
	Finally, in this case, \(\bm{\lambda}_0^\top\mathbf{A}=\mathbf{U}_\mathbf{A}[1,\cdot]\mathbf{A}=\mathbf{R}_\mathbf{A}[1,\cdot]=\mathbf{e}_1^\top\).
\end{proof}

\begin{proof}[Proof of \Cref{Lemma:DisjointBlocksRankLeftover}]
	Every \(r+1\) columns of \(\mathbf{H}'\) are linearly dependent over \(\mathbb{Q}\), and clearing denominators in a nonzero rational kernel vector gives a nonzero vector in \(\ker_\mathbb{Z}(\mathbf{H}'_{(T)})\) for every \(T\subseteq[n]\) with \(|T|=r+1\).
	Partitioning \([n]\) into \(b:=\lfloor n/(r+1)\rfloor\) pairwise disjoint sets \(T_1,\dots,T_b\) of size \(r+1\) and a leftover set \(R\) with \(|R|\leq r\) proves the claim.
\end{proof}

\paragraph{Complexity of \Cref{Algorithm:Construct-g-disjoint-approach}}

The baseline construction using supports of size \(t_\mathrm{max}=r+1\), where \(r:=\mathrm{rank}_\mathbb{Q}(\mathbf{H}')\), processes at most \(\lfloor n/(r+1)\rfloor\) supports and uses polynomial-time integer linear-algebra and generalized-B\'ezout computations.
The optional search for smaller supports examines at most \(\mathit{maxTrials}\) candidates for each size \(t\in[t_0,t_\mathrm{max}-1]\) in every iteration.
For fixed \(t_0\) and \(\mathit{maxTrials}\), the number of rank tests in any execution is \(O(b \cdot \mathit{maxTrials} \cdot(t_\mathrm{max}-t_0))\), where \(b:=|\mathcal{T}|\).
Each rank or kernel computation is polynomial in \(m\), \(t\), and the bit length of the entries of \(\mathbf{H}'\).
Thus, the construction has polynomial bit complexity for fixed search parameters.

\subsection{Reconstruction Coverage}

\paragraph{Deferred Proofs}

\begin{proof}[Proof of \Cref{Corollary:Correctness-LB-T,Corollary:Correctness-LB-Witness}]
	For each \(i\in[b]\), \(\Pr[T_i\subseteq S]=\binom{n-t_i}{s-t_i}/\binom{n}{s}\), and, since the supports are pairwise disjoint, for every \(\emptyset\neq J\subseteq[b]\), \(\Pr[\bigcap_{i\in J}(T_i\subseteq S)]=\binom{n-\sum_{i\in J}t_i}{s-\sum_{i\in J}t_i}/\binom{n}{s}\).
	Inclusion-exclusion gives the union probability in \eqref{Eq:IE-Correctness-Hypergeometric}, and the second-order Bonferroni inequality gives \eqref{Eq:IE-Correctness-Bonferroni}.
    For the witness version, suppose that \(W_i\subseteq S\). 
    Restrict \(\bm{\lambda}_{T_i}\) to \(W_i\) and extend it by zero to the remaining coordinates of \(S\).
	The resulting vector \(\bm{\lambda}_S\in\mathbb{Z}^{|S|}\) satisfies \(\bm{\lambda}_S\in\ker_\mathbb{Z}(\mathbf{H}'_{(S)})\) and \(\bm{\lambda}_S^\top(\mathbf{g})_S=1\).
    Therefore, \(W_i\subseteq S\) implies \(\mathbf{e}_1^\top\in\mathrm{Row}_\mathbb{Z}(\mathbf{G}_S)\).
    Since the sets \(T_i\) are pairwise disjoint and \(W_i\subseteq T_i\), the witness supports \(W_1,\dots,W_b\) are also pairwise disjoint.
	Hence the same inclusion-exclusion and Bonferroni calculations apply with \(t_i\) replaced by \(w_i:=|W_i|\).
\end{proof}

\paragraph{Heuristic Design Rules}

If all planted supports have a common size \(t_i=t\), write \(p(t):=\binom{n-t}{s-t}/\binom{n}{s}\) and \(p(2t):=\binom{n-2t}{s-2t}/\binom{n}{s}\).
Then \eqref{Eq:IE-Correctness-Bonferroni} gives \(\Pr[\bigcup_{i=1}^b(T_i\subseteq S)]\geq bp(t)-\binom{b}{2}p(2t)\).
For parameter-selection intuition, when \(t\ll n\), we have \(p(t)=(s)_t/(n)_t\approx(s/n)^t\).
If the containment events are treated heuristically as independent, then \(\Pr[\bigcup_{i=1}^b(T_i\subseteq S)]\approx 1-(1-p(t))^b\approx 1-(1-(s/n)^t)^b\).
Accordingly, a target reconstruction error \(\epsilon_c\) suggests \(\epsilon_c\gtrapprox(1-(s/n)^t)^b\).
Setting \(s=k_c\) gives \(k_c\gtrapprox(1-\epsilon_c^{1/b})^{1/t}n\).
When \(\epsilon_c\) is small and \(b\) is moderately large, \(k_c\gtrapprox(\ln(1/\epsilon_c)/b)^{1/t}n\).
For witness supports of a common size \(w_i=w\), the same heuristic gives the rules with \(t\) replaced by \(w\).
These independence-based expressions are heuristic approximations used only for parameter-selection intuition; the inclusion-exclusion and Bonferroni expressions are rigorous lower bounds.

\subsection{Linearity and Public-Map Compatibility}

\begin{proof}[Proof of \Cref{Lemma:Linear-Combination-Rec,Lemma:Matrix-Multiplication-Rec}]
	Let \(\mathbf{X}_T,\mathbf{Y}_T\in\mathbb{Z}^{|T|\times D}\) have rows \(\mathbf{x}_i^\top\) and \(\mathbf{y}_i^\top\), ordered according to the indices in \(T\).
	The matrix with rows \((\alpha\mathbf{x}_i+\mathbf{y}_i)^\top\) is \(\alpha\mathbf{X}_T+\mathbf{Y}_T\), and the matrix with rows \((\mathbf{M}\mathbf{x}_i)^\top=\mathbf{x}_i^\top\mathbf{M}^\top\) is \(\mathbf{X}_T\mathbf{M}^\top\).
	Hence \((\bm{\lambda}_T^\top(\alpha\mathbf{X}_T+\mathbf{Y}_T))^\top = \alpha(\bm{\lambda}_T^\top\mathbf{X}_T)^\top + (\bm{\lambda}_T^\top\mathbf{Y}_T)^\top\) and \((\bm{\lambda}_T^\top\mathbf{X}_T\mathbf{M}^\top)^\top = \mathbf{M}(\bm{\lambda}_T^\top\mathbf{X}_T)^\top\), which are the two claims.
\end{proof}

\section{Babai Decoding and Greedy Refinement}\label{Appendix:Babai-Greedy}

This appendix specifies the procedure \(\mathsf{BabaiGreedy}\) (\Cref{Algorithm:Construct-small-lambda-Babai-greedy}) invoked by \Cref{Algorithm:Construct-Small-Lambda} and justifies its correctness, complexity, and determinism.

\begin{algorithm}[!htp]\caption{Babai decoding and greedy \(\ell_\infty\)-refinement for a reconstruction vector.}
	\label{Algorithm:Construct-small-lambda-Babai-greedy}
    \small
	\begin{algorithmic}[1]
	\Require A vector \(\bm{\lambda}_{T,0}\in\mathbb{Z}^{|T|}\); an \(\mathsf{LLL}_{\delta_\mathsf{LLL}}\)-reduced basis \(\mathbf{B}_K'\in\mathbb{Z}^{r\times|T|}\) of \(K_T:=\ker_\mathbb{Z}(\mathbf{G}_T^\top)\); and fixed small parameters \(\mathit{passes},\mathit{window}\in\mathbb{Z}^+\).
	\Ensure A vector \(\bm{\lambda}_T\in\bm{\lambda}_{T,0}+K_T\) satisfying \(\|\bm{\lambda}_T\|_\infty\leq\|\bm{\lambda}_\mathrm{dec}\|_\infty\).
	\Statex \textbf{Stage 1: Babai nearest-plane decoding.}
	\State Let \(\mathbf{b}_i:=(\mathbf{B}_K'[i,\cdot])^\top\) for \(i\in[r]\), and compute the Gram-Schmidt vectors \(\mathbf{b}_1^*,\dots,\mathbf{b}_r^*\) in exact rational arithmetic.
	\State Set \(\bm{\lambda}:=\bm{\lambda}_{T,0}\).
	\For{\(i=r\) \textbf{ down to } \(1\)}
		\State Set \(c_i:=\langle\bm{\lambda},\mathbf{b}_i^*\rangle/\|\mathbf{b}_i^*\|_2^2\).
		\State Set \(\widehat{c}_i:=\lfloor c_i+\tfrac{1}{2}\rfloor\). \Comment{half-integers are rounded toward \(+\infty\)}
		\State Set \(\bm{\lambda}:=\bm{\lambda}-\widehat{c}_i\mathbf{b}_i\).
	\EndFor
	\State Set \(\bm{\lambda}_\mathrm{dec}:=\bm{\lambda}\).
	\Statex \textbf{Stage 2: Greedy \(\ell_\infty\)-local refinement.}
	\For{\(p=1\) \textbf{ to } \(\mathit{passes}\)}
		\State Set \(\mathit{improved}:=\mathbf{False}\).
		\For{\(i=1\) \textbf{ to } \(r\)}
			\State Let \(j\) be the smallest index satisfying \(|\bm{\lambda}[j]|=\|\bm{\lambda}\|_\infty\).
			\If{\(\mathbf{b}_i[j]=0\)}
				\State \textbf{continue}.
			\EndIf
			\State Set \(\alpha_0:=\left\lfloor-\bm{\lambda}[j]/\mathbf{b}_i[j]+\tfrac{1}{2}\right\rfloor\).
			\State Set \(\bm{\lambda}_\mathrm{best}:=\bm{\lambda}\) and \(w_\mathrm{best}:=\|\bm{\lambda}\|_\infty\).
			\For{\(\alpha=\alpha_0-\mathit{window}\) \textbf{ to } \(\alpha_0+\mathit{window}\)}
				\State Set \(\bm{\lambda}_\alpha:=\bm{\lambda}+\alpha\mathbf{b}_i\) and \(w_\alpha:=\|\bm{\lambda}_\alpha\|_\infty\).
				\If{\(w_\alpha<w_\mathrm{best}\)}
					\State Set \(\bm{\lambda}_\mathrm{best}:=\bm{\lambda}_\alpha\) and \(w_\mathrm{best}:=w_\alpha\).
				\EndIf
			\EndFor
			\If{\(w_\mathrm{best}<\|\bm{\lambda}\|_\infty\)}
				\State Set \(\bm{\lambda}:=\bm{\lambda}_\mathrm{best}\) and \(\mathit{improved}:=\mathbf{True}\).
			\EndIf
		\EndFor
		\If{\textbf{not} \(\mathit{improved}\)}
			\State \textbf{break}.
		\EndIf
	\EndFor
	\State \Return \(\bm{\lambda}\).
	\end{algorithmic}
\end{algorithm}

The first stage of \Cref{Algorithm:Construct-small-lambda-Babai-greedy} subtracts an element of \(K_T\) from \(\bm{\lambda}_{T,0}\), and hence returns \(\bm{\lambda}_\mathrm{dec}\in\bm{\lambda}_{T,0}+K_T\).
Every update in the second stage also adds an element of \(K_T\), so the final output \(\bm{\lambda}_T\) remains in the same affine coset.
Consequently, if \(\bm{\lambda}_{T,0}^\top\mathbf{G}_T=\mathbf{e}_1^\top\), then \(\bm{\lambda}_T^\top\mathbf{G}_T=\mathbf{e}_1^\top\).
The greedy procedure accepts only updates that strictly decrease \(\|\bm{\lambda}\|_\infty\).
Since this norm is a nonnegative integer and the number of passes is fixed, Stage 2 terminates and returns a vector satisfying \(\|\bm{\lambda}_T\|_\infty\leq\|\bm{\lambda}_\mathrm{dec}\|_\infty\).
The exact certification step in \Cref{Algorithm:Construct-Small-Lambda} verifies the reconstruction equation before the vector is used by \(\mathsf{Rec}_\mathsf{pp}\).

The Gram-Schmidt orthogonalization and Babai decoding in Stage 1 require \(O(r^2|T|)\) arithmetic operations.
In Stage 2, each of the \(\mathit{passes}\) passes examines \(r\) basis vectors and at most \(2\cdot\mathit{window}+1\) candidate multiples per basis vector, while evaluating each candidate requires \(O(|T|)\) arithmetic operations.
Since \(\mathit{passes}\) and \(\mathit{window}\) are fixed small parameters, the overall arithmetic complexity is \(O(r^2|T|)\).
Its bit complexity is polynomial in \(r\), \(|T|\), and the bit lengths of the rational and integer values encountered during the computation.

It remains to justify the deterministic-selection claim in \Cref{Remark:Canonicity-of-Construct-Small-Lambda}. 
Different transformation matrices \(\mathbf{U}_{\mathbf{G}_T}\) may yield different particular solutions \(\bm{\lambda}_{T,0}\), but any two such solutions differ by an element of \(K_T\) and therefore belong to the same affine coset. 
Let \(V_T:=\mathrm{span}_\mathbb{R}(K_T)\), and let \(\mathbf{b}_1^*,\dots,\mathbf{b}_r^*\) be the Gram-Schmidt vectors of the ordered reduced basis \(\mathbf{B}_K'\). 
Define \(\mathcal{P}_{1/2}(\mathbf{B}_K'^*) := \left\{ \sum_{i=1}^r \alpha_i\mathbf{b}_i^* : \alpha_i\in[-1/2,1/2)\right\} \subseteq V_T\).
For \(\mathbf{t}\in\mathbb{R}^{|T|}\), denote by \(\mathsf{NP}_{\mathbf{B}_K'}(\mathbf{t})\) the residual returned by Stage~1 on input \(\mathbf{t}\). 
With the rounding convention \(\widehat{x}:=\lfloor x+1/2\rfloor\), Babai's nearest-plane procedure, applied within \(V_T\), selects the unique residual satisfying \( \mathsf{NP}_{\mathbf{B}_K'}(\mathbf{t}) \in (\mathbf{t}+K_T) \cap \left( \mathcal{P}_{1/2}(\mathbf{B}_K'^*)+V_T^\perp \right)\) \cite{babai1986lovasz,hhan2026hybrid}. 
Consequently, if \(\mathbf{a}-\mathbf{b}\in K_T\), then \(\mathbf{a}+K_T=\mathbf{b}+K_T\), and hence \( \mathsf{NP}_{\mathbf{B}_K'}(\mathbf{a}) = \mathsf{NP}_{\mathbf{B}_K'}(\mathbf{b})\). 
Therefore, \(\bm{\lambda}_\mathrm{dec}\) is independent of the transformation matrix used to obtain \(\bm{\lambda}_{T,0}\). 
Since the row-HNF basis \(\mathbf{B}_K\), the exact-arithmetic LLL conventions, and all iteration and tie-breaking rules in \Cref{Algorithm:Construct-small-lambda-Babai-greedy} are fixed, the final output is a deterministic function of \(\mathbf{G}_T\) and the stated public parameters.

\section{Heuristic Estimates for Reconstruction Coefficients}\label{Appendix:Heuristic-Coefficients}

This appendix heuristically relates the certificate in \eqref{Eq:Certified-Coefficient-Bound} to the sparsity of \(\mathbf{G}\). 
These estimates are not used in any proof.

\subsection{Kernel-Lattice Scale from Sparsity}

The certified bound depends on the conditioning of \(\mathbf{G}_T\) and the Gram-Schmidt lengths of the reduced basis of \(K_T:=\ker_\mathbb{Z}(\mathbf{G}_T^\top)\). 
We first relate the determinant scale of \(K_T\) to the sparsity of \(\mathbf{G}\).

Assume that \(\|\mathbf{g}\|_\infty\leq\beta_\mathbf{g}\). 
Since every row of \(\mathbf{\hat{G}}\) has exactly \(c_\mathrm{exp}\) nonzero entries in \(\{-1,1\}\), \(\|\mathbf{G}[i,\cdot]\|_2^2 \leq \beta_\mathbf{g}^2+c_\mathrm{exp}\). 
Define \(R_\mathbf{G} := \sqrt{\beta_\mathbf{g}^2+c_\mathrm{exp}}\). 
Let \(r_T:=\mathrm{rank}_\mathbb{R}(\mathbf{A}_T)\), where \(\mathbf{A}_T:=\mathbf{G}_T^\top\).
Choose \(I_T\subseteq[k]\) with \(|I_T|=r_T\) such that \(\mathbf{A}_T' := \mathbf{A}_T[I_T,\cdot] \in \mathbb{Z}^{r_T\times|T|}\) has full row rank. 
Because the rows of \(\mathbf{A}_T'\) span \(\mathrm{Row}_\mathbb{Q}(\mathbf{A}_T)\), \(\ker_\mathbb{Z}(\mathbf{A}_T') = \ker_\mathbb{Z}(\mathbf{A}_T) = K_T\).
Let \(\Delta_T\) be the greatest common divisor of the \(r_T\times r_T\) minors of \(\mathbf{A}_T'\). 
By the standard determinant formula for integer kernel lattices, 
\begin{align}\label{Eq:det_KT} 
	\det(K_T) = \frac{ \sqrt{\det(\mathbf{A}_T'\mathbf{A}_T'^\top)} }{\Delta_T}. 
\end{align} 

By Cauchy-Binet, the squared numerator in \eqref{Eq:det_KT} is the sum of the squares of the \(r_T\times r_T\) minors of \(\mathbf{A}_T'\). 
For every \(J\subseteq T\) with \(|J|=r_T\), the minor \(\det(\mathbf{A}_T'[\cdot,J])\) is the determinant of the transpose of \(\mathbf{G}_T[J,I_T]\), whose rows have Euclidean norm at most \(R_\mathbf{G}\).
Hence, Hadamard's inequality gives \(|\det(\mathbf{A}_T'[\cdot,J])| \leq R_\mathbf{G}^{r_T}\).
There are \(\binom{|T|}{r_T}\) such minors, and therefore 
\begin{align}\label{Eq:det-KT_bound} 
	\det(K_T) \leq \frac{\binom{|T|}{r_T}^{1/2} R_\mathbf{G}^{r_T} }{\Delta_T }. 
\end{align} 
Dropping \(\Delta_T\geq1\) gives a conservative bound, whereas \(\Delta_T>1\) makes the kernel lattice denser than the corresponding Euclidean minor scale alone suggests.

If \(\mathbf{G}_T\) has full column rank, then \(r_T=k\), so \(I_T=[k]\), \(\mathbf{A}_T'=\mathbf{A}_T\), and \(\det(\mathbf{A}_T'\mathbf{A}_T'^\top)=\det(\mathbf{G}_T^\top\mathbf{G}_T)\).
Hadamard's inequality for the positive semidefinite Gram matrix \(\mathbf{G}_T^\top\mathbf{G}_T\) then bounds this determinant by the product of the squared column norms of \(\mathbf{G}_T\).
The first column satisfies \(\|(\mathbf{g})_T\|_2^2\leq\beta_\mathbf{g}^2|T|\).
For \(j\in[m]\), where \(m=k-1\), let \(\rho_j:=\|\mathbf{\hat{G}}_T[\cdot,j]\|_2^2\) be the squared norm of the \(j\)-th masking column, which equals its number of nonzero entries.
Since every row of \(\mathbf{\hat{G}}_T\) has exactly \(c_\mathrm{exp}\) nonzero entries in \(\{-1,1\}\), we have \(\sum_{j=1}^m\rho_j=c_\mathrm{exp}|T|\).
The inequality of arithmetic and geometric means then gives \(\prod_{j=1}^m\rho_j\leq\big(\frac{1}{m}\sum_{j=1}^m\rho_j\big)^m=(c_\mathrm{exp}|T|/m)^m\), and hence \(\det(\mathbf{G}_T^\top\mathbf{G}_T)\leq\beta_\mathbf{g}^2|T|\,(c_\mathrm{exp}|T|/m)^m\).
Substituting this bound into \eqref{Eq:det_KT} yields
\begin{align}\label{Eq:det-KT_bound-full-rank}
	\det(K_T) \leq \frac{\beta_\mathbf{g}\sqrt{|T|}\,\left(c_\mathrm{exp}|T|/m\right)^{m/2}}{\Delta_T},
\end{align}
which avoids the binomial factor in the full-rank case \(\det(K_T)\leq\binom{|T|}{k}^{1/2}R_\mathbf{G}^{k}/\Delta_T\) of \eqref{Eq:det-KT_bound}.

For \(d_T:=|T|-r_T>0\), define the determinant scale \(D_T:=\det(K_T)^{1/d_T}\). 
Thus, sparsity controls a coarse average scale of \(K_T\) through \(R_\mathbf{G}\), or through \(c_\mathrm{exp}|T|/m\) in the full-rank case, but does not determine the shape of its reduced basis or the \(\ell_\infty\)-norm of the decoded reconstruction vector.

\subsection{Heuristic Gram-Schmidt Profile}

To estimate the Babai term from the determinant scale, we use the geometric-series assumption for the Gram-Schmidt profile of the reduced kernel basis. 
Assume that \(d_T>0\), and let \(\delta_\mathrm{RHF}>1\) be the modeled root-Hermite factor of the LLL-reduced basis. 
Then 
\begin{align}\label{Eq:GS-norm-profile} 
	\|\mathbf{b}_i^*\|_2 \approx \delta_\mathrm{RHF}^{\,d_T-2i+1}D_T, \qquad i\in[d_T]. 
\end{align} 
Substituting \eqref{Eq:GS-norm-profile} into \eqref{Eq:decoding-certificate} and \eqref{Eq:Babai-radius} gives 
\begin{align}\label{Eq:Babai-bound-estimate-LLL} 
	B_{\bm{\lambda}}(T) \lesssim \left( \|\bm{\lambda}_{T,\mathbb{R}}\|_2^2 + \frac{1}{4}\det(K_T)^{2/d_T} \sum_{i=1}^{d_T} \delta_\mathrm{RHF}^{\,2(d_T-2i+1)} \right)^{1/2}. 
\end{align}

If the reduced basis is approximately balanced, then \(\sum_{i=1}^{d_T} \delta_\mathrm{RHF}^{\,2(d_T-2i+1)} \approx d_T\), and hence \(R_\mathsf{Babai}(\mathbf{B}_K') \approx \frac{\sqrt{d_T}}{2}D_T\) and \(B_{\bm{\lambda}}(T) \lesssim \left( \|\bm{\lambda}_{T,\mathbb{R}}\|_2^2 + \frac{d_T}{4}D_T^2 \right)^{1/2}\). 
This approximation is most relevant when \(d_T=|T|-\mathrm{rank}(\mathbf{G}_T)\) is small.
The exponents in \eqref{Eq:GS-norm-profile} decrease symmetrically from \(d_T-1\) at \(i=1\) to \(-(d_T-1)\) at \(i=d_T\), so pairing symmetric terms shows that the sum is at least \(d_T\), with equality only if \(\delta_\mathrm{RHF}=1\), and its deviation from \(d_T\) is governed by the largest term \(\delta_\mathrm{RHF}^{2(d_T-1)}\).
Explicitly, \(\sum_{i=1}^{d_T}\delta_\mathrm{RHF}^{2(d_T-2i+1)}=\big(\delta_\mathrm{RHF}^{2d_T}-\delta_\mathrm{RHF}^{-2d_T}\big)/\big(\delta_\mathrm{RHF}^{2}-\delta_\mathrm{RHF}^{-2}\big)\), which remains close to \(d_T\) only while \(\delta_\mathrm{RHF}^{2(d_T-1)}\approx1\).

For LLL with Lov\'asz parameter \(\delta_\mathsf{LLL}\), the first basis vector provably satisfies \(\|\mathbf{b}_1\|_2\leq(\delta_\mathsf{LLL}-1/4)^{-(d_T-1)/4}\det(K_T)^{1/d_T}\) \cite{lenstra1982factoring}, so the root-Hermite factor is at most \((\delta_\mathsf{LLL}-1/4)^{-1/4}\approx1.078\) for \(\delta_\mathsf{LLL}=0.99\) (as in \Cref{Section:SEC-Reconstruction-Coefficient-Evaluation}), and at most \((4/3)^{1/4}\approx1.075\) in the limit \(\delta_\mathsf{LLL} \to 1\).
In practice, LLL performs considerably better on average, with \(\delta_\mathrm{RHF}\approx1.02\), as first observed experimentally by Nguyen and Stehl\'e \cite{nguyen2006lll} and later confirmed by Gama and Nguyen \cite{gama2008predicting}.
These values concern the first basis vector; the geometric-series profile itself remains a heuristic.
This estimate provides parameter-selection intuition but does not replace the instance-specific certificate in \eqref{Eq:Certified-Coefficient-Bound} or the observed values returned by \Cref{Algorithm:Construct-Small-Lambda}. 
For the concrete instantiations, we select \(\beta_{\bm{\lambda}}\) from the observed values \(B_{\bm{\lambda}}(T)\) using the safety margin specified in \Cref{Section:SEC-Observed-Reconstruction-Coefficient-Bounds}.

\section{Structural Noise-Cancellation Analysis}\label{Appendix:SEC-Structural-Obstructions}

This appendix supports the structural setup condition in \Cref{Section:Structural-Obstructions}. 
We first characterize exact noise cancellation over \(\mathbb{Z}\) and \(\mathbb{Q}\), and then show that failure of integral column-module membership yields a cancellation attack after reduction modulo an adversarially chosen modulus. 
For a fixed private coalition \(U\in\Gamma_0\), recall that the restricted share matrix is
\begin{align}\label{Eq:SEC-Restricted-Share-Matrix}
	\mathbf{S}_U = (\mathbf{g})_U\mathbf{s}^\top + \mathbf{\hat{G}}_U\mathbf{R}.
\end{align}

\subsection{Exact Noise Cancellation over \(\mathbb{Z}\) and \(\mathbb{Q}\)}

Fix public parameters \(\mathsf{pp}\) and a private coalition \(U\in\Gamma_0\).
The following lemma characterizes an exact cancellation witness.

\begin{lemma}\label{Lemma:Noise-Cancellation-SEC}
	Let \(\Sigma\in\{\mathbb{Z},\mathbb{Q}\}\).
	Suppose that there exists \(\bm{\alpha}_U\in\Sigma^{|U|}\) such that \(\mathbf{\hat{G}}_U^\top\bm{\alpha}_U=\mathbf{0}\) and \(c:=\bm{\alpha}_U^\top(\mathbf{g})_U\neq0\).
	Then \(\bm{\alpha}_U^\top\mathbf{S}_U = c\mathbf{s}^\top\).
	Thus, the shares indexed by \(U\) reveal a noiseless nonzero scaled secret \(c\mathbf{s}\).
	Over \(\mathbb{Q}\), the secret is recoverable as \(\mathbf{s} = c^{-1}\mathbf{S}_U^\top\bm{\alpha}_U\).
\end{lemma}

\begin{proof}
	Multiplying \eqref{Eq:SEC-Restricted-Share-Matrix} by \(\bm{\alpha}_U^\top\) gives \(\bm{\alpha}_U^\top\mathbf{S}_U = \bm{\alpha}_U^\top(\mathbf{g})_U\mathbf{s}^\top + \bm{\alpha}_U^\top\mathbf{\hat{G}}_U\mathbf{R} = c\mathbf{s}^\top\).
	If the computation is performed over \(\mathbb{Q}\), then \(c\neq0\) is invertible, giving the stated recovery expression.
\end{proof}

Since \(\mathbf{G}_U = \begin{bmatrix} (\mathbf{g})_U \ \mathbf{\hat{G}}_U \end{bmatrix}\), we always have \(\ker_\Sigma(\mathbf{G}_U^\top)\subseteq \ker_\Sigma(\mathbf{\hat{G}}_U^\top)\).
Strict containment is equivalent to the existence of a cancellation witness as in \Cref{Lemma:Noise-Cancellation-SEC}.

\begin{corollary}\label{Corollary:Kernel-Condition-SEC}
	Let \(\Sigma\in\{\mathbb{Z},\mathbb{Q}\}\).
	If \(\ker_\Sigma(\mathbf{G}_U^\top) \subsetneq \ker_\Sigma(\mathbf{\hat{G}}_U^\top)\), then an exact noise-cancellation witness exists over \(\Sigma\).
	In particular, the shares indexed by \(U\) reveal a noiseless nonzero scaled secret.
\end{corollary}

We next express rational cancellation through column-space separation.

\begin{lemma}[Column-space separation]\label{Lemma:SeparationLemma-SEC}
    Let \(\Sigma\) be a field, let \(W\subseteq\Sigma^r\) be a subspace, and let \(\mathbf{v}\in\Sigma^r\setminus W\).
    Then there exists \(\mathbf{x}\in\Sigma^r\) such that \(\langle\mathbf{x},\mathbf{w}\rangle=0\) for every \(\mathbf{w}\in W\), but \(\langle\mathbf{x},\mathbf{v}\rangle\neq0\).
\end{lemma}

\begin{proof}
	The double-annihilator property gives \((W^\bot)^\bot=W\).
	If every \(\mathbf{x}\in W^\bot\) were also orthogonal to \(\mathbf{v}\), then \(\mathbf{v}\in(W^\bot)^\bot=W\), contradicting \(\mathbf{v}\notin W\).
\end{proof}

\begin{lemma}[Rational cancellation criterion]\label{Lemma:ComputationalPrivacy-SEC-Column-Q} 
	Rational noise cancellation is impossible for \(U\) if and only if \((\mathbf{g})_U \in \mathrm{Col}_\mathbb{Q}(\mathbf{\hat{G}}_U)\). 
\end{lemma}

\begin{proof} 
	If \((\mathbf{g})_U\notin \mathrm{Col}_\mathbb{Q}(\mathbf{\hat{G}}_U)\), \Cref{Lemma:SeparationLemma-SEC} gives \(\bm{\alpha}_U\in \ker_\mathbb{Q}(\mathbf{\hat{G}}_U^\top)\) with \(\bm{\alpha}_U^\top(\mathbf{g})_U\neq0\), and \Cref{Lemma:Noise-Cancellation-SEC} gives a cancellation attack. 
	Conversely, if \((\mathbf{g})_U=\mathbf{\hat{G}}_U\mathbf{a}\) for some \(\mathbf{a}\in\mathbb{Q}^{k-1}\), then every \(\mathbf{x}\in \ker_\mathbb{Q}(\mathbf{\hat{G}}_U^\top)\) satisfies \(\mathbf{x}^\top(\mathbf{g})_U = \mathbf{x}^\top\mathbf{\hat{G}}_U\mathbf{a} = 0\). 
	Hence no rational cancellation witness exists. 
\end{proof}

\subsection{Modular Cancellation under Adversarial Post-Processing}

For an integer matrix or vector, an overline denotes component-wise reduction modulo the modulus under consideration.
We use the Smith normal form to show that failure of integral column-module membership is detected modulo some integer.

\begin{lemma}\label{Lemma:SNF-membership}
	Let \(\mathbf{A}\in\mathbb{Z}^{r\times m}\), let \(\mathbf{v}\in\mathbb{Z}^r\), and let \(\mathbf{P}\mathbf{A}\mathbf{Q} = \begin{bmatrix} d_1\\ &\ddots\\ &&d_\rho\\ &&&\mathbf{0} \end{bmatrix}\) be a Smith normal form of \(\mathbf{A}\), where \(\mathbf{P}\in GL_r(\mathbb{Z})\), \(\mathbf{Q}\in GL_m(\mathbb{Z})\), \(0<d_1\mid\cdots\mid d_\rho\), and \(0\leq\rho\leq\min\{r,m\}\).
	Write \(\mathbf{v}':=\mathbf{P}\mathbf{v}\).
	Then \(\mathbf{v}\in\mathrm{Col}_\mathbb{Z}(\mathbf{A}) \) if and only if \(d_i\mid\mathbf{v}'[i]\) for every \(i\in[\rho]\) and \(\mathbf{v}'[i]=0\) for every \(i>\rho\).
\end{lemma}

\begin{proof}
	Since \(\mathbf{P}\) and \(\mathbf{Q}\) are unimodular, they induce isomorphisms of the relevant integer modules.
	The claim therefore follows by solving the diagonal Smith-normal-form system.
\end{proof}

\begin{lemma}\label{Lemma:SNF-separating-modulus}
	Let \(\mathbf{A}\in\mathbb{Z}^{r\times m}\) and \(\mathbf{v}\in\mathbb{Z}^r\).
	If \( \mathbf{v} \notin \mathrm{Col}_\mathbb{Z}(\mathbf{A})\), then there exists \(q\geq2\) such that \(\overline{\mathbf{v}} \notin \mathrm{Col}_{\mathbb{Z}_q} (\overline{\mathbf{A}})\).
\end{lemma}

\begin{proof}
	Use the notation of \Cref{Lemma:SNF-membership}.
	Integral non-membership implies either that \(\mathbf{v}'[i]\neq0\) for some \(i>\rho\), or that \(d_i\nmid\mathbf{v}'[i]\) for some \(i\leq\rho\).
	In the first case, choose a prime \(q\) not dividing \(\mathbf{v}'[i]\).
	The \(i\)-th row of the Smith normal form is zero modulo \(q\), whereas \(\mathbf{v}'[i]\not\equiv0\pmod q\).
	In the second case, choose a prime \(p\) for which \(v_p(\mathbf{v}'[i])<v_p(d_i)\), where \(v_p(\cdot)\) denotes the \(p\)-adic valuation function, and put \(q:=p^{v_p(\mathbf{v}'[i])+1}\).
	Then \(d_i\equiv0\pmod q\), while \(\mathbf{v}'[i]\not\equiv0\pmod q\).
	Thus, in either case, \(\overline{\mathbf{v}'}\) does not lie in the modular column module of the reduced Smith normal form.
	Since \(\mathbf{P}\) and \(\mathbf{Q}\) remain invertible modulo \(q\), the same conclusion holds for \(\mathbf{v}\) and \(\mathbf{A}\).
\end{proof}

\begin{lemma}[Noise-cancellation attack modulo \(q\)]\label{Lemma:ComputationalPrivacy-SEC-Column-Z}
	If \((\mathbf{g})_U \notin \mathrm{Col}_\mathbb{Z}(\mathbf{\hat{G}}_U)\), then there exist \(q\geq2\) and a nonzero vector \(\overline{\mathbf{x}}\in\mathbb{Z}_q^{|U|}\) such that \(\overline{\mathbf{x}}^\top \overline{\mathbf{\hat{G}}_U} = \mathbf{0}^\top\) and \( \overline{c} := \overline{\mathbf{x}}^\top \overline{(\mathbf{g})_U} \neq0\).
	Consequently, \(\overline{\mathbf{x}}^\top \overline{\mathbf{S}_U} = \overline{c}\, \overline{\mathbf{s}}^\top \pmod q\).
	Thus, any two secrets having different images under multiplication by \(\overline{c}\) induce perfectly distinguishable modular share views.
	If \(\overline{c}\) is a unit in \(\mathbb{Z}_q\), then \(\overline{\mathbf{s}}\) is fully recoverable.
\end{lemma}

\begin{proof}
	By \Cref{Lemma:SNF-separating-modulus}, there exists \(q\geq2\) such that \(\overline{(\mathbf{g})_U} \notin \mathrm{Col}_{\mathbb{Z}_q} (\overline{\mathbf{\hat{G}}_U})\).
	The double-annihilator property over the finite Frobenius ring \(\mathbb{Z}_q\) (see, e.g., \cite{wood1999duality} and \cite[Corollary 3.2]{dougherty2017algebraic}) gives a vector \(\overline{\mathbf{x}}\in\mathbb{Z}_q^{|U|}\) that annihilates every column of \(\overline{\mathbf{\hat{G}}_U}\) but not \(\overline{(\mathbf{g})_U}\).
	Reducing \eqref{Eq:SEC-Restricted-Share-Matrix} modulo \(q\) and multiplying by \(\overline{\mathbf{x}}^\top\) gives
	\begin{align}
		\overline{\mathbf{x}}^\top \overline{\mathbf{S}_U} &= \overline{\mathbf{x}}^\top \overline{(\mathbf{g})_U} \overline{\mathbf{s}}^\top + \overline{\mathbf{x}}^\top \overline{\mathbf{\hat{G}}_U} \overline{\mathbf{R}} =\overline{c}\, \overline{\mathbf{s}}^\top \pmod q. \label{Eq:Noise-Cancellation-modulo-q}
	\end{align}
	The distinguishability and recovery statements follow immediately.
\end{proof}

\subsection{Proof of the Structural Proposition}

\begin{proof}[Proof of \Cref{Proposition:No-Noise-Cancellation-Summary}]
	Suppose that \((\mathbf{g})_U \in \mathrm{Col}_\mathbb{Z}(\mathbf{\hat{G}}_U)\).
	Then there exists \(\mathbf{a}\in\mathbb{Z}^{k-1}\) such that \((\mathbf{g})_U = \mathbf{\hat{G}}_U\mathbf{a}\).
	First, the inclusion \(\mathrm{Col}_\mathbb{Z}(\mathbf{\hat{G}}_U) \subseteq \mathrm{Col}_\mathbb{Q}(\mathbf{\hat{G}}_U)\) implies \((\mathbf{g})_U \in \mathrm{Col}_\mathbb{Q}(\mathbf{\hat{G}}_U)\).
	Rational noise cancellation is therefore impossible by \Cref{Lemma:ComputationalPrivacy-SEC-Column-Q}.

	Second, \(\mathbf{G}_U = \begin{bmatrix} (\mathbf{g})_U \ \mathbf{\hat{G}}_U \end{bmatrix} = \mathbf{\hat{G}}_U \begin{bmatrix} \mathbf{a} \ \mathbf{I}_{k-1} \end{bmatrix}\).
	Hence \(\mathrm{Col}_\mathbb{Z}(\mathbf{G}_U) = \mathrm{Col}_\mathbb{Z}(\mathbf{\hat{G}}_U)\), the reverse inclusion being immediate.
	Moreover, \(\ker_\mathbb{Z}(\mathbf{G}_U^\top) \subseteq \ker_\mathbb{Z}(\mathbf{\hat{G}}_U^\top)\) always holds, and conversely, if \(\mathbf{x}\in \ker_\mathbb{Z}(\mathbf{\hat{G}}_U^\top)\), then \((\mathbf{g})_U^\top\mathbf{x} = \mathbf{a}^\top \mathbf{\hat{G}}_U^\top\mathbf{x} = 0\), and hence \(\mathbf{x}\in\ker_\mathbb{Z}(\mathbf{G}_U^\top)\).
	Therefore, \(\ker_\mathbb{Z}(\mathbf{G}_U^\top) = \ker_\mathbb{Z}(\mathbf{\hat{G}}_U^\top)\), which excludes integer noise cancellation.

	Finally, reducing \((\mathbf{g})_U=\mathbf{\hat{G}}_U\mathbf{a}\) modulo any \(q'\geq2\) gives \(\overline{(\mathbf{g})_U} = \overline{\mathbf{\hat{G}}_U}\, \overline{\mathbf{a}}\), so \(\overline{(\mathbf{g})_U}\in \mathrm{Col}_{\mathbb{Z}_{q'}} (\overline{\mathbf{\hat{G}}_U})\).
	Consequently, every vector annihilating \(\overline{\mathbf{\hat{G}}_U}\) also annihilates \(\overline{(\mathbf{g})_U}\), and modular noise cancellation is impossible.
\end{proof}

\section{Deferred Proofs for the Privacy Analysis}\label{Appendix:Privacy-Proofs}

This appendix contains the proofs omitted from \Cref{Section:Secret-Sharing-Privacy-Analysis}.
Throughout, \((\mathcal{U},\mathbf{c})\) is a fixed admissible public realization, the notation is that of \Cref{Table:Privacy-Notation}, and empty block rows and columns are omitted.

\subsection{Joint Precision Bound}\label{Appendix:Privacy-Proofs-Joint}

\begin{proof}[Proof of \Cref{Lemma:Block-Diagonal-Reduction}]
	By \eqref{Eq:Transformed-M-and-y}, \(\mathbf{\widetilde{y}}_{\mathcal{U},\mathbf{c}}\) has blocks \(\mathbf{y}_\mathrm{raw}\), \(\mathbf{r}_\mathrm{in}\), and \(\mathbf{y}_\mathrm{out}\).
	The masks in \(\mathbf{r}_\mathrm{in}\) are independent of \(\mathbf{r}_\mathbf{R}\) and of the distinct masks appearing in \(\mathbf{y}_\mathrm{out}\), so all parameter blocks between the \(\mathcal{S}_\mathrm{in}\) coordinates and the remaining coordinates vanish, and the \(\mathcal{S}_\mathrm{in}\) block equals \(\sigma_\mathbf{r}^2\mathbf{I}\).
	Let \(\mathbf{P}_\mathrm{cor}\) be the permutation that groups the raw-share and \(\mathcal{S}_\mathrm{out}\) coordinates first.
	Then \(\mathbf{P}_\mathrm{cor} \mathbf{\Sigma}_{\mathbf{\widetilde{y}}_{\mathcal{U},\mathbf{c}}} \mathbf{P}_\mathrm{cor}^\top = \mathbf{\Sigma}_\mathrm{cor} \oplus \sigma_\mathbf{r}^2 \mathbf{I}\).
	Both diagonal blocks are invertible, so \(\mathbf{P}_\mathrm{cor} \mathbf{\Sigma}_{\mathbf{\widetilde{y}}_{\mathcal{U},\mathbf{c}}}^{-1} \mathbf{P}_\mathrm{cor}^\top = \mathbf{\Sigma}_\mathrm{cor}^{-1} \oplus \sigma_\mathbf{r}^{-2} \mathbf{I}\), and undoing the permutation gives \eqref{Eq:Sigma-ytilde-inverse-unpermuted}.
	The norm identity holds because permutations preserve the spectral norm and the spectral norm of a block-diagonal positive-definite matrix is the maximum over its diagonal blocks.
\end{proof}

\begin{proof}[Proof of \Cref{Lemma:Exact-Reduction-Mcor}]
	By \eqref{Eq:Precision-Invariance-under-W}, the left-hand side of \eqref{Eq:Exact-Reduction-Mcor} equals \(\mathbf{\widetilde{M}}_{\mathcal{U},\mathbf{c}}^\top \mathbf{\Sigma}_{\mathbf{\widetilde{y}}_{\mathcal{U},\mathbf{c}}}^{-1} \mathbf{\widetilde{M}}_{\mathcal{U},\mathbf{c}}\).
	Substituting \eqref{Eq:Transformed-M-and-y} and \eqref{Eq:Sigma-ytilde-inverse-unpermuted}, every term involving an \(\mathcal{S}_\mathrm{in}\) row or column is multiplied by the zero block of \(\mathbf{\widetilde{M}}_{\mathcal{U},\mathbf{c}}\) and vanishes.
	The remaining blocks give \(\mathbf{\widetilde{M}}_{\mathcal{U},\mathbf{c}}^\top \mathbf{\Sigma}_{\mathbf{\widetilde{y}}_{\mathcal{U},\mathbf{c}}}^{-1} \mathbf{\widetilde{M}}_{\mathcal{U},\mathbf{c}} = \begin{bmatrix} \mathbf{M}_\mathrm{raw}\\ \mathbf{M}_\mathrm{out} \end{bmatrix}^{\!\top} \mathbf{\Sigma}_\mathrm{cor}^{-1} \begin{bmatrix} \mathbf{M}_\mathrm{raw}\\ \mathbf{M}_\mathrm{out} \end{bmatrix}\), which proves \eqref{Eq:Exact-Reduction-Mcor}.
\end{proof}

\begin{proof}[Proof of \Cref{Lemma:Splitting-M-cor}]
	By \eqref{Eq:B-cross-definition} and \eqref{Eq:C-Schur-complement-definition}, \(\mathbf{\Sigma}_\mathrm{raw}\mathbf{B} = \mathbf{\Sigma}_{\mathrm{raw},\mathrm{out}}\) and \(\mathbf{B}^\top\mathbf{\Sigma}_\mathrm{raw}\mathbf{B} + \mathbf{C} = \mathbf{\Sigma}_\mathrm{out}\), and hence
	\begin{align*}
		\mathbf{\Sigma}_\mathrm{cor} = \begin{bmatrix} \mathbf{I} & \mathbf{O}\\ \mathbf{B}^\top & \mathbf{I} \end{bmatrix} \begin{bmatrix} \mathbf{\Sigma}_\mathrm{raw} & \mathbf{O}\\ \mathbf{O} & \mathbf{C} \end{bmatrix} \begin{bmatrix} \mathbf{I} & \mathbf{B}\\ \mathbf{O} & \mathbf{I} \end{bmatrix}.
	\end{align*}
	Taking inverses gives \(\mathbf{\Sigma}_\mathrm{cor}^{-1} = \mathbf{K}^\top ( \mathbf{\Sigma}_\mathrm{raw}^{-1} \oplus \mathbf{C}^{-1} ) \mathbf{K}\) with \(\mathbf{K} := \begin{bmatrix} \mathbf{I} & \mathbf{O}\\ -\mathbf{B}^\top & \mathbf{I} \end{bmatrix}\).
	Since \(\mathbf{K}\mathbf{M}_\mathrm{cor} = \begin{bmatrix} \mathbf{M}_\mathrm{raw}^\top & (\mathbf{M}_\mathrm{out} - \mathbf{B}^\top\mathbf{M}_\mathrm{raw})^\top \end{bmatrix}^\top\), expanding the block-diagonal quadratic form proves \eqref{Eq:Splitting-M-cor}.
\end{proof}

\begin{proof}[Proof of \Cref{Lemma:Sigma-raw-Tout-B-bound}]
	By \eqref{Eq:Covariance-Block-Identities-raw}, \(\mathbf{\Sigma}_\mathrm{raw} = \sigma_\mathbf{R}^2 ( \mathbf{\hat{G}}_{U_\mathrm{raw}} \mathbf{\hat{G}}_{U_\mathrm{raw}}^\top ) \otimes \mathbf{I}_D\).
	Full row rank of \(\mathbf{\hat{G}}_{U_\mathrm{raw}}\) gives \(\mathbf{\Sigma}_\mathrm{raw}\succ0\) with \( \lambda_{\min}(\mathbf{\Sigma}_\mathrm{raw}) = \sigma_\mathbf{R}^2 \sigma_{\min}^2 (\mathbf{\hat{G}}_{U_\mathrm{raw}})\), which proves the inverse-norm identity.

	Suppose that \(\mathcal{S}_\mathrm{out}\neq\emptyset\).
	By \eqref{Eq:B-cross-definition}, \eqref{Eq:Covariance-Block-Identities-raw}, and \eqref{Eq:Covariance-Block-Identities-rawout}, \(\mathbf{B} = ( \mathbf{\hat{G}}_\mathrm{raw}^{+} )^\top \mathbf{T}_\mathrm{out}^\top\).
	Let \(\mathbf{\hat{G}}_{\mathcal{S}_\mathrm{out}}\) stack the rows \(\mathbf{\hat{g}}_i^\top\) over \((s,i)\in\mathcal{S}_\mathrm{out}\), with one copy per occurrence of \(i\), and let \(\mathbf{L}_{\mathcal{S}_\mathrm{out}}\) be block diagonal with block \(\mathbf{L}_s\) at coordinate \((s,i)\).
	Then
	\begin{equation}\label{Eq:T-out-Factorization}
		\mathbf{T}_\mathrm{out} = \mathbf{L}_{\mathcal{S}_\mathrm{out}} ( \mathbf{\hat{G}}_{\mathcal{S}_\mathrm{out}} \otimes \mathbf{I}_D ),
	\end{equation}
	with \(\|\mathbf{L}_{\mathcal{S}_\mathrm{out}}\|_2\leq\omega_c\) and \(\|\mathbf{\hat{G}}_{\mathcal{S}_\mathrm{out}}\|_2 \leq \sqrt{Q}\|\mathbf{\hat{G}}\|_2\); the last bound holds because \(\mathbf{\hat{G}}_{\mathcal{S}_\mathrm{out}} = \mathbf{E}_\mathrm{out}\mathbf{\hat{G}}\) for a row-selection matrix \(\mathbf{E}_\mathrm{out}\) in which each row of \(\mathbf{\hat{G}}\) is selected at most \(Q\) times, so \(\|\mathbf{E}_\mathrm{out}\|_2 \leq \sqrt{Q}\).
	Hence \(\|\mathbf{B}\|_2 \leq \|\mathbf{L}_{\mathcal{S}_\mathrm{out}}\|_2 \|\mathbf{\hat{G}}_{\mathcal{S}_\mathrm{out}}\|_2 \|\mathbf{\hat{G}}_{U_\mathrm{raw}}^+\|_2 \leq \omega_c\sqrt{Q}\|\mathbf{\hat{G}}\|_2/\sigma_{\min} (\mathbf{\hat{G}}_{U_\mathrm{raw}})\).
	Finally, \(\|\mathbf{\hat{G}}\|_2 = \|\mathbf{P}\mathbf{H}\|_2 = \|\mathbf{H}\|_2 \leq \|\mathbf{H}\|_F = \sqrt{c_\mathrm{exp}n}\).
\end{proof}

\begin{proof}[Proof of \Cref{Lemma:C-inverse-bound}]
	Substituting \eqref{Eq:Covariance-Block-Identities-raw}, \eqref{Eq:Covariance-Block-Identities-rawout}, and \eqref{Eq:Covariance-Block-Identities-out} into the definition of \(\mathbf{C}\) in \eqref{Eq:C-Schur-complement-definition} gives
	\begin{align*}
		\mathbf{C} = \sigma_\mathbf{R}^2 \mathbf{T}_\mathrm{out} \left( \mathbf{I} - \mathbf{\hat{G}}_\mathrm{raw}^+ \mathbf{\hat{G}}_\mathrm{raw} \right) \mathbf{T}_\mathrm{out}^\top + \sigma_\mathbf{r}^2 \mathbf{I}.
	\end{align*}
	Since \(\mathbf{\hat{G}}_\mathrm{raw}\) has full row rank, \(\mathbf{\hat{G}}_\mathrm{raw}^+ \mathbf{\hat{G}}_\mathrm{raw}\) is the orthogonal projector onto its row space, so \(\mathbf{I}-\mathbf{\hat{G}}_\mathrm{raw}^+\mathbf{\hat{G}}_\mathrm{raw} \succeq 0\), and congruence by \(\mathbf{T}_\mathrm{out}\) preserves this.
    Thus, \(\mathbf{C}\succeq\sigma_\mathbf{r}^2\mathbf{I}\succ0\) and \(\|\mathbf{C}^{-1}\|_2 =1/\lambda_{\min}(\mathbf{C}) \leq\sigma_\mathbf{r}^{-2}\).
\end{proof}

\begin{proof}[Proof of \Cref{Lemma:M-raw-out-norms}]
	Since \(\mathbf{M}_\mathrm{raw} = (\mathbf{g})_{U_\mathrm{raw}}\otimes\mathbf{I}_D\), we have \(\|\mathbf{M}_\mathrm{raw}\|_2 = \|(\mathbf{g})_{U_\mathrm{raw}}\|_2 \leq \beta_\mathbf{g}\sqrt{k_p}\).
	The matrix \(\mathbf{M}_\mathrm{out}\) stacks \(\mathbf{g}[i]\mathbf{L}_s\) over \((s,i)\in\mathcal{S}_\mathrm{out}\), so \(\|\mathbf{M}_\mathrm{out}\|_2^2 \leq \sum_{(s,i)\in\mathcal{S}_\mathrm{out}} \mathbf{g}[i]^2\|\mathbf{L}_s\|_2^2 \leq \beta_\mathbf{g}^2 \omega_c^2 |\mathcal{S}_\mathrm{out}| \leq \beta_\mathbf{g}^2 \omega_c^2 nQ\).
	Finally, \(\| \mathbf{M}_\mathrm{out} - \mathbf{B}^\top\mathbf{M}_\mathrm{raw} \|_2 \leq \|\mathbf{M}_\mathrm{out}\|_2 + \|\mathbf{B}\|_2 \|\mathbf{M}_\mathrm{raw}\|_2\), and \Cref{Lemma:Sigma-raw-Tout-B-bound} gives \eqref{Eq:Residual-M-bound}.
\end{proof}

\begin{proof}[Proof of \Cref{Lemma:Final-Joint-Bound}]
	We first verify the hypotheses of \Cref{Lemma:Block-Diagonal-Reduction,Lemma:Exact-Reduction-Mcor,Lemma:Splitting-M-cor}.
	By \Cref{Lemma:Sigma-raw-Tout-B-bound}, \(\mathbf{\Sigma}_\mathrm{raw}\succ0\).
	If \(\mathcal{S}_\mathrm{out}=\emptyset\), then \(\mathbf{\Sigma}_\mathrm{cor}=\mathbf{\Sigma}_\mathrm{raw}\succ0\).
	Otherwise, \Cref{Lemma:C-inverse-bound} gives \(\mathbf{C}\succ0\), and the block factorization of \(\mathbf{\Sigma}_\mathrm{cor}\) in the proof of \Cref{Lemma:Splitting-M-cor}, a congruence of \(\mathbf{\Sigma}_\mathrm{raw}\oplus\mathbf{C}\) by an invertible matrix, gives \(\mathbf{\Sigma}_\mathrm{cor}\succ0\).
	Together with \(\sigma_\mathbf{r}>0\), these are the required hypotheses.

	If \(\mathcal{S}_\mathrm{out}=\emptyset\), \Cref{Lemma:Exact-Reduction-Mcor} leaves only the raw-share coordinates, so the precision term equals \(\mathbf{M}_\mathrm{raw}^\top \mathbf{\Sigma}_\mathrm{raw}^{-1} \mathbf{M}_\mathrm{raw}\), whose norm is at most \(\|\mathbf{M}_\mathrm{raw}\|_2^2 \|\mathbf{\Sigma}_\mathrm{raw}^{-1}\|_2 \leq \beta_\mathbf{g}^2k_p / ( \sigma_\mathbf{R}^2 \sigma_{\min}^2 (\mathbf{\hat{G}}_{U_\mathrm{raw}}) )\) by \Cref{Lemma:Sigma-raw-Tout-B-bound,Lemma:M-raw-out-norms}; the second term of \eqref{Eq:Final-Joint-Precision-Bound} is nonnegative.
	If \(\mathcal{S}_\mathrm{out}\neq\emptyset\), \Cref{Lemma:Exact-Reduction-Mcor,Lemma:Splitting-M-cor} and the triangle inequality give
	\begin{align*}
		\left\| \mathbf{M}_{\mathcal{U},\mathbf{c}}^\top \mathbf{\Sigma}_{\mathbf{y}_{\mathcal{U},\mathbf{c}}}^{-1} \mathbf{M}_{\mathcal{U},\mathbf{c}} \right\|_2 &\leq \|\mathbf{M}_\mathrm{raw}\|_2^2 \|\mathbf{\Sigma}_\mathrm{raw}^{-1}\|_2 + \left\| \mathbf{M}_\mathrm{out} - \mathbf{B}^\top\mathbf{M}_\mathrm{raw} \right\|_2^2 \|\mathbf{C}^{-1}\|_2.
	\end{align*}
	The first term is bounded as before, and the second by \Cref{Lemma:C-inverse-bound} and \eqref{Eq:Residual-M-bound}.
\end{proof}

\paragraph{Verification of \Cref{Remark:Boundary-Leakage-Pattern}}

If \(U_\mathrm{raw}=\emptyset\) and \(\mathcal{S}\neq\emptyset\), then \(\mathcal{S}_\mathrm{in}=\emptyset\) and \(\mathcal{S}_\mathrm{out}=\mathcal{S}\), so there are no raw-share or \(\mathcal{S}_\mathrm{in}\) blocks, \(\mathbf{M}_{\mathcal{U},\mathbf{c}}=\mathbf{M}_\mathrm{out}\), and \(\mathbf{\Sigma}_{\mathbf{y}_{\mathcal{U},\mathbf{c}}}=\mathbf{\Sigma}_\mathrm{out} = \sigma_\mathbf{R}^2 \mathbf{T}_\mathrm{out} \mathbf{T}_\mathrm{out}^\top + \sigma_\mathbf{r}^2\mathbf{I} \succeq \sigma_\mathbf{r}^2\mathbf{I}\).
\Cref{Lemma:M-raw-out-norms} assumes \(|U_\mathrm{raw}|\geq1\), but the argument for its bound \(\|\mathbf{M}_\mathrm{out}\|_2^2 \leq \beta_\mathbf{g}^2 \omega_c^2 |\mathcal{S}_\mathrm{out}|\) uses only \(\|\mathbf{g}\|_\infty\leq\beta_\mathbf{g}\) and \(\|\mathbf{L}_s\|_2\leq\omega_c\), and therefore also applies here.
Hence the precision term is at most \(\|\mathbf{M}_\mathrm{out}\|_2^2 \|\mathbf{\Sigma}_\mathrm{out}^{-1}\|_2 \leq \beta_\mathbf{g}^2 \omega_c^2 nQ / \sigma_\mathbf{r}^2\).
If \(U_\mathrm{raw}\neq\emptyset\) and \(\mathcal{S}_\mathrm{out}=\emptyset\), including \(\mathcal{S}=\emptyset\), every response block is eliminated by \(\mathbf{W}\), and the first case of the proof of \Cref{Lemma:Final-Joint-Bound} applies. 
If \(U_\mathrm{raw}=\emptyset\) and \(\mathcal{S}=\emptyset\), then \(\mathbf{M}_{\mathcal{U},\mathbf{c}}\) and \(\mathbf{y}_{\mathcal{U},\mathbf{c}}\) are empty, and the precision term is \(0\) by convention. 
All remaining patterns are covered by \Cref{Lemma:Final-Joint-Bound}.

\subsection{Uniform Parameters}\label{Appendix:Privacy-Proofs-Uniform}

\begin{proof}[Proof of \Cref{Lemma:alpha-kp-uniform}]
	Each \(\mathbf{\hat{G}}_V\) with \(|V|=k_p\) has full row rank, so \(\alpha_{k_p}\) is a minimum of finitely many positive numbers.
	For a nonempty \(U\) with \(|U|\leq k_p\), choose \(V\supseteq U\) with \(|V|=k_p\).
	Since \(\mathbf{\hat{G}}_U\mathbf{\hat{G}}_U^\top\) is a principal submatrix of \(\mathbf{\hat{G}}_V\mathbf{\hat{G}}_V^\top\), eigenvalue interlacing gives \(\sigma_{\min}^2(\mathbf{\hat{G}}_U) \geq \sigma_{\min}^2(\mathbf{\hat{G}}_V) \geq \alpha_{k_p}^2\).
\end{proof}

\begin{proof}[Proof of \Cref{Lemma:Uniform-MQ-Precision}]
	If \(U_\mathrm{raw}\neq\emptyset\), then \(1/\sigma_{\min}(\mathbf{\hat{G}}_{U_\mathrm{raw}}) \leq 1/\alpha_{k_p}\) by \Cref{Lemma:alpha-kp-uniform}, and substituting into \Cref{Lemma:Final-Joint-Bound} gives \(\Xi_\mathrm{MQ}\).
	If \(U_\mathrm{raw}=\emptyset\) and \(\mathcal{S}\neq\emptyset\), \Cref{Remark:Boundary-Leakage-Pattern} gives the bound \(\beta_\mathbf{g}^2\omega_c^2nQ/\sigma_\mathbf{r}^2\), which is at most the second term of \(\Xi_\mathrm{MQ}\) because in \eqref{Eq:Xi-MQ} we have \(\left( \sqrt{n} + \sqrt{k_p} \frac{\|\mathbf{\hat{G}}\|_2}{\alpha_{k_p}} \right)^2 \geq n\).
	The empty view is immediate.
	The bound does not depend on \(\mathbf{c}\), since each \(c_s\) enters only through \(\|\mathbf{L}_s\|_2\leq\omega_c\).
\end{proof}

\begin{proof}[Proof of \Cref{Lemma:Phi-tilde-Surjective}]
	The inclusion \(\mathbf{\widetilde{\Phi}}_{\mathcal{U},\mathbf{c}} \mathbb{Z}^{M} \subseteq \mathbb{Z}^{M'}\) is immediate.
	Suppose that \(U_\mathrm{raw}\neq\emptyset\).
	By \Cref{Lemma:Column-Space-Ghat-U-Expander}, \(\mathbf{\hat{G}}_{U_\mathrm{raw}}\mathbb{Z}^{k-1} = \mathbb{Z}^{|U_\mathrm{raw}|}\), and tensoring with \(\mathbf{I}_D\) gives \(\mathbf{\hat{G}}_\mathrm{raw}\mathbb{Z}^{(k-1)D} = \mathbb{Z}^{|U_\mathrm{raw}|D}\).
	Given an arbitrary \(\begin{bmatrix}\mathbf{a}^\top \ \mathbf{b}_\mathrm{in}^\top \ \mathbf{b}_\mathrm{out}^\top\end{bmatrix}^\top \in \mathbb{Z}^{M'}\), choose \(\mathbf{v}\) integral with \(\mathbf{\hat{G}}_\mathrm{raw}\mathbf{v}=\mathbf{a}\); then the source vector \(\begin{bmatrix}\mathbf{v}^\top \ \mathbf{b}_\mathrm{in}^\top \ (\mathbf{b}_\mathrm{out}-\mathbf{T}_\mathrm{out}\mathbf{v})^\top\end{bmatrix}^\top\) maps to it.
	If \(U_\mathrm{raw}=\emptyset\), then \(\mathcal{S}_\mathrm{in}=\emptyset\), and the output block contains \(\mathbf{I}_{|\mathcal{S}|D}\).
\end{proof}

\begin{proof}[Proof of \Cref{Lemma:Whitened-Kernel-Smoothing}]
	The proof has three steps: we bound the Gram-Schmidt parameter of \(\ker_\mathbb{Z}(\mathbf{\hat{G}}_\mathrm{raw})\), transfer this bound to the whitened kernel through a linear map of norm at most \(1/\sigma_\mathrm{eff}\), and apply a basis-dependent smoothing estimate.

	\noindent\emph{Step 1: the unwhitened kernel.}
	Coefficient expansion does not increase the Gram-Schmidt parameter: for every \(U\subseteq[n]\),
	\begin{equation}\label{Eq:Kernel-Coefficient-Expansion-GS}
		\widetilde{bl} \left( \ker_\mathbb{Z} ( \mathbf{\hat{G}}_U\otimes\mathbf{I}_D ) \right) \leq \widetilde{bl} \left( \ker_\mathbb{Z}(\mathbf{\hat{G}}_U) \right).
	\end{equation}
	Indeed, if \((\mathbf{b}_1,\dots,\mathbf{b}_r)\) is an ordered basis of \(\ker_\mathbb{Z}(\mathbf{\hat{G}}_U)\) and \((\mathbf{f}_c)_{c\in[D]}\) is the standard basis of \(\mathbb{Z}^D\), then, up to a coordinate permutation, \(\{ \mathbf{b}_j\otimes\mathbf{f}_c\}\) is a basis of \(\ker_\mathbb{Z}(\mathbf{\hat{G}}_U\otimes\mathbf{I}_D)\); ordering it by \(c\) and then by \(j\) gives \(D\) mutually orthogonal copies of the original Gram-Schmidt process.
	If \(U_\mathrm{raw}\neq\emptyset\), then \eqref{Eq:Kernel-Coefficient-Expansion-GS} with \(U=U_\mathrm{raw}\) and \eqref{Eq:gamma-kernel-definition} give \(\widetilde{bl} ( \ker_\mathbb{Z}(\mathbf{\hat{G}}_\mathrm{raw}) ) \leq \gamma_{k_p}^{\ker}\).
	If \(U_\mathrm{raw}=\emptyset\), then \(\mathbf{\hat{G}}_\mathrm{raw}\) has no rows, so \(\ker_\mathbb{Z}(\mathbf{\hat{G}}_\mathrm{raw})=\mathbb{Z}^{(k-1)D}\), whose standard basis gives \(\widetilde{bl}(\mathbb{Z}^{(k-1)D})=1\leq\gamma_{k_p}^{\ker}\) by the outer maximum with \(1\) in \eqref{Eq:gamma-kernel-definition}.
	In both cases,
	\begin{equation}\label{Eq:Kernel-Raw-GS-gamma}
		\widetilde{bl} \left( \ker_\mathbb{Z}(\mathbf{\hat{G}}_\mathrm{raw}) \right) \leq \gamma_{k_p}^{\ker}.
	\end{equation}

	\noindent\emph{Step 2: the whitened kernel.}
	Define the injective map \(\mathbf{F}_{\mathcal{U},\mathbf{c}}\mathbf{v} := \begin{bmatrix}\sigma_\mathbf{R}^{-1}\mathbf{v}^\top \ \mathbf{0}^\top \ -\sigma_\mathbf{r}^{-1} \mathbf{v}^\top\mathbf{T}_\mathrm{out}^\top\end{bmatrix}^\top\).
	Since \(\mathbf{\Sigma}_0^{-1/2}\) scales the source blocks by \(\sigma_\mathbf{R}^{-1}\) and \(\sigma_\mathbf{r}^{-1}\), \eqref{Eq:Phi-tilde-kernel} gives \(\mathbf{\Sigma}_0^{-1/2} \ker_\mathbb{Z} ( \mathbf{\widetilde{\Phi}}_{\mathcal{U},\mathbf{c}} ) = \mathbf{F}_{\mathcal{U},\mathbf{c}} \ker_\mathbb{Z} ( \mathbf{\hat{G}}_\mathrm{raw} )\).
	For every \(\mathbf{v}\), \(\|\mathbf{F}_{\mathcal{U},\mathbf{c}}\mathbf{v}\|_2^2 \leq ( \sigma_\mathbf{R}^{-2} + \sigma_\mathbf{r}^{-2}\|\mathbf{T}_\mathrm{out}\|_2^2 ) \|\mathbf{v}\|_2^2\).
	The factorization \eqref{Eq:T-out-Factorization}, together with the bounds \(\|\mathbf{L}_{\mathcal{S}_\mathrm{out}}\|_2\leq\omega_c\) and \(\|\mathbf{\hat{G}}_{\mathcal{S}_\mathrm{out}}\|_2 \leq \sqrt{Q}\|\mathbf{\hat{G}}\|_2\) established after it in the proof of \Cref{Lemma:Sigma-raw-Tout-B-bound} (an argument that does not use \(U_\mathrm{raw}\)), gives \(\|\mathbf{T}_\mathrm{out}\|_2^2 \leq \omega_c^2Q \|\mathbf{\hat{G}}\|_2^2\); if \(\mathcal{S}_\mathrm{out}=\emptyset\), then \(\mathbf{T}_\mathrm{out}\) is empty and the bound is trivial.
	By the definition of \(\sigma_\mathrm{eff}\) in \eqref{Eq:sigma-eff-definition}, \(\|\mathbf{F}_{\mathcal{U},\mathbf{c}}\|_2 \leq 1/\sigma_\mathrm{eff}\).
	For every lattice \(\Lambda\) and every linear map \(\mathbf{F}\) injective on \(\mathrm{span}_\mathbb{R}(\Lambda)\), \(\widetilde{bl}(\mathbf{F}\Lambda) \leq\|\mathbf{F}\|_2\widetilde{bl}(\Lambda)\). 
    Indeed, for an ordered basis \((\mathbf{b}_j)\) of \(\Lambda\) and \(V_{j-1}:=\mathrm{span}_\mathbb{R} \{\mathbf{b}_1,\dots,\mathbf{b}_{j-1}\}\), the \(j\)-th Gram-Schmidt norm of \((\mathbf{F}\mathbf{b}_j)\) is \(\mathrm{dist}(\mathbf{F}\mathbf{b}_j,\mathbf{F}V_{j-1}) \leq \|\mathbf{F}\|_2 \mathrm{dist}(\mathbf{b}_j,V_{j-1})\), where \(\mathrm{dist}(\mathbf{x},V):= \min_{\mathbf{v}\in V}\|\mathbf{x}-\mathbf{v}\|_2\).
	Applying this with \(\mathbf{F}=\mathbf{F}_{\mathcal{U},\mathbf{c}}\) and \eqref{Eq:Kernel-Raw-GS-gamma} gives
	\begin{equation}\label{Eq:Whitened-Kernel-GS-Bound}
		\widetilde{bl} \left( \mathbf{\Sigma}_0^{-1/2} \ker_\mathbb{Z} \left( \mathbf{\widetilde{\Phi}}_{\mathcal{U},\mathbf{c}} \right) \right) \leq \frac{\gamma_{k_p}^{\ker}}{\sigma_\mathrm{eff}}.
	\end{equation}

	\noindent\emph{Step 3: smoothing.}
	By \Cref{Lemma:Phi-tilde-Surjective}, the kernel has rank \(M-M' = (k-1-|U_\mathrm{raw}|)D \leq (k-1)D\).
	The basis-dependent smoothing estimate of \cite[Lemma 2.8]{genise2020improved}, rescaled by \(1/\sqrt{2\pi}\) to the normalization \eqref{Eq:Smoothing-Parameter-SD}, together with \eqref{Eq:Whitened-Kernel-GS-Bound}, gives
	\begin{align*}
		\eta_{\epsilon_0} \left( \mathbf{\Sigma}_0^{-1/2} \ker_\mathbb{Z} \left( \mathbf{\widetilde{\Phi}}_{\mathcal{U},\mathbf{c}} \right) \right) \leq \frac{\gamma_{k_p}^{\ker}}{\sigma_\mathrm{eff}} \sqrt{ \frac{ \ln \left( 2(k-1)D \left( 1+1/\epsilon_0 \right) \right) }{ 2\pi^2 } }.
	\end{align*}
	Since \(1+1/\epsilon_0 = 2+1/\epsilon'\), the condition \eqref{Eq:sigma-eff-condition} implies \eqref{Eq:Whitened-Kernel-Smoothing-Condition}.
\end{proof}

\begin{proof}[Proof of \Cref{Lemma:Kernel-GS-Bound}]
	If \(t=0\), then \(\mathbf{\hat{G}}_U=\mathbf{M}_U\) is invertible and the kernel is trivial.
	Suppose \(t\geq1\), and write \(\mathbf{b}'_a = \mathbf{P}_U^{-1}\mathbf{b}_a\).
	Since \(\mathbf{M}_U\) is unimodular, \(\mathbf{M}_U^{-1}\) is integral, so \(\mathbf{b}'_a\in\mathbb{Z}^r\), and \(\begin{bmatrix} \mathbf{M}_U \ \mathbf{N}_U \end{bmatrix} \mathbf{b}'_a = \mathbf{0}\).
	Conversely, if \(\begin{bmatrix} \mathbf{M}_U \ \mathbf{N}_U \end{bmatrix}\begin{bmatrix}\mathbf{y}\\\mathbf{x}\end{bmatrix} = \mathbf{0}\) with \(\mathbf{y}\in\mathbb{Z}^{|U|}\) and \(\mathbf{x}\in\mathbb{Z}^t\), then \(\mathbf{y} = -\mathbf{M}_U^{-1}\mathbf{N}_U\mathbf{x}\), and hence \(\begin{bmatrix}\mathbf{y}\\\mathbf{x}\end{bmatrix} = \begin{bmatrix}-\mathbf{M}_U^{-1}\mathbf{N}_U\mathbf{x}\\\mathbf{x}\end{bmatrix} = \sum_{a=1}^t\mathbf{x}[a]\begin{bmatrix}-\mathbf{M}_U^{-1}\mathbf{N}_U\mathbf{e}_a\\\mathbf{e}_a\end{bmatrix}=\sum_{a=1}^{t} \mathbf{x}[a]\mathbf{b}'_a\).
	The vectors \(\mathbf{b}'_a\) are linearly independent because their last \(t\) coordinates are \(\mathbf{e}_1,\dots,\mathbf{e}_t\), so they form a basis of the permuted kernel; since \(\mathbf{P}_U\) is unimodular, the \(\mathbf{b}_a\) form a basis of \(\ker_\mathbb{Z}(\mathbf{\hat{G}}_U)\).
	Since \(\mathbf{P}_U\) is an isometry,
	\begin{align*}
		\|\mathbf{b}_a\|_2^2 = 1+ \left\| \mathbf{M}_U^{-1} \mathbf{N}_U\mathbf{e}_a \right\|_2^2 \leq 1+ \frac{ \left\| \mathbf{N}_U\mathbf{e}_a \right\|_2^2 }{ \sigma_{\min}^2(\mathbf{M}_U) } \leq 1+ \frac{ \min\{k_p,\beta_{k_p}^2\} }{ \gamma_{k_p}^2 },
	\end{align*}
	because \(\mathbf{N}_U\mathbf{e}_a\) is a column of \(\mathbf{\hat{G}}_U\in\{-1,0,1\}^{|U|\times r}\), whose squared norm is at most \(\min\{|U|,\|\mathbf{\hat{G}}_U\|_2^2\}\).
	Each Gram-Schmidt vector is no longer than the corresponding basis vector, and the right-hand side is at least \(1\), which proves \eqref{Eq:Kernel-GS-Bound-Explicit}.
\end{proof}

\begin{proof}[Proof of \Cref{Lemma:Uniform-Gaussian-Joint-Perturbation}]
	By \eqref{Eq:Phi-tilde-definition}, \(\mathbf{\widetilde{y}}_{\mathcal{U},\mathbf{c}} = \mathbf{\widetilde{\Phi}}_{\mathcal{U},\mathbf{c}} \mathbf{r}_0\) with \(\mathbf{r}_0 \gets \mathcal{D}_{\mathbb{Z}^{M},\sqrt{\mathbf{\Sigma}_0}}\).

	\noindent\emph{Trivial kernel.}
	If \(\ker_\mathbb{Z} ( \mathbf{\widetilde{\Phi}}_{\mathcal{U},\mathbf{c}} ) = \{\mathbf{0}\}\), then \Cref{Lemma:Phi-tilde-Surjective} gives \(M=M'\) and \(\mathbf{\widetilde{\Phi}}_{\mathcal{U},\mathbf{c}} \in GL_M(\mathbb{Z})\), so the pushforward is exactly \(\mathcal{D}_{ \mathbb{Z}^{M'}, \sqrt{ \mathbf{\Sigma}_{ \mathbf{\widetilde{y}}_{\mathcal{U},\mathbf{c}} } } }\) by \eqref{Eq:Sigma-ytilde-Phi-tilde}.

	\noindent\emph{Non-trivial kernel.}
	Apply \cite[Theorem 3.1]{genise2020improved} with source lattice \(\mathbb{Z}^{M}\), linear map \(\mathbf{\widetilde{\Phi}}_{\mathcal{U},\mathbf{c}}\), and parameter \(\mathbf{S} := \sqrt{2\pi}\, \mathbf{\Sigma}_0^{1/2}\).
	Under the convention of that theorem, the source mass at \(\mathbf{x}\) is proportional to \(\exp ( -\pi \| \mathbf{S}^{-1}\mathbf{x} \|_2^2 ) = \exp ( -\frac{1}{2} \mathbf{x}^\top \mathbf{\Sigma}_0^{-1} \mathbf{x} )\), so the source distribution is \(\mathcal{D}_{\mathbb{Z}^{M},\sqrt{\mathbf{\Sigma}_0}}\) in our convention.
	Since \(\mathbf{S}^{-1} = (2\pi)^{-1/2}\mathbf{\Sigma}_0^{-1/2}\), its smoothing hypothesis \(\eta^\pi_{\epsilon_0} ( \mathbf{S}^{-1}\ker_\mathbb{Z} ( \mathbf{\widetilde{\Phi}}_{\mathcal{U},\mathbf{c}} ) ) \leq 1\) is exactly \eqref{Eq:Whitened-Kernel-Smoothing-Condition}, which holds by \Cref{Lemma:Whitened-Kernel-Smoothing}; the image lattice is \(\mathbb{Z}^{M'}\) by \Cref{Lemma:Phi-tilde-Surjective}.
	The theorem gives \(\Delta ( \mathbf{\widetilde{y}}_{\mathcal{U},\mathbf{c}}, \mathcal{D}_{ \mathbb{Z}^{M'}, \sqrt{ \mathbf{\Sigma}_{ \mathbf{\widetilde{y}}_{\mathcal{U},\mathbf{c}} } } } ) \leq \epsilon_0/(1-\epsilon_0) = \epsilon'\).

	\noindent\emph{Transfer.}
	The map \(\mathbf{x}\mapsto\mathbf{W}^{-1}\mathbf{x}\) is a bijection of \(\mathbb{Z}^{M'}\) and preserves statistical distance.
	It maps \(\mathbf{\widetilde{y}}_{\mathcal{U},\mathbf{c}}\) to \(\mathbf{y}_{\mathcal{U},\mathbf{c}}\) and, by \eqref{Eq:Sigma-ytilde-Phi-tilde}, \(\mathcal{D}_{ \mathbb{Z}^{M'}, \sqrt{ \mathbf{\Sigma}_{ \mathbf{\widetilde{y}}_{\mathcal{U},\mathbf{c}} } } }\) to \(\mathcal{D}_{ \mathbb{Z}^{M'}, \sqrt{ \mathbf{\Sigma}_{ \mathbf{y}_{\mathcal{U},\mathbf{c}} } } }\).
	All quantities in \eqref{Eq:sigma-eff-condition} are independent of \((\mathcal{U},\mathbf{c})\), so the bound is uniform.
\end{proof}

\subsection{Reductions}\label{Appendix:Privacy-Proofs-Reduction}

\begin{proof}[Proof of \Cref{Lemma:HCLeak-to-MHMLWE-multi}]
	For \(b\in\{0,1\}\), let \(H_b\) be branch \(b\) of the fixed-pattern \(\mathsf{HC\text{-}Leak}\) experiment, whose auxiliary leakage is \(\mathbf{z}_{\mathcal{U},\mathbf{c}} = \mathbf{M}_{\mathcal{U},\mathbf{c}}\mathbf{s} + \mathbf{y}_{\mathcal{U},\mathbf{c}}\), and let \(\breve{H}_b\) be obtained by replacing \(\mathbf{y}_{\mathcal{U},\mathbf{c}}\) with \(\mathbf{\breve{y}}_{\mathcal{U},\mathbf{c}}\).
	Both perturbations are independent of the secret and all other variables are generated identically, so applying the deterministic map \((\mathbf{s},\mathbf{y}) \mapsto \mathbf{M}_{\mathcal{U},\mathbf{c}}\mathbf{s} + \mathbf{y}\) and \Cref{Lemma:Uniform-Gaussian-Joint-Perturbation} gives \(\Delta(H_b,\breve{H}_b) \leq \epsilon'\).
	The Gaussianized branches are the one-hint Matrix Hint-MLWE distributions \(( \bm{A}', \bm{A}\bm{s}, \mathbf{M}_{\mathcal{U},\mathbf{c}}, \mathbf{\breve{z}}_{\mathcal{U},\mathbf{c}} )\) and \(( \bm{A}', \bm{u}, \mathbf{M}_{\mathcal{U},\mathbf{c}}, \mathbf{\breve{z}}_{\mathcal{U},\mathbf{c}} )\) with \(\bm{u} \stackrel{\$}{\gets} \mathcal{R}_q^p\), where the integer hint dimension \(M'\) corresponds to module dimension \((|U_\mathrm{raw}|+|\mathcal{S}|)\ell\).
	On a challenge \(( \bm{A}', \bm{v}, \mathbf{M}_{\mathcal{U},\mathbf{c}}, \mathbf{\breve{z}}_{\mathcal{U},\mathbf{c}} )\), the adversary \(\mathcal{B}\) parses \(\mathbf{\breve{z}}_{\mathcal{U},\mathbf{c}}\) according to \eqref{Eq:z-Uc-Structured-Encoding}, reshaping the first \(|U_\mathrm{raw}|D\) coordinates into the raw-share block and the rest into the response blocks indexed by \(\mathcal{S}\); since the ordering and dimensions are public, this recovers the structured Gaussianized view exactly.
	It gives this view, the public realization, and \(\bm{v}\) to \(\mathcal{A}\) and returns its output.
	The simulation is distributed as \(\breve{H}_1\) when \(\bm{v}=\bm{A}\bm{s}\bmod q\) and as \(\breve{H}_0\) when \(\bm{v}\) is uniform, so \(| \Pr[\mathcal{A}(\breve{H}_1)=1] - \Pr[\mathcal{A}(\breve{H}_0)=1] | = \mathsf{Adv}_\mathcal{B}^\mathsf{MHMLWE}(\kappa)\).
	The triangle inequality over the two branches completes the proof.
\end{proof}

\begin{proof}[Proof of \Cref{Lemma:Multi-HCLeak-to-MLWE}]
	By \Cref{Lemma:HCLeak-to-MHMLWE-multi}, there is \(\mathcal{B}_1\) with \(\mathsf{Adv}_{\mathcal{A},\mathsf{HC\text{-}Leak}}^{(\mathsf{pp},\mathcal{U},\mathbf{c})}(\kappa)\leq \mathsf{Adv}_{\mathcal{B}_1}^\mathsf{MHMLWE}(\kappa)+2\epsilon'\).
    The lower bound \(\sqrt{2}\eta_\epsilon(\mathbb{Z}^D)\leq\sigma_\mathrm{MQ}\) is the posterior-resampling condition required by \Cref{Corollary:Hardness-MHMLWE-Tight-Smaller-Sigma}.
	The upper bound \(\sigma_\mathrm{MQ}\leq\sigma_{\mathcal{U},\mathbf{c}}\) follows by comparing \eqref{Eq:Sigma-Pattern-Dependent-MQ} with \eqref{Eq:Sigma-MQ-Uniform-Parameter}, since \Cref{Lemma:Uniform-MQ-Precision} bounds the precision term in \eqref{Eq:Sigma-Pattern-Dependent-MQ} by \(\Xi_\mathrm{MQ}\).
	Hence \(\sqrt{2}\eta_\epsilon(\mathbb{Z}^D)\leq\sigma_\mathrm{MQ}\leq\sigma_{\mathcal{U},\mathbf{c}}\).
	Applying \Cref{Corollary:Hardness-MHMLWE-Tight-Smaller-Sigma} with \(n=p\), \(m = \ell-p\), \(d(n+m)=D\), \(\sigma'=\sigma_\mathrm{MQ}\), and \(\sigma=\sigma_{\mathcal{U},\mathbf{c}}\) gives \(\mathcal{B}\) with \(\mathsf{Adv}_{\mathcal{B}_1}^\mathsf{MHMLWE}(\kappa)\leq \mathsf{Adv}_{\mathcal{B},\mathsf{MLWE}_{p,\ell-p,q,d,\sigma_\mathrm{MQ}}}(\kappa)+2\epsilon\).
    Composing the last inequality with the result from \Cref{Lemma:HCLeak-to-MHMLWE-multi} concludes the lemma.
\end{proof}

\begin{proof}[Proof of \Cref{Theorem:SEC-HCLLR-multi}]
	Fix \(\kappa\), a setup output \(\mathsf{pp}\in\mathsf{Good}_\kappa\), a PPT adversary \(\mathcal{A}\), and a pattern \(\mathcal{U}\in\mathcal{U}_{k_p,Q}\).
	Since \(\mathsf{Good}_\mathrm{MQ}(\mathsf{pp})\) holds, the conclusions of \Cref{Section:Uniform-Parameters-Multi-Query-Privacy} hold for all admissible patterns, so the hypotheses of \Cref{Lemma:Multi-HCLeak-to-MLWE} hold for every \(\mathbf{c}\in\mathcal{C}^Q\) whenever the auxiliary view is nonempty.

	\noindent\emph{Empty auxiliary view.}
	If \(U_\mathrm{raw} = \emptyset\) and \(\mathcal{S} = \emptyset\), then, apart from public information generated identically in both branches, the branches are \(( \bm{A}', \bm{A} \bm{s} )\) and \( (\bm{A}',\bm{u})\), the real and random distributions of \(\mathsf{MLWE}_{p,\ell-p,q,d,\sigma_\mathbf{s}}\).
	In this case, \(\mathcal{B}_\mathsf{pp}\) samples \(\mathbf{c} \stackrel{\$}{\gets} \mathcal{C}^{Q}\) and the remaining public information itself, so \( \mathsf{Adv}_{ \mathcal{A}, \mathsf{HC\text{-}Leak} }^{(\mathsf{pp},\mathcal{U})}(\kappa) \leq \mathsf{Adv}_{ \mathcal{B}_\mathsf{pp}, \mathsf{MLWE}_{p,\ell-p,q,d,\sigma_\mathbf{s}} }(\kappa)\).

	\noindent\emph{Nonempty auxiliary view.}
	On its \(\mathsf{MLWE}\) challenge, \(\mathcal{B}_\mathsf{pp}\) samples \(\mathbf{c} \stackrel{\$}{\gets} \mathcal{C}^{Q}\) exactly as in the experiment and runs the reduction of \Cref{Lemma:Multi-HCLeak-to-MLWE} for \((\mathcal{U},\mathbf{c})\).
	This is well-defined for every sampled \(\mathbf{c}\) because \(\|\mathbf{L}_s\|_2 \leq \omega_c\) and the smoothing and posterior-resampling conditions hold by assumption, and the width \(\sigma_\mathrm{MQ}\) and losses are uniform over \(\mathbf{c}\).
	Hence \(\mathsf{Adv}_{ \mathcal{A}, \mathsf{HC\text{-}Leak} }^{(\mathsf{pp},\mathcal{U})}(\kappa) \leq \mathsf{Adv}_{ \mathcal{B}_\mathsf{pp}, \mathsf{MLWE}_{p,\ell-p,q,d,\sigma_\mathrm{MQ}} }(\kappa) + 2\epsilon' + 2\epsilon\).

	\noindent\emph{Conclusion.}
	In both cases, \(\mathsf{Adv}_{ \mathcal{A}, \mathsf{HC\text{-}Leak} }^{(\mathsf{pp},\mathcal{U})}(\kappa) \leq \mathsf{Adv}_{\mathcal{B}_\mathsf{pp},\mathsf{MLWE}}(\kappa)+2\epsilon'+2\epsilon \leq \epsilon_\mathrm{MQ}(\kappa,\mathcal{A},\mathcal{U})\), where the last inequality holds because \(\epsilon_\mathrm{MQ}\) is the maximum over \(\mathsf{Good}_\kappa\). 
	This holds for every \(\mathsf{pp}\in\mathsf{Good}_\kappa\), every PPT adversary \(\mathcal{A}\), and every \(\mathcal{U}\in\mathcal{U}_{k_p,Q}\), and \(\mathsf{Good}_\kappa\) depends neither on \(\mathcal{A}\) nor on \(\mathcal{U}\). 
	Therefore, the event in \Cref{Theorem:SEC-HCLLR-multi} contains \(\mathsf{Good}_\kappa\), and its probability is at least \(\Pr_{\mathsf{pp}\gets\mathsf{Setup}(1^\kappa)}[\mathsf{pp}\in\mathsf{Good}_\kappa] = 1-\delta_\mathrm{MQ}\) by \eqref{Eq:delta-MQ}. 
	Finally, sampling \(\bm{A}'\) uniformly and setting \(\bm{A}:=\begin{bmatrix}\bm{I}_p \ \bm{A}'\end{bmatrix}\) induces the stated distribution on \(\mathcal{F}_\mathrm{MQ}\), and the real and random branches above are exactly those of \Cref{Definition:HC-Leak} for \(f_{\bm{A}}\). 
	The preceding probability and advantage bounds therefore establish the \((\delta_\mathrm{MQ},\epsilon_\mathrm{MQ})\)-privacy required by \Cref{Definition:HC-LLR}. 
	Under the third assumption of the theorem, \(\epsilon_\mathrm{MQ}(\kappa,\mathcal{A},\mathcal{U})\) is negligible in \(\kappa\) for every PPT adversary \(\mathcal{A}\) and every \(\mathcal{U}\), which gives \(\delta_\mathrm{MQ}\)-privacy.
\end{proof}

\section{Gaussian Tail Bounds and Size Estimates}\label{Appendix:Gaussian-Tail-Size}

This appendix proves the tail bounds of \Cref{Section:SEC-Gaussian-Tail} and derives the size estimates reported in \Cref{Table:SEC-Selected-Sizes}.

\subsection{Discrete Gaussian Tail Bound}

\begin{proof}[Proof of \Cref{Lemma:Discrete-Gaussian-Tail-IA}]
	The bound \eqref{Eq:Discrete-Gaussian-Tail} is \cite[Lemma 4.4]{lyubashevsky2012lattice}, whose Gaussian function \(\exp(-\|\mathbf{x}\|_2^2/(2\sigma^2))\) agrees with \eqref{Eq:DiscreteGaussianDist} for \(\mathbf{\Sigma}=\sigma^2\mathbf{I}_N\).
	Write \(\theta=1+x\) with \(x>0\).
	Then \(-\ln C_\mathrm{tail}(\theta) = x+\frac{x^2}{2}-\ln(1+x) \geq \frac{x^2}{2} > 0\), since \(\ln(1+x)\leq x\); in particular, \(C_\mathrm{tail}(\theta)<1\).
	If, moreover, \(x \geq \sqrt{(2\kappa_\mathrm{tail}\ln 2)/N}\), then \(-\ln C_\mathrm{tail}(\theta) \geq \frac{x^2}{2} \geq \frac{\kappa_\mathrm{tail}\ln 2}{N}\), and hence \(C_\mathrm{tail}(\theta)^N\leq 2^{-\kappa_\mathrm{tail}}\).
\end{proof}

\subsection{Failure Probability of the Good-Tail Event}

We apply \Cref{Lemma:Discrete-Gaussian-Tail-IA} to the three families of Gaussian samples in \Cref{Definition:Good-tail}.

\paragraph{Secret}

Under the coefficient embedding, \(\bm{s}\gets\mathcal{D}_S=\mathcal{D}_{\sigma_{\bm{s}}}^D\) is a sample of \(\mathcal{D}_{\mathbb{Z}^D,\sigma_{\bm{s}}}\).
Applying \Cref{Lemma:Discrete-Gaussian-Tail-IA} with \(N=D\) gives \(\Pr [ \|\bm{s}\|_2 > \theta_{\bm{s}}\sigma_{\bm{s}}\sqrt{D} ] \leq C_\mathrm{tail}(\theta_{\bm{s}})^D \leq 2^{-\kappa}\) by \eqref{Eq:theta_s}.
Since \(\|\bm{s}\|_\infty\leq\|\bm{s}\|_2\) and \(\beta_{\bm{s}} = \theta_{\bm{s}}\sigma_{\bm{s}}\sqrt{D}\), this proves \eqref{Eq:Good-tail-bound-kappa-s}.

\paragraph{Masking Matrix}

Each entry of \(\mathbf{R} \in \mathbb{Z}^{(k-1)\times D} \) is sampled independently from \(\mathcal{D}_{\sigma_\mathbf{R}}\).
For a fixed \((i',j')\), applying \Cref{Lemma:Discrete-Gaussian-Tail-IA} with \(N=1\) gives \(\Pr [ \lvert\mathbf{R}[i',j']\rvert > \theta_\mathbf{R}\sigma_\mathbf{R} ] \leq C_\mathrm{tail}(\theta_\mathbf{R}) \leq 2^{-\kappa_\mathbf{R}}\) by \eqref{Eq:theta_R-single}.
A union bound over the \((k-1)D\) entries gives a failure probability of at most \((k-1)D\cdot 2^{-\kappa_\mathbf{R}} \leq 2^{-\kappa}\), since \(\kappa_\mathbf{R} = \kappa + \lceil \log_2((k-1)D) \rceil\).
This proves \eqref{Eq:Good-tail-bound-kappa_bigR}.

\paragraph{Leakage Masks}

Assume \(Q\geq1\).
For a fixed pair \((i,s)\), the mask \(\bm{r}_{i,s}\) is a sample of \(\mathcal{D}_{\mathbb{Z}^D,\sigma_{\bm{r}}}\), and \Cref{Lemma:Discrete-Gaussian-Tail-IA} with \(N=D\) gives \(\Pr [ \|\bm{r}_{i,s}\|_2 > \theta_{\bm{r}}\sigma_{\bm{r}}\sqrt{D} ] \leq C_\mathrm{tail}(\theta_{\bm{r}})^D \leq 2^{-\kappa_{\bm{r}}}\) by \eqref{Eq:theta_r}.
A union bound over at most \(nQ\) pairs gives a failure probability of at most \(nQ\cdot 2^{-\kappa_{\bm{r}}} \leq 2^{-\kappa}\), since \(\kappa_{\bm{r}} = \kappa + \lceil \log_2(nQ) \rceil\).
This proves \eqref{Eq:Good-tail-bound-kappa_smallr}.
If \(Q\geq1\), a union bound over the secret, masking-matrix, and leakage-mask conditions gives \(\Pr[\neg\mathsf{Good}_\mathsf{tail}] \leq 3\cdot2^{-\kappa}\).  
If \(Q=0\), the leakage-mask condition is vacuous, and a union bound over the remaining two conditions gives \(\Pr[\neg\mathsf{Good}_\mathsf{tail}] \leq 2\cdot2^{-\kappa}\). 
Together, they prove \Cref{Lemma:Good-tail-Failure}.

\subsection{Size Estimates}

We report ideal entropy-coded sizes.
Arithmetic coding approaches the entropy of a stationary memoryless source up to a constant coding overhead \cite[Theorem 2]{saarinen2018arithmetic}.
As the ideal source model for a coefficient of standard deviation \(\sigma\), we use a variance-matched discrete Gaussian, whose entropy is approximately \(\lg\sigma + \frac{1}{2}\lg(2\pi e) \approx \lg\sigma + 2.047\) bits \cite[Section 5.1]{espitau2022shorter}; since the Gaussian maximizes entropy for a given variance, this is an upper estimate whenever the coefficients are independent.

\paragraph{Secret}

The secret \(\bm{s}\in\mathcal{R}^\ell\) has \(D=d\ell\) independent coefficients sampled from \(\mathcal{D}_{\sigma_{\bm{s}}}\), so its size is approximately \(D(\lg\sigma_{\bm{s}} + 2.047)\) bits.

\paragraph{Share}

By \eqref{Eq:s_i-linear-function-s+R}, coefficient \(j\) of \(\bm{s}_i\) is \(\mathbf{g}[i]\mathbf{s}[j] + \sum_{t=1}^{k-1}\mathbf{\hat{G}}[i,t]\mathbf{R}[t,j]\).
For a fixed \(i\), the coefficients are independent across \(j\), because they involve disjoint coordinates of \(\mathbf{s}\) and disjoint columns of \(\mathbf{R}\).
Every row of \(\mathbf{\hat{G}}\) has exactly \(c_\mathrm{exp}\) nonzero entries in \(\{-1,1\}\), so, approximating the variance of \(\mathcal{D}_\sigma\) by \(\sigma^2\), each coefficient has variance at most \(\sigma_\mathsf{sh}^2 := \beta_\mathbf{g}^2\sigma_{\bm{s}}^2 + c_\mathrm{exp}\sigma_\mathbf{R}^2\).
The size of one share is therefore approximately \(D(\lg\sigma_\mathsf{sh} + 2.047)\) bits, and every participant stores exactly one such share.
The values in \Cref{Table:SEC-Selected-Sizes} are obtained by dividing these bit estimates by \(8\cdot2^{10}\).
These estimates exclude implementation-specific coding, framing, finite-block, and byte-alignment overheads.

\section{Comparison with Other Short-Share Secret Sharing Schemes}\label{Appendix:Short-Share-Comparison} 

This appendix compares the ideal entropy-coded secret size and long-term per-participant secret-share storage of \(\mathsf{SEC}\) with those of the replicated and ramp \(\mathsf{sDKG}\) constructions of \cite{del2024shortly} and the Vandermonde secret-sharing construction used by Hermine \cite{borin2026hermine}. 
For \(\mathsf{SEC}\), we use the application-driven parameter sets with \(Q=2^{64}\), since both comparison schemes are instantiated for threshold-signature applications. 
We compare only secret and long-term secret-share sizes, excluding public auxiliary keys, verification material, signing communication, and other components specific to the surrounding threshold-signature schemes.

\subsection{Secret Size}\label{Appendix:Short-Share-Secret-Size}

We use the ideal entropy model of Appendix~\ref{Appendix:Gaussian-Tail-Size}. 
For a secret \(\bm{s}\in\mathcal{R}^{\ell}\) whose \(D=d\ell\) coefficients are independently sampled from \(\mathcal{D}_{\sigma_{\bm{s}}}\), its ideal entropy-coded size is approximately \(D\left(\lg\sigma_{\bm{s}}+2.047\right)\) bits. 
All schemes use the same standard-deviation convention, so their reported Gaussian widths can be compared directly.

\subsubsection{Secret Size in \(\mathsf{sDKG}\)}\label{Appendix:sDKG-Secret-Size}

The threshold-signature parameters of \cite[Table 2]{del2024shortly} comprise two parameter sets at \(\kappa=128\) and one at \(\kappa=256\). 
The first \(\kappa=128\) set has \((d,p,\ell)=(512,4,4)\) and secret width \(\sigma_{\bm{s}}=2^{10}\), and the \(\kappa=256\) set has \((d,p,\ell)=(512,7,7)\) and \(\sigma_{\bm{s}}=2^{10}\); both support up to \(16\) participants. 
The second \(\kappa=128\) set has \((d,p,\ell)=(512,5,5)\) and \(\sigma_{\bm{s}}=2^{8}\), and it is the only reported set supporting more than \(16\) participants, namely up to \(1024\) participants for ramp \(\mathsf{sDKG}\). 
The paper does not report a parameter set for \(\kappa=192\). 

Each \(\mathsf{sDKG}\) sub-share is a pair \((\bm{s}_\mathsf{sub},\bm{e}_\mathsf{sub})\in\mathcal{R}^\ell\times\mathcal{R}^p\), so we measure the shared secret as the corresponding pair \((\bm{s},\bm{e})\in\mathcal{R}^\ell\times\mathcal{R}^p\). 
This is also the like-for-like comparison with \(\mathsf{SEC}\), whose secret contains both parts of the systematic \(\mathsf{MLWE}\) instance. 
Under the common ideal entropy model mentioned earlier, the \(\mathsf{sDKG}\) secret size is \(|(\bm{s},\bm{e})|_\mathsf{sDKG} \approx d(p+\ell)\left(\lg\sigma_{\bm{s}}+2.047\right)\) bits. 
This gives approximately \(6.024\) and \(6.279\) KiB for the first and second \(\kappa=128\) sets, respectively, and \(10.541\) KiB for \(\kappa=256\). 
\Cref{Table:sDKG-SEC-Secret-Size} compares these values with the application-driven \(\mathsf{SEC}\) parameters for \(Q=2^{64}\).

\begin{table}[!htp] 
	\caption{Ideal entropy-coded secret sizes, in KiB, in \(\mathsf{sDKG}\) and \(\mathsf{SEC}\)} 
	\label{Table:sDKG-SEC-Secret-Size}% 
	\begin{tabular}{@{}ccccccccc@{}} 
		\toprule 
		& \multicolumn{4}{c}{\(\mathsf{sDKG}\)} & \multicolumn{2}{c}{\(\mathsf{SEC}\), \(n=40\)} & \multicolumn{2}{c}{\(\mathsf{SEC}\), \(n=60\)} \\ 
		\cmidrule(lr){2-5} \cmidrule(lr){6-7} \cmidrule(l){8-9} 
		\(\kappa\) & \((d,p,\ell)\) & \(\lg\sigma_{\bm{s}}\) & \(n_{\max}\) & Secret & Secret & Ratio & Secret & Ratio \\ 
		\midrule 
		\(128\) & \((512,4,4)\) & \(10\) & \(16\) & \(6.024\) & \(4.333\) & \(1.390\) & \(3.538\) & \(1.703\) \\ 
		\(128\) & \((512,5,5)\) & \(8\) & \(1024\) & \(6.279\) & \(4.333\) & \(1.449\) & \(3.538\) & \(1.775\) \\ 
		\(256\) & \((512,7,7)\) & \(10\) & \(16\) & \(10.541\) & \(7.880\) & \(1.338\) & \(6.486\) & \(1.625\) \\ 
		\botrule 
	\end{tabular}
	\footnotetext{Note: the ratios are \(|(\bm{s},\bm{e})|_\mathsf{sDKG}/|\bm{s}|_\mathsf{SEC}\). 
	\(n_{\max}\) is the largest supported number of participants; the value \(1024\) applies to ramp \(\mathsf{sDKG}\) only. 
	The \(\mathsf{SEC}\) values use the application-driven parameter sets with \(Q=2^{64}\). 
	The \(\mathsf{sDKG}\) paper reports no parameter set for \(\kappa=192\). 
	All sizes use the ideal entropy model of Appendix~\ref{Appendix:Gaussian-Tail-Size}.}
\end{table}

\subsubsection{Secret Size in Vandermonde Secret Sharing}\label{Appendix:Vandermonde-Secret-Size}

We consider the Vandermonde secret-sharing construction used by Hermine \cite{borin2026hermine}. 
For the threshold-signature parameter sets in \cite[Table 3]{borin2026hermine}, the ring degree is \(d=512\), the secret-module ranks are \(\ell=4\), \(6\), and \(8\) for \(\kappa=128\), \(192\), and \(256\), respectively, and \(u_\mathbf{s}:=\lceil\lg\sigma_\mathbf{s}\rceil=13\). 
Since \(\lg\sigma_\mathbf{s}\leq u_\mathbf{s}\), replacing the exact logarithmic width by \(u_\mathbf{s}\) gives a conservative ideal entropy estimate.

Under the common ideal entropy model mentioned earlier, the secret size is bounded approximately by \(|\bm{s}|_\mathsf{Vand} \lesssim d\ell\left(u_{\mathbf{s}}+2.047\right) = 15.047 d\ell\) bits. 
This gives approximately \(3.762\), \(5.643\), and \(7.524\) KiB for \(\kappa=128\), \(192\), and \(256\), respectively. 
\Cref{Table:Vandermonde-SEC-Secret-Size} compares these estimates with the application-driven \(\mathsf{SEC}\) parameters for \(Q=2^{64}\).

\begin{table}[!htp]
	\caption{Ideal entropy-coded secret-size upper estimates, in KiB, in Vandermonde secret sharing and \(\mathsf{SEC}\)} 
	\label{Table:Vandermonde-SEC-Secret-Size}% 
	\begin{tabular}{@{}cccccccc@{}} 
		\toprule
		& \multicolumn{3}{c}{Vandermonde secret sharing} & \multicolumn{2}{c}{\(\mathsf{SEC}\), \(n=40\)} & \multicolumn{2}{c}{\(\mathsf{SEC}\), \(n=60\)} \\ 
		\cmidrule(lr){2-4} \cmidrule(lr){5-6} \cmidrule(l){7-8} \(\kappa\) & \((d,\ell)\) & \(u_\mathbf{s}\) & Secret & Secret & Ratio & Secret & Ratio \\ 
		\midrule 
		\(128\) & \((512,4)\) & \(13\) & \(3.762\) & \(4.333\) & \(0.868\) & \(3.538\) & \(1.063\) \\ 
		\(192\) & \((512,6)\) & \(13\) & \(5.643\) & \(6.087\) & \(0.927\) & \(4.991\) & \(1.131\) \\ 
		\(256\) & \((512,8)\) & \(13\) & \(7.524\) & \(7.880\) & \(0.955\) & \(6.486\) & \(1.160\) \\ 
		\botrule 
	\end{tabular} 
	\footnotetext{Note: the Vandermonde values conservatively replace \(\lg\sigma_\mathbf{s}\) by \(u_\mathbf{s}=\lceil\lg\sigma_\mathbf{s}\rceil=13\). 
	The ratios are \(|\bm{s}|_\mathsf{Vand}/|\bm{s}|_\mathsf{SEC}\). 
	The \(\mathsf{SEC}\) values use the application-driven parameter sets with \(Q=2^{64}\). 
	All sizes use the ideal entropy model of Appendix~\ref{Appendix:Gaussian-Tail-Size}.} 
\end{table}

Measured consistently, the \(\mathsf{sDKG}\) secret is \(1.34\) to \(1.78\) times as large as the \(\mathsf{SEC}\) secret, whereas the Vandermonde upper estimate is comparable to it: slightly smaller for \(n=40\) and slightly larger for \(n=60\). 
The more substantial storage difference arises from the number of sub-shares held by each participant, as analyzed in Appendix~\ref{Appendix:Short-Share-Storage}.

\subsection{Long-Term Secret-Share Size}\label{Appendix:Short-Share-Storage}

Since \(\mathsf{SEC}\) has distinct privacy and empirical reconstruction thresholds \(k_p\) and \(k_c\), no perfect \(t\)-out-of-\(n\) access structure simultaneously matches both thresholds. 
We therefore compare \(\mathsf{SEC}\) with two profiles: \(t_\mathsf{pm}:=k_p+1\) and \(t_\mathsf{qm}:=k_c\). 
The privacy-matched profile \(t_\mathsf{pm}\) tolerates the same number \(k_p\) of corrupted participants, whereas the quorum-matched profile \(t_\mathsf{qm}\) matches the empirical reconstruction threshold of \(\mathsf{SEC}\). 
More generally, every \(t\in[t_\mathsf{pm},t_\mathsf{qm}]\) provides at least the stated privacy of \(\mathsf{SEC}\) without requiring more participants than its empirical reconstruction threshold. 
The two endpoint profiles are used when counting the sub-shares stored in the replicated and ramp \(\mathsf{sDKG}\) constructions and in the Vandermonde construction used by Hermine.
We consider \((n,t_\mathsf{pm},t_\mathsf{qm}) \in \{(40,6,26),(60,7,46)\}\) in our comparative analysis. 

\subsubsection{Secret-Share Size in \(\mathsf{sDKG}\)}\label{Appendix:sDKG-Share-Size}

Each short sub-share in the replicated and ramp \(\mathsf{sDKG}\) constructions of \cite{del2024shortly} is a pair \((\bm{s}_\mathsf{sub},\bm{e}_\mathsf{sub})\in\mathcal{R}^\ell \times \mathcal{R}^p\). 
For a coefficient width \(\sigma\), we estimate its ideal entropy-coded size by \(\mathsf{ShareSize}_{p,\ell}(\sigma):=d(p+\ell)\left(\lg\sigma+2.047\right)\) bits. 
The two constructions assign different numbers of sub-shares to each participant and use different reported parameter sets. 
For replicated \(\mathsf{sDKG}\), we use the first \(\kappa=128\) set of Appendix~\ref{Appendix:sDKG-Secret-Size}, which gives the smallest sub-shares among the reported sets; for ramp \(\mathsf{sDKG}\), we use the second \(\kappa=128\) set, the only one supporting more than \(16\) participants.
We compare their per-participant storage with the application-driven \(\mathsf{SEC}\) share sizes for \(Q=2^{64}\) in \Cref{Table:SEC-Selected-Sizes}. 
The \(\mathsf{sDKG}\) paper reports parameter sets for \(\kappa=128\) and \(256\), but not for \(\kappa=192\).

\paragraph{Replicated \(\mathsf{sDKG}\)}

In a \(t\)-out-of-\(n\) replicated \(\mathsf{sDKG}\), there are \(\binom{n}{t-1}\) short sub-shares in total, and each participant stores \(k_\mathsf{Repl}(n,t) := \binom{n-1}{t-1}\) of them. 
Hence, the estimated per-participant storage is \(|\mathsf{sk}_i^\mathsf{Repl}| \approx \binom{n-1}{t-1} \mathsf{ShareSize}_{p,\ell}(\sigma_{\bm{s}})\).
At \(\kappa=128\), the reported replicated-\(\mathsf{sDKG}\) parameters are \((d,p,\ell)=(512,4,4)\) and \(\sigma_{\bm{s}}=2^{10}\), giving \(\mathsf{ShareSize}_{4,4}(2^{10}) = 512\cdot8\cdot(10+2.047) \approx 6.024\ \mathrm{KiB}\). 
At \(\kappa=256\), the corresponding parameters are \((d,p,\ell)=(512,7,7)\) with the same Gaussian width, so the individual sub-share size increases by the factor \(7/4\) to approximately \(10.541\) KiB.

\Cref{Table:sDKG-SEC-Replicated-Secret-Share-Storage} reports the resulting per-participant long-term secret-share storage for the aforementioned privacy- and quorum-matched profiles.
The reported replicated-\(\mathsf{sDKG}\) parameters support up to \(16\) participants, so the \(n=40\) and \(n=60\) values are extrapolations of the replicated construction.

\begin{table}[!htp] 
	\caption{Estimated per-participant long-term secret-share storage for replicated \(\mathsf{sDKG}\), compared with \(\mathsf{SEC}\)} 
	\label{Table:sDKG-SEC-Replicated-Secret-Share-Storage}% 
	\setlength{\tabcolsep}{3.5pt} 
	\begin{tabular}{@{}ccccrrrr@{}} 
		\toprule 
		\(\kappa\) & \(n\) & \(t\) & Profile & \(\#\) sub-shares & \(\mathsf{sDKG}\) storage & \(\mathsf{SEC}\) share & Ratio \\ 
		\midrule 
		\(128\) & \(40\) & \(6\) & Privacy & \(575{,}757\) & \(3.307\ \mathrm{GiB}\) & \(6.068\ \mathrm{KiB}\) & \(571{,}534.655\) \\ 
		\(128\) & \(40\) & \(26\) & Quorum & \(15{,}084{,}504{,}396\) & \(86{,}652.291\ \mathrm{GiB}\) & \(6.068\ \mathrm{KiB}\) & \(14{,}973{,}881{,}382.549\) \\ 
		\(128\) & \(60\) & \(7\) & Privacy & \(45{,}057{,}474\) & \(258.831\ \mathrm{GiB}\) & \(6.416\ \mathrm{KiB}\) & \(42{,}301{,}074.601\) \\ 
		\(128\) & \(60\) & \(46\) & Quorum & \(13{,}298{,}522{,}298{,}180\) & \(76{,}392{,}792.762\ \mathrm{GiB}\) & \(6.416\ \mathrm{KiB}\) & \(12{,}484{,}982{,}709{,}334.045\) \\ 
		\midrule 
		\(256\) & \(40\) & \(6\) & Privacy & \(575{,}757\) & \(5.788\ \mathrm{GiB}\) & \(10.674\ \mathrm{KiB}\) & \(568{,}589.705\) \\ 
		\(256\) & \(40\) & \(26\) & Quorum & \(15{,}084{,}504{,}396\) & \(151{,}641.508\ \mathrm{GiB}\) & \(10.674\ \mathrm{KiB}\) & \(14{,}896{,}725{,}351.441\) \\ 
		\(256\) & \(60\) & \(7\) & Privacy & \(45{,}057{,}474\) & \(452.954\ \mathrm{GiB}\) & \(11.252\ \mathrm{KiB}\) & \(42{,}210{,}848.349\) \\ 
		\(256\) & \(60\) & \(46\) & Quorum & \(13{,}298{,}522{,}298{,}180\) & \(133{,}687{,}387.333\ \mathrm{GiB}\) & \(11.252\ \mathrm{KiB}\) & \(12{,}458{,}352{,}813{,}757.791\) \\ 
		\botrule 
	\end{tabular}
    \footnotetext{Note: the ratios are \(|\mathsf{sk}_i^\mathsf{Repl}|/|\bm{s}_i^\mathsf{SEC}|\). 
    The \(\mathsf{SEC}\) values use the application-driven parameter sets with \(Q=2^{64}\). 
    The \(\mathsf{sDKG}\) values measure the ideal entropy-coded storage of all secret sub-shares held by one participant and exclude indexing, framing, implementation overheads, and the seed-and-tree compression proposed in \cite{del2024shortly}. 
    The \(n=40\) and \(n=60\) profiles are extrapolated beyond the reported participant range of replicated \(\mathsf{sDKG}\).} 
\end{table}

The number of replicated sub-shares grows combinatorially with the matched threshold. 
Consequently, even the privacy-matched profiles require per-participant storage more than \(5\cdot10^5\) times that of one \(\mathsf{SEC}\) share, while the quorum-matched profiles are substantially larger.

\paragraph{Ramp \(\mathsf{sDKG}\)}

In the basic ramp \(\mathsf{sDKG}\), let \(\nu\) be the number of additive share buckets and \(\mu\) the number of parallel sharings. 
Each party stores one short sub-share from every parallel sharing, and hence stores \(\mu\) sub-shares in total. 
Following the correctness-amplified construction of \cite{del2024shortly}, we set \(\mu:=\kappa\), giving the estimated per-participant storage \(|\mathsf{sk}_i^\mathsf{Ramp}| \approx \kappa \cdot \mathsf{ShareSize}_{p,\ell}(\sigma_\mathsf{share})\).
At \(\kappa=128\), we use the reported ramp-\(\mathsf{sDKG}\) parameter set supporting the larger participant range, where \((d,p,\ell)=(512,5,5)\) and \(\sigma_\mathsf{share}=2^{10}\). 
Thus, one ordinary sub-share has ideal entropy-coded size \(\mathsf{ShareSize}_{5,5}(2^{10}) = 512\cdot10 \left( 10+\frac{1}{2}\lg(2\pi e) \right) \approx 7.529\ \mathrm{KiB}\), and the total per-participant storage is approximately \(128\cdot7.529=963.768\) KiB. 
At \(\kappa=256\), the corresponding parameters are \((d,p,\ell)=(512,7,7)\) with the same Gaussian width. 
The individual sub-share size increases by a factor of \(7/5\), while the number of parallel sharings doubles from \(128\) to \(256\). 
The total storage therefore increases by a factor of \(256/128\cdot7/5 = 14/5\) to approximately \(2698.549\) KiB.

To obtain an \(\mathsf{sDKG}\)-favorable estimate, we assume that every sub-share has width \(\sigma_\mathsf{share}\). 
This omits the larger dependent bucket in each additive sharing and therefore gives an optimistic estimate for the ramp construction. 
\Cref{Table:sDKG-SEC-Ramp-Secret-Share-Storage} compares these values with the application-driven \(\mathsf{SEC}\) share sizes for \(Q=2^{64}\).

\begin{table}[!htp] 
	\caption{Optimistic ideal entropy-coded per-participant long-term secret-share storage, in KiB, for ramp \(\mathsf{sDKG}\), compared with \(\mathsf{SEC}\)} 
	\label{Table:sDKG-SEC-Ramp-Secret-Share-Storage}% 
	\begin{tabular}{@{}ccrrrrr@{}} 
		\toprule 
		\(\kappa\) & \(n\) & \(\mu\) & One sub-share & \(\mathsf{sDKG}\) storage & \(\mathsf{SEC}\) share & Ratio \\ 
		\midrule 
		\(128\) & \(40\) & \(128\) & \(7.529\) & \(963.768\) & \(6.068\) & \(158.828\) \\ 
		\(128\) & \(60\) & \(128\) & \(7.529\) & \(963.768\) & \(6.416\) & \(150.213\) \\ 
		\midrule 
		\(256\) & \(40\) & \(256\) & \(10.541\) & \(2698.549\) & \(10.674\) & \(252.815\) \\ 
		\(256\) & \(60\) & \(256\) & \(10.541\) & \(2698.549\) & \(11.252\) & \(239.828\) \\ 
		\botrule 
	\end{tabular} 
	\footnotetext{Note: the ratios are \(|\mathsf{sk}_i^\mathsf{Ramp}|/ |\bm{s}_i^\mathsf{SEC}|\). 
	The \(\mathsf{SEC}\) values use the application-driven parameter sets with \(Q=2^{64}\). 
	The ramp-\(\mathsf{sDKG}\) estimates charge every sub-share at the ordinary width \(\sigma_\mathsf{share}\) and omit the larger dependent bucket and implementation-specific overheads.} 
\end{table}

The ramp construction avoids the combinatorial sub-share growth of replicated \(\mathsf{sDKG}\), but the correctness-amplified instantiation stores \(\mu=\kappa\) short sub-shares per participant. 
Even under the optimistic ordinary-sub-share estimate, its per-participant storage is approximately \(150\)--\(159\) times that of one \(\mathsf{SEC}\) share at \(\kappa=128\) and \(240\)--\(253\) times at \(\kappa=256\).

\subsubsection{Secret-Share Size in Vandermonde Secret Sharing}\label{Appendix:Vandermonde-Share-Size}

\paragraph{Number of Vandermonde Sub-Shares}

The Vandermonde construction used by Hermine recursively partitions the participants, so an unbalanced partition tree may assign different numbers of sub-shares to different parties \cite{battagliola2025enhancing,borin2026hermine}.
Let \(\mathbf{u}(n,t):=(u_1(n,t),\dots,u_n(n,t))\in\mathbb{Z}_{\geq0}^n\), where \(u_i(n,t)\) denotes the number of sub-shares assigned to party \(i\) in a \(t\)-out-of-\(n\) sharing. 
Define \(L(n,t):=\min_{i\in[n]}u_i(n,t)\), \(U(n,t):=\max_{i\in[n]}u_i(n,t)\), and \(S(n,t):=\sum_{i=1}^n u_i(n,t)\) as the minimum number of sub-shares per party, the maximum number of sub-shares per party, and the total number of sub-shares across all parties, respectively.

Incorporating Hermine's \(t=n\) optimization, set \(\mathbf{u}(n,1)=\mathbf{u}(n,n)=\mathbf{1}_n\), and set \(\mathbf{u}(n,t)=\mathbf{0}_n\) whenever \(t\leq0\) or \(t>n\).
For \(t\in[2,n-1]\), let \(n_\mathsf{L}:=\lfloor n/2\rfloor\), \(n_\mathsf{R}:=\lceil n/2\rceil\), and define \(\mathcal{T}_\mathsf{L}(n,t):=\left[\max\{1,t-n_\mathsf{R}\},\min\{t,n_\mathsf{L}\}\right]\) and \(\mathcal{T}_\mathsf{R}(n,t):=\left[\max\{1,t-n_\mathsf{L}\},\min\{t,n_\mathsf{R}\}\right]\).
The participant-wise count vector satisfies
\begin{align}\label{Eq:u(nt)-vector-concatenation}
	\mathbf{u}(n,t) =
	\left(\sum_{t_\mathsf{L}\in\mathcal{T}_\mathsf{L}(n,t)}\mathbf{u}(n_\mathsf{L},t_\mathsf{L})\right)
	\mathbin{\Vert}
	\left(\sum_{t_\mathsf{R}\in\mathcal{T}_\mathsf{R}(n,t)}\mathbf{u}(n_\mathsf{R},t_\mathsf{R})\right),
\end{align}
where \(\Vert\) denotes vector concatenation in the left-right order of the recursive partition.\footnote{For odd \(n\), the upper endpoint of the right sub-threshold range is \(\min\{t,\lceil n/2\rceil\}\), because the right subtree contains \(\lceil n/2\rceil\) participants. 
This corrects the expression \(\min\{t,\lfloor n/2\rfloor\}\) in \cite[p.~15]{battagliola2025enhancing}; the two expressions coincide when \(n\) is even.}
Recurrence \eqref{Eq:u(nt)-vector-concatenation} determines \(L(n,t)\), \(U(n,t)\), and \(S(n,t)\) exactly. 
The minimum, average, and maximum per-party sub-share counts are \(L(n,t)\), \(S(n,t)/n\), and \(U(n,t)\), respectively. 
We evaluate these quantities for the privacy- and quorum-matched profiles \((40,6)\), \((40,26)\), \((60,7)\), and \((60,46)\) in the subsequent storage comparison.

\paragraph{Size of an Individual Vandermonde Sub-Share}

Each Vandermonde index \(\mathsf{idx}\) corresponds to a pair \((\bm{s}_\mathsf{idx},\bm{e}_\mathsf{idx})\in\mathcal{R}^\ell\times\mathcal{R}^p\), but the long-term signing share stores only \(\bm{s}_\mathsf{idx}\in\mathcal{R}^\ell\).
The component \(\bm{e}_\mathsf{idx}\) is used to construct the indexed partial verification key and is excluded consistently with the scope of this appendix. 
A party stores one such vector for each index assigned to it, and we measure the initial storage before any key refresh.

Following the convolution argument in the proof of \cite[Lemma 6.1]{borin2026hermine}, each coefficient of an initial sub-share is modeled as a signed sum of at most \(1+\lceil\lg n\rceil\) independent Gaussian contributions with standard deviation \(\sigma_\mathbf{s}\). 
We therefore use the uniform variance-matched width \(\sigma_\mathsf{sub}:=\sigma_\mathbf{s}\sqrt{1+\lceil\lg n\rceil}\). 
Since \(\bm{s}_\mathsf{idx}\) has \(d\ell\) coefficients, its ideal entropy-coded size is approximately \(|\bm{s}_\mathsf{idx}|\approx d\ell\big(\lg\sigma_\mathbf{s}+\frac{1}{2}\lg(1+\lceil\lg n\rceil)+2.047\big)\leq d\ell\big(u_\mathbf{s}+\frac{1}{2}\lg(1+\lceil\lg n\rceil)+2.047\big)\) bits, where \(u_\mathbf{s}:=\lceil\lg\sigma_\mathbf{s}\rceil=13\) for the parameter sets considered.
This uniform estimate may overestimate sub-shares arising from shorter or degenerate recursive paths, while replacing \(\lg\sigma_\mathbf{s}\) by \(u_\mathbf{s}\) adds less than one bit per coefficient. 
Measuring the initial sharing before any refresh is favorable to Hermine.

\paragraph{Secret-Share Storage Comparison} 

Since \cite{borin2026hermine} does not provide concrete parameter sets for \(n\in\{40,60\}\), we extrapolate the Vandermonde sub-share counts and sizes using the parameter family with \(\lceil\lg q\rceil=48\) in \cite[Table 3]{borin2026hermine}, which supports the larger reported signing-query budgets. 
For \(\kappa\in\{128,192,256\}\), this family uses \((p,\ell)\in\{(4,4),(6,6),(7,8)\}\), respectively, with \(d=512\) and \(u_\mathbf{s}=13\). 
Since \(\lceil\lg40\rceil=\lceil\lg60\rceil=6\), the uniform estimate of the previous discussion gives \(|\bm{s}_\mathsf{idx}| \leq 512\ell \left( 13+\frac{1}{2}\lg7+2.047 \right)\) bits.
These values extrapolate Hermine beyond its analyzed range \(n\leq16\) and \(t\leq8\), and hence quantify the storage induced by its Vandermonde construction without constituting security-validated Hermine parameter sets for \(n\in\{40,60\}\). 
For a \(t\)-out-of-\(n\) profile, the minimum, average, and maximum per-party long-term secret-share storage are \(L(n,t)|\bm{s}_\mathsf{idx}|\), \(S(n,t)|\bm{s}_\mathsf{idx}|/n\), and \(U(n,t)|\bm{s}_\mathsf{idx}|\), respectively. 
\Cref{Table:Vandermonde-SEC-Secret-Share-Storage} reports these values at \(\kappa=128\) for the privacy- and quorum-matched profiles mentioned earlier, together with the ratio between the minimum Vandermonde storage and the application-driven \(\mathsf{SEC}\) share size for \(Q=2^{64}\). 
Neither profile is strictly like-for-like: privacy matching allows reconstruction by fewer parties than the empirical \(\mathsf{SEC}\) threshold, whereas quorum matching gives the Vandermonde scheme a stronger privacy threshold than \(\mathsf{SEC}\).

\begin{table}[!htp]
	\caption{Estimated initial per-party long-term secret-share storage, in KiB, in Vandermonde secret sharing at \(\kappa=128\), compared with \(\mathsf{SEC}\)} 
	\label{Table:Vandermonde-SEC-Secret-Share-Storage}% 
	\setlength{\tabcolsep}{3.5pt} 
	\begin{tabular}{@{}ccccrrrrr@{}} 
		\toprule \(n\) & \(t\) & Profile & \(\#\) sub-shares & \multicolumn{3}{c}{Vandermonde storage} & \(\mathsf{SEC}\) share & Ratio \\ 
		\cmidrule(lr){5-7} 
		& & & Min./Avg./Max. & Min. & Avg. & Max. & & Min. Vand./\(\mathsf{SEC}\) \\ 
		\midrule 
		\(40\) & \(6\) & Privacy & \(86/104.4/127\) & \(353.692\) & \(429.365\) & \(522.312\) & \(6.068\) & \(58.288\) \\ 
		\(40\) & \(26\) & Quorum & \(474/581.6/713\) & \(1949.417\) & \(2391.942\) & \(2932.350\) & \(6.068\) & \(321.262\) \\ 
		\midrule 
		\(60\) & \(7\) & Privacy & \(170/261.6/280\) & \(699.158\) & \(1075.881\) & \(1151.554\) & \(6.416\) & \(108.971\) \\ 
		\(60\) & \(46\) & Quorum & \(1080/1748.8/1920\) & \(4441.709\) & \(7192.278\) & \(7896.371\) & \(6.416\) & \(692.286\) \\ 
		\botrule 
	\end{tabular} 
	\footnotetext{Note: the sub-share-count column reports \(L(n,t)\), \(S(n,t)/n\), and \(U(n,t)\), respectively. 
	The ratio is the minimum Vandermonde per-party storage divided by the application-driven \(\mathsf{SEC}\) share size with \(Q=2^{64}\). 
	The Vandermonde values are extrapolated beyond Hermine's analyzed range \(n\leq16\) and \(t\leq8\) and measure the initial sharing before any refresh.} 
\end{table}

Even for the party with the smallest estimated storage, the extrapolated Vandermonde construction requires approximately \(58\)--\(692\) times the storage of one \(\mathsf{SEC}\) share at \(\kappa=128\), with the larger ratios arising under quorum matching. 
This difference is because Vandermonde sharing stores many indexed sub-shares per party, whereas \(\mathsf{SEC}\) stores one share per party.

\end{appendices}


\end{document}